\documentclass[11pt]{book}

\newcommand{\be}{\begin{equation}}
\newcommand{\ee}{\end{equation}}
\newcommand{\ba}{\begin{eqnarray}}
\newcommand{\ea}{\end{eqnarray}}
\newtheorem{definizione}{Definition}[section]
\newcommand{\bd}{\begin{definizione}}
\newcommand{\ed}{\end{definizione}}
\newtheorem{teorema}{Theorem}[section]
\newcommand{\bth}{\begin{teorema}}
\newcommand{\eth}{\end{teorema}}
\newtheorem{statement}{Statement}[section]
\newcommand{\bst}{\begin{statement}}
\newcommand{\est}{\end{statement}}
\newtheorem{general}{Generalization}[section]
\newcommand{\bge}{\begin{general}}
\newcommand{\ege}{\end{general}}
\newtheorem{lemma}{Lemma}[section]
\newcommand{\blem}{\begin{lemma}}
\newcommand{\elem}{\end{lemma}}
\newcommand{\brr}{\begin{array}}
\newcommand{\err}{\end{array}}
\newcommand{\nn}{\nonumber}
\newtheorem{corollario}{Corollary}[section]
\newcommand{\bcorol}{\begin{corollario}}
\newcommand{\ecorol}{\end{corollario}}

\newcommand{\eqa}{\begin{eqnarray}}
\newcommand{\en}{\end{equation}}
\newcommand{\ena}{\end{eqnarray}}
\newcommand{\enn}{\nonumber \end{equation}}

\newcommand{\eqn}[1]{(\ref{#1})}

\newsavebox{\uuunit}
\sbox{\uuunit}
    {\setlength{\unitlength}{0.825em}
     \begin{picture}(0.6,0.7)
        \thinlines
        \put(0,0){\line(1,0){0.5}}
        \put(0.15,0){\line(0,1){0.7}}
        \put(0.35,0){\line(0,1){0.8}}
       \multiput(0.3,0.8)(-0.04,-0.02){12}{\rule{0.5pt}{0.5pt}}
     \end {picture}}

\def\Im{{\rm Im ~}}
\def\Re{{\rm Re ~}}
\def\IP{\relax{\rm I\kern-.18em P}}

\def\mun{\underline m}

\def\Eb{{\bf E}}
\def\bu{\bullet}
\def\we{\wedge}
\font\cmss=cmss10 \font\cmsss=cmss10 at 7pt
\def\twomat#1#2#3#4{\left(\matrix{#1 & #2 \cr #3 & #4}\right)}
\def\inbar{\vrule height1.5ex width.4pt depth0pt}
\def\IC{\relax\,\hbox{$\inbar\kern-.3em{\rm C}$}}
\def\IG{\relax\,\hbox{$\inbar\kern-.3em{\rm G}$}}
\def\IB{\relax{\rm I\kern-.18em B}}
\def\ID{\relax{\rm I\kern-.18em D}}
\def\IL{\relax{\rm I\kern-.18em L}}
\def\IF{\relax{\rm I\kern-.18em F}}
\def\IH{\relax{\rm I\kern-.18em H}}
\def\II{\relax{\rm I\kern-.17em I}}
\def\IN{\relax{\rm I\kern-.18em N}}
\def\IP{\relax{\rm I\kern-.18em P}}
\def\IQ{\relax\,\hbox{$\inbar\kern-.3em{\rm Q}$}}
\def\bfzero{\relax\,\hbox{$\inbar\kern-.3em{\rm 0}$}}
\def\IK{\relax{\rm I\kern-.18em K}}
\def\IG{\relax\,\hbox{$\inbar\kern-.3em{\rm G}$}}
 \font\cmss=cmss10 \font\cmsss=cmss10 at 7pt
\def\IR{\relax{\rm I\kern-.18em R}}
\def\ZZ{\relax\ifmmode\mathchoice
{\hbox{\cmss Z\kern-.4em Z}}{\hbox{\cmss Z\kern-.4em Z}}
{\lower.9pt\hbox{\cmsss Z\kern-.4em Z}}
{\lower1.2pt\hbox{\cmsss Z\kern-.4em Z}}\else{\cmss Z\kern-.4em
Z}\fi}
\def\bfone{\relax{\rm 1\kern-.35em 1}}

\def\om{\omega}

\def\la{\lambda}
\def\be{\beta}
\def\ga{\gamma}
\def\Ga{\Gamma}
\def\de{\delta}
\def\epsi{\varepsilon}
\def\we{\wedge}
\def\part{\partial}
\def\bu{\bullet}
\def\ci{\circ}
\def\square{{\,\lower0.9pt\vbox{\hrule \hbox{\vrule height 0.2 cm
\hskip 0.2 cm \vrule height 0.2 cm}\hrule}\,}}

\def\mun{\underline m}

\def\Rb{{\bf R}}
\def\Eb{{\bf E}}
\def\gb{{\bf g}}
\def\dt{{\tilde d}}
\def\Dt{{\tilde D}}
\def\Dcal{{\cal D}}

\def\Rfu#1#2{ R^{{\underline #1}{\underline #2}} }

\def\omef#1#2{\om^{\underline #1}_{~{\underline #2}}}

\def\omefu#1#2{\om^{{\underline #1} {\underline #2}}}
\def\omefub#1#2{{\omb}^{{\underline #1} {\underline #2}}}
\def\Ef#1{E^{\underline #1}}
\def\Efb#1{{\bf E}^{\underline #1}}
\def\omb{\bf \mbox{\boldmath $\om$}}
\def\bfone{\relax{\rm 1\kern-.35em 1}}
\font\cmss=cmss10 \font\cmsss=cmss10 at 7pt

 \def\cD{{\cal D}}
 \def\cG{{\cal G}}
\def\cH{{\cal H}} 
 \def\cK{{\cal K}}
\def\cL{{\cal L}} \def\cM{{\cal M}}
\def\cN{{\cal N}} 
\def\cP{{\cal P}} 
 
\def\tilde{\widetilde}
\def\bar{\overline}

\def\hat{\widehat}

\def\Coe#1.#2.{{#1\over #2}}

\def\coe#1.#2.{\relax{\textstyle {#1 \over #2}}\displaystyle}

\def\to{\rightarrow}
\def\notin{\hbox{{$\in$}\kern-.51em\hbox{/}}}

\def\del{\partial}

\def\IE{\relax{{\rm I\kern-.18em E}}}

\def\IGam{\relax{{\rm I}\kern-.18em \Gamma}}

\def\inbar{\vrule height1.5ex width.4pt depth0pt}
\def\bfzero{\relax{\rm I\kern-.18em 0}}
\def\bfone{\relax{\rm 1\kern-.35em 1}}
\def\twomat#1#2#3#4{\left(\begin{array}{cc}
 {#1}&{#2}\\ {#3}&{#4}\\
\end{array}
\right)}
\def\twovec#1#2{\left(\begin{array}{c}
{#1}\\ {#2}\\
\end{array}
\right)}
\def\o#1#2{{{#1}\over{#2}}}
\newcommand{\La}{{\Lambda}}
\newcommand{\Si}{{\Sigma}}
\newcommand{\im}{{\rm Im\ }}
 
 \def\cD{{\cal D}}
 \def\cG{{\cal G}}
\def\cH{{\cal H}} 
 \def\cK{{\cal K}}
\def\cL{{\cal L}} \def\cM{{\cal M}}
\def\cN{{\cal N}} 
\def\cP{{\cal P}}

\def\T{T}
\begin{document}
\title{{\bf   BPS BLACK HOLES IN SUPERGRAVITY  } \\
~~~~~\\
 {\it  Duality Groups,  $p$--Branes, Central Charges and the Entropy }}
\author{ { by}\\
~~~~~~\\
  RICCARDO D'AURIA and PIETRO FRE'}
  \maketitle
  \tableofcontents
%
\chapter{INTRODUCTION TO THESE LECTURES}
\label{intro}
\section{Extremal Black Holes from Classical General Relativity
to String Theory}
\markboth{BPS BLACK HOLES IN SUPERGRAVITY: CHAPTER 1}
{1.1 EXTREMAL BLACK HOLES FROM G.R. TO STRINGS}
\setcounter{equation}{0}
\label{intro1}
Black hole physics has many aspects of great interest to physicists
with very different cultural backgrounds. These range
from astrophysics to classical general relativity, to
quantum field theory in curved space--times, particle physics  and finally
string theory and supergravity. This is not surprising since
black--holes are one of the basic consequences of a fundamental
theory, namely Einstein general relativity. Furthermore black--holes
have fascinating thermodynamical properties that seem to encode
the deepest properties of the so far unestablished  fundamental theory
of quantum gravity. Central in this context is the Bekenstein Hawking
entropy:
\begin{equation}
S_{BH} = \frac{k_B}{G \hbar} \, \frac{1}{4} \,
\mbox{Area}_H
\label{bekhaw}
\end{equation}
where $k_B$ is the Boltzman constant, $G$ is Newton's constant,
$\hbar $ is Planck's constant and $\mbox{Area}_H$ denotes the area of
the horizon surface.
\par
This very precise relation between a thermodynamical quantity and
a geometrical quantity such as the horizon area has been for more than
twenty years the source of unextinguishable interest and meditation.
Indeed a microscopic statistical explanation of the area law for the
black hole entropy has been correctly regarded as possible only
within a  solid formulation of quantum gravity. Superstring theory
is the most serious candidate for a theory of quantum gravity and as
such should eventually provide such a microscopic explanation of the
area law. Although superstrings have been around for more than twenty
years a significant progress in this direction came only recently
\cite{strovaf}, after the so called second string revolution (1995).
Indeed black holes are a typical non--perturbative phenomenon and
perturbative string theory could say very little about their entropy:
only non perturbative string theory can have a handle on it and
our handle on non perturbative string theory came after 1995 through
the recognition of the role of string dualities. These dualities
allow to relate the strong coupling regime of one superstring model
to the weak coupling regime of another one and are all encoded in the
symmetry group (the $U$--duality group) of the low energy
{\it supergravity effective action}. Paradoxically this low energy
action  is precisely the handle on the non perturbative aspects
of superstrings.
\par
Since these lectures are addressed to an audience assumed to be
mostly unfamiliar with superstrings, supergravity and superstrings we
have briefly summarized this recent conceptual revolution in order
to put what follows in the right perspective. What we want to
emphasize is that the first instance of a microscopic explanation of
the area law within string theory has been limited to what in the
language of general relativity would be an {\it extremal black hole}.
This is not a capricious choice but has a deep reason. Indeed
the extremality condition, namely the coincidence of two horizons,
obtains, in the context of a supersymmetric theory, a profound
reinterpretation that makes extremal black holes the most interesting
objects to study.
To introduce the concept consider the usual Reissner Nordstrom metric
describing a black--hole of mass $m$ and electric (or magnetic) charge
$q$:
\begin{equation}
ds^2 \, = \, -dt^2 \, \left( 1- \frac{2 m}{\rho}+\frac{q}{\rho^2}
\right) + d\rho^2 \, \left( 1- \frac{2 m}{\rho}+\frac{q}{\rho^2}
\right)^{-1} + \rho^2 \, d\Omega^2
\label{reinor}
\end{equation}
where $d\Omega^2= (d\theta^2 + \sin^2\theta \, d\phi^2)$ is the
metric on a $2$--sphere. As it is well known the metric \eqn{reinor}
admits two killing horizons where the norm of the   killing
vector $\frac{\partial}{\partial t}$  changes sign. The horizons are
at the two roots of the quadratic form $\Delta \equiv  -2  m  \rho
+ q^2 + \rho^2$ namely at:
\begin{equation}
 \rho_\pm = m \pm \sqrt{m^2-q^2}
 \label{2hor}
\end{equation}
If $m < |q|$ the two horizons disappear and we have a naked
singularity. For this reason in the context of classical general
relativity the {\it cosmic censorship} conjecture was advanced
that singularities should always be hidden inside horizon
and this conjecture was formulated as the bound:
\begin{equation}
m \, \ge \, |q|
\label{censur}
\end{equation}
As we shall see, in the context of a supersymmetric theory the bound
\eqn{censur} is always guaranteed by supersymmetry.
As anticipated, of particular interest are the states that saturate
the bound \eqn{censur}. If $m=|q|$ the two horizons coincide and
the metric \eqn{reinor} can be rewritten in a new interesting way.
Setting:
\begin{equation}
m=|q|    \quad ; \quad \rho=r+m \quad ; \quad r^2 = {\vec
x}\, \cdot \, {\vec x}
\label{rhor}
\end{equation}
eq.\eqn{reinor} becomes:
\begin{eqnarray}
ds^2 & =& -dt^2 \, \left( 1+ \frac{q}{r}\right)^{-2} +
\left( 1+ \frac{q}{r}\right)^{ 2} \, \left( dr^2 +r^2 \,
d\Omega^2\right)\nonumber\\
& =& - \, H^{-2}({\vec x})\, dt^2  + H^{2}({\vec x})\, d{\vec x} \,
\cdot \, d{\vec x}
\label{guarda!}
\end{eqnarray}
where by:
\begin{equation}
\label{given}
 H ({\vec x})= \left(1 + \frac{q}{\sqrt{{\vec x}\, \cdot \, {\vec
 x}}}\right)
\end{equation}
we have denoted a harmonic function in a three--dimensional space
spanned by the three cartesian coordinates ${\vec x}$ with
the boundary condition that $H ({\vec x})$  goes to $1$ at infinity.
\par
The metric \eqn{guarda!} already contains all the features of the
black hole metrics we shall consider in these lectures:
\begin{enumerate}
\item It is of the form:
\begin{equation}
ds^2 = -e^{2U(r)} \, dt^2 + e^{-2U(r)} \, d{\vec x}^2
\label{ds2Up}
\end{equation}
where the radial function $U(r)$ is expressed as a linear combination
of harmonic functions of the {\it transverse coordinates } ${\vec x}$:
\begin{equation}
U(r) = \sum_{i=1}^{p} \, \alpha _i \, \log \, H_i({\vec x})
\end{equation}
\item It is asymptotically flat $ \Longrightarrow $ $U(\infty) =0$
\item It is a Maxwell Einstein metric in the sense that it satisfies
Einstein equations with the stress--energy tensor $T_{\mu \nu}$
contributed by a suitable collection of abelian gauge fields
$A^\Lambda_\mu $ ($\Lambda = 1, \dots,{\bar n}$).
\item It saturates the cosmic censorship bound \eqn{censur} in the
sense that its ADM mass is expressed as an
algebraic function of the electric and magnetic charges hidden in the
harmonic functions $H_i({\vec x})$.
\end{enumerate}
In the next section we shall interpret black holes of this form as
BPS saturated states namely as quantum states   filling
special irreducible representations of the supersymmetry algebra,
the so called {\it short supermultiplets}, the shortening condition
being precisely the saturation of the cosmic censorship bound \eqn{censur}.
Indeed such a bound can be restated as the equality of the mass with
the central charge which occurs when a certain fraction of the
supersymmetry charges identically annihilate  the state. The
remaining supercharges applied to the BPS state build up a unitary
irreducible representation of supersymmetry that is shorter than the
typical one since it contains  less states. As we stress in the
next section it is precisely this interpretation what makes extremal
black holes relevant to the string theorist. Indeed these classical
solutions behave as {\it the other half} of the particle  spectrum
of superstring theory, the half not accessible to perturbative string
theory.
\par
When in the above sentence we said {\it classical solutions}
we should have specified of which theory. The answer to this question is
{\it supergravity}, that is the low energy effective field theory of
superstrings. The {\it other half} of the quantum particle spectrum,
the non perturbative one, is actually mostly
composed of classical solutions of supergravity. This miracle, as
we shall further explain in the next section \ref{introgen},
is once again due to the BPS condition which protects
these classical solutions from quantum corrections.
\par
Furthermore we said {\it particles}. In modern parlance particles
are just $0$--branes, namely a particular instance of $p$--dimensional
extended objects (the $p$--branes) that occur as quantum states in
the non perturbative superstring spectrum and can be retrieved as
classical solutions of effective supergravity in $D$ space--time
dimensions. If we confine our attention to space--time dimensions
$D=4$ we have only $0$--branes, but these are in fact extended
$p$--branes that have {\it wrapped} along homology cycles
of the compactified dimensions. We shall
elaborate on this point in chapter \ref{pbrasugra}.
In the $p$--brane picture one divides the space--time coordinates
in two subsets, those belonging to the {\it world volume} spanned by the
moving brane and those {\it transverse} to it. The  $p$--brane metric
depends on space--time coordinates only through {  harmonic
functions} of the {transverse coordinates}. For $0$--branes the world
volume consists only of the time $t$ and the above general feature
is precisely that displayed by the extremal Reissner Nordstrom metric
\eqn{guarda!}.
\par
Hence to our reader  whose background is
classical general relativity or quantum gravity and who approaches
the notion of supergravity $p$--branes for the first time we can just
point out the following
\bst
The extremal Reissner Nordstrom metric \eqn{reinor}
where $m=|q|$, once rewritten as in eq.\eqn{guarda!} is the prototype
of the BPS saturated black holes treated in these lectures.
\est
\section{Extremal Black--Holes as quantum  BPS states}
\markboth{SUPERGRAVITY and BPS BLACK HOLES: Chapter 1} {1.2
EXTREMAL BLACK HOLES AS BPS STATES} \setcounter{equation}{0}
\label{introgen} In the previous section we have reviewed the idea
of extremal black holes as it arises in classical general
relativity. Extremal black--holes have become objects of utmost
relevance in the context of superstrings after {\it the second
string revolution} has taken place in 1995. Indeed supersymmetric
extremal black--holes have been studied in depth in a vast recent
literature \cite{malda,gensugrabh,ortino}. This interest is just
part of a more general interest in the $p$--brane classical
solutions of supergravity theories in all dimensions $4 \le D \le
11$ \cite{mbrastelle,duffrep}. This interest streams from the
interpretation of the classical solutions of supergravity that
preserve a fraction of the original supersymmetries as the BPS non
perturbative states necessary to complete the perturbative string
spectrum and make it invariant under the many conjectured duality
symmetries \cite{schw,sesch,huto2,huto,vasch}. Extremal
black--holes and their parent $p$--branes in higher dimensions are
therefore viewed as additional {\it particle--like} states that
compose the  spectrum of a fundamental quantum theory. The reader
of these notes who has a background in classical general
relativity and astrophysics should be advised that the holes we
are discussing here are neither stellar--mass, nor mini--black
holes: their mass  is typically of the order of the Planck--mass:
\begin{equation}
M_{Black~Hole} \, \sim \, M_{Planck}
\end{equation}
The Schwarzschild radius is therefore microscopic.
\par
Yet, as the monopoles in gauge theories, these  non--perturbative
quantum states originate from  regular solutions of the classical
field equations, the same Einstein equations one deals with in
classical general relativity and astrophysics. The essential new
ingredient, in this respect, is supersymmetry that requires the
presence of {\it vector fields} and {\it scalar fields} in
appropriate proportions. Hence the black--holes we are going to
discuss are solutions of generalized Einstein--Maxwell--dilaton
equations.
\par
The above mentioned
identification between classical $p$--brane solutions
(black--holes are instances of $0$--branes)
and the non--perturbative quantum states
of string theory required by duality
has become quite circumstantial with the advent of
$D$--branes \cite{DbranPolch1} and the possibility raised by them of
a direct construction of the BPS states within the language of
perturbative string theory extended by the choice of Dirichlet
boundary conditions \cite{Polchtasi}.
\par
Basic feature of the non--perturbative states of the string spectrum
is that they can carry Ramond--Ramond charges forbidden
at the perturbative level.
On the other hand, the important observation by  Hull
and Townsend \cite{huto,huto2} is that at the level of the low
energy supergravity lagrangians all fields of both the
Neveu-Schwarz,Neveu-Schwarz (NS-NS) and the Ramond-Ramond
(R-R) sector are unified
by the group of duality transformations $U$ which is also the
isometry group of the homogenous scalar manifold ${\cal
M}_{scalar}=U/H$. At least this is true in theories with sufficiently
large number of supersymmetries that is with $N\ge 3$ in $D=4$ or,
in a dimensional reduction invariant language with
$ \# \mbox{supercharges} \, \ge \, 12 $.
This points out that the distinction between R-R and
NS-NS sectors is just an artifact of perturbative string theory. It
also points out to the fact that the unifying symmetry between the
perturbative and non-perturbative sectors is already known from
supergravity, namely it is the $U$-duality group. Indeed the basic
conjecture of Hull and Townsend is that the restriction to integers
$U(\ZZ)$ of the $U$ Lie group determined by supergravity should be an
exact symmetry of non-perturbative string theory.
\par
From an abstract viewpoint BPS saturated states are characterized by the
fact that they preserve, in modern parlance, $1/2$ (or $1/4$, or $1/8$)
of the original
supersymmetries. What this actually means is that there is a
suitable projection operator $\IP^2_{BPS} =\IP_{BPS}$
acting on the supersymmetry charge $Q_{SUSY}$, such that:
\begin{equation}
 \left(\IP_{BPS} \,Q_{SUSY} \right) \, \vert \, \mbox{BPS state} \, >
 \,=  \, 0
 \label{bstato}
\end{equation}
Since the supersymmetry transformation rules of any supersymmetric
field theory are linear in the first derivatives of the fields
eq.\eqn{bstato} is actually a {\it system of first order differential
equations}. This system has to be combined with the second order
field equations of supergravity and the common solutions to both
system of equations is a classical BPS saturated state. That it is
actually an exact state of non--perturbative string theory follows
from supersymmetry representation theory. The classical BPS state is
by definition an element of a {\it short supermultiplet}
and, if supersymmetry is unbroken, it cannot be renormalized to
a {\it long supermultiplet}.
\par
Translating eq. \eqn{bstato} into an explicit first order
differential system requires knowledge of the supersymmetry
transformation rules of supergravity. These latter have a rich
geometrical structure that is the purpose of the present lectures
to illustrate to an audience assumed to be unfamiliar with
supergravity theory. Indeed the geometrical structure of supergravity
which originates in its scalar sector is transferred into the physics
of extremal black holes by the BPS saturation condition.
\par
In order to grasp the
significance of the above statement let us first rapidly review, as
an example, the algebraic definition of $D=4$ BPS states in a theory with
an even number of supercharges $N=2\nu$. The case with $N=\mbox{odd}$
can be similarly treated but needs some minor modifications due to
the fact the eigenvalues of an antisymmetric matrix in odd dimensions
are $\{ \pm \mbox{\rm i}\lambda_i , 0 \}$.
\subsection{General definition of BPS states in a 4D theory with
$N=2 \, \times \, p$ supersymmetries}
The $D=4$ supersymmetry algebra with $N=2\, \times \, p$
supersymmetry charges  is given by
\begin{eqnarray}
&\left\{ {\bar Q}_{A  \alpha }\, , \,{\bar Q}_{B  \beta}
\right\}\, = \,  {\rm i} \left( {\bf C} \,
\gamma^\mu \right)_{\alpha \beta} \,
P_\mu \, \delta_{AB} \, \, - \, {\bf C}_{\alpha \beta} \,
{\bf Z}_{AB}& \nonumber\\
&\left( A,B = 1,\dots, 2p \right)&
\label{susyeven2}
\end{eqnarray}
where the SUSY charges ${\bar Q}_{A}\equiv Q_{A}^\dagger \gamma_0=
Q^T_{A} \, {\bf C}$ are Majorana spinors, $\bf C$ is the charge conjugation
matrix, $P_\mu$ is the 4--momentum operator and
 the antisymmetric tensor
${\bf Z}_{AB}=-{\bf Z}_{BA}$ is the central charge operator. It
can always be reduced to normal form
\begin{equation}
{\bf Z}_{AB} ~~~~= \pmatrix{\epsilon Z_1&0& \dots& 0\cr 0&\epsilon
Z_2& \dots &0\cr \dots &\dots & \dots & \dots \cr 0&0& \dots
&\epsilon Z_p\cr}
\label{skewZ}
\end{equation}
where $\epsilon$ is the $2\times 2$ antisymmetric matrix,
(every zero is a $2\times2$ zero matrix) and
the $p$ skew eigenvalues $Z_I$ of ${\bf Z}_{AB}$ are the central charges.
\par
If we identify each index $A,B,\dots$ with a pair of indices
\begin{equation}
 A=(a,I) \quad ; \quad a,b,\dots =1,2 \quad ; \quad I,J,\dots =1,
 \dots, p
 \label{indrang}
\end{equation}
then the superalgebra \eqn{susyeven2} can be rewritten as:
\begin{eqnarray}
&\left\{ {\bar Q}_{aI \vert \alpha }\, , \,{\bar Q}_{bJ \vert \beta}
\right\}\, = \,  {\rm i} \left( C \, \gamma^\mu \right)_{\alpha \beta} \,
P_\mu \, \delta_{ab} \, \delta_{IJ} \, - \, C_{\alpha \beta} \,
\epsilon_{ab} \, \times \, \ZZ_{IJ}&
\label{susyeven}
\end{eqnarray}
where the SUSY charges ${\bar Q}_{aI}\equiv Q_{aI}^\dagger \gamma_0=
Q^T_{aI} \, C$ are Majorana spinors, $C$ is the charge conjugation
matrix, $P_\mu$ is the 4--momentum operator, $\epsilon_{ab}$ is the
two--dimensional Levi Civita symbol and the central charge
operator is now represented by the {\it symmetric tensor}
$\ZZ_{IJ}=\ZZ_{JI}$ which
can always be diagonalized $\ZZ_{IJ}=\delta_{IJ} \, Z_J$.
The $p$ eigenvalues $Z_J$ are the skew eigenvalues
introduced in equation \eqn{skewZ}.
\par
The Bogomolny bound on the mass of a generalized monopole state:
\begin{equation}
M \, \ge \, \vert \, Z_I \vert \qquad \forall Z_I \, , \,
I=1,\dots,p
\label{bogobound}
\end{equation}
is an elementary consequence of the supersymmetry algebra and of the
identification between {\it central charges} and {\it topological
charges}. To see this it is convenient to introduce the following
reduced supercharges:
\begin{equation}
{\bar S}^{\pm}_{aI \vert \alpha }=\frac{1}{2} \,
\left( {\bar Q}_{aI} \gamma _0 \pm \mbox{i} \, \epsilon_{ab} \,  {\bar Q}_{bI}\,
\right)_\alpha
\label{redchar}
\end{equation}
They can be regarded as the result of applying
a projection operator to the supersymmetry
charges:
\begin{eqnarray}
{\bar S}^{\pm}_{aI} &=& {\bar Q}_{bI} \, \IP^\pm_{ba} \nonumber\\
 \IP^\pm_{ba}&=&\frac{1}{2}\, \left({\bf 1}\delta_{ba} \pm \mbox{i} \epsilon_{ba}
 \gamma_0 \right)
 \label{projop}
\end{eqnarray}
Combining eq.\eqn{susyeven} with the definition \eqn{redchar} and
choosing the rest frame where the four momentum is $P_\mu$ =$(M,0,0,0)$, we
obtain the algebra:
\begin{equation}
\left\{ {\bar S}^{\pm}_{aI}  \, , \, {\bar S}^{\pm}_{bJ} \right\} =
\pm \epsilon_{ac}\, C \, \IP^\pm_{cb} \, \left( M \mp Z_I \right)\,
\delta_{IJ}
\label{salgeb}
\end{equation}
By positivity of the operator
$\left\{ {\bar S}^{\pm}_{aI}  \, , \, {\bar S}^{\pm}_{bJ} \right\} $
it follows that on a generic state the Bogomolny bound \eqn{bogobound} is
fulfilled. Furthermore it also follows that the states which saturate
the bounds:
\begin{equation}
\left( M\pm Z_I \right) \, \vert \mbox{BPS state,} i\rangle = 0
\label{bpstate1}
\end{equation}
are those which are annihilated by the corresponding reduced supercharges:
\begin{equation}
{\bar S}^{\pm}_{aI}   \, \vert \mbox{BPS state,} i\rangle = 0
\label{susinvbps}
\end{equation}
On one hand eq.\eqn{susinvbps} defines {\sl short multiplet
representations} of the original algebra \eqn{susyeven} in the
following sense: one constructs a linear representation of \eqn{susyeven}
where all states are identically
annihilated by the operators ${\bar S}^{\pm}_{aI}$ for $I=1,\dots,n_{max}$.
If $n_{max}=1$ we have the minimum shortening, if $n_{max}=p$ we
have the maximum shortening. On the other hand eq.\eqn{susinvbps}
can be translated into a first order differential equation on the
bosonic fields of supergravity.
 \par
Indeed, let us consider
a configuration where all the fermionic fields are zero.
Setting the fermionic SUSY rules appropriate to such a background equal
to zero we find the following Killing spinor equation:
\begin{equation}
0=\delta \mbox{fermions} = \mbox{SUSY rule} \left( \mbox{bosons},\epsilon_{AI} \right)
\label{fermboserule}
\end{equation}
where the SUSY parameter satisfies the following conditions
\footnote{$\xi^\mu$ denotes a time--like Killing vector}:
\begin{equation}
\begin{array}{rclcl}
\xi^\mu \, \gamma_\mu \,\epsilon_{aI} &=& \mbox{\rm i}\, \varepsilon_{ab}
\,  \epsilon^{bI}   & ; &   I=1,\dots,n_{max}\\
\epsilon_{aI} &=& 0  &;&   I > n_{max} \\
\end{array}
\label{rollato}
\end{equation}
Hence eq.s \eqn{fermboserule} with a parameter satisfying the
condition \eqn{rollato} will be our operative definition of BPS
states.
\section{The Horizon Area and Central Charges}
\markboth{BPS BLACK HOLES IN SUPERGRAVITY: CHAPTER 1}
{1.3 THE HORIZON AREA AND CENTRAL CHARGES}
\setcounter{equation}{0}
\label{introp}
In the main body of these lectures we are going to
see how eq.s \eqn{fermboserule}
can be translated into explicit differential equations. Solving
such first order equations, together with the second
order Einstein Maxwell equations we
can obtain BPS saturated black hole solutions
of the various versions of supergravity theory.
This programme requires the use of the rich and complex structure of
supergravity lagrangians. In these lectures, however
we do not dwell on the technicalities
of supersymmetry theory and except for the supersymmetry
transformations rules needed to write \eqn{fermboserule} we almost
nowhere else mention the fermionic fields. What we rather explain
in detail   the geometric symplectic structure of the supergravity
lagrangians related to the simultaneous presence of {\it vector}
and {\it scalar} fields, the latter interpreted \'a la $\sigma$--model as
coordinates of a suitable scalar manifold ${\cal M}_{scalar}$.
This symplectic structure which is enforced by supersymmetry and
which allows the definition of  generalized {\it electric--magnetic
duality rotations} fits into  a rather general pattern and it is
responsible for the most fascinating  and most intriguing result in
the analysis of supergravity BPS black holes: the interpretation
of the horizon area appearing in the Beckenstein--Hawking formula
\eqn{bekhaw} as a {\it topological U--duality invariant} depending
only on the magnetic and electric charges of the hole.
\par
If we go back to our prototype metric \eqn{reinor} and we calculate
the area of the horizon, we find:
\begin{equation}
\mbox{Area}_H = \int_{\rho=\rho_+} \, \sqrt{g_{\theta\theta}\,g_{\phi\phi}}
\,d\theta \,d\phi \, = \,  4\pi \,  \rho_+^2 = \,4\pi \,
\left(m+\sqrt{m^2-|q|^2}\right)^2
\label{konig}
\end{equation}
In the case of an extremal black--hole ($m=|q|$) eq.\eqn{konig}
becomes:
\begin{equation}
 \frac{\mbox{Area}_H}{4\pi}  =  |q|^2
 \label{prinz}
\end{equation}
The rather innocent looking formula \eqn{prinz} contains a message of
the utmost relevance. For BPS black holes the horizon area and hence
the entropy is a function solely of the electric and magnetic charges
of the hole. It is also a very specific function, since it is an
algebraic invariant of a duality group. But this is difficult to be
seen in a too simple theory where there is just one electromagnetic
field. Eq.\eqn{prinz} reveals its hidden surprising structure when it
is generalized to a theory containing several vector fields
interacting with gravity and scalars within the symplectic scheme
enforced by supersymmetry and proper to the supergravity lagrangians.
\par
Explaining the generalization of eq.\eqn{prinz} both at a general
level and in explicit examples taken from the maximally extended
supergravity theory ($N=8$) is the main purpose of this series of
lectures.
\chapter{SUPERGRAVITY $p$--BRANES IN HIGHER DIMENSIONS }
\label{pbrasugra}
As we already stressed, in line with the topics of this school,
our main goal  is
to classify BPS  {\it black hole
solutions} of $4D$ supergravity and to unravel the fascinating
group theoretical structure of their Bekenstein Hawking entropy.
\par
Black holes are instances of $0$--branes, namely objects with
zero--dimensional space--like extension that can evolve in time. They
carry {\it quantized electric and magnetic charges}
$\{q_ \Sigma , p^ \Lambda \}$  under the host of gauge--fields $A_ \mu ^
\Lambda$ ($ \Lambda = 1,\dots , {\bar n}$)
 appearing in the spectrum of supergravity, that is  the low energy
effective theory of strings, or M--theory. As we shall see,
the entropy
\begin{equation}
S_{BH}  \, = \, {\cal S}\left(q,p \right)
\label{entro1}
\end{equation}
is a topological invariant that depends solely on such quantized charges.
Actually it is not only a topological invariant insensitive to continuous
deformations of the black--hole solution with respect to the parameters
it depends on ({\it the moduli}) but also an invariant
in the group--theoretical sense. Indeed ${\cal S}\left(q,p \right)$
is an invariant of the U--duality group that unifies the various perturbative
and non--perturbative duality of string theory. Such a complicated structure
of the black hole arises from the compactification of too many of the actual
dimensions   of  space--time.  The  black-hole  appears  to  us  as  a
$0$--brane  only  because its spatial extension has been hidden in the
six  (or  seven  ) dimensional compact manifold ${\cal M}_{comp}$ that
is   not   directly   accessible   at  low  energies.  The  $0$--brane
black--hole   is  actually  a  $p$--brane  (or  intersection  of  many
$p$--branes)  that  are  {\it  wrapped} on the homology $p$--cycles of
the internal manifold ${\cal M}_{comp}$.
\par
So  although  our  main  focus  will  be  on  the  $D=4$  theory,
in order to get a better insight into the meaning of such
complicated objects as we propose to study,  it is
appropriate to start from the vantage point of higher dimensions $D > 4$
where  $p$--branes  become  much  simpler by freely unfolding into non
compact directions.
\section{Definition and general features of dilatonic
$p$--brane solutions in dimension $D$ }
\markboth{SUPERGRAVITY and BPS BLACK HOLES: Chapter 2}
{2.1 DEFINITION OF $p$--BRANE SOLUTIONS IN DIMENSION $D$}
\setcounter{equation}{0}
\label{geafeature}
The  basic  idea  of  a  $p$--brane  solution  can  be  illustrated by
considering a very simple action functional in $D$ space-time dimensions
that contains just only three fields:
\begin{enumerate}
\item the metric $g_{\mu \nu}$, namely the {\it graviton}
\item a scalar field $\phi(X)$, namely the {\it dilaton}
\item  a $p+1$-form gauge field $A_{[p+1]} \equiv A_{M _1  \dots
M _{p+1} } dX^{M_1} \, \wedge \dots \wedge dX^M _{p+1} $.
\end{enumerate}
Explicitly we write :
\begin{equation}
 I_D=\int d^D x \sqrt{-g}~ [R-{1\over 2} \nabla_M \phi \nabla^M \phi-
{1 \over 2 n!} e^{a \phi} F^2_{[p+2]}]
  \label{paction}
\end{equation}
where $F_{[p+2]}\equiv d A_{[p+1]}$ is the field strength of the $p+1$--form
gauge potential  and $a$ is some real number whose profound meaning
will become clear in the later discussion of the solutions.
For various values of
\begin{equation}
n=p+2  \quad \mbox{and } \quad  a
\end{equation}
the functional  ${I}_{D}$ is a consistent truncation of some
supergravity bosonic action $S_D^{SUGRA}$ in dimension $D$.
By consistent truncation we mean that a
subset of the bosonic fields have been put equal to zero but
in such a way that all solutions of the truncated action are also
solutions of the complete one.
For instance if we choose:
\begin{equation}
a=1 \quad \quad n=\cases{3 \cr 7\cr}
\label{het}
\end{equation}
eq.\eqn{paction} corresponds to the bosonic low energy action of $D=10$
heterotic superstring (N=1, supergravity) where the $E_8\times
E_8$ gauge fields have been deleted. The two choices $3$ or $7$ in
eq.\eqn{het} correspond to the two formulations (electric/magnetic)
of the theory. Other choices correspond to truncations of the type IIA or
type IIB action in the various intermediate dimensions $4\le D\le
10$. Since the $n-1$--form $A_{[n-1]}$ couples
to the world volume of an extended object of dimension:
\begin{equation}
p = n-2
\label{interpre}
\end{equation}
namely a $p$--brane, the choice of the truncated action \eqn{paction}
is motivated by the search for $p$--brane solutions of supergravity.
According with the interpretation \eqn{interpre} we set:
\begin{equation}
  n=p+2  \qquad  d=p+1 \qquad
{\tilde d}= D-p-3
\label{wvol}
\end{equation}
where $d$ is the world--volume dimension of an electrically charged
{\it elementary} $p$--brane solution, while ${\tilde d}$ is
the world--volume dimension of a magnetically charged {\it solitonic}
${\tilde p}$--brane with ${\tilde p} = D-p-4$. The distinction between
elementary and solitonic is the following. In the elementary case
the field configuration we shall discuss is a true vacuum solution of
the field equations following from the action \eqn{paction} everywhere in
$D$--dimensional space--time except for a singular locus of dimension
$d$. This locus can be interpreted as the location of an elementary
$p$--brane source that is coupled to supergravity via an electric charge
spread over its own world volume. In the solitonic case, the field
configuration we shall consider is instead a bona--fide solution of
the supergravity field equations everywhere in space--time without
the need to postulate external elementary sources. The field energy
is however concentrated around a locus of dimension ${\tilde p}$.
These  solutions  have  been  derived  and  discussed  thoroughly in the
literature \cite{stellebrane}. Good reviews of such results are
\cite{mbrastelle,mbratownsend}. Defining:
\begin{equation}
\Delta = a^2 +2\, \frac{d {\tilde d} }{ D-2}
\label{deltadef}
\end{equation}
it was shown in \cite{stellebrane} that the action \eqn{paction} admits
the following elementary $p$--brane solution
\begin{eqnarray}
ds^2 & =& \left(1+\frac{k}{ r^{\tilde d} } \right)^{-  \frac {4\, { \tilde d}} {\Delta (D-2)}}
\, dx^\mu \otimes dx^\nu \, \eta_{\mu\nu}
- \left(1+\frac{k}{ r^{\tilde d} } \right)^{ \frac {4\, {d}} {\Delta (D-2)}}
\, dy^m \otimes dy^n \, \delta_{mn}\nonumber \\
F &= &\lambda (-)^{p+1}\epsilon_{\mu_1\dots\mu_{p+1}} dx^{\mu_1}
\wedge \dots \wedge dx^{\mu_{p+1}}
\wedge \, \frac{y^m \, dy^m}{r} \, \left(1+\frac{k}{r^{\tilde
d}}\right )^{-2} \, \frac{1}{r^{{\tilde d}+1}}\nonumber\\
e^{\phi(r)} &=& \left(1+\frac{k}{r^{\tilde d}}\right)^{-\frac {2a}{\Delta}}
\label{elem}
\end{eqnarray}
where the coordinates $X^M$ ($M=0,1\dots , D-1$ have been split into
two subsets:
\begin{itemize}
\item $x^\mu$, $(\mu=0,\dots ,p)$ are the coordinates on the
$p$--brane world--volume,
\item $y^m$, $(m=D-d+1,\dots ,D)$ are the coordinates  transverse to
the brane
\end{itemize}
By  $r \equiv \sqrt{y^m y_m}$ we denote the radial distance from the
brane and by $k$   the value of its electric charge.  Finally, in
eq.\eqn{elem} we have set:
\begin{equation}
\lambda= 2\, \frac{{\tilde d} \, k}{\sqrt{\Delta}}
\end{equation}
The same authors of \cite{stellebrane} show that that the
action \eqn{paction} admits also
the following solitonic ${\tilde p}$--brane solution:
\begin{eqnarray}
ds^2 & =& \left(1+\frac{k}{ r^{d} } \right)^{- \frac {4\, {   d}}
{\Delta (D-2)}}
\, dx^\mu \otimes dx^\nu \, \eta_{\mu\nu}
- \left(1+\frac{k}{ r^{ d} } \right)^{ \frac {4\, {\tilde d}} {\Delta (D-2)}}
\, dy^m \otimes dy^n \, \delta_{mn}\nonumber \\
{\tilde F}_{[D-n]} &= &\lambda  \epsilon_{\mu_1\dots\mu_{{\tilde d}}p} dx^{\mu_1}
\wedge \dots \wedge dx^{\mu_{\tilde d}}\wedge \frac{y^p}{r^{d+2}} \nonumber\\
e^{\phi(r)} &=& \left(1+\frac{k}{r^{d}}\right)^{\frac {2a}{ \Delta}}
\label{solit}
\end{eqnarray}
where the $D-p-2$--form ${\tilde F}_{[D-n]}$
is the dual of $F_{[n]}$, $k$ is now the magnetic charge
and:
\begin{equation}
\lambda= - 2\, \frac{{\tilde d} \, k}{\sqrt{\Delta}}
\end{equation}
These  $p$--brane configurations are solutions of the second order
field equations obtained by varying the action \eqn{paction}.
However, when \eqn{paction} is the truncation of
a supergravity action both \eqn{elem} and \eqn{solit} are also the
solutions of a {\it first order differential system of equations}.
This happens because they are BPS--extremal $p$--branes which
preserve a fraction of the original supersymmetries. As an example
we consider the  $10$--dimensional case.
\subsection{The elementary string solution  of heterotic
supergravity in $D=10$}
Here we have :
\begin{equation}
 D=10 \qquad d=2 \qquad {\tilde d}= 6 \qquad
a=1 \qquad \Delta = 4 \qquad \lambda = \pm 6 k
\label{values}
\end{equation}
so that the elementary string solution reduces to:
\begin{eqnarray}
ds^2 &=& \exp[2 U(r)] \, dx^\mu \otimes dx^\nu -
\exp[-\frac{2 }{3}  U(r)] \, dy^m \otimes dy^m
 \nonumber\\
\exp[2 U(r)]&=& \left(1+\frac{k}{r^6}\right)^{-3/4}\nonumber \\
F &=& 6k \, \epsilon_{\mu\nu} dx^\mu \wedge dx^\nu
\wedge \frac{y^m dy^m}{r} \,
\left( 1+\frac{k}{r^6}\right)^{-2} \, \frac{1}{r^7}\nonumber\\
\exp[\phi(r)] &=&  \left(1+\frac{k}{r^6}\right)^{-1/2}
\label{stringsol3}
\end{eqnarray}
As already pointed out, with the values \eqn{values},
the action \eqn{paction}
is just the truncation of heterotic supergravity where, besides the
fermions, also the $E_8\times E_8$ gauge fields have been set to zero.
In this theory the supersymmetry transformation rules we have
to consider are those of the gravitino and of the dilatino.
They read:
\begin{eqnarray}
\delta \psi _\mu &=& \nabla_\mu \epsilon\, +\, \frac{1}{96} \,
\exp[\frac{1}{2} \phi] \, \Bigl ( \Gamma_{\lambda\rho\sigma\mu}
+\, 9\,  \Gamma_{\lambda\rho} \, g_{\sigma\mu} \Bigr )
\, F^{\lambda\rho\sigma}
\, \epsilon \nonumber\\
\delta \chi &=& \mbox{i}\,
\frac{\sqrt{2}}{4} \, \partial^\mu \phi \,
\Gamma_\mu \epsilon  -\,  \mbox{i}\,
\frac{\sqrt{2}}{24} \, \exp[-\frac{1}{2}\phi ] \,
\Gamma_{\mu\nu\rho} \, \epsilon \, F^{\mu\nu\rho}
\label{susvaria}
\end{eqnarray}
Expressing the  ten dimensional gamma matrices as tensor products of
the  two dimensional gamma matrices $\gamma_\mu$ ($\mu=0,1$) on the
$1$--brane world sheet with the eight dimensional gamma matrices
$\Sigma_m$ ($m=2,\dots, 9$) on the transverse space it is easy to check
that in the background \eqn{stringsol3}
the SUSY variations
\eqn{susvaria} vanish for the following choice of the parameter:
\begin{equation}
\epsilon   =   \left( 1+\frac{k}{r^6}\right)^{-3/16} \, \epsilon_0
\otimes \eta_0
\label{carmenpara}
\end{equation}
where the constant spinors $\epsilon_0$ and $\eta_0$ are respectively
$2$--component and $16$--component and have both positive chirality:
\begin{equation}
\matrix{ \gamma_3 \, \epsilon_0 = \epsilon_0 &
\Sigma_{10} \, \eta_0 = \eta_0}
\label{chiralcondo}
\end{equation}
Hence we conclude that the extremal $p$--brane solutions of all
maximal (and non maximal) supergravities can be
obtained by imposing the supersymmetry invariance of the background
with respect to a projected SUSY parameter of the type
\eqn{carmenpara}.
\section{$M2$--branes in $D=11$ and the issue of the horizon geometry}
\setcounter{equation}{0}
\markboth{SUPERGRAVITY and BPS BLACK HOLES: Chapter 2}
{2.2 $M2$--BRANES IN $D=11$ AND HORIZON GEOMETRY}
To illustrate a very crucial feature
of $p$--brane solutions, namely the structure of their horizon
geometry, we choose another example in $D=11$. The bosonic spectrum of $D=11$
supergravity \cite{castdauriafre} is very simple since besides the metric
$g_{MN}$ it contains only a $3$--form gauge field $A_{[3]}$ and no
scalar field $\phi$. This means that $a=0$ and that there is an
elementary electric $2$--brane solution and a magnetically charged
$5$--brane, the $M2$--brane and the $M5$--brane, respectively. This
universally adopted nomenclature follows from $D=11$ supergravity
being the low energy effective action of M--theory. Accordingly
we have:
\begin{equation}
a = 0 \quad ; \quad d=3 \quad ; \quad {\tilde d}=6 \quad ; \quad
\Delta = 4 \quad ; \quad D=11
\end{equation}
and following eq:\eqn{elem} the metric corresponding to the
elementary $M2$--brane can be written as:
\begin{equation}
ds^2_{11} \, = \, \left(1+
\frac{k}{r^{\tilde d}} \right)^{-\frac{\tilde d}{9}} \,
dx^\mu  \,   \, dx^\nu \eta_{\mu\nu} + \, \, \left(1+
\frac{k}{r^{\tilde d}} \right)^{\frac{  d}{9}} \, dX^I \,  \,
dX^J \, \delta_{IJ}\, .
\label{protyp}
\end{equation}
where
\begin{equation}
\begin{array}{ccccccc}
 d & \equiv & 3 & ; & {\tilde d} & \equiv & 11 - 3 -2 = 6\\
\end{array}
\label{ddtil}
\end{equation}
What we would like to note is that in the {\it bulk} of this solution,
namely for generic values of $r$, the isometry of the
$11$--dimensional metric \eqn{protyp} is:
\begin{equation}
 {\cal I}_{2-brane} \, = \, ISO(1,2) \, \otimes \, SO \left( 8
 \right)
 \label{isometr1}
\end{equation}
yet if we take the limiting value of this metric in the two limits
\begin{enumerate}
\item Infinity: $r \to \infty $
\item Horizon: $r \to 0 $
\end{enumerate}
then the isometry gets enhanced and we respectively obtain:
\begin{eqnarray}
ISO(1,10) &\mbox{at}&   \mbox {
infinity} \nonumber\\
SO(2,3) \times SO(8)& \mbox{at}& \mbox{the horizon}
\end{eqnarray}
The reason for this is simple. As one can realize by inspection, at
$r = \infty$ the metric \eqn{protyp} becomes the flat 11-dimensional
metric which obviously admits the 11D Poincar\'e group as isometry,
while in the vicinity of $r = 0$, the metric \eqn{protyp} is
approximated by the following metric:
\begin{equation}
ds^2\, = \,\rho^2 (-dt^2+d\vec{z}\cdot d\vec{z})+\rho^{-2}d\rho^2 \,
+ \,
d\Omega_{7}^2
\label{bertarob}
\end{equation}
where the last term $d\Omega_{7}^2$ is the $SO(8)$ invariant metric on
a 7--sphere $S^{7}$, while the previous ones correspond to a
particular parametrization of the anti de Sitter metric on $AdS_{4}$,
which, by definition, admits $SO(2,3)$ as isometry. To see that this is
indeed the case it suffices to use polar coordinates for the eight
transverse directions to the brane:
\begin{equation}
\, dy^I \,
dy^J \, \delta_{IJ} \, = \, dr^2 \, + \, r^2 \, d\Omega_{7}^2
\label{polar}
\end{equation}
set $\rho= r^2$ and take the limit $\rho \to 0$ in the metric \eqn{protyp}
\par
The conclusion therefore is that in the vicinity of the horizon
the geometry of the $M2$--brane \eqn{protyp} is:
\begin{equation}
  M^{hor}_2 = AdS_{4} \, \times \, S^7
  \label{ads4s7}
\end{equation}
which, by itself, would be an exact solution of $D=11$ supergravity.
It follows that the $M2$--brane can be seen as an M--theory {\it
soliton}, in the usual sense of soliton theory: it is an exact
solution of the field equations that interpolates smoothly between
two vacua of the theory, 11D Minkowski space at infinity and
the space $AdS_{4} \, \times \, S^7$ at the horizon.
\par
This feature of the $M2$--brane \eqn{protyp} corresponds to a very
general property of $p$--branes and can be generalized in two ways.
First we can make the following:
\bst
Whenever $a=0$ the limiting horizon geometry of an elementary or
solitonic $p$--brane as defined in eq.\eqn{elem} or eq. \eqn{solit} is
\begin{equation}
M^{hor}_{p} = AdS_{p+2} \, \times \, S^{D-p-2}
\end{equation}
\est
Secondly we can pursue the following
\bge
Rather than eq.s\eqn{elem} or \eqn{solit}, as candidate solutions
for the action \eqn{paction} consider the following ansatz:
\begin{eqnarray}
& &ds^2=e^{2A(r)} dx^{\mu} dx^{\nu} \eta_{\mu\nu} + e^{2B(r)}[dr^2
+r^2
\la^{-2} ds^2_{G/H}] \label{Ansatz1}\\
& &A_{\mu_1 ...\mu_d}=\epsilon_{\mu_1...\mu_d} e^{C(r)}
\label{Ansatz2}\\
& &\phi=\phi(r) \label{Ansatz3}
\end{eqnarray}
where
\begin{enumerate}
\item  $ \lambda $ is a constant parameter with
the dimensions of  length.
\item The $D$ coordinates $X^M$ are split as follows:
$X^M =(x^{\mu},r,y^m)$, $\eta^{MN}=\mbox{diag}(-,+++...)$
\item $\mu= 0,\dots ,d-1 $ runs on the  p-brane world-volume
($d=p+1$)
\item
$\bullet$ labels  the $r$ coordinate
\item
$m=d+1,\dots, D-1$ runs on some $D-d-1$-dimensional compact
coset manifold $G/H$, $G$ being a compact Lie group and $H\subset G$
a closed Lie subgroup.
\item $ds^2_{G/H}$ denotes a $G$--invariant metric on the above
mentioned coset manifold.
\end{enumerate}
\ege
The basic difference with respect to the previous case is that we have
replaced the invariant metric $ds^2_{S^{D-d-1}}$ on a sphere
$S^{D-d-1}$ by the more general coset manifold metric  $ds^2_{G/H}$.
With such an ansatz exact solutions can be found as it was shown in
\cite{GHbrana}. If $a=0$, namely if there is no dilaton,
such $G/H$ $p$--branes have as horizon geometry:
\begin{equation}
M^{hor}_{p} = AdS_{p+2} \, \times \, \left(\frac{G}{H}\right)_{D-p-2}
\end{equation}
and their horizon symmetry is:
\begin{equation}
SO(2,p+1) \, \times \, G
\end{equation}
Applied to the case of $D=11$ supergravity, the electric ansatz
in eq.s \eqn{Ansatz1},\eqn{Ansatz2} produces $G/H$ M2--brane solutions.
For pedagogical purposes we describe the derivation of such solutions
in a little detail.
\subsection{Derivation of the $G/H$ M2-brane solution}
The field equations derived from \eqn{paction} have the following
form:
\begin{eqnarray}
& &R_{MN}={1\over 2} \partial_M \phi \partial_N \phi
+S_{MN}\label{feq1}\\
& &\nabla_{M_1} (e^{a\phi} F^{M_1 ...M_n})=0 \label{feq2}\\
& &\square \phi = {a\over 2n!} F^2 \label{feq3}
\end{eqnarray}
where $S_{MN}$ is the energy-momentum tensor of  the $n$-form $F$:
\begin{equation}
S_{MN}={1 \over 2(n-1)!} e^{a\phi}  [F_{M...} F_{N}^{~...} - {n-1
\over n(D-2)}
F^2 g_{MN}]
\end{equation}
\subsubsection{The Vielbein}
In order to prove that the ansatz \eqn{Ansatz1}-\eqn{Ansatz3}
is a solution of the field equations
it is necessary to calculate the corresponding vielbein,
spin--connection and curvature tensors. We use the
convention that tangent space indices are underlined. Then the
vielbein components relative to the
ansatz (\ref{Ansatz1}) are:
\begin{eqnarray}
& &\Ef{\mu}=e^Adx^{\mu};~\Ef{\bu} =e^Bdr;~~\Ef{m}=e^B r \la^{-1}
\Efb{m}; \\
& &g_{\mu\nu}=e^{2A}
\eta_{\mu\nu};~~g_{\bu\bu}=e^{2B},~~g_{mn}=e^{2B} r^2
\la^{-2}\gb_{mn}
\end{eqnarray}
with $\Eb^{\mun} \equiv G/H$ vielbein and  $\gb_{mn}$ $\equiv $
$G/H$
metric.
\subsubsection{The spin connection}
The Levi--Civita spin--connection on our $D$--dimensional manifold
is defined as the solution of the vanishing torsion equation:
\begin{equation}
d\Ef{M} + \omef{M}{N} \we \Ef{N}=0
\label{zerotors}
\end{equation}
Solving eq.\eqn{zerotors} explicitly we obtain the spin--connection
components:
\begin{equation}
\omefu{\mu}{\nu}=0,~~\omefu{\mu}{\bu}=e^{-B} A'
\Ef{\mu},~~\omefu{\mu}{n}=0,~~
\omefu{m}{n}=\omefub{m}{n},~~\omefu{m}{\bu}=\exp[-B] \,
(B'+r^{-1})\Ef{m}.
\end{equation}
where $A' \equiv \part_\bu A$ etc.  and $\omefub{m}{n}$ is the spin
connection of the $G/H$ manifold.
\subsubsection{The Ricci tensor}
 From the definition of the curvature 2-form :
\begin{equation}
\Rfu{M}{N}=d\omefu{M}{N}+\omef{M}{S} \we \omefu{S}{N}
\end{equation}
we find the Ricci tensor components:
\begin{eqnarray}
& &R_{\mu\nu}=-{1\over 2} \eta_{\mu\nu} e^{2(A-B)}  [A''+d
(A')^2+\dt
A'B' +
(\dt+1) r^{-1} A']\\
& &R_{\bu\bu}=-{1\over 2}
[d(A''+(A')^2-A'B')+(\dt+1)(B''+r^{-1}B')]\\
& &R_{mn}=-{1\over 2} \gb_{mn}  {r^2\over \la^2}
[dA'(B'+r^{-1})+r^{-1}B'+B''+\dt (B'+r^{-1})^2]+ \Rb_{mn}
\end{eqnarray}
where $\Rb_{mn}$ is the Ricci tensor of $G/H$ manifold, and $\dt
\equiv
D-d-2$.
\subsubsection{The field equations}
Inserting the electric ansatz into the field eq.s \eqn{feq1} yields:
\begin{eqnarray}
& & A''+d(A')^2+\dt A'B' + (\dt+1) A' r^{-1} = {\dt \over 2(D-2)}
S^2 \label{einstein1}\\
& & d[A''+(A')^2-A'B']+(\dt+1)[B'' + r^{-1}B']={\dt \over 2(D-2)}
S^2-{
(\phi')^2\over 2} \label{einstein2}\\
& & \gb_{mn} [dA'(B'+r^{-1})+r^{-1}B'+B''+\dt (B'+r^{-1})^2]-2
\Rb_{mn}=\nonumber\\
& & ~~~~~~~~~~~~~~~~~~~~~~~~~~~~~~~~~~~~~~~
{}~~~~~~~~~~~~~~~~~~~-{d\over 2(D-2)} \gb_{mn} S^2 \label{einstein3}
\end{eqnarray}
while eq.s (\ref{feq2})-(\ref{feq3}) become:
\begin{eqnarray}
C''+ (\dt+1) r^{-1} C' + (\dt B'-d A'+C'+a \phi') C'=0
\label{maxwell}\\
\phi''+ (\dt+1) r^{-1} \phi' + [dA'+\dt B'] \phi'= -{a\over 2} S^2
\label{scalar}
\end{eqnarray}
with
\begin{equation}
S \equiv e^{{1\over 2} a \phi +C -dA} C'
\end{equation}
\subsection{Construction of the BPS Killing spinors in the case
of $D=11$ supergravity}
\label{BPSsusy}
At this point we specialize our analysis to the case
of $D=11$ supergravity, whose action in the bosonic sector reads:
\begin{equation}
I_{11}=\int d^{11}x \sqrt{-g} ~(R-{1\over 48} F^2_{[4]} ) + {1\over
6} \int
F_{[4]} \we
F_{[4]} \we A_{[3]}
\label{elevenaction}
\end{equation}
and we look for the further restrictions imposed on the electric
ansatz by the requirement that the solutions should preserve a
certain amount of supersymmetry. This is essential for our goal
since we are interested in $G/H$ M--branes that are BPS saturated
states and the  BPS condition requires the existence of Killing
spinors.
\par
As discussed  in ref. \cite{mbrastelle},  the above action does not
fall exactly in the general class of actions of type  \eqn{paction}.
Nevertheless, the results of sections 2.1 and 2.2 still apply:
indeed it is straightforward to verify that the FFA term in  the
action
\eqn{elevenaction} gives no contribution to the field equations
once the electric or magnetic ansatz are implemented.
Moreover no scalar fields are present in \eqn{elevenaction}:
this we handle by simply  setting  to zero the scalar coupling
parameter $a$.
\par
Imposing that the ansatz solution admits Killing spinors
allows to simplify the field equations drastically.
 \par
We recall the supersymmetry
transformation for the gravitino:
\begin{equation}
\de \psi_M = \Dt_M \epsilon
\label{gravitin1}
\end{equation}
with
\begin{equation}
\Dt_M = \part_M + {1\over 4} \om_M^{~~AB} \Ga_{AB} -
{1 \over 288} [\Ga^{PQRS}_{~~~~M} + 8 \Ga^{PQR} \de^S_M] F_{PQRS}
\label{gravitin2}
\end{equation}
Requiring that setting $\psi_M=0$  be consistent with the
existence
of residual supersymmetry yields:
\begin{equation}
{\de \psi_M }_{|\psi=0}=\Dt_M \epsilon=0
\label{Kspinor}
\end{equation}
Solutions $\epsilon (x,r,y)$ of the above equation are {\sl Killing
spinor} fields on the bosonic background described by our ansatz.
\par
In order to discuss the solutions of \eqn{Kspinor}
we adopt the following tensor product realization
of the ($32 \times 32$) $SO(1,10)$ gamma matrices:
\begin{equation}
\Ga_A=[\ga_{\mu} \otimes \bfone_8, \ga_3 \otimes \bfone_8, \ga_5
\otimes
\Ga_m]
\label{gambas}
\end{equation}
The above basis \eqn{gambas} is well adapted to our (3+1+7) ansatz.
The $\ga_{\mu} ~(\mu=0,1,2,3)$ are usual $SO(1,3)$ gamma
matrices,  $\ga_5 = i\ga_0 \ga_1 \ga_2 \ga_3$, while $\Ga_m$ are
$8\times 8$
gamma matrices realizing the  Clifford algebra
of $SO(7)$.  Thus for example $\Ga_{\bu}=\ga_3 \otimes \bfone_8$.
 \par
Correspondingly, we split the D=11  spinor $\epsilon$ as follows
\begin{equation}
\epsilon = \varepsilon  \, \otimes \, \eta (r,y)
\end{equation}
where $ \varepsilon $ is an $SO(1,3)$ constant spinor, while the
$SO(7)$ spinor $\eta $, besides the dependence on the internal $G/H$
coordinates
$y^m$, is assumed to depend also on the radial coordinate $r$.
Note the difference with respect to Kaluza Klein supersymmetric
compactifications where $\eta$ depends only  on $y^m$.
Computing $\Dt$ in the ansatz background yields:
\begin{eqnarray}
& & \Dt_{\mu}=\part_{\mu} + {1\over 2} e^{-B-2A}\ga_{\mu} \ga_3
[e^{3A} A' -
 {i\over 3} e^C C' \ga_3 \ga_5] \otimes \bfone_8 \nonumber \\
& &\Dt_{\bu} = \part_r + {i\over 6} e^{-3A} C' e^C \ga_3 \ga_5
\otimes \bfone_8 \nonumber \\
& &\Dt_m=\Dcal^{G/H}_m+{r\over 2 \la} [(B' + r^{-1}) i\ga_3\ga_5+
{1\over 6} e^{C-3A} C']\otimes \Ga_m
\label{splinter}
\end{eqnarray}
where all $\ga_{\mu}$, $\Ga_m$ have tangent space indices.
The Killing spinor equation  $\Dt_{\mu} \epsilon=0$  becomes
equivalent to:
\begin{equation}
(1_4-i \ga_3\ga_5)\epsi = 0 ; ~~~3e^{3A} A'=e^CC'
\label{condominio}
\end{equation}
Thus half of the components of the 4-dim spinor $\epsi$ are
projected
out.
Moreover the second equation is solved by $C=3A$.
Considering next $\Dt_{\bu} \epsilon=0$ leads to the equation (where
we have used  $C=3A$):
\begin{equation}
\part_r \eta+{1\over 6} C' \eta=0
\end{equation}
whose solution is
\begin{equation}
\eta(r,y)=e^{-C(r)/6} \eta_{\ci}(y)
\end{equation}
Finally, $\Dt_m \epsilon=0$ implies
\begin{eqnarray}
& & B=-{1\over 6} C+const.\\
& & [\Dcal^{G/H}_m + {1\over 2\la}  \Ga_m] \eta_{\ci}=0
\label{meraviglia}
\end{eqnarray}
Eq.\eqn{meraviglia} deserves attentive consideration. If we identify
the Freund--Rubin parameter as:
\begin{equation}
 e \equiv {1\over 2\la}
\end{equation}
then eq.\eqn{meraviglia} is nothing else but the Killing spinor
equation for a $G/H$ spinor that one encounters while discussing
the residual supersymmetry of Freund--Rubin vacua.
Freund--Rubin vacua \cite{freundrub} are exact solutions od D=11
supergravity where the $11$--dimensional space is:
\begin{equation}
{\cal M}_{11} \, = \, AdS_{4} \, \times \,
\left(\frac{G}{H}\right)_7
\label{4p7}
\end{equation}
having denoted by $AdS_{D}=SO(2,D-1)/SO(1,D-1)$ anti de Sitter space
in dimension $D$ and by $\left(\frac{G}{H}\right)_7$ a
$7$--dimensional
coset manifold equipped with a $G$--invariant Einstein metric.
Such a metric solves the field
equations under the condition that the $4$--form field strength take
a constant $SO(1,3)$-invariant  vev:
\begin{equation}
F_{\mu _1\mu _2\mu _3 \mu _4} = e \, \epsilon_{\mu _1\mu _2\mu _3
\mu
_4}
\label{f4p7}
\end{equation}
on $AdS_4$.
The Freund Rubin vacua
have been exhaustively studied in  the old literature on
Kaluza--Klein supergravity (see \cite{castdauriafre} for a
comprehensive review) and are all known. Furthermore it is also known
how many killing spinors each of them admits: such a number is
denoted $N_{G/H}$.
\par
In this way we have explicitly verified that the number of
$BPS$ Killing spinors
admitted by the $G/H$ M--brane solution is $N_{G/H}$, i.e.
the number of Killing spinors admitted by the corresponding
Freund--Rubin vacuum.
\subsection{M-brane solution}
To be precise the Killing spinors of the previous section are
admitted by a configuration
that has still to be shown to be a complete solution of the field
equations. To prove this is immediate.
Setting $D=11,d=3,\dt=6$, the scalar coupling parameter $a=0$, and
using the
relations
$C=3A,B=-C/6+const.=-A/2+const.$ we have just deduced,  the field
equations
\eqn{einstein1}, \eqn{einstein2}, \eqn{einstein3}  become:
\begin{eqnarray}
& &A''+ 7 r^{-1} A'={1\over 3} S^2 \\
& & (A')^2={1\over 6} S^2\\
& &\Rb_{mn} = {3\over  \la^2} \gb_{mn} \label{Einstein}
\end{eqnarray}
Combining the first two equations to eliminate $S^2$ yields:
\begin{equation}
\nabla^2 A - 3 (A')^2 \equiv A''+{7 \over r} A' - 3 (A')^2=0
\label{afieldeq}
\end{equation}
or:
\begin{equation}
\nabla^2 e^{-3A} = 0 \label{nablaA}
\end{equation}
whose solution is:
\begin{equation}
e^{-3A(r)}= H(r)=1+{k\over r^6}
\end{equation}
We have chosen the integration constant such that $A(\infty)=0$.
The functions $B(r)$ and $C(r)$ are then given by  $B=-A/2$ (so that
$B(\infty)=0$) and $C=3A$.
Finally, after use of $C=3A$, the F-field equation
\eqn{maxwell} becomes equivalent to  \eqn{afieldeq}.
The equation \eqn{nablaA} determining the radial dependence of the
function
$A(r)$
(and consequently of $B(r)$ and $C(r)$)
 is the same here as in the case
of ordinary branes, while to solve eq.\eqn{Einstein} it suffices to
choose for the manifold $G/H$ the $G$--invariant Einstein metric.
Each of the Freund--Rubin cosets admits such an Einstein metric
which was also constructed in the old Kaluza--Klein supergravity
literature
(see \cite{KKwarncastel,castdauriafre})
\par
Summarizing:  for  $D=11$ supergravity  the field equations are
solved by the ansatz \eqn{Ansatz1}, \eqn{Ansatz2}
where the $A,B,C$ functions are
\begin{eqnarray}
  A(r) & = & -\frac{\dt}{18} \, \ln
  \left(1+\frac{k}{r^{\dt}}\right) = -{1\over 3} \ln
\left(1+\frac{k}{r^{6}}\right)\nonumber \\
  B(r) & = & \frac{d}{18} \, \ln
  \left(1+\frac{k}{r^{\dt}}\right) = {1\over 6}  \ln
\left(1+\frac{k}{r^{6}}\right)\nonumber \\
  C(r) &=& 3 \, A(r)
\end{eqnarray}
 displaying the same $r$--dependence as the ordinary M--brane
solution
\eqn{protyp}.
\par
In this way we have illustrated the existence of $G/H$
M--brane solutions (see. \cite{duffetal}). Table \ref{frubinet} displays the
Freund--Rubin cosets with non vanishing $N_{G/H}$. Each of them
is associated to a BPS saturated M--brane. The notations are as in
ref.s \cite{KKwarncastel}, \cite{castdauriafre}.
\par
\vskip 0.3cm
\begin{table}[ht]
\label{frubinet}
\begin{center}
\caption{Supersymmetric Freund Rubin Cosets with Killing spinors}
\label{tavola}
\vskip .3cm
 \begin{tabular}{|c|c|c|c|}\hline
  G/H &   $G$     &   $H$   &   $N_{G/H}$    \\
 \hline
 \hline
{}&{}&{}&{}\\
 $S^7$ &  $SO(8)$ &   SO(7) &   8             \\
 \hline
 squashed $S^7$ & $SO(5) \times SO(3)$ & $ SO(3)\times SO(3)$& 1\\
 \hline
 $M^{ppr}$ & $SU(3) \times SU(2) \times U(1)$ & $SU(2) \times
U(1)^2$
& 2\\
\hline
$N^{010}$ & $SU(3) \times SU(2)$ & $ SU(2)\times U(1)$& 3\\
\hline
$N^{pqr}$ & $SU(3) \times U(1)$ & $U(1)^2$& 1\\
\hline
$Q^{ppp}$ & $SU(2)^3 $ & $U(1)^3$& 2\\
\hline
$B_{irred}^7$ & $SO(5)$ & $ SO(3)_{max}$& 1\\
\hline
$V_{5,2}$ & $SO(5)) \times U(1)$ & $ SO(3)\times U(1)$& 2\\
\hline
 \end{tabular}
 \end{center}
\end{table}
\chapter{THE SYMPLECTIC STRUCTURE OF 4D SUPERGRAVITY
AND N=2 BPS BLACK HOLES }
\label{n2bholes}
\section{Introduction to tetradimensional BPS black--holes
and the general form of the supergravity action}
\setcounter{equation}{0}
\markboth{SUPERGRAVITY and BPS BLACK HOLES: Chapter 3}
{3.1 INTRODUCTION TO $D=4$ BPS BLACK HOLES}
\label{gen4d}
In this chapter we begin the study of BPS black--hole solutions in
four space-time dimensions. We  mainly   focus on the case
of $N=2$ supergravity, leaving the higher $N$--cases, in particular
that of $N=8$ supergravity to later chapters.  Apart from its intrinsic
interest we want to use the $N=2$ example to  illustrate the profound
relation between the black-hole structure, in particular its
entropy, and the scalar geometry of supergravity,
together with its symplectic embedding that realizes covariance with
respect to electric-magnetic duality rotations. We approach this
intricate and fascinating relation starting from a general
description of the nature of BPS black--hole solutions.
These are field configurations satisfying
the equations of motions derived from the bosonic part of the
supergravity action that are characterized by the following
defining properties:
\begin{enumerate}
\item {The metric has the form of a $0$--brane solution admitting
$\IR \, \times \, SO(3)$  as isometry group:
\begin{equation}
ds^2 = e^{2U(r)} \, dt^2 - e^{-2U(r)} \, d{\vec x}^2
\quad ; \quad \left( r^2 = {\vec x}^2 \right)
\label{ds2U}
\end{equation}  }
\item The ${\bar n}$ gauge  fields of the theory $A^ \Lambda  $ carry both
electric and magnetic charges and are in a $0$--brane configuration,
namely their field strengths $F ^\Lambda$ are of the form:
\begin{equation}
\label{flambda}
F^{\Lambda}\,=\,\frac{p^{\Lambda}}{2r^3}\epsilon_{abc}
x^a dx^b\wedge dx^c-\frac{\ell^{\Lambda}(r)}{r^3}e^{2\cal {U}}dt
\wedge \vec{x}\cdot d\vec{x}
\end{equation}
\item The $m$ scalar fields of the theory $\phi^I$
have only radial dependence:
\begin{equation}
\phi^I = \phi^I (r) \quad ; \quad (I= 1, \dots \, m)
\label{phijey}
\end{equation}
\item The field configuration preserves a certain fraction
of the original $N$--extended supersymmetry:
\begin{equation}
      { \textstyle  f}_{SUSY} = \frac{N }{2\,n_{max}} \quad ; \quad
 \left({ \textstyle  f}_{SUSY}\right)= \quad \frac{1}{2}
 \quad \mbox{or} \quad \frac{1}{4}
 \quad \mbox{or} \quad \frac{1}{8}
\label{susfrac}
\end{equation}
in the sense that there exists a BPS Killing
spinor   $\xi_A(x)$ subject to the   condition:
\begin{eqnarray}
\gamma^0 \,\xi_{A} & =&  \mbox{i}\, \IC_{AB}
\,  \xi^{B} \quad ; \quad A,B=1,\dots ,n_{max} \quad ; \quad
2 \le  \, n_{max} \, \le N \nonumber \\
\xi_{A} & =& 0 \quad ; \quad A=n_{max}+1,\dots ,N  \nonumber\\
\IC_{AB} & = & - \IC_{AB} \quad = \quad \mbox{antisymmetric matrix of rank
$n_{max}$}
\label{kispiro}
\end{eqnarray}
  such that in the considered bosonic background the
supersymmetry transformation rule of all the fermions of the theory
vanishes along the Killing spinor \eqn{kispiro}:
\begin{equation}
\delta^{SUSY}_\epsilon \, \mbox{  fermions} \, =  0 \quad
\mbox{if SUSY parameter}\, \,  \epsilon_A = \xi _A \, =\mbox{killing spinor}
\label{kvano}
\end{equation}
\end{enumerate}
Let us illustrate the above definition of BPS black--holes by
comparison with the results of chapter \ref{intro} on higher
dimensional $p$--branes. To make such a comparison we have to assume
that only one (say $A_{[1]} \equiv A^0$) of the ${\bar n}$
gauge fields $A^\Lambda$ and only one (say $\phi = \phi^1$) of the
$m$ scalar fields $\phi^I$ are switched on in our solution.
Furthermore we have to assume that upon truncation to such fields
the supergravity lagrangian takes the form \eqn{paction}. Under
these conditions a comparison is possible and it reveals the meaning
of the definition of BPS black--holes we have adopted. It will
become clear that the much richer structure of 4D black--holes is due
to the presence of many scalar and many vector fields and to the geometric
structure of their mutual interactions completely fixed by
supersymmetry.
\par
So we begin by comparing the ansatz \eqn{ds2U} for the 4D--metric with
the general form of the metric in either an elementary or solitonic
$p$--brane solution as given in eq.s\eqn{elem} or \eqn{solit}.
In this comparison we have also to recall eq.s \eqn{interpre},\eqn{wvol}
and \eqn{deltadef}. Since the black--hole is a $0$--brane in $D=4$,
we have:
\begin{equation}
\begin{array}{lcrclcr}
   d & = & 1 & ; & {\tilde d} & = & 1 \cr
 p & = & 0 & ; & \Delta &= & a^2 +1 \cr
\end{array}
\label{4dchoice}
\end{equation}
and for a solution coupled to a single vector field we would expect a
metric of the form:
\begin{equation}
ds^2 = \left(1+\frac{k}{r}\right)^{-\frac{2}{a^2+1}} \, dt^2
- \left(1+\frac{k}{r}\right)^{ \frac{2}{a^2+1}} \, d{\vec x}^2
\label{metexpec}
\end{equation}
that is precisely of the form \eqn{ds2U} with a function $U$
given by:
\begin{equation}
 U(r) = - \frac{1}{a^2+1} \, \log \, H(r)
\label{Uharm}
\end{equation}
where:
\begin{eqnarray}
 H(r)& \equiv & \left( 1+ \frac{k}{r} \right)\nonumber \\
 \Delta_3  \, H(r) & = & \sum_{i=1}^{3} \, \frac{\partial^2}
 {\partial x_{i}^2} \, H(r) \, = \, 0
 \label{harmf}
 \end{eqnarray}
is a harmonic function depending on the charge $k$ carried by
the single vector field we have considered in this comparison.
\par
Similarly for the single scalar field we would expect:
\begin{equation}
\phi(r) = \cases{ - \, \frac{2 \, a}{a^2 +1}\, \log \, H(r) \quad
; \quad \mbox{electric $0$-brane} \cr
 \quad \, \frac{2 \, a}{a^2 +1}\, \log \, H(r) \quad
; \quad \mbox{magnetic solitonic $0$-brane} ;}
\label{phiexp}
\end{equation}
and for the field strength of the single vector field we would
expect:
\begin{equation}
\label{fexpec}
\begin{array}{rclcc}
 F & = & -2 \, \frac{k}{\sqrt{a^2+1 }}\, \frac{1}{r^3}
 \left(1+\frac{k}{r}\right)^{-2} dt \, \wedge \, {\vec x} \, \cdot \, {\vec
 x}  & ; & \mbox{electric $0$-brane} \\
 {  F} &=& - 2 \, \frac{k}{\sqrt {a^2+1}} \, \frac{1}{r^3} \,
 \epsilon_{abc} \, x^a \, dx^b \, \wedge \, dx^c & ; &
 \mbox{magnetic solitonic $0$-brane}\\
\end{array}
\end{equation}
Comparison of eq.s \eqn{fexpec} with eq.s\eqn{flambda} shows that
the ansatz we adopted is indeed consistent with the $0$--brane interpretation
of the black--hole upon the identifications:
\begin{equation}
\begin{array}{rclcc}
 \ell(r) & = &r^3 \,  \frac{2 \, \sqrt{a^2+1}}{a^2-1}\, \frac{d}{dr} \,
 \left[H(r)\right]^{\frac{1-a^2}{1+a^2}}& ; & \mbox{electric $0$-brane} \\
 p &=&    - \frac{ 4}{\sqrt{a^2+1}} \, k & ; &
 \mbox{magnetic solitonic $0$-brane}\\
\end{array}
\end{equation}
At this point we must observe that for this one--vector,
one--scalar coupling there are critical values of the $a$--parameter,
namely:
\begin{equation}
a= \cases{ {\begin{array}{lcccccccl}
 \sqrt{3} & \Longrightarrow & \Delta = 4 & ; & U =-\frac{1}{4}\,
 \log H(r) &;& \phi = \mp \frac{\sqrt{3}}{2} \, \log H &;&
 \cases{ \ell = 2 \, r^3 \, \frac{d}{dr}H^{-\frac{1}{2}} \cr
 p= -k \cr }\\
 1 & \Longrightarrow & \Delta = 2 & ; &
 U =-\frac{1}{2}\,
 \log H(r) &;& \phi = \mp   \log H &;&
 \cases{ \ell = -2 \, r^3 \, \frac{d}{dr}H^{-1}
 \cr p= -2\sqrt{2}\, k \cr} \\
 0 & \Longrightarrow & \Delta = 1 & ; &
 U =- \,
 \log H(r) &;& \phi = 0   &;&
 \cases{
 \ell = - 2 \, r^3 \, \frac{d}{dr}H  \cr
 p = -4 \, k \cr} \\
\end{array}} \cr}
\label{critval1}
\end{equation}
The special property of these critical values of $a$ is
that   for all of them we can write:
\begin{equation}
\Delta =  4 \, \times \, 2 \, \times \, {\textstyle  f}_{SUSY} \quad
; \quad \cases{{\textstyle  f}_{SUSY}=\frac{1}{2} \, \leftrightarrow
\, a= \sqrt{3} \cr
{\textstyle  f}_{SUSY}=\frac{1}{4} \, \leftrightarrow
\, a= 1 \cr
{\textstyle  f}_{SUSY}=\frac{1}{8} \leftrightarrow
\, a= 0 \cr }
\label{critval2}
\end{equation}
This numerical coincidence hints at a relation with the number of
preserved supersymmetries. Indeed  such a relation will be verified
in the context of $N=8$ BPS black--hole solutions where the three
cases of supersymmetry fractions make sense. The other important
observation is that it is only in the case $a=0$ that our black--hole
can be {\it dyonic}, namely that we can
simultaneously assign both an electric and a magnetic part to our field
strength. This is evident from eq.\eqn{critval1}. In the other cases
where $\phi(r)$ is non--vanishing it cannot be simultaneously equal to
$\mbox{const}\times \log H(r)$ and to minus the same expression.
For the value $a=0$ the 1-vector BPS black--hole simply
coincides with the extremum Reissner-Nordstrom black-hole
\begin{equation}
ds_{RN}^2 = \left(1+\frac{k}{r}\right)^{-2} \, dt^2 -
\left(1+\frac{k}{r}\right)^2 \, d{\vec x}^2
\label{rnord}
\end{equation}
and for small values of $r$ it is approximated by the horizon
geometry:
\begin{equation}
AdS_2 \, \times \, S^2
\label{ads2s2}
\end{equation}
that corresponds to the Bertotti--Robinson metric \cite{berrobin}:
\begin{equation}
ds^2_{BR}=\frac{1}{m_{BR}^2} r^2 dt^2 - m_{BR}^2 \,
\frac{dr^2}{r^2} -  m_{BR}^2 \, \left( \sin^2(\theta) \, d\phi^2 +
d\theta^2 \right)
\label{bertrob}
\end{equation}
the parameter:
\begin{equation}
 m_{BR} = |k|
\end{equation}
being named the Bertotti--Robinson mass.
\par
Our comparison between the higher dimensional $p$--brane solutions
and the ansatz defining BPS black--holes in $D=4$ is now complete. By
means of this comparison we have understood the form \eqn{ds2U}
assigned to metric. It must reduce to eq.s \eqn{metexpec},
\eqn{Uharm} when there is only one vector and only one scalar
switched on. However, in the general case we are not allowed to
identify directly  the function $U(r)$ with the logarithm of
the harmonic function $H(r)$ as in eq.\eqn{Uharm}. This is so because
there are many vector fields in the theory and to
each of them we can associate a different charge and hence a
different harmonic function. Similarly the presence of many scalar
fields does not allow to know, a priori, whether a given field
strength $F^\Lambda$ is electric, magnetic or dyonic. Correspondingly
the general ansatz for $F^\Lambda$ is that of eq.\eqn{flambda}, the
electric and magnetic parts having the form expected for $0$--branes.
\par
Finally we have to discuss why the  condition defining the
BPS killing spinor is given by eq.\eqn{kispiro}. This is nothing
else but the transcription, on the supersymmetry parameters of the
projection operator:
\begin{equation}
\IP^{\pm}_{AB}=\frac{1}{2} \left( \bfone \, \delta_{A}^B \pm \mbox{\rm i}\,
\IC_{AB} \, \gamma_0 \right)
\label{projet}
\end{equation}
that acts on the supersymmetry generators in the abstract
description \eqn{projop}.
\par
Having illustrated the general form of the ansatz for BPS black--holes
in $D=4$ space--time dimensions, in order to proceed one needs to
insert such an ansatz into
\begin{itemize}
\item The second order bosonic field  equations of supergravity
\item The Killing spinor equations \eqn{kvano} that follow from the
supersymmetry transformations rules of the fermions and
are linear in the derivatives of the bosonic fields
\end{itemize}
In this way one arrives at a coupled system of first and second order
differential equations that depends crucially on two ingredients:
 \begin{itemize}
 \item {\bf a)} The self interaction of the scalar fields, that is on the
 metric $g_{IJ}(\phi)$ of the scalar manifold of which the $\phi^I$
 are interpreted as coordinates
 \item {\bf b)} The non--minimal coupling between the scalars $\phi^I$
 and the vector fields $A^ \Lambda $ of which the exponential coupling
 $ \exp [a \, \phi ] \, F_{[p+2]}^2$ in the action \eqn{paction} is just the
 simplest example.
 \end{itemize}
Indeed for all tetradimensional supergravities theories
the bosonic action takes the following general form:
\begin{eqnarray}
\label{genact}
{\cal L}&=&\int\sqrt{-g}\, d^4x\left(2R+\Im{\cal
N}_{\Lambda \Gamma}F_{\mu\nu}^{~~\Lambda }F^{\Gamma |\mu\nu}+
{1 \over 6}g_{IJ}(\phi) \partial_{\mu}
\phi^{I}\partial^{\mu}\phi^{J}+\right.\nonumber\\
&+&\left.{1 \over 2}\Re{\cal N}_{\Lambda \Gamma  }
{\epsilon^{\mu\nu\rho\sigma}\over\sqrt{-g}}
F_{\mu\nu}^{~~\Lambda }F^{\Gamma }_{~~\rho\sigma}\right)
\end{eqnarray}
where, as already stated, $g_{IJ}(\phi)$ is the scalar metric on the
$m$--dimensional scalar manifold ${\cal M}_{scalar}$ and
\begin{equation}
{\cal N}_{\Lambda\Sigma}(\phi)
\label{permat}
\end{equation}
is a complex, symmetric,  ${\bar n} \, \times \, {\bar n}$ matrix
depending on the scalar fields which we name
the {\it period matrix}. What varies, depending on the number $N$ of
supersymmetries is
\begin{enumerate}
\item The relative and total number  of vectors and scalars, that is
${\bar n}$ and $m$
\item The geometry $g_{IJ}(\phi)$ and the isometry group $G$
of the scalar manifold ${\cal M}_{scalar}$.
\end{enumerate}
Yet the relation between this scalar geometry and the period matrix
${\cal N}$ has a very general and universal form. Indeed it is related
to the solution of a general problem, namely {\it how to lift the
action of the scalar manifold isometries from the scalar  to the
vector fields}. Such a lift is necessary because of supersymmetry
since scalars and vectors generically belong to the same
supermultiplet and must rotate coherently under symmetry operations.
Such a problem is solved by considering a deep property
inherent to lagrangians of type \eqn{genact}: their possible
covariance with respect to generalized {\it electro--magnetic
duality rotations}. In the next section we address this general
property and we show how enforcing covariance with respect to such duality
rotations leads to a determination of the period matrix ${\bar N}$.
It goes without saying that the structure of ${\bar N}$ enters  the
black--hole equations in a crucial way so that, not too surprisingly,
at the end of the day, the  topological invariant  associated with
the hole, that is its {\it entropy}, is an invariant of the group of
electro-magnetic duality rotations, the U--duality group.
In table \ref{topotable} we summarize the geometries of the scalar
manifolds for the various values of $N$. In such a table the
symplectic embedding of these geometries is already mentioned. What
this means will become clear by working through the next section.
{\footnotesize
\begin{table*}
\begin{center}
\caption{\sl Scalar Manifolds of Extended Supergravities}
\label{topotable}
\begin{tabular}{|c||c|c|c||c|c||c||c| }
\hline
\hline
~ & $\#$ scal. & $\#$ scal. & $\#$ scal. & $\#$ vect. &
 $\#$ vect. &~ & $~ $ \\
N & in & in & in & in  &
 in  &$\Gamma_{cont}$ & ${\cal M}_{scalar}$   \\
 ~ & scal.m. & vec. m. & grav. m. & vec. m. & grav. m. & ~ &~
\\
\hline
\hline
~    &~    &~   &~   &~  &~  & ~ & ~ \\
$1$  & 2 m &~   & ~  & n &~  &  ${\cal I}$  & ~   \\
~    &~    &~   &~   &~  &~  &  $\subset Sp(2n,\IR)$ & K\"ahler \\
~    &~    &~   &~   &~  &~  & ~ & ~ \\
\hline
~    &~    &~   &~   &~  &~  & ~ & ~ \\
$2$  & 4 m & 2 n& ~  & n & 1 &  ${\cal I}$ & Quaternionic $\otimes$
\\
~    &~    &~   &~   &~  &~  &  $\subset Sp(2n+2,\IR)$ & Special K\"ahler \\
~    &~    &~   &~   &~  &~  & ~ & ~ \\
\hline
~    &~    &~   &~   &~  &~  & ~ & ~ \\
$3$  & ~   & 6 n& ~  & n & 3 &  $SU(3,n)$ &~  \\
~    &~    &~   &~   &~  &~  & $\subset Sp(2n+6,\IR)$ & $\frac{SU(3,n)}
{S(U(3)\times U(n))}$ \\
~    &~    & ~  &~   &~  &~  & ~ & ~ \\
\hline
~    &~    &~   &~   &~  &~  & ~ & ~ \\
$4$  & ~   & 6 n& 2  & n & 6 &  $SU(1,1)\otimes SO(6,n)$ &
$\frac{SU(1,1)}{U(1)} \otimes $ \\
~    &~    &~   &~   &~  &~  & $\subset Sp(2n+12,\IR)$ &
$\frac{SO(6,n)}{SO(6)\times SO(n)}$ \\
~    &~    &~   &~   &~  &~  & ~ & ~ \\
\hline
~    &~    &~   &~   &~  &~  & ~ & ~ \\
$5$  & ~   & ~  & 10 & ~ & 10 & $SU(1,5)$ & ~  \\
~    &~    &~   &~   &~  &~  & $\subset Sp(20,\IR)$ & $\frac{SU(1,5)}
{S(U(1)\times U(5))}$ \\
~    &~    &~   &~   &~  &~  & ~ & ~ \\
\hline
~    &~    &~   &~   &~  &~  & ~ & ~ \\
$6$  & ~   & ~  & 30 & ~ & 16 & $SO^\star(12)$ & ~ \\
~    &~    &~   &~   &~  &~  & $\subset Sp(32,\IR)$ &
$\frac{SO^\star(12)}{U(1)\times SU(6)}$ \\
~    &~    &~   &~   &~  &~  & ~ & ~ \\
\hline
~    &~    &~   &~   &~  &~  & ~ & ~ \\
$7,8$& ~   & ~  & 70 & ~ & 56 & $E_{7(-7)}$  & ~ \\
~    &~    &~   &~   &~  &~  & $\subset Sp(128,\IR)$ &
$\frac{ E_{7(-7)} }{SU(8)}$ \\
~    &~    &~   &~   &~  &~  & ~ & ~ \\
\hline
 \hline
\end{tabular}
\end{center}
\end{table*}
}
\section{Duality Rotations and Symplectic Covariance}
\label{LL1}
\markboth{SUPERGRAVITY and BPS BLACK HOLES: Chapter 3}
{3.2 DUALITY ROTATIONS AND SYMPLECTIC COVARIANCE}
\setcounter{equation}{0}
In this section, relying on the motivations given above ,
we review the general structure of an
abelian theory of vectors and scalars displaying covariance
under a group of duality rotations.
The basic reference is the 1981 paper by Gaillard and Zumino
\cite{gaizum}. A general presentation in $D=2p$ dimensions was
recently given in \cite{pietrolectures}. Here we fix
 $D=4$.
\par
We consider a theory of  $\bar n$ gauge fields $A^\Lambda_{\mu}$,
in a $D=4$  space--time with Lorentz signature.
They  correspond to a set of  $\bar n$
differential $1$--forms
\begin{equation}
A^\Lambda ~ \equiv ~
A^\Lambda_{\mu} \, dx^{\mu} \quad \quad
 \left ( \Lambda = 1,
\dots , {\bar n} \right )
\end{equation}
The corresponding field strengths and their Hodge duals are defined
by
\begin{eqnarray}
{ F}^\Lambda & \equiv & d \, A^\Lambda \,  \equiv   \,
   {\cal F}^\Lambda_{\mu  \nu} \,
dx^{\mu } \, \wedge \, dx^{\nu} \nonumber\\
{\cal F}^\Lambda_{\mu \nu} & \equiv & {{1}\over{2 }} \,\left ( \partial_{\mu }
A^\Lambda_{\nu} \, - \, \partial_{\nu }
A^\Lambda_{\mu} \right )
\nonumber\\ ^{\star}{ F}^{\Lambda} & \equiv &\, {\tilde
{\cal F}}^\Lambda_{\mu \nu} \, dx^{\mu } \, \wedge \,
 dx^{\nu} \nonumber\\
{\tilde {\cal
F}}^\Lambda_{\mu \nu} & \equiv &{{1}\over{2}}
\varepsilon_{\mu  \nu \rho\sigma}\, {\cal F}^{\Lambda
\vert \rho \sigma}
\label{campfort}
\end{eqnarray}
Defining the space--time integration volume as
\begin{equation}
\mbox{d}^4 x \, \equiv \, -{{1}\over{4!}} \, \varepsilon_{\mu_1\dots
\mu_4} \, dx^{\mu_1} \, \wedge \, \dots \, \wedge dx^{\mu_{4}}\ ,
\label{volume}
\end{equation}
we obtain
\begin{equation}
F^\Lambda \, \wedge \, F^\Sigma \,  =
  \varepsilon^{\mu  \nu \rho \sigma }\,
{\cal F}^\Lambda_{\mu\nu} \, {\cal
F}^\Sigma_{\rho\sigma}\,d^4 x \qquad ; \qquad
F^\Lambda \, \wedge
\, ^{\star}F^{\Sigma}   =   - 2 \,
{\cal F}^\Lambda_{\mu\nu} \, {\cal F}^{\Sigma \vert
\mu\nu} d^4 x\ .
\label{cinetici}
\end{equation}
In addition
to the gauge fields let us also introduce a set of real scalar
fields $\phi^I$ ( $I=1,\dots , m$) spanning an ${\bar
m}$--dimensional manifold ${\cal M}_{scalar}$ \footnotemark
\footnotetext{Whether the $\phi^I$ can be arranged into complex
fields is not relevant at this level of the discussion. } endowed
with a metric $g_{IJ}(\phi)$. Utilizing the above field
content we can write the following action functional:
\begin{equation}
{\cal S}\, = {1\over 2}\,  \int \, \left \{ \left [ \,
  \gamma_{\Lambda\Sigma}(\phi) \, F^\Lambda \, \wedge
\,\star F^{\Sigma} \, +  \,
\theta_{\Lambda\Sigma}(\phi) \, F^\Lambda \, \wedge \, F^{\Sigma}  \,
\right  ]
     \,  + \,
g_{IJ}(\phi) \, \partial_\mu \phi^I \, \partial^\mu \phi^J   \,
\mbox{d}^4 x \,  \right \}\ ,
\label{gaiazuma}
\end{equation}
where the scalar fields
dependent ${\bar n} \times {\bar n}$ matrix
$\gamma_{\Lambda\Sigma}(\phi)$  generalizes the inverse of the
squared coupling constant $\o{1}{g^2}$ appearing in ordinary
 gauge theories. The field dependent matrix
$\theta_{\Lambda\Sigma}(\phi)$ is instead a generalization of the
$theta$--angle of quantum chromodynamics. Both $\gamma$ and $\theta$
are symmetric matrices.
Introducing a formal operator $j$ that maps a field
strength into its Hodge dual
\begin{equation} \left ( j \, {\cal
F}^\Lambda \right )_{\mu \nu} \, \equiv \,
{{1}\over{ 2 }} \, \epsilon_{\mu\nu \rho \sigma} \,
{\cal F}^{\Lambda \vert\rho\sigma}
\label{opjei}
\end{equation}
and a formal scalar product
\begin{equation} \left (
G \, , \, K \right ) \equiv G^T K \, \equiv \,
\sum_{\Lambda=1}^{\bar n} G^{\Lambda}_{\mu\nu} K^{\Lambda
\vert \mu\nu }
\label{formprod}
\end{equation} the total
Lagrangian of eq.~\eqn{gaiazuma} can be rewritten as
\begin{equation}
\cL^{(tot)}\,  = \,  {\cal F}^T \, \left ( -\gamma
\otimes \bfone + \theta \otimes j \right ) {\cal F} \, + \,
\o{1}{2} \, g_{IJ}(\phi) \, \partial_\mu \phi^I \,
\partial^\mu \phi^J
\label{gaiazumadue}
\end{equation}
The operator $j$ satisfies $j^2 \, = \, - \, \bfone$ so that its
eigenvalues are $\pm {\rm i}$.
Introducing self--dual and antiself--dual combinations
\begin{eqnarray}
  {\cal F}^{\pm} &=& {1\over 2}\left({\cal F}\, \pm {\rm i} \, j
{\cal F}\right) \nonumber \\
j \, {\cal F}^{\pm}& =& \mp \mbox{i} {\cal F}^{\pm}
\label{selfduals}
\end{eqnarray}
and the field--dependent symmetric matrices
\begin{eqnarray}
  {\cal N} & = & \theta \, - \,
\mbox{i} \gamma \nonumber\\
{\bar {\cal N}} & = & \theta + \mbox{i} \gamma\ ,
\label{scripten}
\end{eqnarray}
the
vector part of the Lagrangian ~\eqn{gaiazumadue} can be rewritten as
\begin{equation}
{\cal L}_{vec} \,  = \, {\mbox{i}} \, \left [{\cal F}^{-T} {\bar {\cal N}}
{\cal F}^{-}-
{\cal F}^{+T} {\cal N} {\cal F}^{+} \right]
\label{lagrapm}
\end{equation}
Introducing the new tensors
\begin{equation}
  {\tilde{\cal
G}}^\Lambda_{\mu\nu} \, \equiv \,  {1 \over 2 } { {\partial {\cal
L}}\over{\partial {\cal F}^\Lambda_{\mu\nu}}}\leftrightarrow  {\cal
G}^{\mp \Lambda}_{\mu\nu} \, \equiv \,  \mp{{\rm i} \over 2 } { {\partial {\cal
L}}\over{\partial {\cal F}^{\mp \Lambda}_{\mu\nu}}}
\label{gtensor}
\end{equation}
which, in matrix notation, corresponds to
\begin{equation}
j \, {\cal G} \, \equiv \,{1 \over 2 }  \,
{{\partial {\cal L}}\over{\partial {\cal F}^T}} \,  = \, -
\, \left ( \gamma\otimes\bfone -\theta\otimes j \right )
\, {\cal F}
\label{ggmatnot}
\end{equation}
the Bianchi identities and field
equations associated with the Lagrangian ~\eqn{gaiazuma} can be
written as
\ba
\partial^{\mu }{\tilde {\cal F}}^{\Lambda}_{\mu\nu} &=& 0 \\
\partial^{\mu }{\tilde {\cal G}}^{\Lambda}_{\mu\nu} &=& 0
\label{biafieq}
\ea
or equivalently
\ba
 \partial^{\mu }
{\rm Im}{\cal F}^{\pm \Lambda}_{\mu\nu} &=& 0 \\
\partial^{\mu } {\rm Im}{\cal G}^{\pm \Lambda}_{\mu\nu} &=&
0 \ .
\label{biafieqpm}
\ea
This suggests that we introduce the $2{\bar
n}$ column vector
\begin{equation}
{\bf V} \, \equiv \, \left ( \matrix
{ j \, {\cal F}\cr
j \, {\cal G}\cr}\right )
\label{sympvec}
\end{equation}
and that we consider general linear transformations on such a vector
\begin{equation}
\left ( \matrix
{ j \, {\cal F}\cr
j \, {\cal G}\cr}\right )^\prime \, =\,
\left (\matrix{ A & B \cr C & D \cr} \right )
\left ( \matrix
{ j \, {\cal F}\cr
j \, {\cal G}\cr}\right )
\label{dualrot}
\end{equation}
For any matrix $\left (\matrix{ A & B \cr C & D \cr} \right ) \, \in
\, GL(2{\bar n},\IR )$ the new vector ${\bf V}^\prime$ of {\it
magnetic and electric} field--strengths satisfies the same
equations ~\eqn{biafieq} as the old one. In a condensed notation
we can write
\begin{equation}
\partial \, {\bf V}\, = \, 0 \quad \Longleftrightarrow \quad
\partial \, {\bf V}^\prime \, = \, 0
\label{dualdue}
\end{equation}
Separating the self--dual and anti--self--dual parts
\begin{equation}
{\cal F}=\left ({\cal F}^+ +{\cal F}^- \right ) \qquad ;
\qquad
{\cal G}=\left ({\cal G}^+ +{\cal G}^- \right )
\label{divorzio}
\end{equation}
and taking into account that   we have
\begin{equation}
{\cal G}^+ \, = \, {\cal N}{\cal F}^+  \quad
{\cal G}^- \, = \, {\bar {\cal N}}{\cal F}^-
\label{gigiuno}
\end{equation}
the duality
rotation of eq.~\eqn{dualrot} can be rewritten as
\begin{equation}
\left ( \matrix
{   {\cal F}^+ \cr
 {\cal G}^+\cr}\right )^\prime  \, = \,
\left (\matrix{ A & B \cr C & D \cr} \right )
\left ( \matrix
{   {\cal F}^+\cr
{\cal N} {\cal F}^+\cr}\right ) \qquad ; \qquad
\left ( \matrix
{   {\cal F}^- \cr
 {\cal G}^-\cr}\right )^\prime \, = \,
\left (\matrix{ A & B \cr C & D \cr} \right )
\left ( \matrix
{   {\cal F}^-\cr
{\bar {\cal N}} {\cal F}^-\cr}\right )
\label{trasform}
\end{equation}
The problem is that the transformation rule
~\eqn{trasform} of ${\cal G}^\pm$ must be consistent with the definition
of the latter
as variation of the Lagrangian with respect
to ${\cal F}^\pm$ (see eq.~\eqn{gtensor}). This request
restricts the form of the matrix
$\Lambda =\left (\matrix{ A & B \cr C & D \cr} \right )$.
As we are going to show, $\Lambda$ must belong
to the symplectic subgroup  of the general linear group
\begin{equation}
\Lambda \equiv \left (\matrix{ A & B \cr C & D \cr} \right ) \,  \in \,
Sp(2\bar n,\IR)
\,\subset
\, GL(2\bar n ,\IR )
\label{distinguo}
\end{equation}
the   subgroup $Sp(2\bar n,\IR)$ being defined as the set of $2\bar n \times
2\bar n$ matrices that satisfy the condition
\begin{equation}
\Lambda \in Sp(2\bar n,\IR) ~ \longrightarrow ~
   \Lambda^T  \,
\left (\matrix{ {\bf 0}_{} &
\bfone_{} \cr
-\bfone_{} & {\bf 0}_{}
\cr }\right )
 \, \Lambda \,  = \,
   \left (\matrix{ {\bf 0}_{} &
\bfone_{} \cr
-\bfone_{} & {\bf 0}_{}
\cr }\right )
\label{ortosymp}
\end{equation}
that is, using $ n \otimes n$ block components
\begin{equation}
A^T C-C^T A=
B^T D - D^T B  =0 \quad\quad
A^T D - C^T B =1   \label {ortocomp}
\end{equation}
To prove the statement we just made, we calculate the transformed
Lagrangian ${\cal L}^\prime$ and then we compare its variation
${\o{\partial {\cal L}^\prime}{\partial {\cal F}^{\prime T}}}$
with  ${\cal G}^{\pm\prime}$ as it follows from the postulated
transformation rule ~\eqn{trasform}. To perform such a calculation
we rely on the following basic idea. While the
duality rotation~\eqn{trasform} is performed on the field strengths
and on their duals, also the scalar fields are transformed by the action
of some diffeomorphism ${\xi }\,  \in \,
{\rm Diff}\left ( {\cal M}_{scalar}\right )$ of the scalar manifold
and, as a consequence of that, also the matrix ${\cal N}$ changes.
In other words given the scalar manifold ${\cal M}_{scalar}$ we
assume that   there exists a
homomorphism of the  form
\begin{equation}
\iota _{\delta} : \,  {\rm Diff}\left ( {\cal M}_{scalar}\right )
\, \longrightarrow \, GL(2\bar n,\IR)
\label{immersione}
\end{equation}
so that
\begin{eqnarray}
\forall &  \xi   &\in \, {\rm Diff}\left ( {\cal M}_{scalar}\right ) \, :
\, \phi^I \, \stackrel{\xi}{\longrightarrow} \,  \phi^{I\prime}
\nonumber\\
\exists  & \iota _{\delta}(\xi) & = \left (\matrix{ A_\xi & B_\xi \cr
C_\xi & D_\xi \cr }\right ) \, \in \,  GL(2\bar n,\IR)
\label{apnea}
\end{eqnarray}
(In the sequel the suffix $\xi$ will be  omitted when no confusion
can arise and be reinstalled when necessary for clarity. )
\par
Using such a homomorphism
we can define the simultaneous action of $\xi$ on
all the fields of our theory by setting
\begin{equation}
\xi \, : \, \cases{   \phi \,
\longrightarrow \, \xi (\phi) \cr
{\bf V} \,
\longrightarrow \, \iota _{\delta}(\xi) \, {\bf V} \cr
{\cal N}(\phi) \, \longrightarrow \, {\cal N}^\prime (\xi (\phi)) \cr }
\end{equation}
where the notation~\eqn{sympvec} has been utilized.
In the gauge sector the transformed Lagrangian is
\begin{equation}
  {\cal L}^{\prime}_{vec}  \, =  \,
   {\rm i} \, \Bigl [{\cal F}^{-T}
\, \bigl ( A + B {\bar {\cal N}} \bigr )^T {\bar {\cal N}}^\prime
( A + B {\bar {\cal N}} \bigr ) {\cal F}^{-} \, -  \, {\cal F}^{+T}
\, \bigl ( A + B {\cal N} \bigr )^T {\cal N}^\prime
( A + B {\cal N} \bigr ) {\cal F}^{+}
\Bigr ]
\label{elleprima}
\end{equation}
Consistency with
the definition of ${\cal G}^+$ requires  that
\begin{equation}
{\cal N}^\prime \, \equiv \, {\cal N}^\prime (\xi(\phi)) \, = \,
 \left ( C  + D  {\cal N}(\phi) \right )   \left ( A  +
B  {\cal N}(\phi)\right )^{-1}
\label{Ntrasform}
\end{equation}
while consistency with the definition of ${\cal G}^-$ imposes
the transformation rule
\begin{equation}
{\bar {\cal N}}^\prime  \, \equiv \, {\bar {\cal N}}^\prime (\xi(\phi)) \,
= \,
\left ( C  + D  {\bar {\cal N}}(\phi ) \right )   \left ( A  +
B  {\bar {\cal N}}(\phi)\right )^{-1}
\label{Nbtrasform}
\end{equation}
 It is from the transformation rules~\eqn{Ntrasform} and ~\eqn{Nbtrasform}
that we derive a restriction on the form of the
duality rotation matrix $\Lambda \equiv \iota_\delta(\xi)$.
Indeed by requiring that the transformed matrix ${\cal N}^\prime$ be again
symmetric one easily finds that ${\Lambda}$ must obey eq. (\ref {ortosymp}),
namely $\Lambda \in Sp(2\bar n ,\IR)$.
Consequently the
homomorphism of eq.~\eqn{immersione} specializes
as
 \begin{equation}
\iota _{\delta} : \,  {\rm Diff}\left ( {\cal M}_{scalar}\right )
\, \longrightarrow \, Sp(2\bar n,\IR)
\label{spaccoindue}
\end{equation}
Clearly, since   $Sp(2\bar n,\IR)$   is
a finite dimensional Lie group, while
${\rm Diff}\left ( {\cal M}_{scalar}\right )$ is
infinite--dimensional, the homomorphism  $\iota _{\delta}$ can
never be an isomorphism. Defining the Torelli group of the
scalar manifold as
\begin{equation}
{\rm Diff}\left ( {\cal M}_{scalar}\right ) \, \supset \,
\mbox{Tor} \left ({\cal M}_{scalar} \right ) \, \equiv \,
\mbox{ker} \, \iota_\delta
\label{torellus}
\end{equation}
we always have
\begin{equation}
\mbox{dim} \, \mbox{Tor} \left ({\cal M}_{scalar} \right ) \, = \,
\infty
\label{infitor}
\end{equation}
The reason why we have given the name of Torelli to the group defined
by eq.~\eqn{torellus} is because of its similarity with the
Torelli group that occurs in algebraic geometry.
\par
What should
be clear from the above discussion is that a family of Lagrangians
as in eq.~\eqn{gaiazuma} will admit a group of
duality--rotations/field--redefinitions that will map elements
of the family into each other, as long as a {\it kinetic matrix}
${\cal N}_{\Lambda\Sigma}$ can be constructed  that transforms as
in eq.~\eqn{Ntrasform}. A way to obtain such an object is to identify
it with the {\it period matrix} occurring in problems of algebraic
geometry. At the level of the present discussion, however, this
identification is by no means essential: any construction of
${\cal N}_{\Lambda\Sigma}$ with the appropriate transformation
properties is acceptable.
Note also that so far we have used the
words {\it duality--rotations/field--redefinitions} and not the word
duality symmetry. Indeed the diffeomorphisms of the scalar manifold
we have considered were quite general and, as such had no pretension
to be symmetries of the action, or of the theory. Indeed the question
we have answered is the following: what are the appropriate
transformation properties of the tensor gauge fields and of the generalized
coupling constants under diffeomorphisms of the scalar manifold?
The next question is obviously that of duality symmetries.
\par
As it is the case with the difference between general covariance and
isometries in the context of general relativity, duality symmetries
correspond to the subset of duality transformations for which
we obtain an invariance in form of the theory.
In this respect, however, we have to stress that what is invariant
in form cannot be the Lagrangian but only the set of field equations
plus Bianchi identities.
Indeed, while any $\Lambda \in Sp(2\bar n,\IR)$ can, in principle,
be an invariance
in form of eqs.~\eqn{biafieqpm}, the same is  not true for the
Lagrangian.  One can easily find that the vector kinetic part
of this latter transforms as follows:
\begin{eqnarray}
\mbox{Im} {\cal F}^{-\Lambda}{\bar{\cal N}}_{\Lambda\Sigma}
{\cal F}^{-\Sigma}
& \rightarrow & \mbox{Im} {\tilde{\cal F}}^{-\Lambda}\,
{\tilde{\cal G}}^{-}_{\Sigma}\nonumber\\
& \null &= \mbox{Im} \Bigl ( {\cal F}^{-\Lambda} {\cal G}^{-}_{\Lambda} +
2 {\cal F}^{-\Lambda}\, \bigl ( C^{T} B
\bigr )_{\Lambda}^{\phantom{\Lambda}\Sigma}  \,
{\cal G}^{-}_{\Sigma} \nonumber\\
& \null & + {\cal F}^{-\Lambda}\, \bigl ( C^{T} A
\bigr )_{\Lambda \Sigma} {\cal F}^{-\Sigma} +
{\cal G}^{-}_{\Lambda} \, \bigl ( D^{T} B
\bigr )^{\Lambda \Sigma} \,  {\cal G}^{-}_{\Sigma}
\Bigr )
\label{lagtrasf}
\end{eqnarray}
whence we conclude that proper symmetries of the Lagrangian are
to be looked for only among matrices with $ C=B=0 $.
If $ C \ne  0 $ and  $B=0$, the Lagrangian varies through the
addition of a topological density.
  Elements of $Sp(2{\bar n}, \IR)$ with
$B \ne  0$, cannot be symmetries of the classical action under any
circumstance.
\par
The scalar part of the Lagrangian, on the other hand, is invariant
under all those diffeomorphisms of the scalar manifolds that
are
{\it isometries} of the scalar metric
$g_{IJ}$. Naming
$\xi^\star : \, T{\cal M}_{scalar} \, \rightarrow \, T{\cal M}_{scalar}$
the push--forward of $\xi$, this means that
\begin{eqnarray}
&\forall\,  X,Y \, \in \, T{\cal M}_{scalar}& \nonumber\\
& g\left ( X, Y \right
)\, = \, g \left ( \xi^\star X, \xi^\star Y \right )&
\label{isom}
\end{eqnarray}
and $\xi$ is an exact global symmetry of the scalar  part of the
Lagrangian in eq.~\eqn{gaiazuma}. In view of our previous discussion
these symmetries of the scalar sector are not guaranteed to admit an
extension to symmetries of the complete action. Yet we can insist
that they extend to symmetries of the field
equations plus Bianchi identities, namely  to  duality symmetries
in the sense defined above. This requires
that the group of isometries of the scalar metric
${\cal I} ({\cal M}_{scalar})$ be suitably embedded into
the duality group $Sp(2\bar n,\IR)$ and that the kinetic matrix ${\cal
N}_{\Lambda\Sigma}$ satisfies the covariance law:
\begin{equation}
{\cal N}\left ( \xi (\phi)\right ) \, = \,
\left ( C_\xi + D_\xi {\cal N}(\phi) \right )
\left ( A_\xi + B_\xi {\cal N}( \phi )\right )^{-1}\ .
\label{covarianza}
\end{equation}
\section{Symplectic embeddings of homogenous spaces}
\label{LL2}
\markboth{BPS BLACK HOLES IN SUPERGRAVITY: CHAPTER 3}
{3.3 SYMPLECTIC EMBEDDINGS OF HOMOGENEOUS SPACES}
\setcounter{equation}{0}
\par
A general construction of the kinetic coupling matrix $ \cal N$
 can be derived in the
case where the scalar manifold is taken to be a homogeneous space
${\cal G}/{\cal H}$.
This is what happens in all
extended supergravities for $N \ge 3$ and also in specific
instances of N=2 theories. For this reason we shortly review
the construction of the {\it kinetic
period matrix} ${\cal N}$ in the case of homogeneous spaces.
\par
The relevant homomorphism $\iota_\delta$ (see
eq.~\eqn{spaccoindue}) becomes:
\begin{equation}
\iota_\delta : \, \mbox{Diff}\left ({{\cal G}\over{\cal
H}} \right ) \, \longrightarrow \, Sp(2\bar n, \IR)
\label{embeddif}
\end{equation}
In particular, focusing on the isometry group of the canonical metric
defined on ${{\cal G}\over{\cal H}}$\footnotemark
\footnotetext{Actually, in order to be true, the
equation ${\cal I}({\o{\cal G}{\cal H}})={\cal G}$ requires
that that the normaliser of ${\cal H}$ in ${\cal G}$ be the
identity group, a condition that is verified in all the relevant examples}:
$ {\cal I} \left ({{\cal G}\over{\cal H}}\right ) \, = \, {\cal G}$
we must consider the embedding:
\begin{equation}
\iota_\delta : \,  {\cal G}  \, \longrightarrow \, Sp(2\bar n, \IR)
\label{embediso}
\end{equation}
That in eq.~\eqn{embeddif} is a homomorphism of finite dimensional
Lie groups and as such it constitutes a problem that can be solved
in explicit form. What we just need to know is the dimension of the
symplectic group, namely the number $\bar n$ of gauge fields appearing
in the theory. Without supersymmetry the dimension $m$ of the scalar
manifold (namely the possible choices of ${{\cal G}\over{\cal H}}$) and
the number of vectors $\bar n$ are unrelated so that the
possibilities covered by eq.~\eqn{embediso} are infinitely many.
In supersymmetric theories, instead, the two numbers $m$ and $\bar n$
are related, so that there are finitely many cases to be studied
corresponding to the possible  embeddings of given groups
${\cal G}$ into a symplectic group $Sp(2\bar n, \IR)$ of fixed dimension
$\bar n$. Actually taking into account further conditions on
the holonomy of the scalar manifold that are also imposed by
supersymmetry, the solution for the symplectic embedding problem
is unique for all extended supergravities with $N \ge 3$
(see for instance \cite{castdauriafre}).
\par
Apart from the details of the specific case considered
once a symplectic embedding is given there is a general
formula one can write down for the {\it period matrix}
${\cal N}$ that guarantees symmetry (${\cal N}^T = {\cal N}$)
and the required transformation property~\eqn{covarianza}.
This is the result we want to review.
\par
The real symplectic group $Sp(2\bar n ,\IR)$ is defined as the set
of all {\it real} $2\bar n \times 2\bar n$ matrices
$ \Lambda \, = \, \left ( \matrix{ A & B \cr C & D \cr } \right ) $
satisfying equation ~\eqn{ortosymp}, namely
\begin{equation}
\Lambda^T \, \IC \, \Lambda \, = \, \IC
\label{condiziona}
\end{equation}
 where
$ \IC  \, \equiv \, \left ( \matrix{ {\bf 0} & \bfone \cr -\bfone &
{\bf 0} \cr } \right ) $
If we relax the condition that the matrix should be real but we
still impose eq.~\eqn{condiziona} we obtain the definition
of the complex symplectic group $Sp(2\bar n, \IC)$. It is
a well known fact that the following isomorphism is true:
\begin{equation}
Sp(2\bar n, \IR)  \sim  Usp(\bar n , \bar n)   \equiv
Sp(2\bar n, \IC)   \cap   U(\bar n , \bar n)
\label{usplet}
\end{equation}
By definition an element ${\cal S}\,\in \, Usp(\bar n , \bar n)$
is a complex matrix that satisfies simultaneously eq.~\eqn{condiziona}
and a pseudo--unitarity condition, that is:
\begin{equation}
{\cal S}^T \, \IC \, {\cal S} \, = \,  \IC \quad \quad \quad
; \quad \quad \quad
{\cal S}^\dagger \, \IH \, {\cal S} \, =\,  \IH
\label{uspcondo}
\end{equation}
where
$\IH \,  \equiv \, \left ( \matrix{ \bfone & {\bf 0} \cr {\bf 0} & -\bfone
 \cr } \right )$.
The general block form of the matrix ${\cal S}$ is:
\begin{equation}
{\cal S}\, = \, \left ( \matrix{ T & V^\star \cr V & T^\star \cr } \right )
\label{blocusplet}
\end{equation}
and eq.s~\eqn{uspcondo} are equivalent to:
\begin{equation}
T^\dagger \, T \, - \, V^\dagger \, V \, =\, \bfone \quad\quad ; \quad\quad
T^\dagger \, V^\star  \, - \,  V^\dagger \, T^\star \, =\, {\bf 0}
\label{relazie}
\end{equation}
The isomorphism of eq.~\eqn{usplet} is explicitly realized by
the so called Cayley matrix:
\begin{equation}
{\cal C} \, \equiv \, {\o{1}{\sqrt{2}}} \,
\left ( \matrix{ \bfone & {\rm i}\bfone \cr \bfone & -{\rm i}\bfone
 \cr } \right )
\label{cayley}
\end{equation}
via the relation:
\begin{equation}
{\cal S}\, = \, {\cal C} \, \Lambda \, {\cal C}^{-1}
\label{isomorfo}
\end{equation}
which yields:
\begin{equation}
T \, =\,  {\o{1}{2}}\, \left ( A + D \right ) -
{\o{\rm i}{2}}\, \left ( B-C \right ) \quad \quad ; \quad
\quad
V \, =\, {\o{1}{2}}\, \left ( A - D \right ) -
{\o{\rm i}{2}}\, \left ( B+C \right )
\label{mappetta}
\end{equation}
When we set $V=0$ we obtain the subgroup $U(\bar n) \subset Usp (\bar
n , \bar n)$, that in the real basis is given by the subset of
symplectic matrices of the form
$\left ( \matrix{ A & B \cr -B & A
 \cr } \right )$. The basic idea, to obtain the
general formula for the period matrix, is that the symplectic embedding
of the isometry group ${\cal G}$ will be such that the isotropy
subgroup ${\cal H}\subset {\cal G}$ gets embedded into the maximal
compact subgroup $U(\bar n)$, namely:
\begin{equation}
{\cal G} \,  {\stackrel{\iota_\delta}{\longrightarrow}} \,  Usp (\bar
n , \bar n) \quad\quad ; \quad\quad
{\cal G} \supset {\cal H} \,  {\stackrel{\iota_\delta}{\longrightarrow}} \,
U(\bar n) \subset Usp (\bar n , \bar n)
\label{gruppino}
\end{equation}
If this condition is realized let $L(\phi)$ be a parametrization of
the coset ${\cal G}/{\cal H}$ by means of coset representatives.
Relying on the symplectic embedding of eq.~\eqn{gruppino} we obtain
a map:
\begin{equation}
  L(\phi)  \, \longrightarrow  {\cal O}(\phi)\, =  \,
 \left ( \matrix{ U_0(\phi) & U^\star_1(\phi) \cr U_1(\phi)
& U^\star_0(\phi) \cr } \right )\,  \in  \, Usp(\bar n , \bar n)
\label{darstel}
\end{equation}
that associates to $L(\phi)$ a coset representative of $Usp(\bar n ,
\bar n)/U(\bar n)$. By construction if $\phi^\prime \ne \phi$
{\it no} unitary $\bar n \times \bar n$ matrix $W$ {\it can exist}
such that:
\begin{equation}
 {\cal O}(\phi^\prime)  =  {\cal O}(\phi) \,
 \left ( \matrix{ W & {\bf 0} \cr {\bf 0}
& W^\star \cr } \right )
\end{equation}
On the other hand let $\xi \in {\cal G}$ be an element of the
isometry group of ${{\cal G}/{\cal H}}$. Via the symplectic embedding
of eq.~\eqn{gruppino} we obtain a $Usp(\bar n, \bar n)$ matrix
\begin{equation}
{\cal S}_ \xi \, = \,
\left ( \matrix{ T_\xi & V^\star_\xi \cr V_\xi & T^\star_\xi \cr } \right )
\label{uspimag}
\end{equation}
such that
\begin{equation}
{\cal S}_ \xi \,{\cal O}(\phi) \, = \, {\cal O}(\xi(\phi)) \,
\left ( \matrix{ W(\xi,\phi) & {\bf 0} \cr {\bf 0}
& W^\star(\xi,\phi) \cr } \right )
\label{cosettone}
\end{equation}
where $\xi(\phi)$ denotes the image of the point
$\phi \in  {{\cal G}/{\cal H}}$ through $\xi$ and $W(\xi,\phi)$ is
a suitable $U(\bar n)$ compensator depending both on $\xi$ and
$\phi$.
Combining eq.s~\eqn{cosettone},~\eqn{darstel}, with eq.s~\eqn{mappetta}
we immediately obtain:
\begin{eqnarray}
 U_0^\dagger \left( \xi(\phi) \right ) +
U^\dagger_1 \left (\xi(\phi) \right)  & = &
  W  \left [ U_0^\dagger \left( \phi \right )   \left (
A^T + {\rm i}B^T \right ) + U_1^\dagger \left( \phi \right )   \left (
A^T - {\rm i}B^T \right ) \right ]   \nonumber\\
 U_0^\dagger \left( \xi(\phi) \right ) -
U^\dagger_1 \left (\xi(\phi) \right)  & = &
 W \, \left [ U_0^\dagger \left( \phi \right )   \left (
D^T - {\rm i}C^T \right ) - U_1^\dagger \left( \phi \right )   \left (
D^T + {\rm i}C^T \right ) \right ]
\label{semitrasform}
\end{eqnarray}
Setting:
\begin{equation}
{\cal N} \, \equiv \, {\rm i} \left [ U_0^\dagger + U_1^\dagger \right
]^{-1} \, \left [ U_0^\dagger - U_1^\dagger \right ]
\label{masterformula}
\end{equation}
and using the result of eq.~\eqn{semitrasform} one checks
that the transformation rule~\eqn{covarianza} is verified.
It is also an immediate consequence of the analogue of
eq.s~\eqn{relazie} satisfied by $U_0$ and $U_1$ that the matrix
in eq.~\eqn{masterformula} is symmetric
\begin{equation}
{\cal N}^T \, = \, {\cal N}
\label{massi}
\end{equation}
Eq.~\eqn{masterformula} is the master formula derived in 1981 by
Gaillard and Zumino \cite{gaizum}.
It explains the structure of the gauge field
kinetic terms in all $N\ge 3$ extended supergravity theories where
the scalar manifold is always a homogeneous coset manifold. (See
table \ref{topotable}).  In these cases the bosonic lagrangian
\eqn{genact} is completely fixed from the information provided by
table \ref{topotable}. The choice of the coset scalar manifold
suffices to determine the scalar kinetic term, while the choice of
the symplectic embedding plus the use of the master formula
\eqn{masterformula}  determines the period matrix
${\cal N}_{\Lambda\Sigma}$.
In the case of $N=2$ supergravity the scalar manifold is not
necessarily a coset manifold. The requirement imposed by
supersymmetry is that:
\begin{equation}
{\cal M}_{scalar} = \mbox{Special K\"ahler manifold} \, {\cal SM}
\label{statem}
\end{equation}
and for this class of manifolds there exists another formula that
determines the {\it period matrix} ${\cal N}$. In the next section we
shall discuss Special K\"ahler Geometry and derive such a formula.
In the class of special K\"ahler manifolds there is a subclass that is
composed of homogeneous cosets ${\cal G}/{\cal H}$.
For this subclass the construction
of the period matrix through the relations of special geometry and
the construction through the Gaillard--Zumino master formula
\eqn{masterformula} coincide.
 \par
 In the next subsection we pause to illustrate the master formula
 \eqn{masterformula} with an example that is of particular relevance
 in all superstring related supergravities.
\subsection{Symplectic embedding of the ${\cal ST}\left [ m,n \right ]$
homogeneous manifolds}
Because of their relevance in superstring compactifications let us
illustrate the general procedure with the following class of
homogeneous manifolds:
\begin{equation}
 {\cal ST}\left [ m,n \right ] \,  \equiv \,
{\o{SU(1,1)}{U(1)}} \, \otimes \, {\o{SO(m,n)}{SO(m)\otimes SO(n)}}
\label{stmanif}
\end{equation}
The isometry group of the ${\cal ST}\left [ m,n \right ]$
manifolds defined in eq.~\eqn{stmanif} contains a factor ($SU(1,1)$)
whose transformations act as non--perturbative $S$--dualities and
another factor $(SO(m,n))$ whose transformations act as
$T$--dualities,
holding true at each order in string perturbation theory. The field
$S$ is obtained by combining together the {\it dilaton} $D$ and
the {\it axion} ${\cal A}$:
\begin{eqnarray}
S & = & {\cal A} - {\rm i} \mbox{exp}[D] \nonumber\\
\partial^\mu {\cal A} & \equiv & \varepsilon^{\mu\nu\rho\sigma} \,
\partial_\nu \, B_{\rho\sigma}
\label{scampo}
\end{eqnarray}
while $t^i$ is the name usually given to the moduli--fields of the
compactified target space. Now in string and supergravity
applications $S$ will be identified with the complex coordinate
on the manifold ${\o{SU(1,1)}{U(1)}}$, while  $t^i$  will be
the coordinates of the coset space  ${\o{SO(m,n)}{SO(m)\otimes SO(n)}}$.
The case ${\cal ST}[6,n]$ is the scalar manifold in $N=4$
supergravity, while the case ${\cal ST}[2,n]$ is a very interesting
instance of special K\"ahler manifold appearing in superstring
compactifications.
Although as differentiable and metric manifolds
the spaces ${\cal ST}\left [ m,n \right ]$ are just direct products
of two factors (corresponding to the above mentioned different
physical interpretation of the coordinates $S$ and $t^i$), from the
point of view of the symplectic embedding and duality rotations
they have to be regarded as a single entity. This is even more
evident in the case $m=2,n=\mbox{arbitrary}$,
where the following theorem has been proven by
Ferrara and Van Proeyen \cite{ferratoine}:
${\cal ST}\left [ 2,n \right ]$ are the only special K\"ahler
manifolds with a direct product structure. The definition
of special K\"ahler manifolds is given in the next section,
yet the anticipation of this result
should make clear that the special K\"ahler structure (encoding the
duality rotations in the $N=2$ case)
is not a property of the individual factors
but of the product as a whole. Neither factor
is by itself a special manifold although the product is.
\par
At this point comes  the question
of the correct symplectic embedding. Such a question has two aspects:
\begin{enumerate}
\item{Intrinsically inequivalent embeddings}
\item{Symplectically equivalent embeddings that become inequivalent
after gauging}
\end{enumerate}
 The first issue in the above list is group--theoretical in nature.
When we say that the group ${\cal G}$ is embedded into $Sp(2\bar
n,\IR)$ we must specify how this is done from the point of view
of irreducible representations. Group--theoretically the matter is
settled by specifying how the fundamental representation of
$Sp(2\bar n)$ splits into irreducible representations of ${\cal G}$:
\begin{eqnarray}
& {\bf {2 \bar n}} \, {\stackrel{{\cal G}}{\longrightarrow}}
\oplus_{i=1}^{\ell} \, {\bf D}_i &
\label{splitsplit}
\end{eqnarray}
Once eq.~\eqn{splitsplit} is given (in supersymmetric theories
such information is provided by supersymmetry ) the only arbitrariness
which is left is that of conjugation by arbitrary $Sp(2\bar n,\IR)$
matrices. Suppose we have determined an embedding $\iota_\delta$ that
obeys the law in eq.~\eqn{splitsplit}, then:
\begin{equation}
\forall \, {\cal S} \, \in \, Sp(2\bar n,\IR) \, : \,
\iota_\delta^\prime \, \equiv \, {\cal S} \circ  \iota_\delta \circ
{\cal S}^{-1}
\label{matrim}
\end{equation}
will obey the same law. That in eq.~\eqn{matrim} is a symplectic
transformation that corresponds to an allowed
duality--rotation/field--redefinition in the abelian theory of
type in eq.~\eqn{gaiazuma} discussed in the previous subsection. Therefore
all abelian Lagrangians related by such transformations are physically
equivalent.
\par
The matter changes in presence of {\it gauging}. When we switch
on the gauge coupling constant and the electric charges, symplectic
transformations cease to yield physically equivalent theories. This
is the second issue in the above list. The choice of a symplectic
gauge becomes physically significant.
The construction of supergravity theories proceeds in
two steps. In the first step,
one constructs the abelian theory: at that level the only relevant
constraint is that encoded in eq.~\eqn{splitsplit} and the choice of
a symplectic gauge is immaterial. Actually one can write the entire
theory in such a way that {\it symplectic covariance} is manifest.
In the second step one {\it gauges} the theory. This {\it breaks
symplectic covariance} and the choice of the correct symplectic gauge
becomes a physical issue. This issue has been recently emphasized
by the results in \cite{FGP1} where it has been shown that
whether N=2 supersymmetry can be spontaneously broken to N=1 or
not depends on the symplectic gauge.
\par
In the applications of supergravity to the issue of BPS black--holes
what matters is the  {\it abelian ungauged theory}, so we will not
further emphasize the aspects of the theory related to the gauging.
\par
These facts being cleared we proceed to discuss the symplectic
embedding of the ${\cal ST}\left [ m,n \right ]$ manifolds.
\par
Let $\eta$ be the symmetric flat metric with signature
$(m,n)$ that defines the $SO(m,n)$ group, via the relation
\begin{equation}
L \, \in \, SO(m,n) \, \Longleftrightarrow \, L^T \, \eta L \, = \,
\eta
\label{ortogruppo}
\end{equation}
Both in the $N=4$ and in the $N=2$ theory, the number of gauge fields
in the theory is given by:
\begin{equation}
\# \mbox{vector fields} \, = \, m \oplus n
\label{vectornum}
\end{equation}
$m$ being the number of {\it graviphotons} and $n$ the number of
{\it vector multiplets}. Hence we have to embed $SO(m,n)$ into
$Sp(2m+2n,\IR)$ and the explicit form of the decomposition in
eq.~\eqn{splitsplit} required by supersymmetry is:
\begin{equation}
{\bf {2m+2n}} \, {\stackrel{SO(m,n)}{\longrightarrow}} \, {\bf { m+n}}
\oplus  {\bf { m+n}}
\label{ortosplitsplit}
\end{equation}
where ${\bf { m+n}}$ denotes the fundamental representation of
$SO(m,n)$. Eq.~\eqn{ortosplitsplit} is easily understood in physical
terms. $SO(m,n)$ must be a T--duality group, namely a symmetry
 holding true order by order in perturbation theory. As such it must
 rotate electric  field strengths into electric field strengths and
 magnetic field strengths into magnetic field field strengths. The
 two irreducible representations into which the   fundamental
 representation of the symplectic group decomposes when reduced to
 $SO(m,n)$ correspond precisely to electric and magnetic sectors,
 respectively.
In the {\it simplest  gauge} the symplectic embedding satisfying
eq.~\eqn{ortosplitsplit} is block--diagonal and takes the form:
\begin{equation}
\forall \,  L \, \in \, SO(m,n) \quad {\stackrel{\iota_\delta}
{\hookrightarrow}} \quad
\left ( \matrix{ L & {\bf 0}\cr {\bf 0} & (L^T)^{-1} \cr } \right )
\, \in \, Sp(2m+2n,\IR)
\label{ortoletto}
\end{equation}
Consider instead the group $SU(1,1) \sim SL(2,\IR)$. This is the
factor in the isometry group of ${\cal ST}[m,n]$
that is going to act by means of S--duality non perturbative
rotations. Typically it will rotate each electric field strength into
its homologous magnetic one. Correspondingly supersymmetry implies
that its embedding into the symplectic group must satisfy the
following condition:
\begin{equation}
{\bf {2m+2n}} \, {\stackrel{SL(2,\IR)}{\longrightarrow}} \,
\oplus_{i=1}^{m+n} \, {\bf 2}
\label{simposplisplit}
\end{equation}
where  ${\bf 2}$ denotes the fundamental representation of
$SL(2,\IR)$. In addition it must commute with the embedding
of $SO(m,n)$ in eq.~\eqn{ortoletto} . Both conditions are fulfilled
by setting:
\begin{equation}
\forall \,   \left ( \matrix{a & b \cr  c &d \cr }\right )
\, \in \, SL(2,\IR) \quad {\stackrel{\iota_\delta}
{\hookrightarrow}} \quad
\left ( \matrix{ a \, \bfone & b \, \eta \cr c \, \eta &
d \, \bfone \cr } \right )
\, \in \, Sp(2m+2n,\IR)
\label{ortolettodue}
\end{equation}
Utilizing eq.s~\eqn{isomorfo} the corresponding  embeddings into
the group $Usp(m+n,m+n)$ are immediately derived:
\begin{eqnarray}
\forall \,  L \, \in \, SO(m,n) & {\stackrel{\iota_\delta}
{\hookrightarrow}} & \left ( \matrix{ {\o{1}{2}}  \left ( L+
\eta L \eta \right ) & {\o{1}{2}}  \left ( L-
\eta L \eta \right )\cr {\o{1}{2}}  \left ( L -
\eta L \eta \right ) & {\o{1}{2}}  \left ( L+
\eta L \eta \right ) \cr } \right )  \,
  \, \in \, Usp(m+n,m+n)  \nonumber\\
  \forall \,   \left ( \matrix{t & v^\star \cr  v &t^\star \cr }\right )
\, \in \, SU(1,1) & {\stackrel{\iota_\delta}
{\hookrightarrow}} &  \left ( \matrix{ {\rm Re}t \bfone +{\rm i}{\rm Im}t\eta &
{\rm Re}v \bfone -{\rm i}{\rm Im}v \eta   \cr
{\rm Re}v \bfone +{\rm i}{\rm Im}v\eta &
{\rm Re}t \bfone - {\rm i}{\rm Im}t\eta \cr } \right )
 \, \in \, Usp(m+n,m+n)  \nonumber\\
\label{uspembed}
\end{eqnarray}
where the relation between the entries of the $SU(1,1)$ matrix
and those of the corresponding $SL(2,\IR)$ matrix are provided
by the relation in eq.~\eqn{mappetta}.
\par
Equipped with these relations we can proceed to derive the explicit
form of the {\it period matrix} ${\cal N}$.
\par
The homogeneous manifold $SU(1,1)/U(1)$ can be conveniently
parametrized in terms of a single complex coordinate $S$, whose
physical interpretation will be that of {\it axion--dilaton},
according to eq.~\eqn{scampo}. The coset parametrization appropriate
for comparison with other constructions (special geometry  or
$N=4$ supergravity) is given
by the family of matrices:
\begin{equation}
  M(S) \, \equiv \, {\o{1}{n(S)} } \, \left (
\matrix{ \bfone & { \o{{\rm i} -S }{ {\rm i} + S } }\cr
{\o{ {\rm i} + {\bar S} }{ {\rm i} -{\bar S} } } & \bfone \cr}
\right )\quad \quad : \quad \quad
n(S) \, \equiv \, \sqrt{ {\o{4 {\rm Im}S } {
 1+\vert S \vert^2 +2 {\rm Im}S } } }
 \label{su11coset}
\end{equation}
To parametrize the coset $SO(m,n)/SO(m)\times SO(n)$ we can instead
take the usual coset representatives
(see for instance~\cite{castdauriafre}):
\begin{equation}
L(X) \, \equiv \, \left (\matrix{ \left ( \bfone + XX^T \right )^{1/2}
& X \cr X^T & \left ( \bfone + X^T X \right )^{1/2}\cr } \right )
\label{somncoset}
\end{equation}
where the $m \times n $ real matrix $X$ provides a set of independent
coordinates. Inserting these matrices into the embedding formulae of
eq.s~\eqn{uspembed} we obtain a matrix:
\begin{equation}
\iota_\delta \left ( M (S) \right ) \circ  \iota_\delta
\left ( L(X) \right ) \,
 = \, \left ( \matrix{ U_0(S,X) & U^\star_1(S,X) \cr
U_1(S,X)
& U^\star_0(S,X) \cr } \right ) \, \in \, Usp(n+m , n+m)
\label{uspuspusp}
\end{equation}
that inserted into the master formula of eq.~\eqn{masterformula}
yields the following result:
\begin{equation}
{\cal N}\, = \, {\rm i} {\rm Im}S \, \eta L(X) L^T(X) \eta
+ {\rm Re}S \, \eta
\label{maestrina}
\end{equation}
Alternatively, remarking that if $L(X)$ is an $SO(m,n)$ matrix
also $L(X)^\prime =\eta L(X) \eta$ is such a matrix and represents
the same equivalence class, we can rewrite ~\eqn{maestrina} in the
simpler form:
\begin{equation}
{\cal N}\, = \, {\rm i} {\rm Im}S \,   L(X)^\prime L^{T\prime}
(X)
+ {\rm Re}S \, \eta
\label{maestrino}
\end{equation}
\section{Special K\"ahler Geometry}
\setcounter{equation}{0}
\markboth{BPS BLACK HOLES IN SUPERGRAVITY: CHAPTER 3}
{3.4 SPECIAL KAHLER GEOMETRY}
As already stressed, in the case of $N=2$ supergravity
the requirements imposed by supersymmetry on the scalar manifold
${\cal M}_{scalar}$ of the theory is that it should be the following
direct product:
\begin{eqnarray}
 {\cal M}_{scalar}&=&{\cal SM}\, \otimes \, {\cal HM}\nonumber\\
 \mbox{dim}_{\bf C} \,{\cal SM}&=& n_v \nonumber\\
  \mbox{dim}_{\bf R} \,{\cal HM}&=& 4 \,n_h \nonumber\\
  \label{scalando}
\end{eqnarray}
where ${\cal SM}$, ${\cal HM}$ are respectively {\it special
K\"ahler} and {\it quaternionic} and $n_v$, $n_h$ are respectively
the number of {\it vector multiplets} and {\it hypermultiplets}
contained in the theory. The direct product structure \eqn{scalando}
imposed by supersymmetry precisely reflects the fact that the
quaternionic and special K\"ahler scalars belong to different
supermultiplets. In the construction of BPS black--holes it turns out
that the hyperscalars are spectators playing no dynamical role. Hence
in this set of lectures we do not discuss
the hypermultiplets any further and we confine our attention to an
$N=2$ supergravity where the graviton multiplet, containing,
besides the graviton $g_{\mu \nu}$, also a graviphoton
$A^0_{\mu}$, is coupled to  $n_v=n$  {\it vector multiplets}.
Such a theory has an action of type \eqn{genact} where the number
of gauge fields is ${\bar n}=1+n$ and the number of scalar fields
is $m=2 \, {n}$. Correspondingly the indices have the
following ranges
\begin{eqnarray}
 \Lambda,\Sigma,\Gamma,\dots & = & 0,1,\dots, n \nonumber\\
 I,I,K &=&1,\dots,2 \, n \nonumber\\
 \label{indin}
\end{eqnarray}
To make the action \eqn{genact} fully explicit, we need to discuss
the geometry of the vector multiplets scalars, namely special
K\"ahler geometry.  This is what we do in the next subsections.
Let us begin by reviewing the notions of K\"ahler and Hodge--K\"ahler
manifolds. The readers interested in the full-fledged structure of
$N=2$ supergravity can read \cite{n2standa} which contains its most
general formulation and explains all the details of its construction.
\subsection{Hodge--K\"ahler manifolds}
\def\mom{{M(k, \IC)}}
Consider a {\sl line bundle}
${\cal L} {\stackrel{\pi}{\longrightarrow}} {\cal M}$ over a K\"ahler
manifold. By definition this is a holomorphic
vector bundle of rank $r=1$. For such bundles the only available
Chern class is the first:
\begin{equation}
c_1 ( {\cal L} ) \, =\, \o{i}{2\pi}
\, {\bar \partial} \,
\left ( \, h^{-1} \, \partial \, h \, \right )\, =
\, \o{i}{2\pi} \,
{\bar \partial} \,\partial \, \mbox{log} \,  h
\label{chernclass23}
\end{equation}
where the 1-component real function $h(z,{\bar z})$ is some hermitian
fibre metric on ${\cal L}$. Let $f (z)$ be a holomorphic section of
the line bundle
${\cal L}$: noting that  under the action of the operator ${\bar
\partial} \,\partial \, $ the term $\mbox{log} \left ({\bar \xi}({\bar z})
\, \xi (z) \right )$ yields a vanishing contribution, we conclude that
the formula in eq.~\eqn{chernclass23}  for the first Chern class can be
re-expressed as follows:
\begin{equation}
c_1 ( {\cal L} ) ~=~\o{i}{2\pi} \,
{\bar \partial} \,\partial \, \mbox{log} \,\parallel \, \xi(z) \, \parallel^2
\label{chernclass24}
\end{equation}
where $\parallel \, \xi(z) \, \parallel^2 ~=~h(z,{\bar z}) \,
{\bar \xi}({\bar z}) \,
\xi (z) $ denotes
the norm of the holomorphic section $\xi (z) $.
\par
Eq.~\eqn{chernclass24} is the starting point for the definition
of Hodge K\"ahler manifolds,
an essential notion in supergravity theory.
\par
A K\"ahler manifold ${\cal M}$ is a Hodge manifold if and
only if there exists
a line
bundle ${\cal L} \, \longrightarrow \, {\cal M}$ such that its
first Chern class equals
the cohomology class of the K\"ahler 2-form K:
\begin{equation}
c_1({\cal L} )~=~\left [ \, K \, \right ]
\label{chernclass25}
\end{equation}
\par
In local terms this means that there is a holomorphic section
$W(z)$ such that we can write
\begin{equation}
K\, =\, \o{i}{2\pi} \, g_{ij^{\star}} \, dz^{i} \, \wedge \,
d{\bar z}^{j^{\star}} \, = \,
\o{i}{2\pi} \, {\bar \partial} \,\partial \, \mbox{log} \,\parallel \,
W(z) \,
\parallel^2
\label{chernclass26}
\end{equation}
Recalling the local expression of the K\"ahler metric
in terms of the K\"ahler potential
$ g_{ij^{\star}} $ = ${\partial}_i \, {\partial}_{j^{\star}}
{\cal K} (z,{\bar z})$,
it follows from eq.\eqn{chernclass26} that if the
manifold ${\cal M}$ is a Hodge manifold,
then the exponential of the K\"ahler potential
can be interpreted as the metric
$h(z,{\bar z})$ = $\exp \left ( {\cal K} (z,{\bar z})\right )$
on an appropriate line bundle ${\cal L}$.
\par
This structure is precisely that advocated by the Lagrangian of
$N=1$ matter coupled supergravity:
the holomorphic section $W(z)$ of the line bundle
${\cal L}$ is what, in N=1 supergravity theory, is
named the superpotential and the logarithm of
its norm  $\mbox{log} \,\parallel \, W(z) \, \parallel^2\, = \,
{\cal K} (z,{\bar z})\, + \, \mbox{log} \, | \, W(z) \, |^2  ~=~ G(z,{\bar z})$
is precisely the  invariant function in terms of which one writes the
potential and Yukawa coupling terms of the supergravity action
(see \cite{sugkgeom_4} and for a review \cite{castdauriafre}).
\par
\subsection{Special K\"ahler Manifolds: general discussion}
\par
There are in fact two kinds
of special K\"ahler geometry: the local and the rigid one.
The former describes the scalar field sector of vector multiplets
in $N=2$ supergravity while the latter describes the same sector
in rigid $N=2$  Yang--Mills theories. Since $N=2$
includes $N=1$ supersymmetry, local and rigid special
K\"ahler manifolds must be compatible with the geometric structures that are
respectively enforced by local and rigid $N=1$ supersymmetry in the
scalar sector. The distinction between the two cases
deals with the first Chern--class of the line--bundle
${\cal L} {\stackrel{\pi}{\longrightarrow}} {\cal M}$, whose sections
are the possible superpotentials.  In the local theory $c_1({\cal L})
=[K]$ and this restricts ${\cal M}$ to be a Hodge--K\"ahler manifold.
In the rigid theory, instead, we have $c_1({\cal L})=0$. At the level
of the Lagrangian this reflects into a different behaviour of the
fermion fields. These latter are sections of ${\cal L}^{1/2}$ and
couple to the canonical hermitian connection defined on ${\cal L}$:
\begin{equation}
\begin{array}{ccccccc}
{\theta}& \equiv & h^{-1} \, \partial  \, h = {\o{1}{h}}\, \partial_i h \,
dz^{i} &; &
{\bar \theta}& \equiv & h^{-1} \, {\bar \partial}  \, h = {\o{1}{h}} \,
\partial_{i^\star} h  \,
d{\bar z}^{i^\star} \cr
\end{array}
\label{canconline}
\end{equation}
In the local case where
\begin{equation}
\left  [ \, {\bar \partial}\,\theta \,  \right ] \, = \,
c_1({\cal L}) \, = \, [K]
\label{curvc1}
\end{equation}
the fibre metric $h$ can be identified with the exponential of the
K\"ahler potential and we obtain:
\begin{equation}
\begin{array}{ccccccc}
{\theta}& = &  \partial  \,{\cal K} =  \partial_i {\cal K}
dz^{i} & ; &
{\bar \theta}& = &   {\bar \partial}  \, {\cal K} =
\partial_{i^\star} {\cal K}
d{\bar z}^{i^\star}\cr
\end{array}
\label{curvconline}
\end{equation}
In the rigid case,  ${\cal L}$ is instead a flat bundle and its
metric is unrelated to the K\"ahler potential. Actually one can choose
a vanishing connection:
\begin{equation}
\theta \,= \, {\bar \theta} \, = \, 0
\label{rigconline}
\end{equation}
The distinction between rigid and local special manifolds is the
$N=2$ generalization of this difference occurring at the
$N=1$ level.
\par
In these lectures, since we are interested in BPS black-holes
and therefore in supergravity,  we discuss only the case of local
special geometry, leaving aside the rigid case that is relevant for
$N=2$ gauge theories. The interested reader can find further details
in \cite{pietrolectures} or in \cite{n2standa} which contain also all
the references to the original papers on special geometry.
\par
In the $N=2$ case, in addition to the line--bundle
${\cal L}$ we need a flat holomorphic vector bundle ${\cal SV}
\, \longrightarrow \, {\cal M}$ whose sections can be identified
with the superspace {\it fermi--fermi} components of electric and magnetic
field--strengths. In this way, according
to the discussion of
previous sections the diffeomorphisms of the scalar manifolds
will be lifted to produce an action on the gauge--field strengths
as well. In a supersymmetric theory where scalars and gauge fields
belong to the same multiplet this is a mandatory condition.
However  this symplectic bundle structure must be made
compatible with the line--bundle structure already requested
by $N=1$ supersymmetry. This leads to the existence of
two kinds of special geometry. Another essential distinction
between the two kind of geometries arises from the different
number of vector fields in the theory. In the rigid case
this number equals that of the vector multiplets so that
\begin{eqnarray}
\# \, \mbox{vector fields}\, \equiv \, {\bar n} & = & n
\nonumber\\
\# \, \mbox{vector multiplets}\equiv n & = &
\mbox{dim}_{\bf C} \, {\cal M}\nonumber\\
\mbox{rank} \, {\cal SV}   \, \equiv \, 2\bar n & = & 2 n
 \label{rigrank}
\end{eqnarray}
On the other hand,
in the local case, in addition to the vector fields arising
from the vector multiplets we have also the graviphoton coming from
the graviton multiplet. Hence we conclude:
\begin{eqnarray}
\# \, \mbox{vector fields}\, \equiv \, {\bar n} & = & n+1
\nonumber\\
\# \, \mbox{vector multiplets}\equiv n & = &
\mbox{dim}_{\bf C} \, {\cal M}\nonumber\\
\mbox{rank} \, {\cal SV}   \, \equiv \, 2\bar n & = & 2 n+2
 \label{locrank}
\end{eqnarray}
In the sequel we make extensive use of covariant derivatives with
respect to the canonical connection of the line--bundle ${\cal L}$.
Let us review its normalization. As it is well known there exists
a correspondence between line--bundles and
$U(1)$--bundles. If $\mbox{exp}[f_{\alpha\beta}(z)]$ is the transition
function between two local trivializations of the line--bundle
${\cal L} \, \longrightarrow \, {\cal M}$, the transition function
in the corresponding principal $U(1)$--bundle $U \,
\longrightarrow {\cal M}$ is just
$\mbox{exp}[{\rm i}{\rm Im}f_{\alpha\beta}(z)]$ and the K\"ahler potentials
in two different charts are related by:
\begin{equation}
{\cal K}_\beta = {\cal K}_\alpha + f_{\alpha\beta}   + {\bar
{f}}_{\alpha\beta}
\label{carte}
\end{equation}
At the level of connections this correspondence is formulated by
setting:
\begin{equation}
\mbox{ $U(1)$--connection}   \equiv   {\cal Q} \,  = \,   \mbox{Im}
\theta = -{\o{\rm i}{2}}   \left ( \theta - {\bar \theta}
\right)
\label{qcon}
\end{equation}
If we apply the above formula to the case of the $U(1)$--bundle
${\cal U} \, \longrightarrow \, {\cal M}$
associated with the line--bundle ${\cal L}$ whose first Chern class equals
the K\"ahler class, we get:
\begin{equation}
{\cal Q}  =    -{\o{\rm i}{2}} \left ( \partial_i {\cal K}
dz^{i} -
\partial_{i^\star} {\cal K}
d{\bar z}^{i^\star} \right )
\label{u1conect}
\end{equation}
 Let now
 $\Phi (z, \bar z)$ be a section of ${\cal U}^p$.  By definition its
covariant derivative is
\begin{equation}
\nabla \Phi = (d + i p {\cal Q}) \Phi
\end{equation}
or, in components,
\begin{equation}
\begin{array}{ccccccc}
\nabla_i \Phi &=&
 (\partial_i + {1\over 2} p \partial_i {\cal K}) \Phi &; &
\nabla_{i^*}\Phi &=&(\partial_{i^*}-{1\over 2} p \partial_{i^*} {\cal K})
\Phi \cr
\end{array}
\label{scrivo2}
\end{equation}
A covariantly holomorphic section of ${\cal U}$ is defined by the equation:
$ \nabla_{i^*} \Phi = 0  $.
We can easily map each  section $\Phi (z, \bar z)$
of ${\cal U}^p$
into a  section of the line--bundle ${\cal L}$ by setting:
\begin{equation}
\tilde{\Phi} = e^{-p {\cal K}/2} \Phi  \,   .
\label{mappuccia}
\end{equation}
  With this position we obtain:
\begin{equation}
\begin{array}{ccccccc}
\nabla_i    \tilde{\Phi}&    =&
(\partial_i   +   p   \partial_i  {\cal K})
\tilde{\Phi}& ; &
\nabla_{i^*}\tilde{\Phi}&=& \partial_{i^*} \tilde{\Phi}\cr
\end{array}
\end{equation}
Under the map of eq.~\eqn{mappuccia} covariantly holomorphic sections
of ${\cal U}$ flow into holomorphic sections of ${\cal L}$
and viceversa.
\subsection{Special K\"ahler manifolds: the local case}
We are now ready to give the definition of local special K\"ahler
manifolds and illustrate their properties.
A first definition that does not  make direct reference to the
symplectic bundle is the following:
\bd
A Hodge K\"ahler manifold is {\bf Special K\"ahler (of the local type)}
if there exists a completely symmetric holomorphic 3-index section $W_{i
j k}$ of $(T^\star{\cal M})^3 \otimes {\cal L}^2$ (and its
antiholomorphic conjugate $W_{i^* j^* k^*}$) such that the following
identity is satisfied by the Riemann tensor of the Levi--Civita
connection:
\begin{eqnarray}
\partial_{m^*}   W_{ijk}& =& 0   \quad   \partial_m  W_{i^*  j^*  k^*}
=0 \nonumber \\
\nabla_{[m}      W_{i]jk}& =&  0
\quad \nabla_{[m}W_{i^*]j^*k^*}= 0 \nonumber \\
{\cal R}_{i^*j\ell^*k}& =&  g_{\ell^*j}g_{ki^*}
+g_{\ell^*k}g_{j i^*} - e^{2 {\cal K}}
W_{i^* \ell^* s^*} W_{t k j} g^{s^*t}
\label{specialone}
\end{eqnarray}
\label{defspecial}
\ed
In the above equations $\nabla$ denotes the covariant derivative with
respect to both the Levi--Civita and the $U(1)$ holomorphic connection
of eq.~\eqn{u1conect}.
In the case of $W_{ijk}$, the $U(1)$ weight is $p = 2$.
\par
The holomorphic sections $W_{ijk}$ have two different physical
interpretations in the case that the special manifold is utilized
as scalar manifold in an N=1 or N=2 theory. In the first case
they correspond to the Yukawa couplings of Fermi families
\cite{petropaolo}. In the second case they provide the coefficients
for the anomalous magnetic moments of the gauginos, since they appear
in the Pauli--terms of the $N=2$ effective action.
Out of the $W_{ijk}$ we can construct covariantly holomorphic
sections of weight 2 and - 2 by setting:
\begin{equation}
C_{ijk}\,=\,W_{ijk}\,e^{  {\cal K}}  \quad ; \quad
C_{i^\star j^\star k^\star}\,=\,W_{i^\star j^\star k^\star}\,e^{  {\cal K}}
\label{specialissimo}
\end{equation}
Next we can give the second more intrinsic definition that relies
on the notion of the flat symplectic bundle.
Let ${\cal L}\, \longrightarrow \,{\cal  M}$ denote the complex
line bundle whose first Chern class equals
the K\"ahler form $K$ of an $n$-dimensional Hodge--K\"ahler
manifold ${\cal M}$. Let ${\cal SV} \, \longrightarrow \,{\cal  M}$
denote a holomorphic flat vector bundle of rank $2n+2$ with structural
group $Sp(2n+2,\IR)$. Consider   tensor bundles of the type
${\cal H}\,=\,{\cal SV} \otimes {\cal L}$.
A typical holomorphic section of such a bundle will be
denoted by ${\Omega}$ and will have the following structure:
\begin{equation}
{\Omega} \, = \, {\twovec{{X}^\Lambda}{{F}_ \Sigma} } \quad
\Lambda,\Sigma =0,1,\dots,n
\label{ololo}
\end{equation}
By definition
the transition functions between two local trivializations
$U_i \subset {\cal M}$ and $U_j \subset {\cal M}$
of the bundle ${\cal H}$ have the following form:
\begin{equation}
{\twovec{X}{ F}}_i=e^{f_{ij}} M_{ij}{\twovec{X}{F}}_j
\end{equation}
where   $f_{ij}$ are holomorphic maps $U_i \cap U_j \, \rightarrow
\,\IC $
while $M_{ij}$ is a constant $Sp(2n+2,\IR)$ matrix. For a consistent
definition of the bundle the transition functions are obviously
subject to the cocycle condition on a triple overlap:
\begin{eqnarray}
e^{f_{ij}+f_{jk}+f_{ki}} &=&1 \ \nn\\
M_{ij} M_{jk} M_{ki} &=& 1 \
\end{eqnarray}
Let $i\langle\ \vert\ \rangle$ be the compatible
hermitian metric on $\cal H$
\begin{equation}
i\langle \Omega \, \vert \, \bar \Omega \rangle \, \equiv \,  -
i \Omega^\T \twomat {0} {\bfone} {-\bfone}{0} {\bar \Omega}
\label{compati}
\end{equation}
\bd
We say that a Hodge--K\"ahler manifold ${\cal M}$
is {\bf special K\"ahler of the local type} if there exists
a bundle ${\cal H}$ of the type described above such that
for some section $\Omega \, \in \, \Gamma({\cal H},{\cal M})$
the K\"ahler two form is given by:
\begin{equation}
K= \o{i}{2\pi}
 \partial \bar \partial \, \mbox{\rm log} \, \left ({\rm i}\langle \Omega \,
 \vert \, \bar \Omega
\rangle \right )
\label{compati1} .
\end{equation}
\ed
From the point of view of local properties, eq.~\eqn{compati1}
implies that we have an expression for the K\"ahler potential
in terms of the holomorphic section $\Omega$:
\begin{equation}
{\cal K}\,  = \,  -\mbox{log}\left ({\rm i}\langle \Omega \,
 \vert \, \bar \Omega
\rangle \right )\,
=\, -\mbox{log}\left [ {\rm i} \left ({\bar X}^\Lambda F_\Lambda -
{\bar F}_\Sigma X^\Sigma \right ) \right ]
\label{specpot}
\end{equation}
The relation between the two definitions of special manifolds is
obtained by introducing a non holomorphic section of the bundle
${\cal H}$ according to:
\begin{equation}
V \, = \, \twovec{L^{\Lambda}}{M_\Sigma} \, \equiv \, e^{{\cal K}/2}\Omega
\,= \, e^{{\cal K}/2} \twovec{X^{\Lambda}}{F_\Sigma}
\label{covholsec}
\end{equation}
so that eq.~\eqn{specpot} becomes:
\begin{equation}
1 \, = \,  {\rm i}\langle V  \,
 \vert \, \bar V
\rangle  \,
= \,   {\rm i} \left ({\bar L}^\Lambda M_\Lambda -
{\bar M}_\Sigma L^\Sigma \right )
\label{specpotuno}
\end{equation}
Since $V$ is related to a holomorphic section by eq.~\eqn{covholsec}
it immediately follows that:
\begin{equation}
\nabla_{i^\star} V \, = \, \left ( \partial_{i^\star} - {\o{1}{2}}
\partial_{i^\star}{\cal K} \right ) \, V \, = \, 0
\label{nonsabeo}
\end{equation}
On the other hand, from eq.~\eqn{specpotuno}, defining:
\begin{equation}
U_i   =   \nabla_i V  =   \left ( \partial_{i} + {\o{1}{2}}
\partial_{i}{\cal K} \right ) \, V   \equiv
\twovec{f^{\Lambda}_{i} }{h_{\Sigma\vert i}}
\label{uvector}
\end{equation}
it follows that:
\begin{equation}
\nabla_i U_j  = {\rm i} C_{ijk} \, g^{k\ell^\star} \, {\bar U}_{\ell^\star}
\label{ctensor}
\end{equation}
where $\nabla_i$ denotes the covariant derivative containing both
the Levi--Civita connection on the bundle ${\cal TM}$ and the
canonical connection $\theta$ on the line bundle ${\cal L}$.
In eq.~\eqn{ctensor} the symbol $C_{ijk}$ denotes a covariantly
holomorphic (
$\nabla_{\ell^\star}C_{ijk}=0$) section of the bundle
${\cal TM}^3\otimes{\cal L}^2$ that is totally symmetric in its indices.
This tensor can be identified with the tensor of eq.~\eqn{specialissimo}
appearing in eq.~\eqn{specialone}.
Alternatively, the set of differential equations:
\begin {eqnarray}
&&\nabla _i V  = U_i\\
 && \nabla _i U_j = {\rm i} C_{ijk} g^{k \ell^\star} U_{\ell^\star}\\
 && \nabla _{i^\star} U_j = g_{{i^\star}j} V\\
 &&\nabla _{i^\star} V =0 \label{defaltern}
\end{eqnarray}
with V satisfying eq.s \eqn{covholsec}, (\ref {specpotuno}) give yet
another definition of special geometry. This is actually what one
obtains from the $N=2$ solution of superspace Bianchi identities.
In particular it is easy to find eq.~\eqn{specialone}
as integrability conditions of~\eqn{defaltern}
The {\it period matrix} is now introduced via the relations:
\begin{equation}
{\bar M}_\Lambda = {\bar {\cal N}}_{\Lambda\Sigma}{\bar L}^\Sigma \quad ;
\quad
h_{\Sigma\vert i} = {\bar {\cal N}}_{\Lambda\Sigma} f^\Sigma_i
\label{etamedia}
\end{equation}
which can be solved introducing the two $(n+1)\times (n+1)$ vectors
\begin{equation}
f^\Lambda_I = \twovec{f^\Lambda_i}{{\bar L}^\Lambda} \quad ; \quad
h_{\Lambda \vert I} =  \twovec{h_{\Lambda \vert i}}{{\bar M}_\Lambda}
\label{nuovivec}
\end{equation}
and setting:
\begin{equation}
{\bar {\cal N}}_{\Lambda\Sigma}= h_{\Lambda \vert I} \circ \left (
f^{-1} \right )^I_{\phantom{I} \Sigma}
\label{intriscripen}
\end{equation}
As a consequence of its definition the matrix ${\cal N}$ transforms,
under diffeomorphisms of the base K\"ahler manifold exactly as it
is requested by the rule in eq.~\eqn{covarianza}.
Indeed this is the very reason
why the structure of special geometry has been introduced. The
existence of the symplectic bundle ${\cal H} \, \longrightarrow \,
{\cal M}$ is required in order to be able to pull--back the action
of the diffeomorphisms on the field
strengths and to construct the kinetic matrix ${\cal N}$.
\par
From the previous formulae it is easy to derive a set of useful
relations among which we quote the following \cite{CDFp}:
\begin{eqnarray}
{\rm Im}{\cal N}_{\Lambda\Sigma}L^\Lambda\bar{L}^\Sigma &=& -{1 \over 2}
\label{normal}\\
\langle V ,U_i \rangle &=& \langle V, U_{i^\star}\rangle = 0
\label{ortogo}\\
 U^{\Lambda\Sigma} \, \equiv \,  f^\Lambda_i \, f^\Sigma_{j^\star} \,
g^{ij^\star} &=&
-{\o{1}{2}} \, \left ( {\rm Im}{\cal N} \right )^{-1 \vert
\Lambda\Sigma} \, -\, {\bar L}^\Lambda L^\Sigma
\label{dsei} \\
g_{ij^\star} &=& -{\rm i} \langle \, U_{i} \, \vert \, {\bar U}_{j^\star}
\, \rangle = -2 f^\La_i \im \cN_{\La\Si} f^\Si_{j^\star}\ ;\\
C_{ijk} &=& \langle \, \nabla_i U_{j} \, \vert \, {  U}_{k} \,
\rangle=f^\La_i \del_j \bar \cN_{\La\Si} f^\Si_k =(\cN-\bar\cN)_{\La\Si}
f^\La_i \del_j f^\Si_k
\label{sympinvloc}
\end{eqnarray}
In particular eq.s \eqn{sympinvloc} express the K\"ahler metric and
the anomalous magnetic moments in terms of symplectic invariants.
It is clear from our discussion that nowhere we have
assumed the base K\"ahler manifold to be a homogeneous space. So,
in general, special manifolds are not homogeneous spaces. Yet
there is a subclass of homogenous special manifolds. The homogeneous
symmetric ones were classified by Cremmer and Van Proeyen in
\cite{cremvanp} and are displayed in
table~ \ref{homospectable}.
\begin{table}
\begin{center}
\caption{\sl Homogeneous Symmetric Special Manifolds  }
\label{homospectable}
\vskip 0.3cm
\begin{tabular}{|c||c||c||c|}
\hline
n  & $G/H$  & $Sp(2n+2)$  & symp rep of G
\\
\hline
~~&~~&~~& ~~\\
$1$ & $\o{SU(1,1)}{U(1)}$ & $Sp(4)$ &
${\underline {\bf 4}}$ \\
~~&~~&~~& ~~\\
\hline
~~&~~&~~& ~~\\
~~&~~&~~& ~~\\
$n$ & $\o{SU(1,n)}{SU(n)\times U(1)}$ & $Sp(2n+2)$ &
${\underline {\bf n+1}}\oplus{\underline {\bf n+1}}$ \\
~~&~~&~~& ~~\\
\hline
~~&~~&~~& ~~\\
$n+1$ & $\o{SU(1,1)}{U(1)}\otimes \o{SO(2,n)}{SO(2)\times SO(n)}$
 & $Sp(2n+4)$ & $
{\underline {\bf 2}}\otimes \left ({\underline {\bf n+2}}
\oplus{\underline {\bf n+2}}\right )  $ \\
~~&~~&~~& ~~\\
\hline
~~&~~&~~& ~~\\
$6$ & $\o{Sp(6,\IR)}{SU(3)\times U(1)}$ & $Sp(14)$ & $
{\underline {\bf 14}}$ \\
~~&~~&~~& ~~\\
\hline
\hline
~~&~~&~~& ~~\\
$9$ & $\o{SU(3,3)}{S(U(3)\times U(3))}$& $Sp(20)$ & $
{\underline {\bf 20}}$ \\
~~&~~&~~& ~~\\
\hline
~~&~~&~~& ~~\\
$15$ & $\o{SO^\star(12)}{SU(6)\times U(1)}$& $Sp(32)$ & $
{\underline {\bf 32}}$ \\
~~&~~&~~& ~~\\
\hline
~~&~~&~~& ~~\\
$27$ & $\o{E_{7(-6)}}{E_6\times SO(2)}$& $Sp(56)$ & $
{\underline {\bf 56}}$ \\
~~&~~&~~& ~~\\
\hline
\end{tabular}
\end{center}
\end{table}
\vfill
\eject
It goes without saying that for homogeneous special manifolds the two
constructions of the period matrix, that provided by the master
formula in eq.~\eqn{masterformula} and that given by eq.~\eqn{intriscripen}
must agree. In   subsection \ref{st2n} we   verify it in the case
of the manifolds
${\cal ST}[2,n]$ that correspond to the second infinite family of
homogeneous special manifolds displayed in table
~ \ref {homospectable}.
\par
Anyhow, since special geometry guarantees the existence of a kinetic
period matrix with the correct covariance property it is evident that
to each special manifold we can associate a duality covariant bosonic
Lagrangian of the type considered in eq.~\eqn{gaiazuma}. However special
geometry contains more structures than just the period matrix ${\cal
N}$ and the scalar metric $g_{ij^\star}$.
All the other items of the construction do have
a place and play an essential role in the supergravity Lagrangian and
the supersymmetry transformation rules. We shall have to manipulate
them also in discussing the BPS black--holes.
\subsection{Special K\"ahler manifolds: the issue of special
coordinates}
 So far no privileged coordinate system has been chosen on the base
 K\"ahler manifold ${\cal M}$ and no mention has been made
 of the holomorphic prepotential $F(X)$ that is ubiquitous in the $N=2$
 literature. The simultaneous avoidance
 of privileged coordinates and of the prepotential is not
 accidental. Indeed, when the definition of special K\"ahler
 manifolds  is given in intrinsic terms, as we did in the previous
 subsection, the holomorphic prepotential $F(X)$ can be dispensed
 of. Whether a prepotential $F(X)$ exists or not
 depends on the choice of a symplectic gauge which is
 immaterial in the abelian theory but not in the gauged one.
 Actually, in the local case, it appears that some  physically
 interesting cases are precisely instances where $F(X)$ does not
 exist. On the contrary the prepotential $F(X)$ seems to be
 a necessary ingredient in the tensor calculus constructions of
 $N=2$ theories that for this reason are not completely general.
 This happens because tensor calculus uses special coordinates
 from the very start. Let us then see how the notion of  $F(X)$
 emerges if we resort to special coordinate systems.
 \par
 Note that under a K\"ahler transformation ${\cal K} \, \to \,
 {\cal K} + f(z) +{\bar f}({\bar z})$ the holomorphic section transforms,
 in the local case, as $\Omega \, \to \, \Omega \, e^{-f}$, so that
 we have $X^\Lambda \, \to \, X^\Lambda \, e^{-f}$. This means that,
 at least locally, the upper half of $\Omega$ (associated with
 the electric field strengths) can be regarded as a set $X^\Lambda$
 of homogeneous coordinates on ${\cal M}$, provided that the
 jacobian matrix
 \begin{equation}
e^I_{i}(z) = \partial_{i} \left ( {\o{X^I}{X^0}}\right ) \quad ;
\quad a=1,\dots ,n
\label{nonsingcoord}
\end{equation}
is invertible. In this case, for the lower part of the symplectic
section $\Omega$ we obtain $F_\Lambda = F_\Lambda(X)$. Recalling
eq.s~(\ref{ortogo}), in particular:
\begin{equation}
0 \, =\,  \langle\,  V \, \vert \, U_i \, \rangle
\, =\, X^\Lambda \, \partial_{i} F_\Lambda - \partial_{i} X^\Lambda \,
F_\Lambda
\label{ortogonalcosa}
\end{equation}
we obtain:
\begin{equation}
X^\Sigma \, \partial_ \Sigma F_ \Lambda (x) \, = \, F_ \Lambda (X)
\label{nipiolbuit}
\end{equation}
so that we can conclude:
\begin{equation}
F_ \Lambda (X) \, = \, {\o{\partial}{\partial X^\Lambda} } F(X)
\label{lattedisoia}
\end{equation}
where $F(X)$ is a homogeneous function of degree 2 of the homogeneous
coordinates $X^\Lambda$. Therefore,when the determinant of the
Jacobian (\ref{nonsingcoord}) is non vanishing,
we can use the {\it special coordinates}:
\begin{equation}
t^I \, \equiv \, {\o{X^I}{X^0}}
\label{speccoord}
\end{equation}
and the whole geometric structure can be derived by a single
holomorphic prepotential:
\begin{equation}
{\cal F}(t) \, \equiv \,  (X^0)^{-2} F(X)
\label{gianduia}
\end{equation}
In particular, eq.(\ref{specpot}) for
the K\"ahler potential becomes
\begin{equation}
{\cal K}(t, \bar t) = -\mbox{log} \,\,  {\rm i}\Bigl [
2 \left ( {\cal F} - \bar {\cal F} \right )  -
\left ( \partial_I {\cal F}
+ \partial_{{I^\star}} \bar {\cal F} \right )\left  ( t^I -\bar t^{I^\star}
\right ) \Bigr ]
\label{vecchiaspec}
\end{equation}
while eq.(\ref{sympinvloc}) for the magnetic moments simplifies into
\begin{equation}
W_{IJK}= \partial_I \partial_J \partial_K {\cal F}(t)
\label{yuktre}
\end{equation}
Finally we note that in the rigid case the Jacobian from a generic
parametrization to special coordinates
\begin{equation}
e^I_{i}(z) = \partial_{i} \left ( {\o{X^I}{X^0}}\right )=A+B{\bar{\cal  N}}
\label{nonsingcoord2}
\end{equation}
 cannot have zero eigenvalues, and therefore the function $F$ always
 exist. In this case the matrix $\bar\cN$ coincides with $\frac{\del^2 F}
{\del X^I \del X^J}$.
\subsection{The special geometry of the ${\cal ST}[2,n]$
manifolds}
\label{st2n}
When we studied the symplectic embeddings of the ${\cal ST}[m,n]$
manifolds, defined by eq.(\ref{stmanif}), a study that lead us to
the general formula in eq.(\ref{maestrino}), we remarked that the
subclass ${\cal ST}[2,n]$ constitutes a family of special K\"ahler
manifolds: actually a quite relevant one. Here we survey the
special geometry of this class.
\par
Consider a standard parametrization of the $SO(2,n)/SO(2)\times
SO(n)$ manifold, like for instance that in eq.(\ref{somncoset}).
In the $m=2$ case we can introduce a canonical complex structure
on the manifold by setting:
\begin{equation}
\Phi^\Lambda (X)\,  \equiv \,{\o{1}{\sqrt{2}}} \, \left (
L^\Lambda_{\phantom{\Lambda}0 } + {\rm i} \,
L^\Lambda_{\phantom{\Lambda}1 } \right )\qquad ;\qquad
 \left ( \Lambda=0,1,a \quad a=2,\dots,n+1  \right )
\label{cicciophi}
\end{equation}
The relations satisfied by the upper two rows of the coset
representative (consequence of $L(X)$ being pseudo--orthogonal with
respect to metric $\eta_{\Lambda\Sigma}={\rm diag}(+,+,-,\dots,-)$):
\begin{equation}
L^\Lambda_{\phantom{\Lambda}0 } \, L^\Sigma_{\phantom{\Lambda}0 } \,
\eta_{\Lambda\Sigma} \, =\,  1  \quad ; \quad
L^\Lambda_{\phantom{\Lambda}0 } \, L^\Sigma_{\phantom{\Lambda}1 } \,
\eta_{\Lambda\Sigma} \, =\,  0  \quad ; \quad
L^\Lambda_{\phantom{\Lambda}1 } \, L^\Sigma_{\phantom{\Lambda}1 } \,
\eta_{\Lambda\Sigma} \,= \, 1
\label{pseudodionigi}
\end{equation}
can be summarized into the complex equations:
\begin{equation}
{\bar \Phi}^\Lambda  \,
\Phi^\Sigma
\eta_{\Lambda\Sigma} \, =\,  1  \quad ; \quad
\Phi^\Sigma
\eta_{\Lambda\Sigma} \, =\,  0
 \label{pseudomichele}
\end{equation}
Eq.s (\ref{pseudomichele}) are solved by posing:
\begin{equation}
\Phi^\Lambda \, = \,{\o{X^\Lambda}{\sqrt{{\bar X}^\Lambda \,
X^\Sigma \, \eta_{\Lambda\Sigma}}}}
\label{fungomarcio}
\end{equation}
where $X^\Lambda$ denotes any set of complex parameters, determined
up to an overall multiplicative constant and satisfying the
constraint:
\begin{equation}
 X^\Lambda  \,
X^\Sigma
\eta_{\Lambda\Sigma} \, = \, 0
\label{ciliegia}
\end{equation}
In this way we have proved the identification, as differentiable
manifolds, of the coset space $G/H$ where $G=SO(2,n)$ and
$H=SO(2)\times SO(n)$ with the vanishing locus of the quadric in
eq.(\ref{ciliegia}). Taking any holomorphic solution of
eq.(\ref{ciliegia}), for instance:
\begin{equation}
 X^{\Lambda}(y)  \, \equiv \,
\left(\begin{array}{c} 1/2\hskip 2pt (1 + y^2) \\
{\rm i}/2\hskip 2pt (1 -
y^2)\\ y^{a}\end{array}\right)
\label{calabivise}
\end{equation}
where $y^a$ is a set of $n$ independent complex coordinates,
inserting it into eq.(\ref{fungomarcio}) and comparing with
eq.(\ref{cicciophi}) we obtain the relation between whatever
coordinates we had previously used to write the coset representative
$L(X)$ and the complex coordinates $y^a$. In other words
we can regard the matrix $L$ as a function of the $y^a$ that
are named the Calabi Visentini coordinates \cite{calvis}.
\par
Consider in addition the {\it axion--dilaton} field  $S$
that parametrizes the $SU(1,1)/U(1)$ coset according with
eq.(\ref{su11coset}). The special geometry of the manifold
${\cal ST}[2,n]$ is completely specified by writing the
holomorphic symplectic section $\Omega$ as follows
(\cite{CDFVP}):
\par
\begin{equation}
\label{so2nssec}
\Omega (y,S) \, = \, \left (\matrix{
X^{\Lambda}  \cr F_{\Lambda}\cr }\right)
\, = \, \left ( \matrix { X^{\Lambda}(y) \cr
{\cal  S }\, \eta_{\Lambda\Sigma}
X^{\Sigma}(y) \cr } \right  )
\end{equation}
Notice that  with the above choice, it is
not possible to describe $F_\Lambda$ as derivatives of any
prepotential. Yet everything else can be calculated utilizing
the formulae we presented in the text.
The K\"ahler potential is:
\begin{equation} \label{so2nkpotskn}
\cK \, =\, \cK_1(S)+\cK_2(y)\,
 =\, -\mbox{log} \, \left[ \, {\rm i} \, (\bar S - S)\right]
 -\log X^T\eta X
\end{equation}
The K\"ahler metric  is  block diagonal:
\begin{equation}
\label{so2n5}
g_{ij^\star} \, = \,
\left(\begin{array}{cc}g_{S\bar S} & {\bf
0}\\ {\bf 0} & g_{a \bar b}\end{array}\right)
\qquad \quad
\left\{\begin{array}{l} g_{S\bar S} = \partial_S \partial_{\bar S} \cK_1 =
{-1\over (\bar S - S)^2}\\ g_{a\bar b}(y)=
\partial_{a}\partial_{\bar b} \cK_2 \end{array}\right.
\end{equation}
as expected.
The anomalous magnetic moments-Yukawa
couplings $C_{ijk}$ ($i=S, a$) have a very simple expression
in the chosen coordinates:
\begin{equation}
\label{so2n6bis} C_{Sab} = -{\rm exp}[{\cal K}] \,
\delta_{ab},
\end{equation}
all the other components being zero.
\par
Using the definition of the {\it period matrix} given
in eq.(\ref{intriscripen}) we obtain
\begin{equation} \label{so2n10} \cN_{\Lambda\Sigma} = (S -
\bar S) {{X_{\Lambda} {\bar X}_{\Sigma} + {\bar X}_{\Lambda} X_{\Sigma}
}\over {{\bar X}^T\eta X}} + \bar S\eta_{\Lambda\Sigma}.
\label{discepolo}
\end{equation}
In order to see that eq.(\ref{discepolo}) just coincides
with eq.(\ref{maestrino}) it suffices to note that as a
consequence of its definition (\ref{cicciophi}) and of the
pseudo--orthogonality of the coset representative $L(X)$,
the vector $\Phi^\Lambda$ satisfies the following
identity:
\begin{equation}
\Phi^\Lambda \, {\bar \Phi}^\Sigma + \Phi^\Sigma \, {\bar
\Phi}^\Lambda  \, = \,
  {\o{1}{2}} \,  L^{\Lambda}_{\phantom{\Lambda}\Gamma} \,
L^{\Sigma}_{\phantom{\Sigma}\Delta} \left ( \delta^{\Gamma\Delta} +
\eta^{\Gamma\Delta} \, \right )
\label{meraviglia2}
\end{equation}
Inserting eq.(\ref{meraviglia2}) into eq.(\ref{discepolo}),
formula (\ref{maestrino}) is retrieved.
\par
This completes the proof that the choice (\ref{so2nssec}) of
the special geometry holomorphic section corresponds to the
symplectic embedding (\ref{ortoletto}) and (\ref{ortolettodue})
of the coset manifold ${\cal ST}[2,n]$. In this symplectic
gauge the symplectic transformations of the isometry group
are the simplest possible ones and  the entire group $SO(2,n)$
is represented by means of {\it classical} transformations
that do not mix electric fields with magnetic fields.
The disadvantage of this basis, if any,
is that there is no holomorphic prepotential. To find
an $F(X)$ it suffices to make a symplectic rotation to
a different basis.
\par
 If we set:
\begin{eqnarray}
X^1 = {{1}\over{2}} (1 + y^2) &=& -{{1}\over{2}} (1 - \eta_{ij} t^{i}
 t^{j})\nonumber\\
X^2 = i {{1}\over{2}} (1 - y^2) &=& t^2\nonumber\\
X^a = y^a &=& t^{2+a} \quad
{a=1,\dots,n-1}\nonumber\\
X^{a=n} = y^{n} &=& {{1}\over{2}} (1+\eta_{ij} t^{i} t^{j})
\end{eqnarray}
where
\begin{equation}
\eta_{ij} = {\rm diag} \left ( + , -, \dots , -\right ) ~
i,j=2,\dots , n +1
\end{equation}
then we can show that $\exists \, {\cal C}\, \in Sp(2 n+2,\IR)$ such that:
\begin{equation}
 {\cal C} \,
\left (
       \matrix {
                 X^\Lambda \cr
                 S\eta_{\Lambda\Sigma} \, X^\Lambda \cr
                }
\right ) \, =\,  \exp[\varphi(t)]\,
\left (
\matrix{
         1 \cr
         S \cr
           t^{i} \cr
          2 \, {\cal F} - t^{i}
         {{\partial}\over{\partial t^{i}}}{\cal F}
          - S{{\partial}\over S}{\cal F} \cr
           S{{\partial}\over S}{\cal F}\cr
       {{\partial}\over{\partial t^{i}}}{\cal F}\cr
      }
\right )
\end{equation}
with
\begin{eqnarray}
{\cal F}(S,t) &=&{{1}\over{2}} \,
S\,  \eta_{ij} t^{i} t^{j}
= {{1}\over{2}} \, d_{IJK} t^{I} t^{J} t^{K}
\nonumber\\
t^{1} &=&S\nonumber\\
d_{IJK}&=&\cases{
d_{1jk}=\eta_{ij}\cr
0 ~~\mbox{otherwise}\cr}
\label{vecchiume}
\end{eqnarray}
and
\begin{equation}
W_{IJK} = d_{IJK} = {{\partial^3{\cal F}(S, t^{i})}
\over{\partial t^{I}\partial
t^{J} \partial t^{K} } }
\end{equation}
This means that in the new basis
the symplectic holomorphic section ${\cal C}\Omega$ can be derived
from the following cubic prepotential:
\begin{equation}
F(X) \, =\,{\o{1}{3!}}\, {\o{ d_{IJK} \, X^I \, X^J \, X^K}{X_0}}
\label{duplocubetto}
\end{equation}
For instance in the case $n=1$  the matrix which does such a job
is:
\begin{equation}
{\cal C}=\left (\matrix{ 1 & 0 & -1 & 0 & 0 & 0 \cr 0 & 0 & 0 & 1
& 0 & 1 \cr 0 & -1
   & 0 & 0 & 0 & 0 \cr 0 & 0 & 0 & {1\over 2} & 0 & -{1\over 2} \cr
  -{1\over 2} & 0 & -{1\over 2} & 0 & 0 & 0 \cr 0 & 0 & 0 & 0 & -1
   & 0 \cr  }\right )
\end{equation}
\markboth{BPS BLACK HOLES IN SUPERGRAVITY: CHAPTER 3}
{3.5 THE BOSONIC LAGRANGIAN OF N=2 SUPERGRAVITY}
\section{The bosonic lagrangian and the supersymmetry transformation
rules of N=2 supergravity}
\setcounter{equation}{0}
\label{n2bosla}
Using   special  K\"ahler geometry as described in the previous
section we can now write both the bosonic action and the
supersymmetry transformation rules of the fermions pertaining to
$N=2$ supergravity, namely the ingredients needed to write down the
equations defining N=2 BPS black--holes.
\par
In the complex notation appropriate to special geometry and using
the definitions \eqn{selfduals} the action can be written as follows
\begin{eqnarray}
{\cal L}_{ungauged}^{SUGRA \vert Bose} & =&
\sqrt{-g}\Bigl  [ \,  R[g]
\, + \, g_{i {j^\star}}(z,\bar z )\, \partial^{\mu} z^i \,
\partial _{\mu} \bar z^{j^\star} \,
\nonumber\\
& & + \,{\rm i} \,\left(
\bar {\cal N}_{\Lambda \Sigma} {\cal F}^{- \Lambda}_{\mu \nu}
{\cal F}^{- \Sigma \vert {\mu \nu}}
\, - \,
{\cal N}_{\Lambda \Sigma}
{\cal F}^{+ \Lambda}_{\mu \nu} {\cal F}^{+ \Sigma \vert {\mu \nu}} \right )
\, \Bigr ]
\label{ungausugra}
\end{eqnarray}
The fermion fields of the theory are of two kinds:
\begin{enumerate}
\item The gravitino $\psi_{A\mu }$
\item The gaugino $\lambda^{iA}$
\end{enumerate}
The gravitino $\psi_{A\mu }$   is an $SU(2)$ doublet ($A=1,2$) of
spin $3/2$ spinor--vectors.
Here the group $SU(2)$ is the automorphism group of the
$N=2$ supersymmetry algebra, usually called R--symmetry. Following
the conventions of \cite{castdauriafre} and \cite{n2standa}, the position
of the $SU(2)$ index, whether upper or lower, is related to the
chirality projection on the fields. For the gravitino doublet we have
left chirality:
\begin{equation}
 \gamma_5 \, \psi_{A\mu } = \psi_{A\mu }
 \label{leftpsi}
\end{equation}
The complex conjugate doublet,
transforming under the conjugate representation
of $SU(2)$ contains the right chirality projection of the same Majorana
field:
\begin{equation}
 \gamma_5 \, \psi^A_{ \mu } = -\psi^A_{\mu }
 \label{rightpsi}
\end{equation}
On the other hand the gauginos $\lambda^{iA}$,   are   $SU(2)$
doublets of spin 1/2 fields that carry also a contravariant world-index
of the special K\"ahler manifold. In other words the gauginos,
in addition to being sections of the spin--bundle on   space--time
are also sections of the holomorphic tangent bundle $T{\cal SM}$.
The chirality of the gaugino with upper $SU(2)$ doublet index is left:
 \begin{equation}
 \gamma_5 \, \lambda^{iA} =  \lambda^{iA}
 \label{rightla}
\end{equation}
The right chirality projection
$\lambda_{A}^{i^\star}=-\gamma_5\,\lambda_{A}^{i^\star} $ transforms
under $SU(2)$ in the same way as the left-projection of the gravitino
but it is a section of the antiholomorphic tangent bundle
${\bar T}{\cal SM}$.
\par
Correspondingly in a purely bosonic background the supersymmetry
transformation rules of the fermion fields are the following ones:
\begin{eqnarray}
\delta\,\Psi _{A \mu} &=& {\cal D}_{\mu}\,\epsilon _A\,
+ \epsilon_{AB} \, T^-_{\mu \nu}\, \gamma^{\nu}\epsilon^B
 \label{trasfgrav} \\
\delta\,\lambda^{iA}&=&
{\rm i}\,    \nabla _ {\mu}\, z^i  \, \gamma^{\mu} \epsilon^A
+G^{-i}_{\mu \nu} \gamma^{\mu \nu} \epsilon _B \epsilon^{AB}
\label{gaugintrasfm}
\end{eqnarray}
where the local supersymmetry parameter is also split into
chirality projections transforming in the doublet or conjugate
doublet representation of $SU(2)$:
\begin{equation}
\gamma_5 \, \epsilon_A = \epsilon_A  \quad  ; \quad
\gamma_5 \, \epsilon^A = -\epsilon^A
\label{eppro}
\end{equation}
and where we have defined:
\begin{eqnarray}
T^-_{\mu\nu} &=& 2{\rm i}
 \left(\im {\cal N}\right)_{\Lambda\Sigma} L^{\Sigma}
 { {F}}_{\mu\nu}^{\Lambda -}  \label{T-def}\\
T^+_{\mu\nu} &=& 2 {\rm i}
\left (\im {\cal N}\right)_{\Lambda\Sigma} {\bar L}^{\Sigma}
 { {F}}_{\mu\nu}^{\Lambda +}
\label{T+def} \\
G^{i-}_{\mu\nu} &=& - g^{i{{j}^\star}}
 \bar f^\Gamma_{{j}^\star}
\left (\im  {\cal N}\right)_{\Gamma\Lambda} \,
  {  {F}}^{\Lambda -}_{\mu\nu}
\label{G-def}\\
G^{{{i}^\star}+}_{\mu\nu} &=& - g^{{{i}^\star}j} f^{\Gamma}_j
\left (\im {\cal N}\right)_{\Gamma\Lambda} \,
  { {F}}^{\Lambda +}_{\mu\nu}  \label{G+def}
\end{eqnarray}
The scalar field dependent combinations of fields strengths appearing
in the fermion supersymmetry transformation rules that we listed above
in eq.s \eqn{T-def}\eqn{T+def},\eqn{G-def},\eqn{G+def} have a
profound meaning and play a key role in the physics of BPS
black--holes.
The combination $T^-_{\mu\nu}$ is named the graviphoton field
strength and its integral over a $2$--sphere at infinity gives the
value of {\it the central charge $Z$} of the $N=2$ supersymmetry algebra.
The combination $G^{i-}_{\mu\nu}$ is named the matter field strength.
Evaluating its integral on a $2$--sphere at infinity one obtains the
so called {\it matter charges} $Z^i$. These are the covariant
derivatives of the central charge $Z$ with respect to the {\it
moduli}, namely the vector multiplet scalars $z^i$.
\par
This will become
clear in our general discussion of the $N=2$ BPS black holes
that we are finally in a position to address, having clarified all
the necessary geometric ingredients.
\section{$N=2$ BPS black holes: general discussion}
\markboth{BPS BLACK HOLES IN SUPERGRAVITY: CHAPTER 3}
{3.6 $N=2$ BPS BLACK HOLES: GENERAL DISCUSSION}
\setcounter{equation}{0}
\label{n2bpsh}
In this section we consider the general properties of   BPS saturated
black--holes in the context of $N=2$ supergravity. As our analysis will
reveal these properties are completely rooted in the special K\"ahler
geometric structure of the mother theory. In particular the entropy
of the black--hole is related to the central charge, namely to the
integral of the graviphoton field strength evaluated for very special
values of the scalar fields $z^i$. These special values, named the
{\it fixed scalars} $z^i_{fix}$, are functions solely of the electric
and magnetic charges $\{ q_ \Sigma,p^\Lambda\}$
of the hole  and are attained by the scalars
$z^i(r)$ at the hole-horizon $r=0$.
\par
To illustrate how this happens let us finally go back to the ansatz
\eqn{ds2U} for the metric and to the ansatz \eqn{flambda}
for the vector field strengths. It is convenient to rephrase the same
ansatz in the complex formalism well-adapted to the $N=2$ theory.
To this effect we begin by constructing a
2--form which is {\it anti--self--dual} in the background of the
metric \eqn{ds2U} and whose integral on the $2$--sphere at
infinity $ S^2_\infty$ is normalized to $ 2 \pi $. A short
calculation yields:
\begin{eqnarray}
E^-   &=&   \mbox{i} \frac{e^{2U(r)}}{r^3} \, dt \wedge {\vec x}\cdot
d{\vec x}  +  \frac{1}{2}   \frac{x^a}{r^3} \, dx^b   \wedge
dx^c   \epsilon_{abc} \nonumber\\
2 \, \pi  &=&    \int_{S^2_\infty} \, E^-
\label{eaself}
\end{eqnarray}
and with a little additional effort one obtains:
\begin{equation}
E^-_{\mu\nu} \, \gamma^{\mu\nu}  = 2 \,\mbox{i} \frac{e^{2U(r)}}{r^3}  \,
\gamma_a x^a \, \gamma_0 \, \frac{1}{2}\left[ {\bf 1}+\gamma_5 \right]
\label{econtr}
\end{equation}
which will prove of great help in the unfolding of the supersymmetry
transformation rules.
\par
Next, introducing the following complex combination of the magnetic
charge $ p^\Lambda $ and of the radial function $\ell^\Sigma(r)$
defined by eq. \eqn{flambda}:
\begin{equation}
\label{tlamb}
t^\Lambda(r)\,=\, 2\pi(p^\Lambda+{\rm i}\ell ^\Lambda (r))\nonumber\\
\end{equation}
we can rewrite the ansatz \eqn{flambda} as:
\begin{eqnarray}
\label{strenghtsans}
F^{-\vert \Lambda}\,&=&\, \frac{t^\Lambda}{4\pi}E^-\nonumber\\
\end{eqnarray}
and we retrieve the original formulae from:
\begin{equation}
\begin{array}{rcccl}
F^{\Lambda}\,&=&\,2{\rm Re}F^{-\vert \Lambda}&= &
\frac{p^{\Lambda}}{2r^3}\epsilon_{abc}
x^a dx^b\wedge dx^c-\frac{\ell^{\Lambda}(r)}{r^3}e^{2\cal {U}}dt
\wedge \vec{x}\cdot d\vec{x} \\
\tilde{F}^{\Lambda}&=&-2{\rm Im}F^{-\vert \Lambda}& = &
-\frac{\ell^{\Lambda}(r)}{2r^3}\epsilon_{abc}
x^a dx^b\wedge dx^c
-\frac{p^{\Lambda}}{r^3}e^{2\cal {U}}dt\wedge \vec{x}\cdot d\vec{x}
\\
\end{array}
\label{fedfd}
\end{equation}
Before proceeding further it is convenient to define the electric and
magnetic charges of the hole as it is appropriate in any
electromagnetic theory. Recalling the general form of the field
equations and of the Bianchi identities as given in \eqn{biafieq} we
see that the field strengths ${\cal F}_{\mu \nu}$ and
${\cal G}_{\mu \nu}$ are both closed 2-forms, since their duals are
divergenceless. Hence we  can invoke Gauss theorem and claim that
their integral on a closed space--like $2$--sphere does not depend on
the radius of the sphere. These integrals are the electric and
magnetic charges of the hole that, in a quantum theory, we expect to
be quantized. We set:
\begin{eqnarray}
q_\Lambda & \equiv & \frac{1}{4 \pi} \,
\int_{S^2_\infty} \,G_{\Lambda\vert \mu\nu}\,
dx^\mu \wedge dx^\nu
\label{holele}\\
p^\Sigma & \equiv & \frac{1}{4 \pi} \,
\int_{S^2_\infty} \,F^{\Sigma}_{ \mu\nu}\,\,
dx^\mu \wedge dx^\nu
\label{holmag}
\end{eqnarray}
If rather than the integral of $G_{\Lambda }$
we were to calculate the integral of ${\tilde F}^\Lambda$
which is not a closed form we would obtain a function of the radius:
\begin{eqnarray}
4\pi \ell^\Lambda (r)\,&=&\,-\int_{S^2_r}\tilde{F}^{\Lambda}\,=\,2
{\rm Im}t^\Lambda
\label{cudierre}
\end{eqnarray}
Consider now the supersymmetry transformations rules \eqn{trasfgrav},
\eqn{gaugintrasfm} and write the BPS condition:
\begin{eqnarray}
0 &=& {\cal D}_{\mu}\,\xi _A\,
+ \epsilon_{AB} \, T^-_{\mu \nu}\, \gamma^{\nu}\xi^B
 \label{kigrav} \\
0 &=&
{\rm i}\,    \nabla _ {\mu}\, z^i  \, \gamma^{\mu} \xi^A
+G^{-i}_{\mu \nu} \gamma^{\mu \nu} \xi _B \epsilon^{AB}
\label{kigaug}
\end{eqnarray}
where the killing spinor $\xi _A(r)$ satisfies eq.\eqn{kispiro}
and is of the form of a single radial function times a constant spinor:
\begin{eqnarray}
\xi_A (r)\,&=&\,e^{f(r)} \chi _A~~~~~~~~~\chi_A=\mbox{constant}\nonumber\\
\gamma_0 \chi_A\,&=&\,\pm {\rm i}\epsilon_{AB}\chi^B
\label{quispi}
\end{eqnarray}
Inserting eq.s \eqn{T-def},\eqn{G-def},\eqn{quispi} into eq.s\eqn{kigrav},
\eqn{kigaug} and using  the result \eqn{econtr}, with a little
work we obtain the first order differential equations:
\begin{eqnarray}
\frac{dz^i}{dr}\, &=&\, \mp \left(\frac{e^{U(r)}}{4\pi r^2}\right)
g^{ij^\star}\bar{f}_{j^\star}^\Lambda
({\cal N}-\bar{{\cal N}})_{\Lambda\Sigma}t^\Sigma\,=\,\nonumber\\
&&\mp \left(\frac{e^{U(r)}}{4\pi r^2}\right) g^{ij^\star}
\nabla_{j^\star}\bar{Z}(z,\bar{z},{p},{q}) \label{zequa}\\
\frac{dU}{dr}\, &=&\,\mp \left(\frac{e^{U(r)}}{r^2}\right)
(M_\Sigma {p}^\Sigma-
L^\Lambda {q}_\Lambda)\,=\,
\mp \left(\frac{e^{U(r)}}{r^2}\right)Z(z,\bar{z},{p},{q})
\label{Uequa}
\end{eqnarray}
where \begin{enumerate}
\item ${\cal N}_{\Lambda\Sigma}(z,\bar{z})$ is the period matrix  of
special geometry defined by eq.\eqn{etamedia}
\item The vector $V= \left(L^\Lambda(z,\bar{z}),
M_\Sigma(z,\bar{z})\right)$ is the
covariantly holomorphic section of
the symplectic bundle entering the definition
of a Special K\"ahler manifold (see eq. \eqn{covholsec}
\item
\begin{equation}
 Z(z,\bar{z},{p},{q}) \equiv \left( M_\Sigma {p}^\Sigma-
L^\Lambda {q}_\Lambda\right)
\label{zentrum}
\end{equation}
is the local realization on the scalar manifold ${\cal SM}$ of
the central charge of the $N=2 $ superalgebra
\item
\begin{equation}
 Z^i(z,\bar{z},{p},{q}) \equiv g^{ij^\star}\nabla_
{j^\star}\bar{Z}(z,\bar{z},{p},{q})
\label{zmatta}
\end{equation}
are the central charges associated with the matter vectors,
the so--called matter central charges
\end{enumerate}
To obtain eqs. (\ref{zequa}),(\ref{Uequa})
we made use of  the following two properties :
\begin{eqnarray}
0\,&=&\, \bar{h}_{j^\star\vert\Lambda}
t^{\star \Sigma}-
\bar{f}_{j^\star}^\Lambda{\cal N}_{\Lambda\Sigma}t^{\star \Sigma}\nonumber\\
0\,&=&\, M_\Sigma t^{\star \Sigma}-L^\Lambda {\cal N}_{\Lambda\Sigma}
t^{\star \Sigma}
\end{eqnarray}
which are a direct consequence of the definition \eqn{etamedia}
of the period matrix.
The electric charges $\ell^\Lambda (r)$ defined
in (\ref{cudierre}) are {\it moduli dependent} charges
which are functions of the radial direction through
the moduli $z^i$. On the other hand, the
{\it moduli independent} electric charges
$q_\Lambda$ in eqs. (\ref{Uequa}),(\ref{zequa}) are those defined
by eq.\eqn{holele} which, together with $p^\Lambda$
fulfil a Dirac quantization condition. Their definition
allows them to be expressed in terms of
of $t^\Lambda (r)$ as follows:
\begin{equation}
q_\Lambda\,=\,
\frac{1}{2\pi}{\rm Re}({\cal N}(z(r),\bar{z}(r))t (r))_\Lambda
\label{ncudierre}
\end{equation}
Equation (\ref{ncudierre}) may be inverted in order
to find the moduli dependence of
$\ell_\Lambda (r)$.
The independence of $q_\Lambda$
on $r$ is a consequence of one of the Maxwell's equations:
\begin{equation}
\partial_a \left(\sqrt{-g}\tilde{G}^{a0\vert \Lambda}(r)\right)\,=\,0\Rightarrow
\partial_r {\rm Re}({\cal N}(z(r),\bar{z}(r))t(r))^\Lambda\,=\,0
\label{prione}
\end{equation}
\subsection{Fixed scalars and the entropy}
\label{fissato}
In this way we have reduced the  condition that the black-hole should be
a BPS saturated state to a pair of first order differential
equations for the metric scale factor $U(r)$ and for the
scalar fields $z^i(r)$. To obtain explicit solutions one should
specify the special K\"ahler manifold one is working with, namely
the specific Lagrangian model. There are, however, some very
general and interesting conclusions that can be drawn in a
model--independent way. They are just consequences of the fact that
these BPS conditions are {\it first order differential equations}.
Because of that there are fixed points
(see the papers \cite{ferkal2,ferkal4,strom3})
namely values either of the metric
or of the scalar fields which, once attained in the evolution
parameter $r$ (= the radial distance ) persist indefinitely. The
fixed point values are just the zeros of the right hand side in
either of the coupled eq.s \eqn{Uequa} and \eqn{zequa}. The fixed
point for the metric equation is $r=\infty $,
which corresponds to its
asymptotic flatness.
The fixed point for the moduli is $r=0$.  So, independently
from the initial data at  $r=\infty$ that determine the details
of the evolution,  the scalar fields flow into
their fixed point values at $r=0$, which, as we will show,
turns out to be a horizon. Indeed in the vicinity of $r=0$ also the
metric takes the universal form of an $AdS_2 \, \times \, S^2$,
Bertotti Robinson metric.
\par
Let us see this more closely.
\par
To begin with we consider the equations determining the fixed point
values for the moduli and the universal form attained by the metric
at the moduli fixed point:
\begin{eqnarray}
0 &=& -g^{ij^\star} \, {\bar f}^\Gamma_{j^\star}
\left( \mbox{Im}{\cal N}\right)_{\Gamma\Lambda} \, t^{\Lambda }(0)
\label{zequato} \\
\frac{dU}{dr}\, & \cong &
\mp \left(\frac{e^{U(r)}}{r^2}\right)
Z\left (z_{fix},{\bar{z}}_{fix} ,{p},{q} \right)
\label{Uequato}
\end{eqnarray}
Multiplying eq.\eqn{zequato} by $f^\Sigma_i$ using the identity
\eqn{dsei} and the definition \eqn{zentrum} of the central charge
 we conclude that
at the fixed point the following condition is true:
\begin{equation}
0=-\frac{1}{2} \, \frac{t^\Lambda}{4\pi} \, -\, \frac{Z_{fix} \,
{\bar L}_{fix}^\Lambda}{8\pi}
\label{passetto}
\end{equation}
In terms of the previously defined
electric and magnetic charges (see eq.s \eqn{holele},\eqn{holmag},
\eqn{ncudierre})
eq.\eqn{passetto} can be rewritten as:
\begin{eqnarray}
p^\Lambda & = & \mbox{i}\left( Z_{fix}\,{\bar L}^\Lambda_{fix}
- {\bar Z}_{fix}\,L^\Lambda_{fix} \right)\\
q_\Sigma & = & \mbox{i}\left( Z_{fix}\,{\bar M}_\Sigma^{fix}
- {\bar Z}_{fix}\,M_\Sigma^{fix} \right)\\
Z_{fix} &=& M_\Sigma^{fix} \, p^\Sigma \, - L^\Lambda_{fix} \, q_\Lambda
\label{minima}
\end{eqnarray}
which can be regarded as algebraic equations determining the value
of the scalar fields at the fixed point as functions of the electric
and magnetic charges $p^\Lambda, q_\Sigma$:
\begin{equation}
L_{fix}^\Lambda = L^\Lambda(p,q) \, \longrightarrow \,
Z_{fix}=Z(p,q)=\mbox{const}
\label{ariacalda}
\end{equation}
\par
In the vicinity of the fixed point the differential equation for the
metric becomes:
\begin{equation}
\pm \, \frac{dU}{dr}=\frac{Z(p,q)}{  \, r^2} \, e^{U(r)}
\end{equation}
which has the approximate solution:
\begin{equation}
\exp[U(r)]\, {\stackrel{r \to 0}{\longrightarrow}}\,
\mbox{const} + \frac{Z(p,q)}{  \, r}
\label{approxima}
\end{equation}
Hence, near $r=0$   the metric \eqn{ds2U}
becomes of the Bertotti Robinson type (see eq.\eqn{bertrob} )
with Bertotti Robinson mass given by:
\begin{equation}
m_{BR}^2 = \vert  {Z(p,q)}  \vert^2
\label{brmass}
\end{equation}
In the metric \eqn{bertrob} the surface $r=0$ is light--like and
corresponds to a horizon since it is the locus where the
Killing vector generating time translations $\frac{\partial}{\partial t} $,
which is time--like at spatial infinity $r=\infty$, becomes
light--like. The horizon $r=0$ has a finite area given by:
\begin{equation}
\mbox{Area}_H = \int_{r=0} \, \sqrt{g_{\theta\theta}\,g_{\phi\phi}}
\,d\theta \,d\phi \, = \,  4\pi \, m_{BR}^2
\label{horiz}
\end{equation}
Hence, independently from the details of the considered model,
the BPS saturated black--holes in an N=2 theory have a
Bekenstein--Hawking entropy given by the following horizon area:
\begin{equation}
 \frac{\mbox{Area}_H}{4\pi} = \,     \vert Z(p,q) \vert^2
 \label{ariafresca}
\end{equation}
the value of the central charge being determined by eq.s
\eqn{minima}. Such equations can also be seen as the variational
equations for the minimization of the horizon area as
given by \eqn{ariafresca}, if the central charge is regarded
as a function of both the scalar fields and the charges:
\begin{eqnarray}
 \mbox{Area}_H (z,{\bar z})&=& \,  4\pi \, \vert Z(z,{\bar z},p,q) \vert^2
 \nonumber\\
 \frac{\delta \mbox{Area}_H }{\delta z}&=&0 \, \longrightarrow \,  z = z_{fix}
\end{eqnarray}
This view point will be pursued in the next chapter where we consider
the general properties of {\it supergravity central charges} and their
extremization
\chapter{The Structure of Supergravity Central Charges and the Black Hole
Entropy}
\label{chargcha}
\section{Introduction}
\markboth{BPS BLACK HOLES IN SUPERGRAVITY: CHAPTER 4}
{4.1 INTRODUCTION TO CENTRAL CHARGES}
\setcounter{equation}{0}
The present one is the central chapter in this series of
lectures. As we explained in the introduction the relevance
of extremal black holes for string theory is that  in the context
of a supersymmetric theory the extremality condition, that is the
coincidence of the two horizons, can be reinterpreted as the
statement that the black hole is a BPS saturated state where
the ADM mass is equal to the modulus of the supersymmetry central charge.
\par
Actually in $N$--extended supersymmetry the central charge is an
antisymmetric tensor $Z_{AB}$ that admits several skew eigenvalues (see
eq.\eqn{skewZ}) so that the discussion of the BPS condition becomes
more involved depending on the structure of these skew eigenvalues.
In this chapter we study the general properties of the supergravity
central charges and their relation with the horizon area, namely
with the black hole entropy. A first instance of such a relation was
already derived in the previous chapter in the context of $N=2$
supergravity where we obtained the fundamental relations
\eqn{ariafresca} and \eqn{ariacalda} expressing the horizon area
as the modulus square of the central charge evaluated  at the
extremum value of the scalars. In the present chapter we put
this $N=2$ result in a much more general perspective and we
illustrate its deep meaning. Indeed, after illustrating the general
structure and properties of the supergravity central charges,
we show that to any lagrangian of type \eqn{genact} we can associate
an effective one--dimensional field theory that governs the radial
dependence of the black hole metric and of the scalar fields. The
interactions in such an effective theory are determined by a
specific potential, {\it the geodesic potential} $V(\phi,{\vec Q})$
that depends on the scalar fields and on the electric and magnetic
charges of the hole. Finite horizon area black holes are
characterized by scalar fields that are regular on the horizon and
reach at $r=0$  extremum values of the geodesic potential. Actually the
horizon area and, hence, the black hole entropy is given by the
extremum value of the geodesic potential. When the theory is
supersymmetric the geodesic potential can be reexpressed as a
quadratic positive definite form in the supersymmetry central
charges. Hence extremizing the potential and finding the black hole
entropy means to extremize such a quadratic form in the central
charges. Using the formal structure of the supergravity central
charges and their invariance under the $U$--duality group
one can derive very general properties of the horizon area. Indeed
we can show that it is a topological invariant depending solely
on the vector ${\vec Q}$ of  {\it quantized electric and magnetic
charges} that is also the square root of a quartic
algebraic invariant of the $U$--duality group. This is probably the
deepest and most significant aspect of black--hole physics in the context
of supergravity. In the next chapter we shall illustrate the concepts
introduced in the present one through the explicit construction of
$N=8$ black holes. Here  we keep our discussion at a very
general level.
\section{Central Charges in $N$--extended $D=4$ supergravity}
\setcounter{equation}{0}
\label{cqnd4}
\markboth{BPS BLACK HOLES IN SUPERGRAVITY: CHAPTER 4}
{4.2 CENTRAL CHARGES IN EXTENDED SUPERGRAVITY}
As we have explained in table \ref{topotable}, for $N \ge 3$ the
scalar manifold of $D=4$ supergravity is always a homogeneous coset
manifold $G/H$. Furthermore, as we discussed in section
\ref{LL2}, the isometry group $G$   is symplectically
embedded into $Sp(2 {\bar n},\IR)$, the symbol ${\bar n}$ denoting
the total number of vector
fields contained in the theory that can either belong to the graviton
multiplet ({\it graviphotons}), or to the matter multiplets, if such
multiplets are allowed ($N \le 4$). Correspondingly
if $L(\phi)$ is the coset representative of $G/H$ in some representation
of $G$,
we denote by
\begin{equation}
\IL_{Sp}(\phi) =\left( \matrix {A & B \cr C & D \cr } \right)
\label{Lspes}
\end{equation}
 the same coset representative embedded in the
fundamental $2{\bar n}$ real representation of
  $Sp(2{\bar n},\IR)$.  Its blocks $A(\phi) ,B(\phi),C(\phi),D(\phi)$ can be constructed
  in terms of the original coset representative $L(\phi)$.
Performing the isomorphism transformation \eqn{isomorfo} we can
rather use a coset representative that is embedded in the fundamental
$2{\bar n}$ representation of the group
  $Usp({\bar n},{\bar n})$.   We name such a matrix $\IL_{Usp}(\phi)$ and
  for convenience in later manipulations we parametrize it as
  follows:
\begin{equation}
\IL_{Usp} = {1 \over \sqrt{2}}\pmatrix{f+{\rm i}h & \bar f+{\rm i}\bar h \cr
f-{\rm i}h &\bar f-{\rm i}\bar h \cr} =
  {\cal C} \IL_{Sp} {\cal C}^{-1}
  \label{defu}
\end{equation}
Explicitly the relation between the ${\bar n} \times {\bar n}$
complex matrices $f$ and $h$ defined by eq.\eqn{defu} and the real
blocks of the symplectic coset representative is written below:
\begin{eqnarray}
  f&=&{1 \over \sqrt{2}} (A-{\rm i} B) \nonumber\\
h &=&{1 \over \sqrt{2}} (C-{\rm i} D) \label{ABCDfh}
\end{eqnarray}
Explicitly the condition $ \IL_{Usp} \in Usp({\bar n}, {\bar n})$ reads as follows:
 \begin{equation}
\left\lbrace\matrix{{\rm i}(f^\dagger h - h^\dagger f) &=& \bfone \cr
(f^t  h - h^t f) &=& 0\cr} \right.
\label{specdef}
\end{equation}
The ${\bar n}\times {\bar n}$ sub--blocks of $\IL_{Usp}$ are submatrices $f,h$ which can be
decomposed with respect to the
isotropy subgroup $H_{Aut} \times H_{matter} \subset G$.
\par
Indeed in all supergravity theories $N \ge 3$, where the scalar
manifold is a homogeneous coset $G/H$, the isotropy subgroup $H$ has
the following direct product structure:
\begin{eqnarray}
H &=&  H_{Aut}  \, \otimes \, H_{matter}\nonumber\\
H_{Aut}& = & SU(N) \, \otimes \, U(1)
\label{HautH}
\end{eqnarray}
where $H_{Aut}$ is the automorphism group of the supersymmetry
$N$--extended algebra, usually called R--symmetry, while $H_{matter}$
is a group related to the structure of the matter multiplets one
couples to supergravity. Actually in $D=4$ there are just two cases:
$H_{matter} = SU(n)$ in $N=3$ and $H_{matter} = SO(n)$ in $N=4$, as
the reader can see by looking at table \ref{topotable}.
The decomposition we mentioned is the following one:
\begin{eqnarray}
  f&=& (f^\Lambda_{AB} , f^\Lambda _I) \nonumber\\
h&=& (h_{\Lambda AB} , h_{\Lambda I})
\label{deffh}
\end{eqnarray}
where $AB$ is a pair of indices in the antisymmetric representation of
$H_{Aut}= SU(N) \times U(1)$ and
$I$ is an index in the fundamental representation of $H_{matter}$.
Upper $SU(N)$ indices label objects in the complex conjugate representation of $SU(N)$,
as we have already emphasized while discussing the $N=2$ case (see for
example formula  \eqn{leftpsi} and following ones):
\begin{equation}
 (f^\Lambda_{AB})^* = f^{\Lambda AB}  \quad \mbox{and similarly for the other cases}
 \label{ABconj}
\end{equation}
Note that we can consider $(f^\Lambda_{AB}, h_{\Lambda AB})$ and
$(f^\Lambda_{I}, h_{\Lambda I})$
as symplectic sections of a $Sp(2{\bar n}, \IR)$ bundle over the $G/H$ base manifold.
We will prove in the following that this bundle is actually flat.
The real embedding given by $\IL_{Sp}$ is appropriate for duality transformations  of
${\cal F}^\pm$ and their duals $\cG^\pm$,
according to equations \eqn{dualrot}, while
the complex embedding in the matrix $\IL_{Usp}$ is appropriate in writing down
the supersymmetry transformation rules of the fermions and of the vectors.
The kinetic matrix $\cN$, given by  the  Gaillard--Zumino master formula
\eqn{masterformula}, can be rewritten as follows:
\begin{equation}
\cN= hf^{-1}, \quad\quad \cN = \cN^t
\label{nfh-1}
\end{equation}
Using (\ref{specdef})and (\ref{nfh-1}) we find
   \begin{equation}
   (f^t)^{-1} = {\rm i} (\cN - \bar \cN)\bar f
\end{equation}
that, more explicitly reads
 \begin{eqnarray}
  f_{AB \Lambda} &\equiv& (f^{-1})_{AB \Lambda} =
{\rm i} (\cN - \bar \cN)_{\Lambda\Sigma}\bar f^\Sigma_{AB}\\
f_{I \Lambda} &\equiv& (f^{-1})_{I \Lambda} =
{\rm i} (\cN - \bar \cN)_{\Lambda\Sigma}\bar f^\Sigma_{I}
\end{eqnarray}
Similarly to the $N=2$ case, the graviphoton field strengths and the
matter field strengths, whose integrals yield the central and matter
charges, are defined by the symplectic invariant, scalar field dressed,
combinations of $F^\Lambda_{\mu\nu}$ fields appearing in the fermion
supersymmetry transformation rules. In other words there exists a
generalization of formulae \eqn{trasfgrav},\eqn{gaugintrasfm} and
\eqn{T-def},\eqn{T+def},\eqn{G-def},\eqn{G+def} that we presently
describe.
In extended supergravity there are three kinds of fermions:
\begin{enumerate}
\item The gravitino $\psi_{A\mu}$ which carries an R--symmetry index
$A$ running in the fundamental of $SU(N)$, the chirality assignments
being as described in eq.s\eqn{leftpsi},\eqn{rightpsi}.
\item The dilatino  fields $\chi_{ABC}$, that are spin $1/2$ members
of the graviton multiplet (for $N \ge 3$) and transform in the three times
antisymmetric representation of $SU(N)$. Their  chirality
assignment is:
\begin{eqnarray}
\gamma_5 \, \chi_{ABC} & = & \chi_{ABC}\nonumber\\
\gamma_5 \, \chi^{ABC} & = & -\chi^{ABC}
\label{dilachir}
\end{eqnarray}
\item The gauginos $\lambda^{I}_A$ that are spin $1/2$ members of the
vector multiplets and transform in the fundamental representation of
the R--symmetry $SU(N)$. For $N=3,4$ they are also in the fundamental
representation of the $H_{matter}$ group, respectively $SU(n)$ or
$SO(n)$. In the $N=2$ case, where the scalar multiplet is not
necessarily a coset manifold $G/H$, the index $I$ is replaced by the
world index $i^\star$, since the gaugino transforms as a section of
the antiholomorphic tangent bundle (see eq.s \eqn{rightla})
\end{enumerate}
The supersymmetry  transformation laws of these fermion fields on bosonic
backgrounds are
\begin{eqnarray}
\label{psisusy}
\delta\psi_{A\mu}&=&\nabla_{\mu}\epsilon_A~
~-\frac{\rm 1}{4}\, T^{(-)}_{AB|\rho\sigma}\gamma^{\rho\sigma}\gamma_{\mu}\epsilon^{B}=0\\
\label{chisusy}
\delta\chi_{ABC}&=&~4 \mbox{\rm i} ~P_{ABCD|i}\partial_{\mu}\Phi^i\gamma^{\mu}
\epsilon^D-3T^{(-)}_{[AB|\rho\sigma}\gamma^{\rho\sigma} \,
\epsilon_{C]}=0
\end{eqnarray}
for the gravitino and dilatino, respectively, and
\begin{eqnarray}
\delta\lambda^I_A &=&a  P^I_{AB,i}\partial_a \phi^i \gamma^a
\epsilon^B + b \, T^{- I}_{ ab} \gamma^{ab}\epsilon_A
\label{lasusy}
\end{eqnarray}
for the gaugino, $a,b$ being some numerical coefficients, depending on the model
and unessential for the purpose of the present discussion.
Furthermore the dressed graviphoton and matter field strengths are
defined by the following equations, completely analogous to equations
( \eqn{T-def} and following ones):
\begin{eqnarray}
T^-_{AB}&=& {\rm i} (\bar f^{-1})_{AB \Lambda}F^{- \Lambda} =
f^\Lambda_{AB}(\cN - \bar \cN)_{\Lambda\Sigma}F^{-\Sigma}=
 h_{\Lambda AB} F^{-\Lambda} -f^\Lambda_{AB} \cG^-_\Lambda  \nonumber\\
  T^-_{I}&=&{\rm i} (\bar f^{-1})_{I\Lambda}F^{- \Lambda} =
  f^\Lambda_{I}(\cN - \bar \cN)_{\Lambda\Sigma}F^{-\Sigma}=
 h_{\Lambda I} F^{-\Lambda} - f^\Lambda_{I} \cG^-_\Lambda \nonumber\\
\bar T^{+ AB} &=& (T^-_{AB})^* \nonumber\\
\bar T^{+ I} &=& (T^-_{I})^* \label{gravi}
\end{eqnarray}
 Obviously, for $N>4$, $f _{\Lambda I}= T_I =0$. Note also that the
 right hand side of eq.s \eqn{gravi} makes it explicit the symplectic
 invariance of the dressed field strengths.
 \par
 Finally, in order to complete our description of the supersymmetry
 transformation rules, let us explain the meaning of the $H_{Aut}$
 antisymmetric tensors
 $P_{ABCD \vert i}, P_{AB \, I \vert i}$ (the index $i$ is a world index labelling
 $G/H$ coordinates).
 As it will be clear in the
 sequel these correspond to a decomposition of the $G/H$ coset manifold   vielbein
 into irreducible representations of the
 isotropy subgroup $H$. What is general in this decomposition is
 that, for all $N$--cases, we always obtain the required
 representations needed to make sense of the supersymmetry
 transformation rules \eqn{psisusy},\eqn{chisusy},\eqn{lasusy}.
 This happens because the
 choice of $G/H$ is completely determined by supersymmetry.
\par
 In  full analogy to what we did in the $N=2$ case we can now
 construct {\it field dependent central and matter charges}, by integrating
 the dressed field strengths $T_{AB} = T^+_{AB} + T^ -_{AB}  $ and  (for $N=3, 4$)
 $T_I = T^+_I + T^ -_I $ on a two--sphere of radius $r$.  As before
 these objects depend on $r$ since the corresponding field
 strength is not conserved. Note that as already discussed in the
 $N=2$ case we always assume spherical symmetry and hence purely
 radial dependence of the scalar and vector fields.
For this purpose we note that
 \begin{eqnarray}
 T^+_{AB} & = &  h_{\Lambda
 AB}F^{+\Lambda} - f^\Lambda_{AB} \cG_\Lambda^+  =0  \label{tiden0}\\
   T^+_I & = &  h_{\Lambda
 I}F^{+\Lambda}-f^\Lambda_{I} \cG_\Lambda^+   =0 \label{tiden}
\end{eqnarray}
as a consequence of eqs. (\ref{nfh-1}) and (\ref{gigiuno}).
Therefore we have:
\begin{eqnarray}
Z_{AB} & = & \int_{S^2} T_{AB} = \int_{S^2} (T^+_{AB} + T^ -_{AB}) = \int_{S^2} T^ -_{AB} =
  h_{\Lambda AB}(r) p^\Lambda- f^\Lambda_{AB}(r) q_\Lambda
\label{zab}\\
Z_I & = & \int_{{S}^2} T_I =
\int_{S^2} (T^+_I + T^ -_I) = \int_{S^2} T^ -_I =
 h_{\Lambda I}(r) p^\Lambda - f^\Lambda_I(r) q_\Lambda   \quad (N\leq 4)
\label{zi}
\end{eqnarray}
where:
\begin{equation}
q_\Lambda = \int_{S^2} \cG_\Lambda , \quad
p^\Lambda = \int_{S^2} F^\Lambda \label{charges}
\end{equation}
are the conserved quantized charges defined as in eq.\eqn{holele} and
\eqn{holmag}.
We see that the  central and matter charges are given  by
symplectic invariants
and that the presence of dyons in $D=4$ is related to the
 symplectic embedding.
 \par
 \subsection{Differential relations obeyed by the dressed charges}
Let $\Gamma = \IL_{Usp}^{-1} d\IL_{Usp}$ be the $Usp({\bar n},{\bar n})$ Lie algebra left invariant 1--form
satisfying the Maurer Cartan equation:
\begin{equation}
  d\Gamma +\Gamma \wedge \Gamma = 0
\label{int}
\end{equation}
Note that the above equation \eqn{int} actually implies the flatness
of the symplectic bundle over $G/H$ we have already mentioned.\par
In terms of the matrices $(f,h)$ the 1--form $\Gamma$ reads:
\begin{equation}
  \label{defgamma}
  \Gamma \equiv \IL_{Usp}^{-1} d\IL_{Usp} =
\pmatrix{{\rm i} (f^\dagger dh - h^\dagger df) & {\rm i} (f^\dagger d\bar h - h^\dagger d\bar f) \cr
-{\rm i} (f^t dh - h^t df) & -{\rm i}(f^t d\bar h - h^t d\bar f) \cr}
\equiv
\pmatrix{\Omega^{(H)}& \bar \cP \cr
\cP & \bar \Omega^{(H)} \cr }
\end{equation}
where the ${\bar n} \times {\bar n}$ sub--blocks $ \Omega^{(H)}$ and  $\cP$
embed the $H$ connection and the $G/H$ vielbein,  respectively.
This identification follows from the Cartan decomposition
of the $Usp({\bar n},{\bar n})$ Lie algebra.
Explicitly, if we define the $H_{Aut} \times H_{matter}$--covariant
derivative of
a vector $V= (V_{AB},V_I)$ as:
\begin{equation}
\nabla V = dV - V\omega , \quad \omega = \pmatrix{\omega^{AB}_{\ \
CD} & 0 \cr 0 & \omega^I_{\ J} \cr }
\label{nablav}
\end{equation}
from  (\ref{defu}) and (\ref{defgamma}),
we obtain the $({\bar n} \times {\bar n})$ matrix equation:
\begin{eqnarray}
  \nabla(\omega) (f+{\rm i} h) &=& (\bar f + {\rm i} \bar h) \cP \nonumber \\
\nabla(\omega) (f-{\rm i} h) &=& (\bar f - {\rm i} \bar h) \cP
\end{eqnarray}
together with their complex conjugates. Using further the definition (\ref{deffh}) we have:
\begin{eqnarray}
  \nabla(\omega) f^\Lambda_{AB} &=&  \bar f^\Lambda_{I}  P^I_{AB} +
  {1 \over 2}\bar f^{\Lambda CD} P_{ABCD} \nonumber \\
\nabla(\omega) f^\Lambda_{I} &=&  {1 \over 2} \bar f^{\Lambda AB}
P_{AB I} +\bar f^{\Lambda J}  P_{JI}
 \label{df}
\end{eqnarray}
where we have decomposed the embedded vielbein $\cP$ as follows:
\begin{equation}
  \label{defp}
  \cP = \pmatrix{P_{ABCD} & P_{AB J} \cr P_{I CD} & P_{IJ}\cr }
\end{equation}
 the sub--blocks being related to the vielbein of $G/H$, $P = L^{-1} \nabla^{(H)} L$,
 written in terms of the indices of $H_{Aut} \times H_{matter}$.
Note that, since $f$ belongs to the unitary matrix $\IL_{Usp}$, we have:
$(  f^\Lambda _{AB},  f^\Lambda_I)^\star = (\bar f^{\Lambda AB},\bar f^{\Lambda I})$.
Obviously, the same differential relations that we wrote for $f$ hold true
for the dual matrix $h$ as well.
\par
Using the definition of the charges (\ref{zab}), (\ref{zi}) we
then get the following differential
relations among charges:
\begin{eqnarray}
  \nabla(\omega) Z_{AB} &=&  \bar Z_{I}  P^I_{AB} +  {1 \over 2} \bar Z^{CD} P_{ABCD}
  \nonumber \\
\nabla(\omega) Z_{I} &=& {1 \over 2}  \bar Z^{AB}  P_{AB I} + \bar Z_{J}  P^J_{I}
 \label{dz1}
\end{eqnarray}
Depending on the coset manifold, some of the blocks of (\ref{defp}) can be actually zero.
For example in $N=3$ the vielbein of $G/H = {SU(3,n) / SU(3)
 \times SU(n)\times U(1)}$ \cite{ccdffm} is $P_{I AB}$ ($AB $ antisymmetric),
 $I=1,\cdots,n;
A,B=1,2,3$ and it turns out
that $P_{ABCD}= P_{IJ} = 0$.
\par
In $N=4$, $G/H= {SU(1,1) / U(1)} \otimes { O(6,n)/ O(6) \times O(n)}$ \cite{bekose},
and we have
$P_{ABCD}= \epsilon_{ABCD} P$, $P_{IJ} = \bar P \delta_{IJ}$, where $P$ is the
K\"ahlerian vielbein of ${SU(1,1)\over
U(1)}$,  ($A,\cdots , D $ $SU(4)$ indices and $I,J$ $O(n)$ indices)
and $P_{I AB}$ is the vielbein of ${ O(6,n)/ O(6) \times O(n)}$.
\par
For $N>4$ (no matter indices) we have that $\cP $ coincides with the vielbein
$P_{ABCD}$ of the relevant $G/H$.
\par
\subsection{Sum rules satisfied by central charges}
Besides the differential relations
(\ref{dz1}), the charges also satisfy a  sum rule  that we presently describe.
\par
The sum rule has the following form:
\begin{equation}
 {1 \over 2} Z_{AB} \bar Z^{AB} + Z_I \bar Z^I =
 -{1\over 2} Q^t \cM (\cN) Q
\label{sumrule}
\end{equation}
where $\cM(\cN)$ and $Q$ are:
\begin{equation}
\cM = \left( \matrix{ \bfone & - \mbox{\rm Re} \cN \cr 0 &\bfone\cr}\right)
\left( \matrix{ \mbox{Im} \cN & 0 \cr 0 &\mbox{Im} \cN^{-1}\cr}\right)
\left( \matrix{ \bfone & 0 \cr - \mbox{Re} \cN & \bfone \cr}\right)
\label{m+}
\end{equation}
\begin{equation}
Q=\left(\matrix{p^\Lambda \cr q_ \Lambda \cr} \right)
\label{eg}
\end{equation}
In order to obtain this result we just need to observe that from the
 fundamental identities \eqn{specdef} and  from the definition of the
 kinetic matrix given in (\ref{nfh-1}) it  follows:
 \begin{eqnarray}
ff^\dagger &= &-{\rm i} \left( \cN - \bar \cN \right)^{-1} \\
hh^\dagger &= &-{\rm i} \left(\bar \cN^{-1} - \cN^{-1} \right)^{-1}\equiv
-{\rm i} \cN \left( \cN - \bar \cN \right) ^{-1}\bar \cN \\
hf^\dagger &= & \cN ff^\dagger  \\
fh^\dagger & = & ff^\dagger \bar \cN
\end{eqnarray}
Note that if we go back to the $N=2$ case treated in the previous
chapter, the sum rule on the central and matter charges \eqn{sumrule}
takes the form:
\begin{equation}
  \vert Z \vert ^2 +  \vert Z_i \vert ^2 \equiv
  \vert Z \vert ^2 +   Z_i g^{i\bar \jmath} \bar Z_{\bar \jmath} =
  -{1\over 2} Q^t \cM   Q
  \label{sumrules}
\end{equation}
with the same right hand side as in the previous discussion.
\section{The geodesic potential}
\setcounter{equation}{0}
\markboth{BPS BLACK HOLES IN SUPERGRAVITY: CHAPTER 4}
{4.3 THE GEODESIC POTENTIAL}
\label{geopoto}
At this point we can use the information we have so far collected on
the central and  matter charges in a different way that makes no
direct  reference to supersymmetry and enlightens the significance of
the scalar field functional:
\begin{equation}
V(\phi, Q)= -\frac{1}{2} \, Q^t \, {\cal M}\left( {\cal N}\right) \, Q
\label{potent}
\end{equation}
introduced in the sum rule \eqn{sumrule},\eqn{sumrules}.
\par
To this effect we consider the general action \eqn{genact} and the
ansatz \eqn{ds2U} for the metric, \eqn{flambda} for the vector fields
and \eqn{phijey} for the scalars. From such an action one can derive
the field equations varying with respect to the metric, the vector
fields and the scalars. Inserting the above ansatz such equations
reduce to a system of second order ordinary differential equations in
the variable $r$ for the unknown functions $U(r)$ and $\phi^I(r)$. We
can think of such equations as the Euler Lagrange equations derived
from an effective action which is quite straightforward to write down.
Such an effective action has the following explicit form:
\begin{eqnarray}
S_{eff} & \equiv & \int \, {\cal L}_{eff}(\tau) \, d\tau \quad ;
\quad \tau = -\frac{1}{r} \nonumber\\
 {\cal L}_{eff}(\tau ) & = & \left( \frac{dU}{d\tau} \right)^2 +
 g_{IJ} \, \frac{d\phi^I}{d\tau} \,  \frac{d\phi^J}{d\tau} + e^{2U}
 \, V(\phi , Q)
 \label{effact}
\end{eqnarray}
where the potential $V(\phi)$ appearing in \eqn{effact} is   the same object
defined in eq. \eqn{potent}.
Actually the field equations from the original action are equivalent
to the variational equations obtained from the effective action
\eqn{effact} provided we add to them also the following constraint:
\begin{equation}
    \left( \frac{dU}{d\tau}\right)^2
 g_{IJ} \, \frac{d\phi^I}{d\tau} \,  \frac{d\phi^J}{d\tau} - e^{2U}
 \, V(\phi , Q) \,  \cong \, 0
 \label{legato}
\end{equation}
Let us first consider a simple case where we assume that the scalar
fields are constants from {\it horizon} $r=0$  to {\it infinity}
$r= \infty $:
\begin{equation}
       \phi^I = \mbox{constant} = \phi^I_\infty
       \label{dubext}
\end{equation}
Extremal black--holes satisfying such an additional simplifying condition \eqn{dubext} are
named {\it double extreme black holes}.
\par
It follows from \eqn{legato}, upon use of \eqn{dubext} that:
\begin{equation}
 \left( \frac{dU}{d\tau}\right)^2
   =e^{2U}
 \, V(\phi , Q)
 \label{soliton}
\end{equation}
At the horizon $\tau \to   -\infty$, the metric \eqn{ds2U}  must approach
the Bertotti Robinson metric \eqn{bertrob}, as we already discussed
in complete generality, so that we have a boundary condition for the
differential equation \eqn{soliton}:
\begin{equation}
e^{2U(\tau)} \, \stackrel{\tau \to \infty}{\longrightarrow } \,
\frac{1}{m_{BR}^2} \, \frac{1}{\tau^2} \, = \, \frac{4\pi}{{\rm Area}_H} \, \frac{1}{\tau^2}
\label{bunder}
\end{equation}
 where $m_{BR}$ is the Bertotti Robinson mass and ${\rm Area}_H$ is the
 area of the horizon, related to the first by eq.\eqn{horiz}.
 Using this information, for double extreme black holes we have:
\begin{equation}
 V(\phi_H, Q)=\frac{\mbox{Area}_H}{4\pi}
 \label{valarea}
\end{equation}
where $\phi_H$ denotes the value of the scalar fields at the horizon,
which in this case is the same as their value at infinity.
Actually the result \eqn{valarea} is more general and it is true also
for generic extremal black--holes where we relax  condition \eqn{dubext} but we
still assume that the kinetic term of the scalars $\phi^I$ (that in the original lagrangian
\eqn{genact} and  not that in the effective action) should be finite at the horizon:
\begin{equation}
\lim_{\tau \to \infty} \,   g_{IJ} \, \frac{d\phi^I}{d\tau} \,  \frac{d\phi^J}{d\tau}
\, e^{2U} \, \tau^4 \, < \, \infty
\end{equation}
Indeed, using again the boundary condition \eqn{bunder} on $U$ we
find:
\begin{equation}
  g_{IJ} \, \frac{d\phi^I}{d\tau} \,  \frac{d\phi^J}{d\tau}
  \, \frac{4\pi}{\mbox{Area}_H} \, \tau^2
  \, \stackrel{\tau \to \infty}{\longrightarrow } \, X^2 \,= \, \mbox{finite quantity}
\end{equation}
However, we must have $X^2=0$ if the moduli $\phi$ are assumed to be finite
at the horizon. Indeed, if $X^2 \ne 0$, near the horizon we can write:
\begin{equation}
\tau \, \frac{d\phi^I}{d\tau} \, \stackrel{\tau \to \infty}{\longrightarrow } \, \mbox{const} \quad
 \rightarrow \quad \phi^I \, \sim  \, \mbox{const}\,  \times \, \log \tau
\end{equation}
 Hence also for generic extremal black--holes the same conclusion
 reached before in eq. \eqn{valarea} holds true.
 \par
 Hence the area of the horizon is expressed in terms of the finite
 value reached by the scalars at the horizon. But what is such a
 value? We can easily show that $\phi^I_H$ is determined by the
 following extremization of the potential \eqn{potent}:
 \begin{equation}
\frac{\partial V}{\partial \phi^I} \, |_H \, = \, 0
\label{horomin}
\end{equation}
Indeed considering the variational equation for the scalar fields
derived from the effective action \eqn{effact} we have:
\begin{equation}
\frac{D^2}{d\tau^2}\, \phi^I \, = \, \frac{1}{2} \, e^{2U} \,
\frac{\partial V}{\partial \phi^I} \, g^{IJ}
\label{fimoto}
\end{equation}
Near the horizon the contribution from the quadratic terms
proportional to the Levi Civita connection $\Gamma^I_{JK} \, d_\tau
\, \phi^J   \, d_\tau \phi^K $ vanishes so that eq.\eqn{fimoto}
reduces to:
\begin{equation}
\frac{d^2}{d\tau^2}\, \phi^I \, \cong \, \frac{1}{2} \,
\frac{\partial V}{\partial \phi^I} \, g^{IJ} \, \frac{4\pi}{\mbox{Area}_H} \, \frac{1}{\tau^2}
\label{fiappr}
\end{equation}
whose solution is:
\begin{equation}
\phi^I \, \sim \, \frac{2\pi}{\mbox{Area}_H} \, \frac{\partial V}{\partial \phi^I} \, g^{IJ}  \,
\log \tau \, + \, \phi^I_H
\label{solfi}
\end{equation}
Invoking once again the finiteness of the scalar fields $\phi$ at the
horizon we conclude that their the extremum condition \eqn{horomin}
must be true in order to be consistent with eq. \eqn{solfi}.
\par
In this way we have reached, for the general case of an extremal black hole  with a finite horizon area,
the same conclusion we had reached in section \ref{fissato}
for such a black hole in the context
of $N=2$ supergravity. Namely the scalars flow at the horizon at a fixed point that is
determined as the extremum of a potential. In the $N=2$  case we
obtained the   equations \eqn{minima} determining the horizon values by starting from the BPS
conditions \eqn{zequa} and \eqn{Uequa} and imposing that we are at
the fixed point of the first order differential equations. We already
anticipated that the same result could be obtained by extremizing a
potential, whose value at the horizon is the horizon area. We can now
verify explicitly such a statement.  To this effect we should stress
that in the supersymmetric case the {\it geodesic potential}
appearing in the effective action \eqn{effact} and defined in
eq.\eqn{potent} is the right hand side of the {\it central and matter
charges sum rule} of eq.\eqn{sumrule} or \eqn{sumrules} for the $N=2$
case. Hence extremizing the geodesic potential, in supersymmetric
cases is the same thing as extremizing the left hand side of the sum
rule. For the $N=2$ case we find:
\begin{eqnarray}
0 & = & \frac{d}{dz^i} \,
\left( Z \, {\bar Z} + g^{ij^\star} \, Z_i \, {\bar
Z}_{j^\star}\right) \nonumber\\
&=& \nabla_i Z \, {\bar Z} + \nabla_i    Z_j  \, {\bar Z}_{j^\star}
\, g^{jk^\star} \, + Z_j \, \nabla_i \, {\bar Z}_{k^\star} \, g^{jk^\star} \,  \nonumber\\
&=& \nabla_i Z \, {\bar Z} + \mbox{i} \, C_{ijm} \, {\bar
Z}_{\ell^\star} \, {\bar Z}_{k^\star} \, g^{\ell^\star m} \,
g^{j k^\star} \, + \, Z_j \,{Z} \, \delta^i_j
\label{contin}
\end{eqnarray}
 where we have used the special geometry identities \eqn{defaltern}
 and the already quoted identification (see \eqn{zentrum}and following ones):
  \begin{equation}
Z_i = \nabla_i Z
\end{equation}
Indeed the special geometry identities holding true on the symplectic
sections and their derivatives are automatically extended to the
central and matter charges that are linear combinations of such
sections with constant coefficients (the quantized charges). From
eq.\eqn{contin} it immediately follows that the extremum condition
translates into eq.\eqn{zequato} since it implies:
\begin{equation}
 Z_i = \nabla_i Z =0 \quad \longrightarrow \quad \partial_i \, |Z |^2
 =0
 \label{0Zi}
\end{equation}
Hence the horizon area is indeed minimum and, as we claimed in
section \ref{fissato}, the fixed values of the scalars are obtained
by extremizing such an area.
\section{General properties of central charges at the extremum}
\markboth{BPS BLACK HOLES IN SUPERGRAVITY: CHAPTER 4}
{4.4 CENTRAL CHARGES AT THE EXTREMUM}
\setcounter{equation}{0}
\label{extrzen}
 The geodesic potential \eqn{potent},
whose extrema correspond to the finite  values reached
by the scalar fields at the horizon of an extremal black--hole,
can be rewritten as the left hand side of
eq.\eqn{sumrule} in the case of a supersymmetric theory.
This statement must be interpreted in the following
way. We are free to consider an arbitrary theory of gravity coupled
to scalar and vector fields described by an action of type \eqn{genact}.
If supersymmetry is not advocated   no special relation exists on
the number of scalars relative to the number of vectors and any
symplectically embedded scalar manifold is allowed. For any such
theory there is a {\it period matrix} ${\cal N}$ and correspondingly
we can construct the potential \eqn{potent}. We are also free to look
for extremal black--solutions of such a theory and, according to the
discussion presented in the previous section, the value
attained by the scalars at the horizon is determined by
eq.\eqn{horomin}, provided we look for {\it finite horizon area}
solutions. Yet it is only in the supersymmetric case that there
exists a concept of central charges of the supersymmetry algebra and
in that case the geodesic potential can be rewritten as a sum of squares
of such charges:
\begin{equation}
\label{suspot}
 V^{SUSY} \left(\phi, {\vec Q}\right) \, =
 \, {1 \over 2} Z_{AB} \bar Z^{AB} + Z_I \bar Z^I
\end{equation}
We can reconsider the extremization of the geodesic potential from
the point of view of its supersymmetric reinterpretation \eqn{suspot}.
Writing the extremum condition as the vanishing of the exterior
differential, we immediately get:
\begin{eqnarray}
0 & = & dV^{SUSY} \left(\phi, {\vec Q}\right)  \nonumber\\
  & = & \frac{1}{2}\, \nabla Z_{AB} \, {\bar Z}^{AB} +
         \frac{1}{2}\,  Z_{AB} \,\nabla {\bar Z}^{AB} +
         \nabla Z_{I} \, {\bar Z}^{I} +
            Z_{I} \,\nabla {\bar Z}^{I}
\label{exdif}
\end{eqnarray}
and  inserting the differential relations \eqn{dz1} into \eqn{exdif}
we obtain the following condition:
\begin{eqnarray}
0 & = & \left[ \frac{1}{2}
\left( {\bar Z}^{I} \, P_{I \vert AB} + \frac{1}{2}
{\bar Z}^{CD} \, P_{ABCD} \right) \, {\bar Z}^{AB} + \mbox{c.c}
\right]\nonumber\\
&& + \left[ \frac{1}{2} \left( {\bar Z}^{AB} \, P_{AB \vert I} +
{\bar Z}^{J} \, P_{J\vert I} \right) \, {\bar Z}^J +\mbox{c.c} \right]
\label{cagnol}
\end{eqnarray}
Since the scalar vielbein provides a frame of independent $1$--forms
on the scalar manifold, eq.\eqn{cagnol} can be true only if the coefficients
of each vielbein component vanishes independently.
This implies that, at the extremum,  the following
conditions on the central and matter charges have to be true:
\begin{eqnarray}
Z_I^{fix} &=& 0 \label{0matc}\\
Z_{[AB}^{fix} \, Z_{CD]}^{fix}& =& 0 \label{0zen}
\end{eqnarray}
Eq.s \eqn{0matc} and \eqn{0zen} are the higher $N$ generalization of
the extremum condition \eqn{0Zi} we found in the case of $N=2$
supergravity. As we see for all values of $N$  the matter charges
$Z_I$ vanish at the horizon. In the $N=2$ case where the matter
charges can be interpreted as the moduli covariant derivatives of
the unique central charge condition \eqn{0matc} suffices to determine
the extremum. In the higher $N$ case where the central charge is an
antisymmetric $N \times N$ matrix,  the extremum is characterized by
the additional algebraic condition \eqn{0zen}. Its meaning becomes
clear if we make a local  $SU(N)$ R--symmetry transformation that
reduces the central charge tensor to its {\it normal frame} where it
is skew--diagonal (see eq.\eqn{skewZ}). Focusing on the $N=2 p = even$
case, in the  normal frame the only non--vanishing entries are
\begin{equation}
Z_1^{fix} \equiv Z_{1\, 2}^{fix} \, , \,Z_2^{fix} \equiv
Z_{3\, 4}^{fix} \, , \, \dots \, , \, Z_{N/2}^{fix} \equiv Z_{N-1 \, N}^{fix}
\label{lista}
\end{equation}
and equation \eqn{0zen} implies that either all
vanish, or at most one of them, say $Z_1^{fix} =Z_{12}^{fix}$,
is non zero while all the other
vanish. Hence the constraint \eqn{0zen} yields two possibilities:
\begin{equation}
\mbox{solutions of the constraint}=
\cases{\mbox{$1^{st}$ solution}: \quad  Z_i^{fix}=0
\quad;\quad   (i=1,\dots,p) \cr
\mbox{$2^{nd}$ solution}: \quad\cases{Z_1^{fix} \ne 0 \cr
Z_i^{fix} \quad \quad;   \quad (i=2,\dots,p)}   }
\label{dueca}
\end{equation}
On the other hand, recalling \eqn{valarea} we have:
\begin{equation}
\label{checa}
\frac{\mbox{Area}_H}{4 \pi} \, = \, \sum_{i=1}^{p} \vert \, Z_i^{fix} \,
\vert^2
\end{equation}
so that if the horizon area ( and hence the entropy) of the
black--hole has to be finite then we can exclude the first
possibility in eq.\eqn{dueca} and we find:
\begin{equation}
\label{sorda}
 \frac{\mbox{Area}_H}{4 \pi} \, = \, \vert \, Z_1^{fix} \,
\vert^2  \, > \, 0
\end{equation}
Eq.\eqn{sorda} is the higher $N$ generalization of eq.\eqn{ariafresca}
holding true in the $N=2$ case. It leads to the following very important
conclusion:
\bst
\label{horzen}
BPS saturated black--holes are classified by the skew--eigenvalue
structure of their central charge $Z_{AB}$ evaluated at the
horizon. The only BPS black--holes
that have a non vanishing entropy ${\mbox{Area}_H}/ {4 \pi} \, > \, 0$
are those admitting a single non-zero skew--eigenvalue $Z_1^{fix}$ while
all the other $Z_i^{fix}$ vanish.
\est
\markboth{BPS BLACK HOLES IN SUPERGRAVITY: CHAPTER 5}
{4.5 THE HORIZON AREA AS A TOPOLOGICAL INVARIANT}
\section{The horizon area as a topological $U$ duality invariant}
\setcounter{equation}{0}
\label{51v56}
As we just saw the horizon area can be found by extremizing the geodesic
potential and then replacing the fixed values of the scalars in eq.
\eqn{valarea}. Since the only parameters appearing in the geodesic
potential \eqn{potent} are the quantized charges $\{p^\Lambda,q_\Sigma\}$,
it follows that the fixed values of the scalars will depend only on such
quantized charges and so will do the horizon area:
\begin{equation}
\frac{\mbox{Area}_H}{4\pi} \, = \, S(p,q)
\label{arg1}
\end{equation}
On the other hand we recall that, by construction the geodesic potential
is a symplectic invariant and hence an invariant under $U$--duality
transformations.  Consider for instance the $N>4$ supersymmetric case where
the geodesic potential is expressed as in eq.\eqn{suspot} with no matter
charges $Z_I$. Introduce the charge vector \eqn{eg}
that transforms in the appropriate
{\it real symplectic representation} of the
$U$ duality group which defines the symplectic
embedding of the coset manifold $U/H$. Then recall the definition
\eqn{zab} of the central charges and eq.\eqn{defu} which relates the
complex $Usp$--realization $\IL_{Usp}(\phi)$
of the coset representative to its real
symplectic version $\IL_{Sp}(\phi)$. By straightforward algebra we
verify that the geodesics potential is nothing else but the
following bi--quadratic form in the charge vector and in the coset
representative:
\begin{eqnarray}
V(\phi,{\vec Q}) & \equiv & \frac{1}{2}\, {\bar Z}^{AB}(\phi) \, Z_{AB}(\phi)
\nonumber\\
&=& \frac{1}{2}\,{\vec Q}^T \, \left[ \IL^{-1}_{Sp}
\left(\phi\right) \right]^T \,
\IL^{-1}_{Sp}\left(\phi\right) \, {\vec Q}
\label{higeo}
\end{eqnarray}
The invariance under the $U$--transformations is verified in the
following way. By definition of coset representative, if $A \, \in \, U
\subset Sp(2{\bar n},\IR)$ we have:
\begin{eqnarray}
\IL_{Sp}\left({\cal A}\phi\right)& = & A \,
\IL_{Sp}\left( \phi\right) \, W_A(\phi)\label{costra}\\
 W_A(\phi)   \in  H \subset U({\bar n}) \, \subset  \,  Sp(2{\bar
 n},\IR) & \Longrightarrow &  \left[W_A(\phi)\right]^T \,W_A(\phi)
 \, = \, \bfone
\label{saldo}
\end{eqnarray}
where ${\cal A}\phi$ denotes the non--linear action of the $U$ group element
$A$ on the scalar fields $\phi$ and $W_A(\phi)$ is the $H$--compensator
that in the symplectic representation is an orthogonal matrix (indeed
we have $U({\bar n}) = SO(2{\bar n}) \cap Sp(2{\bar n},\IR)$).
Relying on eq.\eqn{saldo} and on the representation \eqn{higeo}
of the geodesic potential we find:
\begin{eqnarray}
V({\cal A}\phi,A{\vec Q})&=&{\vec Q}^T \, A^T \,
\left[ \IL^{-1}_{Sp}
\left({\cal A}\phi\right) \right]^T \,
\IL^{-1}_{Sp}\left({\cal A}\phi\right) \, A\, {\vec Q} \nonumber \\
&=&
{\vec Q}^T \, A^T \,[A^{-1}]^T \,\left[ \IL_{Sp}^{-1}
\left( \phi\right) \right]^T \,[W^{-1}_A(\phi)]^T \,
W^{-1}_A(\phi) \IL^{-1}_{Sp}\left(\phi\right)\,  A^{-1} \, A \,
{\vec Q}\nonumber\\
&=& V( \phi, {\vec Q})
\label{novari}
\end{eqnarray}
which explicitly proves the $U$ invariance of of the geodesic
potential. It follows that also the horizon area $S(p,q)$
obtained by substitution of the fixed scalar values in $V(\phi,Q)$
will be a $U$ invariant. Not only that. Eq.\eqn{higeo} shows that
the invariant $I(p,q)$ must be homogenous of order two in the charge
vector ${\vec Q}$. At this point we can just rely on group theory.
The geodesic potential is bi--quadratic in the charge vector and in
the coset representative. Hence $S(p,q)$ cannot be just a quadratic
invariant since it is obtained by substitution of $\phi^{fix} =\phi(p,q)$
into an expression that is already quadratic in the charge vector.
The  other possibility is that $S(p,q)$ be the square root of a
quartic invariant. Hence we can guess the following formula:
\begin{equation}
\label{bella}
\frac{\mbox{Area}_H}{4\pi} \, \equiv \, S(p,q) = \sqrt{Q^\Lambda \,
Q^\Sigma \, Q^\Delta \, Q^\Gamma \, d_{\Lambda\Sigma\Delta\Gamma} }
\end{equation}
where $d_{\Lambda\Sigma\Delta\Gamma}$ is a {\it four-index symmetric
invariant tensor} of the $U$ duality group in the   real
representation that defines the symplectic embedding of the manifold
$U/H$.
For instance in the case of $N=8$ supergravity where $U=E_{7(7)}$ a
four-index symmetric invariant tensor does indeed exist in the
fundamental {\bf 56} representation. The horizon area of an $N=8$
BPS black--hole is correctly given by eq.\eqn{bella}. We   explicitly
verify this statement in the next chapter by solving the appropriate
differential equations and computing the horizon area. Here we pursue
a more abstract discussion. Having established that the square of the
horizon area is a quartic $U$--invariant we can try to construct it
as an appropriate linear combination of $H$--invariants. We start
from the central charges $Z_{AB}(p,q,\phi)$ that are linear in the
charge vector ${\vec Q}$ and that transform covariantly
under the compensating subgroup $H$. We consider the possible
$H$--invariants $I_i \left( Z(p,q,\phi)\right)$ that are
quartic in the central charges
$Z_{AB}$ and we write the ansatz:
\begin{equation}
S^2(p,q) = I  = \sum_{i} \, \alpha _i \, I_i \left( Z(p,q,\phi)\right)
\label{iansa}
\end{equation}
where $ \alpha _i$ are numerical coefficients. For arbitrary $ \alpha
_i$ values $I$ is a function of the scalar fields $ \phi $ through the scalar
field dependence of the central charges. Yet for special choices of
the coefficients $\alpha_i$ the $H$--invariant $I$ can be independent
from the $ \phi^I$. We argue that such a linear combination of
$H$--invariants is the quartic $U$--invariant we look for. To find the
appropriate $\alpha_i $ it suffices to impose:
\begin{equation}
\frac{\partial}{\partial \phi^I} \, I \, = 0
\label{dfI0}
\end{equation}
This is a very much indirect way of constructing $U$--invariants that
has a distinctive advantage. It puts into evidence the relation
between the black--hole horizon area and the eigenvalues of the
central charge. As an example we consider the $N=8$ case.
\subsection{Construction of the quartic $E_{7(7)}$ invariant in the
$N=8$ theory}
\label{5param}
In the $N=8$ case we have $U=E_{7(7)}$, $H=SU(8)$ and the central
charge $Z_{AB}$ is in the complex ${\bf 28}$ representation of
$SU(8)$. There are three quartic $SU(8)$ invariants that we can
write:
\begin{eqnarray}
I_1 & \equiv & \left(Z_{AB} \, {\bar Z}^{BA}\right)^2 \nonumber\\
I_2 & \equiv & Z_{AB} \, {\bar Z}^{BC} \, Z_{CD} \, {\bar Z}^{DA} \nonumber\\
I_3 & \equiv & \frac{1}{2} \left( \frac{1}{2!4!}\epsilon^{ABCDEFGH}
\, Z_{AB} \, Z_{CD} \, Z_{EF} \, Z_{GH} \, +\mbox{c.c} \right)
\label{trilis}
\end{eqnarray}
We can easily determine the coefficients $\alpha _1,\alpha _2,\alpha
_3$ by going to the normal frame. Naming the four skew-eigenvalues
as in eq.\eqn{lista} we obtain:
\begin{eqnarray}
 I_1 & = & 4 \left(\sum_{i=1}^{4} \, \vert Z_i \vert ^2 \right)^2
 \nonumber \\
 I_2 & = & 2  \sum_{i <j} \vert Z_i \vert ^2 \, \vert Z_j \vert ^2
 \nonumber \\
 I_3 & = & \frac{1}{2}\, \left( \prod_{i=1}^{4} Z_i + \prod_{i=1}^{4}
 {\bar Z}_i \right)
 \label{I123}
\end{eqnarray}
On the other hand, naming:
\begin{equation}
P_1 = P_{1234} \quad ; \quad P_2 = P_{1256} \quad ; \quad P_3 =
P_{1278}
\label{P123}
\end{equation}
the differential relations \eqn{dz1} reduce  in the normal frame to:
\begin{eqnarray}
\nabla Z_1 = P_1 \, {\bar Z}_2 + P_2 \, {\bar Z}_3 + P_3 \, {\bar
Z}_4 \nonumber \\
\nabla Z_2 = P_1 \, {\bar Z}_1 + P_3 \, {\bar Z}_3 + P_2 \, {\bar
Z}_4 \nonumber \\
\nabla Z_3 = P_2 \, {\bar Z}_1 + P_3 \, {\bar Z}_2 + P_1 \, {\bar
Z}_4 \nonumber \\
\nabla Z_4 = P_3 \, {\bar Z}_1 + P_2 \, {\bar Z}_2 + P_1 \, {\bar
Z}_3
\label{dZPi}
\end{eqnarray}
Inserting eq.s \eqn{dZPi} and \eqn{I123} in the condition
\begin{equation}
d \left( \sum_{i=1}^{3} \, \alpha_i \, I_i \right) = 0
\end{equation}
and cancelling separately the terms proportional to $P_i$, $(i=1,2,3)$
we obtain $\alpha_1 = 1/4$, $\alpha _2 = -1 $, $\alpha _3 = 8$.
\par
Hence we can finally write:
\begin{eqnarray}
\left(\frac{\mbox{Area}_H}{4 \pi}\right)^2 &=& \frac{1}{4} \, I_1
- I_2 + 8 \, I_3 \nonumber\\
& =& \sum_{i=1}^{4} \,\left( \vert Z_i \vert ^2 \right)^2 -
2  \sum_{i <j} \vert Z_i \vert ^2 \, \vert Z_j \vert ^2
+ 4 \, \left( \prod_{i=1}^{4} Z_i + \prod_{i=1}^{4}
 {\bar Z}_i \right)\nonumber\\
 &=&(\rho_1+\rho_2+\rho_3+\rho_4) \,
 (\rho_1+\rho_2-\rho_3-\rho_4) \, \times \,\nonumber\\
 && \times \, (\rho_1 -\rho_2+\rho_3 -\rho_4) \,
 (\rho_1-\rho_2-\rho_3+\rho_4) \nonumber\\
&& +8 \rho_1 \, \rho_2 \, \rho_3 \, \rho_4 \, \left( \cos \theta
 -1\right)
 \label{funf}
\end{eqnarray}
The last equality in eq.\eqn{funf} has been obtained by writing the four
skew eigenvalues $Z_i$ in the following way:
\begin{equation}
Z_i = \rho_i \, \exp[\mbox{i}\theta] \quad ; \quad \rho_i \in \IR
\label{rhofi}
\end{equation}
where $\theta$ is a common phase. The central charge $Z_{AB}$ can
always be reduced to such a form because after the matrix has been
skew--diagonalized we still have enough $SU(8)$ transformations to
gauge away the three relative phases of the eigenvalues.
An alternative and more intrinsic way of seeing that the truly independent
parameters are ${\bf 5}$, namely the $4$ moduli $\rho_i$ plus the overall
phase $\theta$, is to observe that the stability subgroup of $Z_{AB}$
in the normal form is $\left( SU(2)\right)^4$. Then it suffices to
count the dimensionality of the coset
\begin{equation}
\mbox{dim}_{\bf R}  \, \frac{SU(8)}{\left( SU(2)\right)^4} \, = \, 51
\end{equation}
to conclude that $51$ of the $56$ real components of $Z_{AB}$ can be
gauged away leaving a residual $5$.
\bst
The most general $N=8$ BPS black--solution depends on $5$ intrinsic
parameters and its horizon area is given by eq.\eqn{funf} in terms
of such parameters.
\est
From eq.s\eqn{funf},\eqn{rhofi}  we see that there are three
possible cases:
\begin{equation}
\begin{array}{ccc}
\rho_1 > \rho_2 > \rho_3 > \rho_4 \ge 0 & \mbox{Area}_H > 0 & \frac{1}{8}\,
\mbox{BPS}  \\
\rho_1 = \rho_2 > \rho_3 = \rho_4 \ge 0 & \mbox{Area}_H = 0 & \frac{1}{4}\,
\mbox{BPS}  \\
\rho_1 = \rho_2 = \rho_3 = \rho_4 \ge 0 & \mbox{Area}_H = 0 & \frac{1}{2}\,
\mbox{BPS}  \\
\end{array}
\label{arusp}
\end{equation}
The last column in table \eqn{arusp} is a prediction about the fraction
of preserved supersymmetries that we shall verify in the next
chapter.
\par
Let us finally clarify the meaning of eq.\eqn{funf} in comparison
with our previous conclusion in statement \ref{horzen}
that at the horizon there is at most one non--vanishing central
charge eigenvalue and that the horizon area is essentially the
square of that eigenvalue (see eq.\eqn{sorda}).
Superficially eq.s\eqn{funf} and eq.\eqn{sorda} might seem
to be contradictory, but it is not so.
\bst The numerical value of the horizon area is a {\bf topological}
quantity depending only on the vector of quantized charges ${\vec Q}$
that is also a group theoretical invariant of the $U$--duality group.
This numerical value was named $S(p,q)$ since it is interpreted
as the black hole entropy in black hole thermodynamics and its
quantized charge dependence is given by eq.\eqn{bella}
\est
By means of the construction described in the present section we have tried
to reexpress the topological invariant $S(p,q)$ in terms of the
central charges  skew--eigenvalues $Z_i$. These latter depend on the
scalar fields $\phi^I$ and therefore are not constant in space-time.
In a black hole solution they have a radial dependence: $Z_i =
Z_i(r)$ (see eq. \eqn{zab}). However the combination of central
charges we have constructed is such that its numerical value does not
depend on the scalar fields. Hence eq.\eqn{funf} is true at any point
of space--time and expresses the value of the black--entropy in terms
of the central charges skew eigenvalue evaluated at that point.
Thus it happens that if we go to radial infinity $r=\infty $,
by integrating the
graviphoton field strength on a two sphere of very large radius we
get, in principle,  {\bf 5} different parameters, namely:
\begin{equation}
\rho_1^\infty \, \ne \, \rho_2^\infty \, \ne \, \rho_3^\infty \, \ne \,
\rho_4^\infty \quad ; \quad  \theta^\infty
\label{5dif}
\end{equation}
On the other hand if we evaluate the central charges at the horizon,
for a finite area black hole we find only one parameter:
\begin{equation}
\rho_1^H \, \ne \, 0 \quad ; \quad
\rho_2^H \, = \, \rho_3^H \, = \,
\rho_4^H \, = \, 0 \quad ; \quad  \theta^H \, = \, 0
\label{unsolo}
\end{equation}
Yet whether we use the parameters in eq.\eqn{5dif} or those in eq.
\eqn{unsolo} the topological invariant corresponding to the horizon
area is always expressed by eq.\eqn{funf} the numerical value being
that provided by eq.\eqn{bella}. In particular we obtain the relation:
\begin{eqnarray}
 \left(\rho_1^H\right)^2 & = &(\rho_1^\infty+\rho_2^\infty+
 \rho_3^\infty+\rho_4^\infty) \,
 (\rho_1^\infty+\rho_2^\infty-\rho_3^\infty-\rho_4^\infty)
 \, \times \,\nonumber\\
 && \times \, (\rho_1^\infty -\rho_2^\infty+\rho_3^\infty -\rho_4^\infty) \,
 (\rho_1^\infty-\rho_2^\infty-\rho_3^\infty+\rho_4^\infty) \nonumber\\
&& +8 \rho_1^\infty \, \rho_2^\infty \, \rho_3^\infty \, \rho_4^\infty
\, \left( \cos \theta^\infty
 -1\right)
 \label{0inf}
\end{eqnarray}
that applies to finite area $N=8$ black holes. The {\bf 5} parameters
parametrizing the central charge on a large $2$--sphere depend, besides
the quantized charges ${\vec Q},$ also on the
{\it black hole  moduli}, namely on the values:
\begin{equation}
\mu ^I \equiv \phi^I(\infty)
\end{equation}
attained by the scalar fields at radial infinity. These moduli are
completely arbitrary. Yet the combination \eqn{funf} of central
charges is such that it actually depends only on the charges
${\vec Q}$ and not on the moduli. On the other hand, at the horizon
the scalar fields flow to
their fixed value that depends only on the charges
and is insensitive to the boundary conditions at radial infinity.
\par
The relevance of this discussion is that it puts into evidence that
the $5$ {\it moduli} characterizing the most general finite
area $N=8$ black hole can be parametrized by the skew eigenvalues
of the central charge at radial infinity.
\par
In the next chapter we address the problem of the explicit
construction of $N=8$ BPS black holes.
\chapter{N=8 supergravity: BPS black holes with 1/2, 1/4 and 1/8 of
supersymmetry}
\section{Introduction to $N=8$ BPS Black Holes}
\setcounter{equation}{0}
\markboth{BPS BLACK HOLES IN SUPERGRAVITY: CHAPTER 5}
{5.1 INTRODUCTION TO $N=8$ BPS BLACK HOLES}
In this chapter we consider BPS extremal black--holes in the context
of $N=8$ supergravity.
\par
$N=8$ supergravity is the $4$--dimensional effective
lagrangian  of both type IIA and type IIB superstrings compactified
on a torus $T^6$. Alternatively it can be viewed as the $4D$
effective lagrangian of $11$--dimensional M--theory compactified on a
torus $T^7$. For this reason its $U$--duality group $E_{7(7)}(\ZZ)$,
which is defined as the discrete part of the isometry group of its scalar
manifold:
\begin{equation}
{\cal M}^{(N=8)}_{scalar} \, = \, \frac{E_{7(7)}}{SU(8)} \, ,
\label{e7su8n}
\end{equation}
unifies all superstring dualities relating the various consistent
superstring models. The {\it non perturbative BPS states}
one needs to adjoin  to the string states in order to complete
linear representations of the
$U$--duality group are, generically, {\it BPS black--holes}.
\par
These latter can be viewed as intersections of several $p$--brane
solutions of the higher dimensional theory {\it wrapped} on the
homology cycles of the $T^6$ (or $T^7$) torus.
Depending on how many $p$--branes intersect, the residual
supersymmetry can be:
\begin{enumerate}
\item $ \frac{1}{2}$ of the original supersymmetry
 \item $\frac{1}{4}$ of the original supersymmetry
 \item $\frac{1}{8}$ of the original supersymmetry
\end{enumerate}
  As we have already emphasized in
section \ref{gen4d} and further discussed in chapter \ref{chargcha}
the distinction between these three kinds of BPS solutions can be
considered directly in a $4$--dimensional setup and it is related to
the structure of the central charge eigenvalues and to the behaviour
of the scalar fields at the horizon. BPS black--holes with a finite
horizon area are those for which the scalar fields are regular at the
horizon and reach a fixed value there. These
can only be the $1/8$--type black holes, whose structure is that of
$N=2$ black--holes embedded into the $N=8$ theory. For $1/2$ and
$1/4$ black--holes the scalar fields always diverge at the horizon
and the entropy is zero.
\par
The nice point, in this respect, is that we can make a complete
classification of all BPS black holes belonging to the three possible
types. Indeed the distinction of the solutions into these
three classes can be addressed a priori and, as we are going to see,
corresponds to a classification into different orbits of  the
possible
${\bf 56}$--dimensional vectors $Q=\{ p^\Lambda , q_\Sigma \}$
of magnetic-electric charges of the hole. Indeed
$N=8$ supergravity contains ${\bf 28}$ gauge fields $A^\Lambda_\mu $
and correspondingly the hole can carry ${\bf 28}$ magnetic
$p^\Lambda$ and ${\bf 28}$ electric $q_\Sigma$ charges. Through the
symplectic embedding of the scalar manifold \eqn{e7su8n} it follows
that the field strengths $F^\Lambda_{\mu \nu}$ plus their duals
$G_{\Sigma \vert \mu \nu}$ transform in the fundamental ${\bf 56}$
representation of $E_{7(7)}$ and the same is true of their integrals,
namely the charges.
\par
The Killing spinor equation \eqn{kvano} that imposes preservation of
either $1/2$, or $1/4$, or $1/8$ of the original supersymmetries
enforces two consequences different in the three cases:
\begin{enumerate}
\item a different decomposition of the scalar field manifold into
two sectors:
{\begin{itemize}
\item a sector of {\it dynamical scalar fields} that evolve in the radial
parameter $r$
\item a sector of {\it spectator scalar fields} that do not evolve in
$r$ and are constant in the BPS solution.
\end{itemize}
}
\item a different orbit structure for the charge vector $Q$
\end{enumerate}
Then, up to $U$--duality transformations, for each case one can write
 a fully general {\it generating solution} that contains the
minimal necessary number of excited dynamical fields and the minimal
necessary number of non vanishing charges. All other solutions of the
same supersymmetry type can be obtained form the generating one by
the action of   $E_{7(7)}$--rotations.
\par
Such an analysis is clearly group--theoretical and requires
the use of appropriate techniques. The basic one is provided by the
solvable Lie algebra representation of the scalar manifold.
\section{Solvable Lie algebra description
of the supergravity scalar manifold}
\markboth{BPS BLACK HOLES IN SUPERGRAVITY: CHAPTER 5}
{5.2 SOLVABLE LIE ALGEBRA DESCRIPTION}
\setcounter{equation}{0}
\label{soldesc}
As we already explained in chapter \ref{n2bholes}
it has been known for many years \cite{sase} that
the scalar field manifold of both pure and matter coupled
$N>2$ extended supergravities in $D=10-r$ ($r=6,5,4,3,2,1$)
is a non compact homogenous symmetric manifold $U_{(D,N)} /H_{(D,N)}$,
where $U_{(D,N)}$ (depending on the space--time dimensions
and on the number of supersymmetries) is a non compact
Lie group and $H_{(D,N)}\subset U_{(D,N)}$ is a maximal compact subgroup.
For instance in the physical $D=4$ case the situation was summarized
in table \ref{topotable}. In the case of maximally extended
supergravities in $D=10-r$ dimensions the scalar
manifold has a universal structure:
\begin{equation}
 { U_D\over H_D}  = {E_{r+1(r+1)} \over H_{r+1}}
\label{maximal1}
\end{equation}
where the Lie algebra of the $U_D$--group $E_{r+1(r+1)} $ is the
maximally non compact real section of the
exceptional $E_{r+1}$ series  of the simple complex Lie Algebras
and $H_{r+1}$ is its maximally compact subalgebra \cite{cre}.
This series of homogeneous spaces is summarized in table
\ref{costab}.
\par
\begin{table}[ht]
\begin{center}
\begin{tabular}{|c|c|c|c|}
\hline
$D=9$ &      $E_{2(2)}  \equiv SL(2,\IR)\otimes O(1,1)$ & $H =
O(2) $ &
$\mbox{dim}_{\bf R}\,(U/H) \, =\, 3$ \\
 \hline
$D=8$ &      $E_{3(3)}  \equiv SL(3,\IR)\otimes Sl(2,\IR)$ & $H =
O(2)\otimes O(3) $ &
$\mbox{dim}_{\bf R}\,(U/H) \, =\, 7$ \\
\hline
$D=7$ &      $E_{4(4)}  \equiv SL(5,\IR) $ & $H = O(5) $ &
$\mbox{dim}_{\bf R}\,(U/H) \, =\, 14$ \\
\hline
$D=6$ &      $E_{5(5)}  \equiv O(5,5) $ & $H = O(5)\otimes O(5) $ &
$\mbox{dim}_{\bf R}\,(U/H) \, =\, 25$ \\
\hline
$D=5$ &      $E_{6(6)}$   & $H = Usp(8) $ &
$\mbox{dim}_{\bf R}\,(U/H) \, =\, 42$ \\
\hline
$D=4$ &      $E_{7(7)}$   & $H = SU(8) $ &
$\mbox{dim}_{\bf R}\,(U/H) \, =\, 70$ \\
\hline
$D=3$ &      $E_{8(8)}$   & $H = O(16) $ &
$\mbox{dim}_{\bf R}\,(U/H) \, =\, 128$ \\
\hline
\end{tabular}
\caption{U--duality groups and maximal compact subgroups of maximally
extended supergravities.}
\label{costab}
\end{center}
\end{table}
All the homogeneous coset manifolds appearing
in the various extended supergravity models
share the very important property of being
non--compact Riemannian manifolds
${\cal M}$ admitting a solvable Lie algebra description, i.e.
they can be expressed
as Lie group manifolds generated by a solvable Lie algebra $Solv$:
\begin{equation}
{\cal M}\,=\, \exp{(Solv)}
\label{solvrep}
\end{equation}
This property is of great relevance for the physical interpretation
of supergravity since it allows an intrinsic algebraic
characterization of the scalar fields of these theories  which,
without such a characterization, would simply be undistinguishable
coordinates of a manifold. Furthermore the choice of the solvable
parametrization provides essential technical advantages with respect
to the choice of other parametrizations and plays a crucial role in
the solution of BPS black hole equations. For this reason we devote
the present section to explaining the basic features of such a solvable
Lie algebra description of supergravity scalars.
\subsection{Solvable Lie algebras: the machinery}
\par
Let us start by giving few preliminary definitions.
A {\it solvable } Lie algebra $\IG_s$ is a Lie algebra whose $n^{th}$ order
(for some $n\geq 1$) derivative algebra vanishes:
\begin{eqnarray}
{\cal D}^{(n)}\IG_{s}&=&0 \nonumber \\
\cD\IG_s=[\IG_s,\IG_s]&;&\quad \cD^{(k+1)}\IG_s=[\cD^{(k)}\IG_s,
\cD^{(k)}\IG_s]\nonumber
\end{eqnarray}
A {\it metric} Lie algebra $(\IG,h)$ is a Lie algebra endowed with an
euclidean metric $h$. An important theorem states that if a Riemannian
manifold
$(\cM,g)$ admits a transitive group of isometries $\cG_s$ generated by
a solvable Lie algebra $\IG_s$ of the same dimension as $\cM$, then:
\begin{eqnarray}
\cM\sim \cG_s&=& \exp (\IG_s)\nonumber\\
 g_{|e\in \cM}&=&h \nonumber
\end{eqnarray}
where $h$ is an euclidean metric defined on $\IG_s$.
 Therefore there is a one to one correspondence between
 Riemannian manifolds fulfilling the hypothesis
stated above and solvable metric Lie algebras $(\IG_s,h)$.\\
Consider now a homogeneous coset manifold $\cM=\cG /\cH$,
$\cG$ being a non compact real
 form of a semisimple Lie group and $\cH$ its maximal compact
subgroup. If $\IG$ is the Lie algebra generating $\cG$, the so
called Iwasawa
decomposition ensures the existence of a solvable Lie subalgebra
$\IG_s\subset
\IG$, acting transitively on $\cM$, such that \cite{hel}:
\begin{equation}
\IG=\IH + \IG_s \qquad \mbox{dim }\IG_s=\mbox{dim }\cM
\label{soldeco}
\end{equation}
$\IH$ being the maximal compact subalgebra of $\IG$ generating $\cH$.
Note that the sum \eqn{soldeco} is not required to be an orthogonal
decomposition, namely the elements of $\IG_s$ are not requested to be
orthogonal to the elements of $\IH$ but simply linearly
independent from them.\\
In virtue of the previously stated theorem, $\cM$ may be expressed
 as a solvable
 group manifold generated by $\IG_s$. The algebra $\IG_s$ is constructed
as follows \cite{hel}. Consider the Cartan orthogonal decomposition
\begin{equation}
\IG = \IH \oplus \IK
\label{cartdeco}
\end{equation}
Let us denote by $\cH_K$ the  maximal abelian subspace
of $\IK$ and by $\cH$
the Cartan subalgebra of $\IG$.
It can be proven \cite{hel} that $\cH_K = \cH \cap \IK$,
that is it consists of all non compact elements of $\cH$.
Furthermore let $h_{\alpha_i}$ denote the elements
of $\cH_K$, $\{\alpha_i\}$ being
 a subset of the positive roots of $\IG$ and $\Phi^+$
 the set of positive roots $\beta$ not orthogonal to
all the $\alpha_i$ (i.e. the corresponding ``shift'' operators
$E_\beta$ do not commute with $\cH_K$).
It can be demonstrated that the solvable algebra $\IG_s$
defined by the Iwasawa decomposition
may be expressed in the following way:
\begin{equation}
  \label{iwa}
  \IG_s = \cH_K \oplus \{\sum_{\alpha \in \Phi^+}E_\alpha \cap \IG \}
\end{equation}
where  the intersection with $\IG $ means that $\IG_s$ is generated
by those suitable complex combinations of the ``shift'' operators
which belong to the real form of the isometry algebra $\IG$.
\par
The {\it rank} of a homogeneous  coset manifold is defined as
the maximum number of commuting semisimple
elements of the non compact subspace $\IK$. Therefore it
coincides with the dimension of $\cH_K$,
i.e. the number of non compact Cartan generators of $\IG$.
A  coset manifold is {\it maximally non compact} if
$\cH =\cH_K \subset \IG_s$.
The relevance of maximally non compact coset manifolds relies
on the fact that they are spanned by
the scalar fields in the maximally extended supergravity theories.
In the solvable representation of a manifold the local
coordinates of the manifold are the parameters of the generating Lie algebra,
therefore adopting this parametrization of scalar manifolds in supergravity
implies the definition of a one to one correspondence
between the scalar fields and the
generators of $Solv$ \cite{RR,solv}.
\par
Special K\"ahler manifolds and Quaternionic manifolds
admitting such a description have been
classified in the $70$'s by Alekseevskii \cite{alek}.
\subsection{Solvable Lie algebras: the simplest example}
The simplest example of solvable Lie algebra
parametrization is the case of the two
dimensional manifold ${\cal M}=SL(2,\IR)/SO(2)$ which
may be described as the exponential of the following solvable Lie algebra:
\begin{eqnarray}
SL(2,\IR)/SO(2)\, &=&\,\exp{(Solv)}\nonumber\\
Solv\, &=&\,\{\sigma_3,\sigma_+  \}\nonumber\\
\left[\sigma_3,\sigma_+\right]\, &=&\,2\sigma_+\nonumber\\
\sigma_3\, =\,\left(\matrix{1 & 0\cr 0 & -1}\right)\,\,&;&\,\,\sigma_+\, =\,
\left(\matrix{0 & 1\cr 0 & 0}\right)
\label{key}
\end{eqnarray}
From (\ref{key}) we can see a general feature of $Solv$, i.e. it may always be expressed as the
direct sum  of semisimple
(the non--compact Cartan generators of the isometry group)
 and nilpotent generators, which in a suitable basis are represented
respectively by diagonal and upper triangular matrices. This property, as we
shall see, is one of the advantages of the solvable Lie algebra description
since it allows to express the coset representative of a homogeneous manifold
as a solvable group element which is the product of a diagonal matrix and the
exponential of a nilpotent matrix, which is a polynomial in the parameters.
The simple solvable algebra represented in (\ref{key}) is called {\it key}
algebra and will be denoted by F.
\subsection{Another example: the solvable Lie algebra of the $STU$
model}
\label{another}
As a second example we illustrate the solvable Lie algebra parametrization
of   a coset  manifold that will play a crucial role
in the discussion of $1/8$ preserving BPS black holes.
The coset manifold we refer to is an instance of special K\"ahler
homogeneous manifold, namely
\begin{eqnarray}
 {\cal ST}[2,2] &=& \frac{SU(1,1)}{U(1)}\, \otimes \,
 \frac{SO(2,2)}{SO(2) \times SO(2)} \nonumber\\
 & \sim &
 \frac{SU(1,1)}{U(1)}\, \otimes \, \frac{SU(1,1)}{U(1)}\, \otimes \,
 \frac{SU(1,1)}{U(1)}\nonumber\\
 & \sim & \left(\frac{SL(2,\IR)}{SO(2)}\right)^3\,
 \label{su113}
\end{eqnarray}
This is the scalar manifold of an $N=2$ supergravity theory
containing
three vector multiplets. The corresponding three complex scalar
fields are usually named $S$, $T$ and $U$ and for this reason such a
model is named the $STU$ model. Hence
the scalar manifold of the $STU$ model is
a special K\"ahler manifold generated
by a solvable Lie algebra which is the sum of $3$ commuting
key algebras:
\begin{eqnarray}
{\cal M}_{STU}\,&=&\, \left(\frac{SL(2,\IR)}{SO(2)}\right)^3\,=\,
\exp{(Solv_{STU})}\nonumber\\
Solv_{STU}\,&=&\, F_1\oplus F_2\oplus F_3\nonumber\\
F_i\,=\,\{h_i,g_i\}\qquad&;&\qquad \left[h_i,g_i\right]=2g_i\nonumber\\
\left[F_i,F_j\right]\,&=&\,0
\end{eqnarray}
the parameters of the Cartan generators $h_i$ are the dilatons of the theory,
while the parameters of the nilpotent generators $g_i$ are the axions.
The three $SO(2)$ isotropy groups of the manifold are generated by the three
compact generators $\tilde{g}_i=g_i-g^\dagger_i$. \par
\subsection{A third example: solvable Lie algebra description of the
manifold ${\cal M}_{T^6/Z_3}$ }
\label{trexe}
If we compactify the type IIA superstring on the orbifold of a
six--torus modulo some discrete group we obtain a low energy
effective supergravity that corresponds to a truncation of the
original $N=8$ supergravity, generally with lower supersymmetry.
For instance if the six extra dimensions are compactified on the
orbifold $T^6/\ZZ_3$, the resulting theory is $N=2$ supergravity
coupled to ${\bf 9}$ vector multiplets. The $18$ scalars belonging to
these vector multiplets are the coordinates of a special K\"ahler
manifold  that we name ${\cal M}_{T^6/Z_3}$ since it can be
geometrically interpreted as the moduli space of K\"ahler class
deformations of the orbifold. This space is one of the exceptional
special K\"ahler homogeneous spaces of the Cremmer Van Proeyen
classification: indeed it is $SU(3,3)/SU(3) \, \times \, U(3)$
( see table \eqn{homospectable}). It plays an important role
in deriving the {\it generating solution} of BPS
black holes of the $1/8$ type. For this reason we anticipate here
its solvable Lie algebra description as an illustration of the
general theory by means of a third example.
We find that this  $18$--dimensional Special K\"ahler manifold is generated by a solvable
algebra whose structure is slightly more involved than that considered in the previous
example, indeed it contains the solvable Lie algebra of the $STU$
model plus some additional nilpotent generators. Explicitly we have:
\begin{eqnarray}
{\cal M}_{T^6/Z_3}\, &=&\,\frac{SU(3,3)}{SU(3)\times U(3)}\,=\,  \exp{(Solv)}\nonumber\\
Solv\, &=&\,Solv_{STU} \oplus\, {\bf X}\, \oplus\,
{\bf Y}\, \oplus\, {\bf Z}\nonumber\\
\label{stuembed}
\end{eqnarray}
The $4$ dimensional subspaces ${\bf X},{\bf Y},{\bf Z}$ consist of nilpotent
generators, while the only semisimple generators are the $3$ Cartan generators
contained in $Solv_{STU}$ which define the rank of the manifold.
The algebraic structure of $Solv$ together with the
details of the construction of the $SU(3,3)$ generators in the representation
${\bf 20}$ is what we explain next. This corresponds to the explicit
construction of the symplectic embedding of the coset manifold
implied by special geometry. Such a construction will be
particularly relevant to the construction of the $1/8$ supersymmetry
preserving black--holes. Indeed the ${\bf 20}$ dimensional
representation of $SU(3,3)$ is that which accommodates the ${\bf 10}\oplus{\bf 10}$
electric plus magnetic field strengths of $N=2$ supergravity coupled
to ${\bf 9}$ vector multiplets. In a first step we shall argue that
the $N=8$ black--holes of type $1/8$ are actually solutions of such
an $N=2$ theory embedded into the $N=8$ one. In a second step we
shall even argue that the {\it generating solution} is actually
a solution of a further truncation of this $N=2$ theory to an
$STU$--model.
\par
Anticipating these physical motivations,
in describing the solvable Lie algebra
structure of the ${\cal M}_{T^6/Z_3}$ manifold we pay particular attention to how it
incorporates the $STU$ solvable algebra.
Indeed applying the decomposition (\ref{iwa}) to the manifold ${\cal M}_{T^6/Z_3}$
one obtains:
\begin{eqnarray}
SU(3,3)\, &=&\, \left[SU(3)_1\oplus SU(3)_2 \oplus U(1)\right]\oplus Solv\nonumber\\
Solv\, &=&\, F_1\, \oplus\, F_2\, \oplus\, F_3\, \oplus\, {\bf X}\, \oplus\,
{\bf Y}\, \oplus\, {\bf Z}\nonumber\\
F_i\, &=&\, \{{\rm h}_i\, ,\,{\rm g}_i\}\quad i=1,2,3\nonumber\\
{\bf X}\,=\, {\bf X}^+\, \oplus\, {\bf X}^- \;,\;
{\bf Y}\, &=&\, {\bf Y}^+\, \oplus\, {\bf Y}^-\;,\;
{\bf Z}\, =\, {\bf Z}^+\, \oplus\, {\bf Z}^-\nonumber\\
\left[{\rm h}_i\, ,\,{\rm g}_i\right]\, &=&\,
2{\rm g}_i\quad i=1,2,3\nonumber\\
\left[F_i\, ,\,F_j\right]\, &=&\, 0 \quad i\neq j\nonumber\\
\left[{\rm h}_3\, ,\,{\bf Y}^{\pm}\right]\, &=&\,\pm{\bf Y}^{\pm}\;,\;
\left[{\rm h}_3\, ,\,{\bf X}^{\pm}\right]\, =\,\pm{\bf X}^{\pm}
\nonumber\\
\left[{\rm h}_2\, ,\,{\bf Z}^{\pm}\right]\, &=&\,\pm{\bf Z}^{\pm}\;,\;
\left[{\rm h}_2\, ,\,{\bf X}^{\pm}\right]\, =\,{\bf X}^{\pm}
\nonumber\\
\left[{\rm h}_1\, ,\,{\bf Z}^{\pm}\right]\, &=&\,{\bf Z}^{\pm}\;,\;
\left[{\rm h}_1\, ,\,{\bf Y}^{\pm}\right]\, =\,{\bf Y}^{\pm}
\nonumber\\
\left[{\rm g}_1\, ,\,{\bf X}\right]\, &=&\left[{\rm g}_1\, ,\,{\bf Y}\right]
\,=\,\left[{\rm g}_1\, ,\,{\bf Z} \right]\, =\, 0\nonumber\\
\left[{\rm g}_2\, ,\,{\bf X}\right]\, &=&\left[{\rm g}_2\, ,\,{\bf Y}\right]
\,=\,\left[{\rm g}_2\, ,\,{\bf Z}^{+} \right]\, =\, 0\;,\;
\left[{\rm g}_2\, ,\,{\bf Z}^{-}\right]\, =\,{\bf Z}^+ \nonumber \\
\left[{\rm g}_3\, ,\,{\bf Y}^{+}\right]\, &=&\,\left[{\rm g}_3\, ,\,{\bf X}^{+}\right]
\,=\,\left[{\rm g}_3\, ,\,{\bf Z} \right]\, =\, 0\nonumber\\
\left[{\rm g}_3\, ,\,{\bf Y}^{-}\right]\, &=&\,{\bf Y}^+\, ;\,\,
\left[{\rm g}_3\, ,\,{\bf X}^{-}\right]\, =\,{\bf X}^+\nonumber\\
\left[F_1\, ,\,{\bf X}\right]\, &=&\,\left[F_2\, ,\,{\bf Y}\right]\, =\,
\left[F_3\, ,\,{\bf Z}\right]\, =\, 0\nonumber\\
\left[{\bf X}^-\, ,\,{\bf Z}^{-}\right]\, &=&\, {\bf Y}^-
\label{Alekal}
\end{eqnarray}
where the solvable subalgebra $Solv_{STU}=F_1\oplus F_2
\oplus F_3$ is the solvable algebra generating ${\cal M}_{STU}$.
Denoting by $\alpha_i$, $i=1,2,\dots,5$ the simple roots of $SU(3,3)$,
using the {\it canonical} basis
for the $SU(3,3)$ algebra, the generators in (\ref{Alekal})
have the following form:
\begin{eqnarray}
{\rm h}_1\, &=& \,  H_{\alpha_1}\quad {\rm g}_1\, = \,
{\rm i}E_{\alpha_1}\nonumber\\
{\rm h}_2\, &=& \, H_{\alpha_3}\quad {\rm g}_2\, = \,
{\rm i}E_{\alpha_3}\nonumber\\
{\rm h}_3\, &=& \,H_{\alpha_5}\quad {\rm g}_3\, = \,
{\rm i}E_{\alpha_5}\nonumber\\
{\bf X}^+\, &=&\, \left( \begin{array}{c}
{\bf X}^+_1={\rm i}(E_{-\alpha_4}+E_{\alpha_3+\alpha_4+\alpha_5})\\
{\bf X}^+_2=E_{\alpha_3+\alpha_4+\alpha_5}-E_{-\alpha_4}
\end{array}\right)\,\,\,\nonumber\\
{\bf X}^-\, &=&\, \left( \begin{array}{c}
{\bf X}^-_1={\rm i}(E_{\alpha_3+\alpha_4}+E_{-(\alpha_4+\alpha_5)})\\
{\bf X}^-_2=E_{\alpha_3+\alpha_4}-E_{-(\alpha_4+\alpha_5)}\end{array}\right)\nonumber\\
{\bf Y}^+\, &=&\, \left( \begin{array}{c}
{\bf Y}^+_1={\rm i}(E_{\alpha_1+\alpha_2+\alpha_3+\alpha_4+\alpha_5}+
E_{-(\alpha_2+\alpha_3+\alpha_4)})\\
{\bf Y}^+_2=E_{\alpha_1+\alpha_2+\alpha_3+\alpha_4+\alpha_5}-
E_{-(\alpha_2+\alpha_3+\alpha_4)}\end{array}\right)\nonumber\\
{\bf Y}^-\, &=&\, \left( \begin{array}{c}
{\bf Y}^-_1={\rm i}(E_{\alpha_1+\alpha_2+\alpha_3+\alpha_4}+
E_{-(\alpha_2+\alpha_3+\alpha_4+\alpha_5)})\\
{\bf Y}^-_2=E_{\alpha_1+\alpha_2+\alpha_3+\alpha_4}-
E_{-(\alpha_2+\alpha_3+\alpha_4+\alpha_5)}\end{array}\right)\nonumber\\
{\bf Z}^+\, &=&\, \left( \begin{array}{c}
{\bf Z}^+_1={\rm i}(E_{\alpha_1+\alpha_2+\alpha_3}+E_{-\alpha_2})\\
{\bf Z}^+_2=E_{\alpha_1+\alpha_2+\alpha_3}-E_{-\alpha_2}
\end{array}\right)\nonumber\\
{\bf Z}^-\, &=&\, \left( \begin{array}{c}
{\bf Z}^-_1={\rm i}(E_{\alpha_1+\alpha_2}+E_{-(\alpha_2+\alpha_3)})\\
{\bf Z}^-_2=E_{\alpha_1+\alpha_2}-E_{-(\alpha_2+\alpha_3)}\end{array}\right)
\label{su33struc}
\end{eqnarray}
As far as the embedding of the isotropy
group $SO(2)^3$ of ${\cal M}_{STU}$ inside the ${\cal M}_{T^6/Z_3}$
isotropy group $SU(3)_1\times SU(3)_2\times U(1)$ is concerned, the $3$ generators of
the former ($\{\tilde{g}_1,\tilde{g}_2,\tilde{g}_3\}$ )
are related to the Cartan generators of the latter in the following way:
\begin{eqnarray}
\tilde{g}_1\, &=&\, \frac{1}{2}\left(\lambda +\frac{1}{2}\left(H_{c_1}-H_{d_1}+
H_{c_1+c_2}-H_{d_1+d_2}\right)\right)\nonumber\\
\tilde{g}_2\, &=&\, \frac{1}{2}\left(\lambda +\frac{1}{2}\left(H_{c_1}-H_{d_1}
-2(H_{c_1+c_2}-H_{d_1+d_2})\right)\right)\nonumber\\
\tilde{g}_3\, &=&\, \frac{1}{2}\left(\lambda +\frac{1}{2}
\left(-2(H_{c_1}-H_{d_1})+(H_{c_1+c_2}-H_{d_1+d_2})\right)\right)
\label{relcart}
\end{eqnarray}
where $\{c_i\}, \{d_i\}$, $i=1,2$ are the simple roots of $SU(3)_1$ and
$SU(3)_2$ respectively, while $\lambda$ is the generator of $U(1)$.
\par
In order to perform the  truncation to an $STU$ model,
one needs to know also which of the ${\bf 10}+{\bf 10}$
 field strengths should  be set to zero in order to be left with the
 ${\bf 4}+{\bf 4}$ of $STU$ model.
 This information is provided by the decomposition of the ${\bf 20}$
of $SU(3,3)$  with respect to the isometry group of the $STU$ model,
namely $[SL(2,\IR)]^3$:
\begin{equation}
{\bf 20}\,\stackrel{SL(2,\IR)^3}{\rightarrow}\,{\bf (2,2,2)} \oplus 2\times
\left[{\bf (2,1,1)}\oplus
{\bf (1,2,1)}\oplus {\bf (1,1,2)}\right]
\label{chargedec}
\end{equation}
Then, skew diagonalizing the $5$ Cartan generators
of $SU(3)_1\times SU(3)_2\times U(1)$ on the ${\bf 20}$
we obtain the $10$ positive weights of the
representation as $5$ components vectors $\vec{v}^{\Lambda^\prime}$
($\Lambda^\prime=0,\dots,9$):
\begin{eqnarray}
\{C(n)\}\,&=&\, \{\frac{H_{c_1}}{2},\frac{H_{c_1+c_2}}{2},\frac{H_{d_1}}{2},
\frac{H_{d_1+d_2}}{2},
{\lambda}\}\nonumber\\
C(n)\cdot \vert v^{\Lambda^\prime}_x \rangle \,&=&\, v_{(n)}^{\Lambda^\prime} \vert
v^{\Lambda^\prime}_y \rangle \nonumber\\
C(n)\cdot \vert v^{\Lambda^\prime}_y \rangle \,&=&\, -v_{(n)}^{\Lambda^\prime} \vert
v^{\Lambda^\prime}_x \rangle
\end{eqnarray}
Using the relation (\ref{relcart}) we compute the value of the weights $v^{\Lambda^\prime}$
on the three generators $\tilde{g}_i$ and find out which are the
$4$ positive weights $\vec{v}^\Lambda$ ($\Lambda=0,\dots,3$)
of the ${\bf (2,2,2)}$ in (\ref{chargedec}). The complete list of weights
$\vec{v}^{\Lambda^\prime}$ and their
eigenvectors $\vert v^{\Lambda^\prime}_{x,y} \rangle$ are worked out below.\par
Evaluated on the Cartan subalgebra ${\cal H}$ of $SU(3)_1\oplus SU(3)_2\oplus U(1)$
the weights $\vec{v}^{\Lambda^\prime}$ read as follows:
\begin{eqnarray}
\vec{v}^{\Lambda^\prime}\, &=&\, v^{\Lambda^\prime}(\frac{H_{c_1}}{2},\frac{H_{c_1+c_2}}{2},
\frac{H_{d_1}}{2},\frac{H_{d_1+d_2}}{2},{\lambda})\nonumber\\
v^{0}\, &=&\,\{ 0,0,0,0,{\frac{3}{2}}\} \nonumber\\
v^{1}\, &=&\,\{ {\frac{1}{2}},{\frac{1}{2}},-{\frac{1}{2}},-{\frac{1}{2}},
-{\frac{1}{2}}\}\nonumber\\
v^{2}\, &=&\,\{ 0,{\frac{1}{2}},0,-{\frac{1}{2}},{\frac{1}{2}}\} \nonumber\\
v^{3}\, &=&\,\{ {\frac{1}{2}},0,-{\frac{1}{2}},0,{\frac{1}{2}}\}\nonumber\\
v^{4}\, &=&\,\{ {\frac{1}{2}},0,0,-{\frac{1}{2}},{\frac{1}{2}}\}\nonumber\\
v^{5}\, &=&\,\{ 0,{\frac{1}{2}},-{\frac{1}{2}},0,{\frac{1}{2}}\}\nonumber\\
v^{6}\, &=&\,\{ {\frac{1}{2}},0,{\frac{1}{2}},{\frac{1}{2}},{\frac{1}{2}}\}
\nonumber\\
v^{7}\, &=&\,\{ {\frac{1}{2}},{\frac{1}{2}},{\frac{1}{2}},0,-{\frac{1}{2}}\}
\nonumber\\
v^{8}\, &=&\,\{ 0,{\frac{1}{2}},{\frac{1}{2}},{\frac{1}{2}},{\frac{1}{2}}\}
\nonumber\\
v^{9}\, &=&\,\{ {\frac{1}{2}},{\frac{1}{2}},0,{\frac{1}{2}},-{\frac{1}{2}}\}
\label{9pesi}
\end{eqnarray}
The weights \eqn{9pesi} have been ordered in such a way that the first four define
the ${\bf (2,2,2)}$ of the subgroup $SL(2,\IR)^3\subset SU(3,3)$ and in the
physical interpretation of this algebraic construction the weight $\vec{v}^0$ will
be associated with the graviphoton since its restriction to the Cartan generators
$H_{c_1},H_{c_1+c_2},H_{d_1},H_{d_1+d_2}$ of the matter isotropy
group $H_{matter}=SU(3)_1\oplus SU(3)_2$ is trivial. \par
As promised, while constructing the ${\cal M}_{T^6/\ZZ_3}$ we have
completely defined the way it incorporates the $STU$ model.
Indeed we obtained an algebraic recipe to perform the truncation to the $STU$ model:
setting to zero all the scalars parametrizing the $12$ generators ${\bf X}\, \oplus\,{\bf Y}\,
\oplus\, {\bf Z}$ in (\ref{stuembed}) and the $6$ vector fields corresponding to the weights
$v^{\Lambda^\prime}$, $\Lambda^\prime=4,\dots,9$. Restricting the action of the $[SL(2,\IR)]^3$
generators ($h_i,g_i,\tilde{g}_i$) inside $SU(3,3)$ to the $8$ eigenvectors
$\vert v^{\Lambda}_{x,y}\rangle $($\Lambda=0,\dots,3$) the embedding of  $[SL(2,\IR)]^3$ in
$Sp(8)$ is automatically obtained
\footnote{In the $Sp(8)$ representation of the U--duality group
$[SL(2,\IR)]^3$ we shall use the non--compact Cartan generators
$h_i$ are diagonal. Such a representation will be
denoted by $Sp(8)_D$, where the subscript ``D'' stands for ``Dynkin''. This notation has been
introduced in \cite{mp1} to distinguish the representation $Sp(8)_D$ from $Sp(8)_Y$
(``Y'' standing for ``Young'') where on the contrary the Cartan generators of the compact isotropy
group (in our case $\tilde{g}_i$) are diagonal.
The two representations are related by an
orthogonal transformation.}.
\par
After performing the restriction to the $Sp(8)_D $ representation of $[SL(2,\IR)]^3$
we have just described,
the orthonormal basis $\vert v^\Lambda_{x,y}\rangle$ ($\Lambda =0,1,2,3$) is:
\begin{eqnarray}
\vert v^{1}_x\rangle\, &=&\,\{ 0,0,0,0,{1\over 2},-{1\over 2},{1\over 2},-{1\over 2}\} \nonumber \\
\vert v^{2}_x\rangle\, &=&\,\{ 0,0,0,0,{1\over 2},{1\over 2},{1\over 2},{1\over 2}\}\nonumber \\
\vert v^{3}_x\rangle\, &=&\,\{ -{1\over 2},{1\over 2},{1\over 2},-{1\over 2},0,0,0,0\} \nonumber \\
\vert v^{4}_x\rangle\, &=&\,\{ {1\over 2},{1\over 2},-{1\over 2},-{1\over 2},0,0,0,0\}\nonumber \\
\vert v^{1}_y\rangle\, &=&\,\{ {1\over 2},-{1\over 2},{1\over 2},-{1\over 2},0,0,0,0\}\nonumber \\
\vert v^{2}_y\rangle\, &=&\,\{ {1\over 2},{1\over 2},{1\over 2},{1\over 2},0,0,0,0\} \nonumber \\
\vert v^{3}_y\rangle\, &=&\,\{ 0,0,0,0,-{1\over 2},{1\over 2},{1\over 2},-{1\over 2}\}\nonumber \\
\vert v^{4}_y\rangle\, &=&\,\{ 0,0,0,0,{1\over 2},{1\over 2},-{1\over 2},-{1\over 2}\}
\end{eqnarray}
The $Sp(8)_D$ representation of the generators of $Solv_{STU}$ reads as:
{\small
\begin{eqnarray}
h_1\,&=&\,\frac{1}{2}\left(\matrix{ 1 & 0 & 0 & 0 & 0 & 0 & 0 & 0 \cr 0 &
    -1 & 0 & 0 & 0 & 0 & 0 & 0 \cr 0 & 0 & 1 & 0 & 0 & 0 & 0 &
   0 \cr 0 & 0 & 0 & -1 & 0 & 0 & 0 & 0 \cr 0 & 0 & 0 & 0 &
    -1 & 0 & 0 & 0 \cr 0 & 0 & 0 & 0 & 0 & 1 & 0 & 0 \cr 0 &
   0 & 0 & 0 & 0 & 0 &
    -1 & 0 \cr 0 & 0 & 0 & 0 & 0 & 0 & 0 & 1 \cr  }\right)\quad ;\quad g_1\,=\,\frac{1}{2}\left(\matrix{
   0 & 0 & 0 & 0 & 0 & 0 & 1 & 0 \cr 0 & 0 & 0 & 0 & 0 & 0 &
   0 & 0 \cr 0 & 0 & 0 & 0 & 1 & 0 & 0 & 0 \cr 0 & 0 & 0 & 0 &
   0 & 0 & 0 & 0 \cr 0 & 0 & 0 & 0 & 0 & 0 & 0 & 0 \cr 0 & 0 &
   0 &
   -1 & 0 & 0 & 0 & 0 \cr 0 & 0 & 0 & 0 & 0 & 0 & 0 & 0 \cr 0 &
   -1 & 0 & 0 & 0 & 0 & 0 & 0 \cr  }\right)\nonumber\\
h_2\,&=&\,\frac{1}{2}\left(\matrix{ -1 & 0 & 0 & 0 & 0 & 0 & 0 & 0 \cr 0 &
    -1 & 0 & 0 & 0 & 0 & 0 & 0 \cr 0 & 0 & 1 & 0 & 0 & 0 & 0 &
   0 \cr 0 & 0 & 0 & 1 & 0 & 0 & 0 & 0 \cr 0 & 0 & 0 & 0 & 1 &
   0 & 0 & 0 \cr 0 & 0 & 0 & 0 & 0 & 1 & 0 & 0 \cr 0 & 0 & 0 &
   0 & 0 & 0 & -1 & 0 \cr 0 & 0 & 0 & 0 & 0 & 0 & 0 & -1 \cr  }\right)\quad ;\quad
g_2\,=\,\frac{1}{2}\left(\matrix{
   0 & 0 & 0 & 0 & 0 & 0 & 0 & 0 \cr 0 & 0 & 0 & 0 & 0 & 0 &
   0 & 0 \cr 0 & 0 & 0 & 0 & 0 & 0 & 0 &
    -1 \cr 0 & 0 & 0 & 0 & 0 & 0 &
    -1 & 0 \cr 0 & 1 & 0 & 0 & 0 & 0 & 0 & 0 \cr 1 & 0 & 0 &
   0 & 0 & 0 & 0 & 0 \cr 0 & 0 & 0 & 0 & 0 & 0 & 0 & 0 \cr 0 &
   0 & 0 & 0 & 0 & 0 & 0 & 0 \cr  }\right)\nonumber\\
h_3\,&=&\,\frac{1}{2}\left(\matrix{ 1 & 0 & 0 & 0 & 0 & 0 & 0 & 0 \cr 0 &
    -1 & 0 & 0 & 0 & 0 & 0 & 0 \cr 0 & 0 &
    -1 & 0 & 0 & 0 & 0 & 0 \cr 0 & 0 & 0 & 1 & 0 & 0 & 0 &
   0 \cr 0 & 0 & 0 & 0 &
    -1 & 0 & 0 & 0 \cr 0 & 0 & 0 & 0 & 0 & 1 & 0 & 0 \cr 0 &
   0 & 0 & 0 & 0 & 0 & 1 & 0 \cr 0 & 0 & 0 & 0 & 0 & 0 & 0 &
    -1 \cr  }\right)\quad ;\quad g_3\,=\,\frac{1}{2}\left(\matrix{ 0 & 0 & 0 & 0 & 0 & 0 & 0 &
    -1 \cr 0 & 0 & 0 & 0 & 0 & 0 & 0 & 0 \cr 0 & 0 & 0 & 0 &
   0 & 0 & 0 & 0 \cr 0 & 0 & 0 & 0 &
    -1 & 0 & 0 & 0 \cr 0 & 0 & 0 & 0 & 0 & 0 & 0 & 0 \cr 0 &
   0 & 1 & 0 & 0 & 0 & 0 & 0 \cr 0 & 1 & 0 & 0 & 0 & 0 & 0 &
   0 \cr 0 & 0 & 0 & 0 & 0 & 0 & 0 & 0 \cr  }\right)\nonumber
\end{eqnarray}
}
\subsection{Solvable Lie algebras:
R-R and NS-NS scalars in maximally extended supergravities}
\label{RRNS}
 At the beginning of this section we stated that every non compact homogeneous space ${\cal G}/{\cal H}$
is actually  a solvable group manifold and that its generating solvable Lie
algebra $Solv \left( {\cal G}/{\cal H} \right)$ can be constructed
utilizing roots and Dynkin diagram techniques. As we stressed this is important
in string theory since it
offers the   possibility of introducing an
intrinsic algebraic characterization of the supergravity scalars.
In particular this yields
a group--theoretical definition of Ramond--Ramond  and Neveu--Schwarz scalars.
It goes as follows. The same supergravity lagrangian admits different
interpretations as low energy theory of different superstrings
related by duality transformations or of M--theory. The
identification of the Ramond and Neveu Schwarz sectors is different in the
different interpretations. Algebraically this corresponds to
inequivalent decompositions of the solvable Lie algebra
$Solv \left( {\cal G}/{\cal H} \right)$ with respect to different subalgebras.
Each string theory admits a $T$--duality and an $S$--duality group
whose product $S \otimes   T$ constitutes a subgroup of the
$U$--duality group, namely of the isometry group $U \equiv {\cal G}$
of the homogeneous scalar manifold ${\cal G}/{\cal H}$. Physically
$S$ is a non perturbative symmetry acting on the {\it dilaton} while $T$ is
a perturbative symmetry acting on the "{\it radii}" of the
compactification. There exist also two compact subgroups ${\cal H}_S
\subset S$ and ${\cal H}_T \subset T$ whose product ${\cal H}_S \otimes
{\cal H}_T \,\subset  \, H$ is contained in the maximal compact
subgroup ${\cal H} \subset U$ such that we can write:
\begin{equation}
Solv \left( {\cal U}/{\cal H} \right) = Solv \left( { S}/{\cal H}_S \right)
\, \oplus  \,  Solv \left( { T}/{\cal H}_T \right) \,
\oplus \, {\cal W}
\label{stw}
\end{equation}
the three addends being all subalgebras of
$Solv \left( {\cal U}/{\cal H} \right)$. The first two addends
constitute the Neveu Schwarz sector while the last subalgebra ${\cal W}$
which is not only solvable but also {\it nilpotent} constitutes the
Ramond sector  relative to the chosen superstring interpretation.
\par
An example of this way of reasoning is provided by maximal
supergravities in $D=10-r$ dimensions. For such lagrangians the
scalar sector is given by ${\cal M}_{scalar} = E_{r+1(r+1)}/{\cal H}_{r+1}$
where the group $E_{r+1(r+1)}$ is obtained exponentiating
the maximally non compact real form of the exceptional rank $r+1$ Lie
algebra $E_{r+1}$ and ${\cal H}_{r+1}$ is the corresponding maximal
compact subgroup (see table \ref{costab}). If we interpret supergravity as the low energy
theory of Type IIA superstring compactified on a torus $T^r$, then
the appropriate $S$-duality group is $O(1,1)$ and the appropriate
$T$--duality group is $SO(r,r)$. Correspondingly we obtain the
decomposition:
\begin{equation}
Solv \left(  E_{r+1(r+1)}/ {\cal H}_{r+1} \right )
= O(1,1)
\, \oplus  \,  Solv \left(   \frac {SO(r,r)}{SO(r)\times SO(r)  }  \right) \,
\oplus \, {\cal W}_{n_{r+1}}
\label{equata}
\end{equation}
where the Ramond subalgebra ${\cal W}_{n_{r+1}}\equiv  spin[r,r]  $
is nothing else but the
chiral spinor representation of   $SO(r,r)$.
In the four dimensional case $r=6$ equation
\eqn{equata} takes the exceptional form:
\begin{equation}
Solv \left( \frac{E_{7(7)}}{ SU(8)}   \right )
= Solv \left(\frac{SL(2,R)}{O(2)}   \right )
\, \oplus  \,  Solv \left(   \frac {SO(6,6)}{SO(6)\times SO(6)  }  \right) \,
\oplus \, {\cal W}_{32}
\label{equadue}
\end{equation}
The 38 Neveu Schwarz scalars are given by the first two addends in
\eqn{equadue}, while the 32 Ramond scalars in the algebra ${\cal W}_{32}$
transform in the spinor representation of $SO(6,6)$ as in
all the other cases.
\par
Alternatively we can interpret maximal supergravity in $D=10-r$ as
the compactification on a torus $T^r$ of Type IIB superstring. In
this case the ST--duality group is different. We just have:
\begin{equation}
S \, \otimes  \, T \, = \, O(1,1) \, \otimes \, GL(r) \label{equatre}
\end{equation}
Correspondingly we write the solvable Lie algebra decomposition:
\begin{equation}
Solv \left(  E_{r+1(r+1)}/ {\cal H}_{r+1} \right )
= O(1,1)
\, \oplus  \,  Solv \left(   \frac {GL(r)}{SO(r) }  \right) \,
\oplus \, {\tilde {\cal W}}_{n_{r+1}}
\label{equaquat}
\end{equation}
where ${\tilde {\cal W}}_{r+1}$ is the new algebra of Ramond scalars
with respect to the Type IIB interpretation. Actually, as it is
well known, Type IIB theory already admits an $SL(2,R)$ U--duality symmetry
in ten dimensions that mixes Ramond and Neveu Schwarz states. The proper
S--duality group $O(1,1)$ is just a maximal subgroup of such $SL(2,R)$.
Correspondingly eq. \eqn{equaquat} can be restated as:
\begin{equation}
Solv_{r+1}\, \equiv \,Solv \left(  E_{r+1(r+1)}/ {\cal H}_{r+1} \right )
= Solv \left(  SL(2,R)/ O(2) \right )
\, \oplus  \,  Solv \left(   \frac {GL(r)}{SO(r) }  \right) \,
\oplus \, {\bar {\cal W}}_{n_{r+1}}
\label{equacinq}
\end{equation}
Finally a third decomposition of the same solvable Lie algebra can be
written if the same supergravity lagrangian is interpreted as
compactification on a torus $T^7$ of M--theory. For the details on
this and other decompositions of the scalar sector
that keep track of the sequential compactifications on multiple torii
we refer the reader to the original papers \cite{RR,solv}.
\section{Summary of $N=8$ supergravity.}
\label{41}
\setcounter{equation}{0}
\markboth{BPS BLACK HOLES IN SUPERGRAVITY: CHAPTER 5}
{5.3 SUMMARY OF $N=8$ SUPERGRAVITY}
Having reviewed the solvable Lie algebra description of the
supergravity scalar manifolds we next proceed to summarize the
structure of $N=8$ supergravity. The action is of the general form
\eqn{genact} with $g_{IJ}$ being the invariant metric of $E_{7(7)}/SU(8)$
and the period matrix ${\cal N}$ being determined from the Gaillard
Zumino master formula \eqn{masterformula} via the appropriate
symplectic embedding of $E_{7(7)}$ into $Sp(56,\IR)$ (see table
\ref{topotable}).
Hence, according to the general formalism discussed in
chapter \ref{chargcha} and to eq.\eqn{defu} we introduce the coset representative $\IL$ of
$E_{7(7)}\over SU\left(8\right)$ in the ${\bf 56}$ representation
of $E_{7(7)}$:
\begin{equation}
\label{Lcoset}
\IL={1 \over\sqrt{2}}\left(\begin{array}{c|c}
                             f+\mbox{\rm i} h & \bar{f}+\mbox{\rm i}\bar{h}\\
                             \hline\\
                             f-\mbox{\rm i} h & \bar{f}-\mbox{\rm i}\bar{h}
                             \end{array}
                     \right)
\end{equation}
where the submatrices $\left(h,f\right)$ are $28\times28$ matrices
labeled by antisymmetric pairs $\Lambda,\Sigma, A,B$ (
$\Lambda,\Sigma=1,\dots,8$, $A,B=1,\dots,8 $) the first pair
transforming under $E_{7(7)}$ and the second one under
$SU\left(8\right)$:
\begin{equation}
\left(h,f\right)=\left(h_{\Lambda\Sigma|AB},
f^{\Lambda\Sigma}_{~~AB}\right)
\end{equation}
As expected from the general formalism we have $\IL\in Usp\left(28,28\right)$.
The vielbein $P_{ABCD}$ and the $SU\left(8\right)$ connection
$\Omega_{A}^{~B}$ of $E_{7(7)}\over SU\left(8\right)$ are
computed from the left invariant $1$-form $\IL^{-1}d\IL$:
\begin{equation}
\label{L-1dL}
\IL^{-1}d\IL=\left(\begin{array}{c|c}
                             \delta^{[A}_{~~[C}\Omega^{B]}_{~~D]} & \bar{P}^{ABCD}\\
                             \hline\\
                             P_{ABCD} & \delta_{[A}^{~~[C}\bar{\Omega}_{B]}^{~~D]}\end{array}
                     \right)
\end{equation}
where $P_{ABCD}\equiv
P_{ABCD, i}d\Phi^{i}~~~\left(i=1,\dots,70\right)$ is
completely antisymmetric and satisfies the reality condition
\begin{equation}
\label{realcondP}
P_{ABCD}={1 \over 24}\epsilon_{ABCDEFGH}\bar P^{EFGH}
\end{equation}
The bosonic lagrangian of $N=8$ supergravity is \cite{cre}
\begin{eqnarray}
\label{lN=8}
{\cal L}&=&\int\sqrt{-g}\, d^4x\left(2R+\Im{\cal
N}_{\Lambda\Sigma|\Gamma\Delta}F_{\mu\nu}^{~~\Lambda\Sigma}F^{\Gamma\Delta|\mu\nu}+
{1 \over 6}P_{ABCD,i}\bar{P}^{ABCD}_{j}\partial_{\mu}
\Phi^{i}\partial^{\mu}\Phi^{j}+\right.\nonumber\\
&+&\left.{1 \over 2}\Re{\cal N}_{\Lambda\Sigma|\Gamma\Delta}
{\epsilon^{\mu\nu\rho\sigma}\over\sqrt{-g}}
F_{\mu\nu}^{~~\Lambda\Sigma}F^{\Gamma\Delta}_{~~\rho\sigma}\right)
\end{eqnarray}
where the curvature two-form is defined as
\begin{equation}
R^{ab}=d\omega^{ab}-\omega^a_{~c}\wedge\omega^{cb}.
\end{equation}
and the kinetic matrix ${\cal N}_{\Lambda\Sigma|\Gamma\Delta}$ is
given by the usual general formula:
\begin{equation}
\label{defN}
{\cal N}=hf^{-1}~~\rightarrow~~ {\cal N}_{\Lambda\Sigma|\Gamma\Delta}=
h_{\Lambda\Sigma|AB}f^{-1~AB}_{~~~~~\Gamma\Delta}.
\end{equation}
The same matrix relates the (anti)self-dual electric and magnetic
$2$-form field strengths, namely, setting
\begin{equation}
\label{defFpm}
F^{\pm~\Lambda\Sigma}={1 \over 2}\left(F\pm \mbox{\rm i}~\star
F\right)^{\Lambda\Sigma}
\end{equation}
according to the general formulae \eqn{gigiuno} one has
\begin{eqnarray}
G^{-}_{\Lambda\Sigma}&=&\bar{{\cal N}}_{\Lambda\Sigma|\Gamma\Delta}F^{-~\Gamma\Delta}
\nonumber\\
\label{defG}
G^{+}_{\Lambda\Sigma}&=&{\cal N}_{\Lambda\Sigma|\Gamma\Delta}F^{+~\Gamma\Delta}
\end{eqnarray}
where the ''dual'' field strengths $G^{\pm}_{\Lambda\Sigma}$, according to
the general formalism (see eq.\eqn{gtensor}) are
defined as $G^{\pm}_{\Lambda\Sigma}={i \over 2}{\delta{\cal L} \over
\delta F^{\pm~\Lambda\Sigma}}$.
Note that the $56$ dimensional (anti)self-dual vector $\left(F^{\pm~\Lambda\Sigma},G^{\pm}_{\Lambda\Sigma}\right)$
transforms covariantly under $U\in Sp\left(56,\IR\right)$
\begin{eqnarray}
&&U\left(\begin{array}{c} F \\ G\end{array}\right)=\left(\begin{array}{c} F' \\ G'\end{array}\right)~~;~~
U=\pmatrix{A&B\cr C&D\cr}\nonumber\\
&&A^tC-C^tA=0\nonumber\\
&&B^tD-D^tB=0\nonumber\\
&&A^tD-C^tB={\bf 1}
\end{eqnarray}
The matrix transforming the coset representative $\IL$ from the
$Usp\left(28,28\right)$ basis, eq.\eqn{Lcoset}, to the real
$Sp\left(56,\IR\right)$ basis is the Cayley matrix:
\begin{equation}
\IL_{Usp}={\cal C}\IL_{Sp}{\cal C}^{-1}~~~~~{\cal C}=\pmatrix{ \bfone &
\mbox{\rm i} \bfone
\cr \bfone &-\mbox{\rm i}\bfone \cr}
\label{caylone}
\end{equation}
implying eq.s\eqn{ABCDfh}.
Having established our definitions and notations, let us now write
down the Killing spinor equations obtained by equating to zero the
SUSY transformation laws of the gravitino $\psi_{A\mu}$ and dilatino $\chi_{ABC}$ fields
of $N=8$ supergravity in a purely bosonic background:
\begin{eqnarray}
\label{2}
\delta\chi_{ABC}&=&~4 \mbox{\rm i} ~P_{ABCD|i}\partial_{\mu}\Phi^i\gamma^{\mu}
\epsilon^D-3T^{(-)}_{[AB|\rho\sigma}\gamma^{\rho\sigma}
\epsilon_{C]}=0\\
\label{psisusy2}
\delta\psi_{A\mu}&=&\nabla_{\mu}\epsilon_A~
~-\frac{\rm 1}{4}\, T^{(-)}_{AB|\rho\sigma}\gamma^{\rho\sigma}\gamma_{\mu}\epsilon^{B}=0
\end{eqnarray}
 where
$\nabla_{\mu}$ denotes the derivative covariant both with respect to
Lorentz and $SU\left(8\right)$ local transformations
\begin{equation}
\nabla_{\mu}\epsilon_A=\partial_{\mu}\epsilon_A-{1\over
4}\gamma_{ab} \, \omega^{ab}\epsilon_A-\Omega_A^{~B}\epsilon_B
\label{delepsi}
\end{equation}
and where $T^{\left(-\right)}_{AB}$ is the ''dressed graviphoton''$2-$form,
  defined according to the general formulae \eqn{gravi}
\begin{equation}
T^{\left(-\right)}_{AB}=\left(h_{\Lambda\Sigma AB}
\left(\Phi\right)F^{-\Lambda\Sigma}-f^{\Lambda\Sigma}_{~~AB}
\left(\Phi\right)G^-_{\Lambda\Sigma}\right)
\label{tab}
\end{equation}
From equations (\ref{defN}), (\ref{defG}) we have the following
identities that are the particular $N=8$ instance of eq.\eqn{tiden0}:
\begin{eqnarray}
&&T^+_{AB}=0\rightarrow T^-_{AB}=T_{AB}\nonumber ~~~~~~
\bar{T}^-_{AB}=0\rightarrow \bar{T}^+_{AB}=\bar{T}_{AB}
\end{eqnarray}
Following the general procedure indicated by eq.\eqn{zab}
we can define the central charge:
\begin{equation}
Z_{AB}=\int_{S^2}{T_{AB}}=h_{\Lambda\Sigma|AB}
p^{\Lambda\Sigma}-f^{\Lambda\Sigma}_{~~AB}q_{\Lambda\Sigma}
\label{zab2}
\end{equation}
which in our case is an antisymmetric tensor transforming in the
${\bf {28}}$ irreducible representation of $SU(8)$. In eq.\eqn{zab2}
 the integral of the two-form $T_{AB}$ is evaluated on a large
two-sphere at infinity and the quantized charges ($p_{\Lambda\Sigma},~q^{\Lambda\Sigma}$)
are defined, following the general eq.s \eqn{charges} by
\begin{eqnarray}
p^{\Lambda\Sigma}&=&\int_{S^2}{F^{\Lambda\Sigma}}\nonumber\\
q_{\Lambda\Sigma}&=&\int_{S^2}{\cal N}_{\Lambda\Sigma|\Gamma\Delta}\star F^{\Gamma\Delta}.
\end{eqnarray}
\section{The Killing spinor equation and its covariance group}
\setcounter{equation}{0}
\markboth{BPS BLACK HOLES IN SUPERGRAVITY: CHAPTER 5}
{5.4 THE KILLING SPINOR EQUATION}
\label{ksecov}
In order to translate eq.\eqn{2} and \eqn{psisusy2}
into first order differential equations
 on the bosonic fields of supergravity we consider
a configuration where all the fermionic fields are zero and
a SUSY parameter that satisfies the following conditions:
\begin{equation}
\begin{array}{rclcl}
\chi^\mu \, \gamma_\mu \,\epsilon_{A} &=& \mbox{i}\, \IC_{AB}

\,  \epsilon^{B}   & ; &   A,B=1,\dots,n_{max}\\
\epsilon_{A} &=& 0  &;&   A> n_{max} \\
\end{array}
\label{kilspieq}
\end{equation}
Here $\chi^\mu$ is a time--like Killing vector for the space--time metric
( in the following we just write
 $\chi^\mu \gamma_{\mu} = \gamma^0$) and
$ \epsilon _{A}, \epsilon^{A}$ denote the two chiral projections of
a single Majorana spinor: $ \gamma _5 \, \epsilon _{A} \, = \, \epsilon _{A} $ ,
$ \gamma _5 \, \epsilon ^{A} \, = - \epsilon ^{A} $. This is just the
particularization to the $N=8$ case of eq.s \eqn{kispiro} defining
the Killing spinor $\xi_A$ and inserting an $\epsilon_A$ with the
properties \eqn{kilspieq} into eq.s \eqn{2},\eqn{psisusy2} we
obtain the $N=8$ instance of the general equation \eqn{kvano}.
We name such an equation the {\sl Killing spinor equation}
and the investigation of its
group--theoretical structure is the main task we face in order to
derive the three possible types of BPS black--holes, those preserving
$1/2$ or $1/4$ or $1/8$ of the original supersymmetry.
To appreciate the distinction among the three types of $N=8$ black--hole solutions
we need to recall the results of \cite{FGun} where a classification was given of the
${\bf 56}$--vectors of quantized electric and magnetic charges ${\vec Q}$ characterizing
such solutions. The basic argument is provided by the reduction of
the central charge skew--symmetric tensor $\ZZ_{AB}$ to normal form.
The reduction can always be obtained by means of local $SU(8)$
transformations, but the structure of the skew eigenvalues depends on
the orbit--type of the $\bf 56$--dimensional charge vector which can be described by means
of its stabilizer subgroup $G_{stab}({\vec Q}) \subset E_{7(7)}$:
\begin{equation}
g \, \in \, G_{stab}({\vec Q}) \subset E_{7(7)} \quad \Longleftrightarrow \quad
g \,  {\vec Q}  = {\vec Q}
\end{equation}
There are three possibilities:
\begin{equation}
\begin{array}{cccc}
\null & \null & \null & \null \\
 \mbox{SUSY} & \mbox{Central Charge} & \mbox{Stabilizer}\equiv G_{stab} &
 \mbox{Normalizer}\equiv G_{norm}\\
  \null & \null & \null & \null \\
   1/2 & Z_1 = Z_2 =Z_3 =Z_4  & E_{6(6)} & O(1,1) \\
 \null & \null & \null & \null \\
 1/4 & Z_1 = Z_2 \neq Z_3=Z_4 & SO(5,5)&
 SL(2,\IR) \times O(1,1) \\
 \null & \null & \null & \null \\
 1/8 & Z_1 \ne Z_2 \ne Z_3 \ne Z_4 & SO(4,4) & SL(2,\IR)^3 \\
 \null & \null & \null & \null \\
\label{classif}
\end{array}
\end{equation}
where the normalizer $G_{norm}({\vec Q})$ is defined as the subgroup
of $E_{7(7)}$ that commutes with the stabilizer:
\begin{equation}
   \left[ G_{norm} \, , \,  G_{stab} \right] = 0
\end{equation}
The main result of \cite{mp1} is that the most general $1/8$
black--hole solution of $N=8$ supergravity is related to the normalizer group
$SL(2,\IR)^3$. In the subsequent paper \cite{mp2}
the $1/2$ and $1/4$ cases were completely worked out. Finally
in \cite{mp3} the explicit form of the generating solutions
was discussed for the $1/8$ case. In these lectures
we review all these results in detailed form.
\par
In all three cases the Killing spinor equation has two features which we want
presently to stress:
\begin{enumerate}
\item{It requires an efficient parametrization of the scalar field
sector}
\item{It breaks the original $SU(8)$ automorphism group of the
supersymmetry algebra to the subgroup $Usp(2 \,n_{max})\times SU(8-2\,n_{max})\times U(1)$}
\end{enumerate}
The first feature is the reason why the use of the rank $7$ solvable Lie
algebra $Solv_7$ associated with $E_{7(7)}/SU(8)$ is of great help in this problem.
The second feature is the
reason why the solvable Lie algebra $Solv_7$ has to be decomposed in
a way appropriate to the decomposition of the isotropy group
$SU(8)$
with respect to the subgroup $Usp(2\,n_{max})\times SU(8-2\,n_{max})\times U(1)$.
\par
This decomposition of the solvable Lie algebra is a close relative of
the decomposition of $N=8$ supergravity into multiplets of the lower
supersymmetry $N^\prime = 2 \, n_{max}$. This is easily understood by
recalling that close to the horizon of the black hole one doubles the supersymmetries
holding in the bulk of the solution. Hence the near horizon
supersymmetry is precisely $N^\prime = 2 \, n_{max}$ and the black
solution can be interpreted as a soliton that interpolates between
{\it ungauged} $N=8$ supergravity at infinity and some form of
$N^\prime$ supergravity at the horizon.
\subsubsection{The $1/2$ SUSY case}
Here we have $n_{max} = 8$ and correspondingly the covariance subgroup
of the Killing spinor equation is $Usp(8)\, \subset \, SU(8)$. Indeed
 condition \eqn{kilspieq} can be rewritten as follows:
 \begin{equation}
\label{1/2killingspin}
\gamma^0 \,\epsilon_{A}  =  \mbox{i}\, \IC_{AB}
\,  \epsilon^{B} \quad ; \quad A,B=1,\dots ,8
\label{urcunmez}
\end{equation}
where $\IC_{AB}= - \IC_{BA}$ denotes an $ 8 \times 8$
antisymmetric matrix satisfying $\IC^2 = -\bfone$. The group
$Usp(8)$ is the subgroup of unimodular, unitary $ 8 \times 8$
matrices that are also symplectic, namely that preserve the matrix
$\IC$. Relying on eq. (\ref{classif}) we see that in the present case $G_
{\it stab}=E_{6(6)}$ and $G_{\it norm}=O(1,1)$. Furthermore we have the
following
decomposition of the ${\bf
70}$ irreducible representation of $SU(8)$ into irreducible
representations of $Usp(8)$:
\begin{equation}
{\bf 70} \, \stackrel{Usp(8)}{\longrightarrow} \, {\bf 42} \, \oplus
\, {\bf 1} \, \oplus \, {\bf 27}
\label{uspdecompo1}
\end{equation}
\par
We are accordingly lead to decompose the solvable Lie algebra as
\begin{eqnarray}
Solv_7 & =& Solv_6 \, \oplus \, O(1,1) \, \oplus \ID_6
\label{decomp1a}\\
70 & = & 42 \, + \, 1 \, + \, 27
\label{decomp1b}
\end{eqnarray}
where, following the notation established in \cite{mp1,RR}:
\begin{eqnarray}
Solv_7 & \equiv & Solv \left(\frac{E_{7(7)}}{SU(8)}\right)
\nonumber\\
Solv_6 & \equiv & Solv \left(\frac{E_{6(6)}}{Usp(8)}\right)
\nonumber\\
\mbox{dim}\, Solv_7 & = & 70 \quad ; \quad \mbox{rank}\, Solv_7 \, =
\, 7 \nonumber\\
\mbox{dim}\, Solv_6 & = & 42 \quad ; \quad \mbox{rank}\, Solv_6 \, =
\, 6 \nonumber\\
\label{defsolv7}
\end{eqnarray}
In eq.\eqn{decomp1a}
$Solv_6$ is the solvable Lie algebra that describes the scalar
sector of $D=5$, $N=8$ supergravity, while the $27$--dimensional
abelian ideal $\ID_6$ corresponds to those $D=4$ scalars that
originate from the $27$--vectors of supergravity one--dimension
above \cite{solv}.
Furthermore, we can also decompose the $\bf 56$ charge representation
of $E_{7(7)}$ with respect to $O(1,1)\times E_{6(6)}$ obtaining
\begin{equation}
{\bf 56}\stackrel{Usp(8)}{\longrightarrow}
({\bf 1,27})\oplus({\bf 1,27})\oplus({\bf 2,1})
\label{visto}
\end{equation}
 \par
In order to  single out the content of the first order Killing
spinor equations we need to decompose them into irreducible $Usp(8)$
representations.
The gravitino equation \eqn{psisusy2} is an ${\bf 8}$ of $SU(8)$ that
remains irreducible under $Usp(8)$ reduction. On the other hand the dilatino
equation \eqn{2} is a ${\bf 56}$ of $SU(8)$ that reduces as follows:
\begin{equation}
{\bf 56} \, \stackrel{Usp(8)}{\longrightarrow} \, {\bf 48} \, \oplus
\, {\bf 8}
\label{uspdecompo2}
\end{equation}
Hence altogether we have that $3$ Killing spinor equations in the
representations ${\bf 8}$,
${\bf 8}^\prime$ , ${\bf 48}$ constraining the scalar fields
 parametrizing  the three subalgebras
${\bf 42}$, ${\bf 1}$ and ${\bf 27}$. Working out
the consequences of these constraints and deciding which scalars are
set to constants, which are instead evolving and how many charges
are different from zero is what we will do in a later
section \ref{det1/2}.    As it will be  explicitly seen there
the
content of the Killing spinor equations after $Usp(8)$ decomposition,
is such as to set to a constant $69$ scalar fields parametrizing
$Solv_6\oplus\ID_6$ thus confirming the SLA analysis discussed in the
above: indeed in this case $G_{norm} = O(1,1)$ and $H_{norm}
=\bf 1$, so that there is just one
 surviving field parametrizing $G_{norm}= O(1,1)$.
Moreover, the same Killing spinor equations
tell us that the $54$ belonging to the two $({\bf 1},{\bf 27})$
representation of eq. (\ref{visto})
are actually zero, leaving only two non--vanishing charges transforming
as a doublet of $O(1,1)$.
\subsubsection{The $1/4$ SUSY case}
Here we have $n_{max} = 4$ and correspondingly the covariance subgroup
of the Killing spinor equation is $Usp(4)\,\times \, SU(4) \,\times \, U(1)
\subset \, SU(8)$. Indeed condition \eqn{kilspieq} can be rewritten as follows:
\begin{eqnarray}
\gamma^0 \,\epsilon_{a} & =&  \mbox{i}\, \IC_{ab}
\,  \epsilon^{b} \quad ; \quad a,b=1,\dots ,4 \nonumber\\
\epsilon_{X} & =& 0; \quad X=5,\dots ,8 \nonumber\\
\label{urcunquart}
\end{eqnarray}
where $\IC_{ab}= - \IC_{ba}$ denotes a  $ 4 \times 4$
antisymmetric matrix satisfying $\IC^2 = -\bfone$. The group
$Usp(4)$ is the subgroup of unimodular, unitary $ 4 \times 4$
matrices that are also symplectic, namely that preserve the matrix
$\IC$.
\par
We are accordingly lead to decompose the solvable Lie algebra in the
way we describe below. Recalling the decomposition with respect to the $S
\otimes T$ duality subgroups given in eq.\eqn{stw} and choosing the
type IIA interpretation of $N=8$ supergravity we can start from eq.
\eqn{equadue}, which we can summarize as follows:
\begin{eqnarray}
Solv_7 & =& Solv_S \, \oplus \, Solv_T \, \oplus \, {\cal W}_{32}
\label{stdecomp1a}\\
70 & = & 2\, + \, 36 \, + \, 32  \ .
\label{stdecomp1b}
\end{eqnarray}
where we have adopted the shorthand notation
\begin{eqnarray}
Solv_S & \equiv & Solv \left(\frac{SL(2,R)}{U(1)}\right)
\nonumber\\
Solv_T & \equiv & Solv \left(\frac{SO(6,6)}{SO(6) \times SO(6)}\right)
\nonumber\\
\mbox{dim}\, Solv_S & = & 2 \quad ; \quad \mbox{rank}\, Solv_S \, =
\, 1 \nonumber\\
\mbox{dim}\, Solv_T & = & 36 \quad ; \quad \mbox{rank}\, Solv_T \, =
\, 6 \nonumber\\
\label{defsolst}
\end{eqnarray}
As discussed in section \ref{RRNS} and more extensively
explained in ref.s\cite{RR},\cite{solv},
the solvable Lie algebras $Solv_S$ and $Solv_T$ describe the dilaton--axion
sector and the six torus moduli, respectively, in the interpretation
of $N=8$ supergravity as the compactification of Type IIA
theory on a six--torus
$T^6$ \cite{solv}. The rank zero abelian subalgebra ${\cal W}_{32}$
is instead composed by  the $32$ Ramond-Ramond scalars.
\par
Introducing the decomposition \eqn{stdecomp1a}, \eqn{stdecomp1b} we have
succeeded in singling out a holonomy subgroup $SU(4) \, \times \,
SU(4) \, \times \, U(1) \, \subset \, SU(8)$. Indeed we have
$SO(6) \, \equiv \, SU(4)$. This is a step forward but it is not yet
the end of the story since we actually need a subgroup $Usp(4) \,
\times \, SU(4)\times U(1)$ corresponding to the invariance group
of the Killing spinor equation \eqn{psisusy2},\eqn{2} with
parameter \eqn{urcunquart} . This means that we must further decompose the
solvable Lie algebra  $Solv_T$. This latter is the manifold of the
scalar fields associated with vector multiplets in an $N=4$
decomposition of the $N=8$ theory. Indeed the decomposition
\eqn{stdecomp1a} with respect to the S--T--duality subalgebra is the
appropriate decomposition of the scalar sector according to $N=4$
multiplets.
\par
The  further SLA decomposition we need is  :
\begin{eqnarray}
Solv_T &=& Solv_{T5} \oplus Solv_{T1}  \nonumber\\
Solv_{T5}& \equiv &Solv \left( \frac{SO(5,6)}{SO(5) \times
SO(6)}\right)\nonumber  \\
Solv_{T1}& \equiv &Solv \left( \frac{SO(1,6)}{
SO(6)}\right)\nonumber  \\
\label{stspilt}
\end{eqnarray}
where we rely on the
isomorphism $Usp(4)\equiv SO(5)$ and we have taken into
account that the $\bf{70}$ irreducible
representation of $SU(8)$ decomposes with respect to
$Usp(4) \, \times \, SU(4)\times U(1)$ as follows
\begin{equation}
{\bf 70} \, \stackrel{Usp(4) \, \times \, SU(4) \, \times
\, U(1)}{\longrightarrow}
\, \left( {\bf 1},{\bf 1},{\bf 1}+{\bar {\bf 1}}\right) \, \oplus
\, \left({\bf 5},{\bf 6}, {\bf 1} \right) \, \oplus \,
\left( {\bf 1},{\bf 6},{\bf 1} \right) \, \oplus \,
\left( {\bf 4} ,{\bf 4}, {\bf 1}\right) \, \oplus \,
\left( {\bf 4},{\bf 4},{\bf 1} \right)
\label{uspdecompo3}
\end{equation}
Hence,
altogether we can write:
\begin{eqnarray}
Solv_7 &=& Solv_S \, \oplus \, Solv_{T5} \, \oplus \, Solv_{T1} \,
\oplus \, {\cal W}_{32} \nonumber\\
70 & = & 2 \, + \, 30 \, + \,  6\, + 32
\end{eqnarray}
Just as in the previous case we should now
single out the content of the first order Killing
spinor equations by decomposing  them into irreducible
$Usp(4) \, \times \, SU(4) \, \times \, U(1)$ representations.
The dilatino equations $\delta \chi_{ABC}=0$, and the gravitino
equation $\delta\psi_A=0$, $A,B,C=1,\ldots,8$  ($SU(8)$ indices),
decompose as follows
\begin{equation}
{\bf 56} \, \stackrel{Usp(4) \, \times \, SU(4) }{\longrightarrow}
\, ( {\bf 4},{\bf 1}) \, \oplus
\, 2({\bf 1},{\bf 4},)\, \oplus \,
( {\bf 5},{\bf 4}) \oplus \,
( {\bf 4} ,{\bf 6})
\label{uspdecomp56}
\end{equation}
\begin{equation}
{\bf 8} \, \stackrel{Usp(4) \, \times \, SU(4) }{\longrightarrow}
\, ( {\bf 4},{\bf 1}) \, \oplus
\, ({\bf 1},{\bf 4},)
\label{usp56}
\end{equation}
As we
shall see explicitly in
section \ref{seckilling1/4}, the content of the reduced Killing
spinor equations is such that only two scalar fields are essentially
dynamical all the other being set to constant up to $U$- duality
transformations. Moreover $52$ charges are set to zero leaving $4$
charges transforming in the $(2,2)$ representation of $Sl(2,\IR)\times
O(1,1)$.
Note that in the present case on the basis of the SLA analysis given
above, one would expect $3$ scalar fields parametrizing
 $ G_{norm}/ H_{norm}$ =
${Sl(2,\IR)\over U(1)} \times O(1,1)$; however the relevant Killing
spinor equation gives an extra reality constraint on the
$Sl(2,\IR)\over U(1)$ field
thus reducing the number of non trivial scalar fields to two.
\subsubsection{The $1/8$ SUSY case}
\label{gen1/8}
Here we have $n_{max} = 2$ and
$Solv_7$ must be decomposed  according to the decomposition
of the isotropy subgroup: $SU(8) \longrightarrow SU(2)\times U(6)$. We
showed in \cite{mp1}  that the corresponding decomposition of the solvable
Lie algebra is the following one:
\begin{equation}
Solv_7   =  Solv_3 \, \oplus \, Solv_4
\label{7in3p4}
\end{equation}
\begin{equation}
\begin{array}{rclrcl}
Solv_3 & \equiv & Solv \left( SO^\star(12)/U(6) \right) & Solv_4 &
\equiv & Solv \left( E_{6(4)}/SU(2)\times SU(6) \right) \\
\mbox{rank }\, Solv_3 & = & 3 & \mbox{rank }\,Solv_4 &
= & 4 \\
\mbox{dim }\, Solv_3 & = & 30 & \mbox{dim }\,Solv_4 &
= & 40 \\
\end{array}
\label{3and4defi}
\end{equation}
The rank three  Lie algebra $Solv_3$ defined above describes the
thirty dimensional scalar sector of $N=6$ supergravity, while the rank four
solvable Lie algebra $Solv_4$ contains the remaining forty scalars
belonging to $N=6$ spin $3/2$ multiplets. It should be noted
that, individually, both manifolds $ \exp \left[ Solv_3 \right]$ and
$ \exp \left[ Solv_4 \right]$ have also an $N=2$ interpretation since we have:
\begin{eqnarray}
\exp \left[ Solv_3 \right] & =& \mbox{homogeneous special K\"ahler}
\nonumber \\
\exp \left[ Solv_4 \right] & =& \mbox{homogeneous quaternionic}
\label{pincpal}
\end{eqnarray}
so that the first manifold can describe the interaction of
$15$ vector multiplets, while the second can describe the interaction
of $10$ hypermultiplets. Indeed if we decompose the $N=8$ graviton
multiplet in $N=2$ representations we find:
\begin{equation}
\mbox{N=8} \, \mbox{\bf spin 2}  \,\stackrel{N=2}{\longrightarrow}\,
 \mbox{\bf spin 2} + 6 \times \mbox{\bf spin 3/2} + 15 \times \mbox
{\bf vect. mult.}
 +
 10 \times \mbox{\bf hypermult.}
 \label{n8n2decompo}
\end{equation}
Introducing the decomposition \eqn{7in3p4}
we found in \cite{mp1} that the $40$ scalars belonging
to $Solv_4$ are constants
independent of the radial variable $r$. Only the $30$ scalars in the
K\"ahler algebra $Solv_3$ can be radial dependent. In fact their
radial dependence is governed by a first order differential equation
that can be extracted from a suitable component of the Killing spinor
equation.The result in this case is that $64$ of the scalar fields are
 actually constant
while $6$ are dynamical. Moreover $48$ charges are annihilated leaving
$6$ nonzero charges transforming in the representation $(2,2,2)$ of
the normalizer $G_{norm} = [Sl(2,\IR)]^3$.
More precisely we obtained the following result.
Up to U--duality transformations the most general $N=8$ black--hole
is actually an $N=2$ black--hole corresponding to a very
specific choice of the special K\"ahler manifold, namely $ \exp[ Solv_3 ]$
as in eq.\eqn{pincpal},\eqn{3and4defi}. Furthermore up to the duality
rotations of $SO^\star(12)$ this general solution is actually
determined by the so called $STU$ model studied in \cite{STUkallosh}
and based on the solvable subalgebra:
\begin{equation}
Solv_{STU} \, \equiv \, Solv \left( \frac{SL(2,\IR)^3}{U(1)^3}
\right) \, \subset \, Solv_3
\label{rilevanti}
\end{equation}
\par
In other words the only truly independent degrees of freedom of the
black hole solution are given by three complex scalar fields,
$S,T,U$. This is the result   we already anticipated in section
\ref{another} where we used the solvable Lie algebra $Solv_{STU}$ as a
preferred example in one illustration of the general concept.
The real parts of the  scalar fields $S,T$ and $U$
correspond to the three Cartan
generators of $Solv_3$ and have the physical interpretation of radii
of the torus compactification from $D=10$ to $D=4$. The imaginary
parts of these complex fields are generalised theta angles.
 \par
A more detailed argument leading to the conclusion that the only relevant
scalar fields in the $1/8$ solution are those
of the $STU$--model goes as follows.
Let
\begin{equation}
{\vec Q} \equiv \left( \matrix { p^{\vec {\Lambda}} \cr q_{\vec {\Sigma}}\cr } \right)
\label{chavecto}
\end{equation}
be the vector of electric and magnetic charges
that transforms in the ${\bf 56}$ dimensional real representation
of the U duality group $E_{7(7)}$.
Through the Cayley matrix we can convert
it to the ${\bf Usp(56)}$ basis namely to:
\begin{equation}
\left(\matrix { t^{{\vec {\Lambda}}_1}=
p^{{\vec {\Lambda}}_1}+ {\rm i} \, q_{{\vec {\Lambda}}_1}\, \cr
{\bar t}_{{\vec {\Lambda}}_1}=
p^{{\vec {\Lambda}}_1}- {\rm i} \, q_{{\vec {\Lambda}}_1}\,\cr
} \right)
\label{uspqvec}
\end{equation}
Acting on ${\vec Q}$ by means of suitable $ E_{7(7)} $
transformations, we can reduce it to a   {\it normal} form:
\begin{equation}
{\vec Q}\rightarrow {\vec Q}^N
\equiv \left(\matrix { t^0_{(1,1,1)}\cr t^1_{(1,1,15)}
\cr t^2_{(1,1,15)}\cr t^3_{(1,1,15)}\cr 0\cr
\dots \cr 0 \cr  \cr {\bar t}^0_{{\bar (1,1,1)}}\cr {\bar t}^1_{{\bar (1,1,15)}
} \cr
{\bar t}^2_{{\bar (1,1,15)}}\cr {\bar t}^3_{{\bar (1,1,15)}}\cr 0 \cr
\dots \cr 0 \cr  } \right)
\label{qnormalf}
\end{equation}
where there are only $4$ complex (alternatively $8$ real) independent
charges and these charges are located in the representation ${\bf (1,1,15)}$
of the subgroup:
\begin{equation}
U(1) \, \times \, SU(2) \,  \times \, SU(6) \, \subset SU(8) \,
\subset E_{7(7)}
\label{orphan}
\end{equation}
Indeed the general decomposition of the fundamental $56$--representation
of $E_{7(7)}$ with respect to the subgroup \eqn{orphan} is:
\begin{equation}
{\bf 56}_{real} = ({\bf  1,1,1})_{comp.}\,  \oplus \,
 ({\bf  1,2,6})_{comp.}  \, \oplus \, ({\bf  1,1,15})_{comp.}
 \label{56real}
\end{equation}
Consequently also the central charge ${\vec Z}\equiv \left( Z^{AB}\, , \,
Z_{CD}\right)$, which depends on ${\vec Q}$ through
the coset representative
in a symplectic--invariant way, through a suitable $SU(8)$
transformation will be brought to its {\it normal} form
where it has only $4+4$ non--vanishing components. Since the
decomposition of the complex ${\bf 28}$ representation of $SU(8)$ under
the subgroup \eqn{orphan} is the same as the decomposition of the
real ${\bf 56}$ of $E_{7(7)}$, namely eq.\eqn{56real} we can also write:
\begin{equation}
{\vec Z}\rightarrow {\vec Z}^N \equiv \left(\matrix { z^0_{(1,1,1)}\cr
z^1_{(1,1,15)}
\cr z^2_{(1,1,15)}\cr z^3_{(1,1,15)}\cr 0\cr
\dots \cr 0 \cr  \cr {\bar z}^0_{{\bar (1,1,1)}}\cr {\bar z}^1_{{\bar (1,1,15)}
} \cr
{\bar z}^2_{{\bar (1,1,15)}}\cr {\bar z}^3_{{\bar (1,1,15)}}\cr 0 \cr
\dots \cr 0 \cr  } \right)
\label{znormalf}
\end{equation}
 It is an easy consequence of elementary group theory, as it was shown in
\cite{lastserg}, \cite{savoy} that ${\vec Q}^N$ is invariant with respect to
the action of an $O(4,4)$ subgroup of $ E_{7(7)}$ and its {\it normalizer}
is an $SL(2,\IR)^3 \subset E_{7(7)}$ commuting with it.
 Indeed it turns out that
the eight real parameters in ${\vec Q}^N$ are singlets with respect to
$O(4,4)$ and in a ${\bf (2,2,2)}$ irreducible representation of
$SL(2,\IR)^3$  as it is shown in the following
decomposition of the ${\bf 56}$ with respect to $O(4,4)\otimes SL(2,\IR)^3$:
\begin{equation}
{\bf 56}\rightarrow {\bf (8_v,2,1,1)\, \oplus \, (8_s,1,2,1)\,\oplus \,
(8_{s^\prime},1,1,2)}\, \oplus\, {\bf (1,2,2,2)}
\label{56normaldec}
\end{equation}
The corresponding subgroup of $SU(8)$ leaving ${\vec Z}^N$ invariant
is therefore $SU(2)^4 \cong SO(4) \times SO(4) $
which is the maximal compact subgroup of $O(4,4)$. The reader should
compare this argument with the discussion in section \ref{5param}
where we have shown that $SU(2)^4$ is the stabilizer of the central charge
written in normal form and where we have argued that the most general
form of the $1/8$ solution should contain ${\bf 5}$ parameters, that
is:
\begin{equation}
{\bf 5} = 56 - \mbox{dim} \, \frac{SU(8)}{SU(2)^4}
\label{conteg}
\end{equation}
Indeed we can write the generic {\bf 56}--dimensional charge vector
${\vec Q}$ in terms
of   five normal frame parameters plus 51 ``angles'' which
parametrize the 51 dimensional compact space $\frac{SU(8)}{SU(2)^4}$,
where $ SU(2)^4 $ is the maximal compact subgroup of the stability group
$O(4,4)$ \cite{cvet}.
This same counting is achieved by arguing at the level of the
normalizer $SL(2,\IR)^3$.
This latter  contains a $U(1)^3 \subset SU(8)$ that
 can be  used to gauge away three of the $4$ phases of the $4$
 complex  charges $t^\Lambda = p^\Lambda + i q_ \Lambda $.
 Independently from the path used to reach it the conclusion is
 \bst
 The generating solution of $1/8$ supersymmetry preserving $N=8$
 BPS black--holes depends on $5$--essential parameters,
 namely four complex numbers with
the same phase.
\est
Consider now the scalar {\it geodesic  potential}
defined in eq.\eqn{sumrule}, which  identically can
be rewritten as it follows:
\begin{eqnarray}
V(\phi) & \equiv & \frac{1}{2}\, {\bar Z}^{AB}(\phi) \, Z_{AB}(\phi)
\nonumber\\
&=& \frac{1}{2}\,{\vec Q}^T \, \left[ \IL^{-1}\left(\phi\right) \right]^T \,
\IL^{-1}\left(\phi\right) \, {\vec Q}
\label{newgeo}
\end{eqnarray}
As explained in section \ref{geopoto} the minimization of the
potential \eqn{newgeo} determines the fixed values of the scalar fields
at the horizon of the black--hole.
Because of its invariance properties
the scalar potential $V(\phi)$ depends on
${\vec Z}$ and therefore on ${\vec Q}$ only  through their normal forms.
Since the fixed scalars at the horizon of the black--hole
are obtained minimizing
$V(\phi)$, it can be inferred that the most general solution of
this kind will depend (modulo duality transformations)
only on those scalar fields associated with the
{\it normalizer} of the normal form ${\vec Q}^N$.
Indeed the dependence of $V(\phi)$ on
a scalar field is achieved by acting on ${\vec Q}$ in the expression
of $V(\phi)$ by means of the transformations in $Solv_7$
associated with that field.
Since at any point of the scalar manifold $V(\phi)$
can be made to depend only on
${\vec Q}^N$, its minimum will be defined only by those scalars
that correspond
to transformations acting on the non--vanishing components
of the normal form ({\it normalizer} of ${\vec Q}^N$). Indeed
all the other isometries were used to rotate  ${\vec Q}$ to the normal
form ${\vec Q}^N$.
Among those scalars which are not determined by the fixed point conditions
there are the {\it flat direction fields} namely
those on which the scalar potential does not  depend  at all:
\begin{equation}
  \mbox{flat direction field } \, q_f \quad \leftrightarrow \quad
  \frac{\partial}{\partial q_f} \, V(\phi) =0
\end{equation}
Some of these fields parametrize $Solv \left( O(4,4)/O(4)
\times O(4)\right)$  since
they are associated with isometries leaving ${\vec Q}^N$ invariant,
and the remaining ones
are  obtained from the latter by means of duality transformations.
In order to identify the scalars which are {\it flat} directions
of $V(\phi)$, let
us consider the way in which  $Solv\left(O(4,4)\right)$
is embedded into $Solv_7$.
To this effect we start by reviewing the algebraic structure of the
solvable Lie algebras  $Solv_3$ and $Solv_4$ defined by eq.s
\eqn{3and4defi}
Since $Solv_3$ and $Solv_4$  respectively define a special K\"ahler and
a quaternionic manifold, it is useful to describe them  in Alekseevski's
formalism \cite{alek}.
\begin{eqnarray}
Solv_3\, :\quad\quad\quad\quad\quad\quad\quad\quad & &\nonumber\\
Solv_3\, &=&\, F_1\, \oplus\, F_2\, \oplus\, F_3\, \oplus\,
{\bf X}\, \oplus\,
{\bf Y}\, \oplus\, {\bf Z}\nonumber\\
F_i\, &=&\, \{{\rm h}_i\, ,\,{\rm g}_i\}\quad i=1,2,3\nonumber\\
{\bf X}\, &=&\, {\bf X}^+\, \oplus\, {\bf X}^-\, =\, {\bf X}_{NS}\,
\oplus\,
{\bf X}_{RR}\nonumber\\
{\bf Y}\, &=&\, {\bf Y}^+\, \oplus\, {\bf Y}^-\, =\, {\bf Y}_{NS}\,
\oplus\,
{\bf Y}_{RR}\nonumber\\
{\bf Z}\, &=&\, {\bf Z}^+\, \oplus\, {\bf Z}^-\, =\, {\bf Z}_{NS}\,
\oplus\,
{\bf Z}_{RR}\nonumber\\
Solv\left(SU(3,3)_1\right)\, &=&\, F_1\, \oplus\, F_2\, \oplus\, F_3\,
\oplus\,
{\bf X}_{NS}\, \oplus\,
{\bf Y}_{NS}\, \oplus\, {\bf Z}_{NS}\nonumber\\
Solv\left(SL(2,\IR)^3\right)\, &=&\, F_1\, \oplus\, F_2\,
\oplus\, F_3\nonumber
\\
{\cal W}_{12}\, &=&\, {\bf X}_{RR}\, \oplus\,{\bf Y}_{RR}\,
\oplus\,{\bf Z}_{RR}
\nonumber\\
\mbox{dim}\left( F_i\right)\, &=&\, 2\, ;\quad \mbox{dim}\left( {\bf X}_{NS/RR}\right)\,
=\,
\mbox{dim}\left( {\bf X}^{\pm}\right)\, =\,4\nonumber \\
 \mbox{dim}\left( {\bf Y}_{NS/RR}\right)\, &=&\,\mbox{dim}\left( {\bf Y}^{\pm}\right)\,
 =\,
\mbox{dim}\left( {\bf Z}_{NS/RR}\right)\, =\,\mbox{dim}\left( {\bf Z}^{\pm}\right)\,
=\,4\nonumber\\
\left[{\rm h}_i\, ,\,{\rm g}_i\right]\, &=&\, {\rm g}_i\quad i=1,2,3
\nonumber\\
\left[F_i\, ,\,F_j\right]\, &=&\, 0 \quad i\neq j\nonumber\\
\left[{\rm h}_3\, ,\,{\bf Y}^{\pm}\right]\, &=&\,\pm\frac{1}{2}{\bf Y}^{\pm}
\nonumber\\
\left[{\rm h}_3\, ,\,{\bf X}^{\pm}\right]\, &=&\,\pm\frac{1}{2}{\bf X}^{\pm}
\nonumber\\
\left[{\rm h}_2\, ,\,{\bf Z}^{\pm}\right]\, &=&\,\pm\frac{1}{2}{\bf Z}^{\pm}
\nonumber\\
\left[{\rm g}_3\, ,\,{\bf Y}^{+}\right]\, &=&\,\left[{\rm g}_2\, ,
\,{\bf Z}^{+}
\right]\, =\,\left[{\rm g}_3\, ,\,{\bf X}^{+}\right]\, =\, 0 \nonumber\\
 \left[{\rm g}_3\, ,\,{\bf Y}^{-}\right]\, &=&\,{\bf Y}^+\, ;\,\,
\left[{\rm g}_2\, ,\,{\bf Z}^{-}\right]\, =\,{\bf Z}^+\, ;\,\,
\left[{\rm g}_3\, ,\,{\bf X}^{-}\right]\, =\,{\bf X}^+\nonumber\\
\left[F_1\, ,\,{\bf X}\right]\, &=&\,\left[F_2\, ,\,{\bf Y}\right]\, =\,
\left[F_3\, ,\,{\bf Z}\right]\, =\, 0 \nonumber\\
\left[{\bf X}^-\, ,\,{\bf Z}^{-}\right]\, &=&\, {\bf Y}^-
\label{Alekal2}
\end{eqnarray}
\begin{eqnarray}
Solv_4\, :\quad\quad\quad\quad\quad\quad\quad\quad\quad\quad\quad\quad\quad
 & &\nonumber\\
Solv_4 & =& F_0\, \oplus  \, F_1^{\prime}\,\oplus \, F_2^{\prime}
\,\oplus \, F_2^{\prime}\,\nonumber\\
 & &\oplus \, {\bf X}_{NS}^{\prime}\,\oplus \,
{\bf Y}_{NS}^{\prime}\,\oplus \, {\bf Z}_{NS}^{\prime}\,\oplus \,
{\cal W}_{20}\nonumber\\
Solv\left(SL(2,\IR)\right)\,\oplus \,Solv\left(SU(3,3)_2\right)\, &=&\,
\left[F_0\right]\,\oplus\, \Biggl [ F_1^{\prime}\,\oplus \, F_2^{\prime}
\,\oplus \, F_2^{\prime} \nonumber \\
 & & \,\oplus \, {\bf X}_{NS}^{\prime}\,\oplus \,\Biggr ]\nonumber\\
F_0\, &=&\, \{{\rm h}_0\, ,\,{\rm g}_0\}\quad \left[{\rm h}_0\, ,\,
{\rm g}_0\right]\, =\, {\rm g}_0\nonumber\\
F_i^\prime\, &=&\, \{{\rm h}_i^\prime\, ,\,{\rm g}_i^\prime\}
\quad i=1,2,3\nonumber\\
\left[F_0\, , \, Solv\left(SU(3,3)_2\right)\right]\,&=& \, 0\, ;\quad
\left[{\rm h}_0\, ,\,{\cal W}_{20}\right]\, =\, \frac{1}{2}{\cal W}_{20}
\nonumber\\
\left[{\rm g}_0\, ,\,{\cal W}_{20}\right]\, &=&\, \left[{\rm g}_0\, , \,
 Solv\left(SU(3,3)_2\right)\right]\,=\, 0\nonumber\\
\left[Solv\left(SL(2,\IR)\right)\,\oplus \,Solv\left(SU(3,3)_2\right)\, ,\,
{\cal W}_{20}\right]\, &=&\, {\cal W}_{20}
\label{Alekquat}
\end{eqnarray}
The operators ${\rm h}_i\quad i=1,2,3$ are the Cartan generators of
$SO^\star(12)$ and ${\rm g}_i$ the corresponding axions which together
with
${\rm h}_i$ complete the solvable algebra
$Solv_{STU} \equiv Solv\left(SL(2,\IR)^3\right)$
Referring to the description of $Solv_4$
given in eqs. (\ref{Alekquat}):
\begin{eqnarray}
Solv\left(O(4,4)\right)&\subset& Solv_4\nonumber\\
Solv\left(O(4,4)\right)\, &=&\,  F_0\,\oplus \, F_1^{\prime}\,\oplus \,
F_2^{\prime}\,\oplus \, F_3^{\prime}\, \oplus \, {\cal W}_{8}
\label{o44}
\end{eqnarray}
where the R--R part ${\cal W}_{8}$
of $Solv\left(O(4,4)\right)$ is the quaternionic image
of $ F_0\,\oplus \, \\F_1^{\prime}\,\oplus \, F_2^{\prime}\,\oplus \, F_3^{\prime}$
in ${\cal W}_{20}$.
Therefore $Solv\left(O(4,4)\right)$
is parametrized by the $4$ {\it hypermultiplets}
containing the Cartan fields of
$Solv\left(E_{6(4)}\right)$.
One finds that the other flat directions are all
the remaining parameters of $Solv_4$, that is all the hyperscalars.
\par
Alternatively we can observe that since the hypermultiplet
scalars are flat directions of the potential,
then we can use the solvable Lie algebra
$ Solv_4$    to set them to zero at the horizon.
Since we know from the Killing spinor equations that these
$40$ scalars  are  constants
it follows that we can safely set them to zero and forget
about their existence (modulo U--duality transformations).
Hence the non zero scalars required for a general
solution have to be looked for among the vector multiplet scalars
that is in the solvable Lie algebra $Solv_3$. In other words
the most general $N=8$ black--hole (up to U--duality rotations) is
given by the most general $N=2$ black--hole based on the $15$--dimensional
special K\"ahler manifold:
\begin{equation}
{\cal SK }_{15}  \, \equiv \, \exp \left[ Solv_3 \right] \, =
\frac{SO^\star(12)}{U(1) \times SU(6)}
\label{mgeneral}
\end{equation}
Having determined the little group of the normal form enables us
to decide which among the above $30$ scalars have to be
kept alive in order to generate the most general BPS black--hole
solution (modulo U--duality).
\par
We argue as follows. The {\it normalizer} of the normal form
is contained in the largest subgroup
of $E_{7(7)}$ commuting with $O(4,4)$.
Indeed, a necessary condition for a group $G^N$ to be the {\it normalizer}
of ${\vec Q}^N$ is to commute with the {\it little group} $G^L=O(4,4)$ of
${\vec Q}^N$:
\begin{eqnarray}
{\vec Q}^{\prime N}\, &=&\, G^N\cdot {\vec Q}^N\quad {\vec Q}^N\, =\,
 G^L\cdot {\vec Q}^{N}\nonumber\\
{\vec Q}^{\prime N}\, &=&\, G^L\cdot {\vec Q}^{\prime N}\Rightarrow
\left[G^N\, ,\, G^L\right]\, =\, 0
\label{gngl}
\end{eqnarray}
As previously mentioned, it was proven that $G^N\, =\, SL(2,\IR)^{3}
\subset SO^{\star}(12)$ whose solvable
algebra is defined by the last
of eqs. (\ref{Alekal2}).
Moreover $G^N$ coincides with the largest subgroup of $Solv_7$
commuting with $G^L$.\\
The duality transformations associated with
the $SL(2,\IR)^{3}$ isometries act only on the eight non
vanishing components
of ${\vec Q}^N$ and therefore belong to ${\bf Sp(8)}$.\par
{\it In conclusion the most general $N=8$ black--hole solution is  described
by the 6 scalars
parametrizing  $Solv_{STU} \equiv Solv\left(SL(2,\IR)^{3}\right)$,
which are the only
ones involved in the
fixed point conditions at the horizon.}
\par
Another way of seeing this is to
notice that all the other $64$ scalars are either the $16$ parameters of
$Solv\left(O(4,4)\right)$
which are flat directions of $V\left(\phi\right)$, or coefficients of the
$48=56-8$ transformations needed to rotate ${\vec Q}$ into
${\vec Q}^N$ that is to set $48$
components of ${\vec Q}$ to zero as shown in eq. (\ref{qnormalf}).
\par
Therefore in section \ref{det1/8} we shall reduce our attention to
the Cartan vector multiplet sector, namely to the 6 vectors
corresponding to the solvable Lie algebra $Solv \left ( SL(2,\IR)
\right )$.
\section{Detailed study of the $1/2$ case}
\setcounter{equation}{0}
\label{det1/2}
\markboth{BPS BLACK HOLES IN SUPERGRAVITY: CHAPTER 5}
{5.5 DETAILED STUDY OF THE $1/2$ CASE}
As established in section \ref{ksecov},
the $N=1/2$ SUSY preserving black hole
solution of $N=8$ supergravity has $4$ equal skew eigenvalues
in the normal frame for the central charges.
The stabilizer of the normal form is $E_{6(6)}$
and the normalizer of this latter in $E_{7(7)}$ is $O\left(1,1\right)$:
\begin{equation}
E_{7(7)}\supset E_{6(6)}\times O\left(1,1\right)
\label{orfan1}
\end{equation}
According to our previous discussion, the relevant subgroup
of the $SU\left(8\right)$ holonomy group is $Usp\left(8\right)$, since
the BPS Killing spinor conditions involve supersymmetry parameters
$\epsilon_A,~\epsilon^A$ satisfying eq.\eqn{urcunmez}.
Relying on this information, we can write the solvable Lie algebra
decomposition \eqn{decomp1a},\eqn{decomp1b}
of the $\sigma-$model scalar coset $E_{7(7)}
\over SU\left(8\right)$.
\par
As discussed in the introduction, it is natural to guess that
modulo $U-$duality
transformations the complete solution is given in terms
of a single scalar field
parametrizing $O\left(1,1\right)$.
\par
Indeed, we can  now demonstrate that according to the previous
discussion there is just one scalar field,
parametrizing the normalizer $O\left(1,1\right)$,
which appears in the
final lagrangian, since the Killing spinor equations imply that
$69$ out of the $70$ scalar fields are actually  constants.
In order to achieve this result, we have to decompose the $SU\left(8\right)$ tensors
appearing in the equations \eqn{2},\eqn{psisusy2} with respect to $Usp\left(
8\right)$ irreducible representations. According to the decompositions
\begin{eqnarray}
\label{70-28decomp1/2}
{\bf 70}& \stackrel{Usp(8)}{=} &{\bf 42}\oplus{\bf 27}\oplus{\bf 1}\nonumber\\
{\bf 28}& \stackrel{Usp(8)}{=} &{\bf 27}\oplus{\bf 1}
\end{eqnarray}
we have
\begin{eqnarray}
\label{usp8dec}
P_{ABCD}&=&\stackrel{\circ}{P}_{ABCD}+{3\over 2}
C_{[AB}\stackrel{\circ}{P}_{CD]}+{1\over 16}C_{[AB}C_{CD]}P\nonumber\\
T_{AB}&=&\stackrel{\circ}{T}_{AB}+{1\over 8}C_{AB}T
\end{eqnarray}
where the notation $ \stackrel{\circ}{t}_{A_1 \dots , A_n}$
means that the antisymmetric tensor is  $Usp\left(8\right)$ irreducible, namely
has vanishing $C$-traces: $C^{A_1 A_2} \,\stackrel{\circ}{t}_{A_1 A_2 \dots , A_n}=0$.
\par
Starting from  equation \eqn{2} and using equation (\ref{1/2killingspin})
we easily find:
\begin{equation}
4  P_{,a}\gamma^a\gamma^0-6T_{ab} \gamma^{ab}=0\, ,
\end{equation}
where we have twice contracted the free $Usp(8)$ indices with the
$Usp(8)$ metric $C_{AB}$. Next, using the decomposition
(\ref{usp8dec}), eq. \eqn{2} reduces to
\begin{equation}
-4  \left(  \stackrel{\circ}{P}_{ABCD,a}+ {3\over2} \stackrel{\circ}{P}_{[CD,a}C_{AB]}\right)
C^{DL}\gamma^a\gamma^0-3
\stackrel{\circ}{T}_{[AB}\delta^L_{C]}\gamma^{ab}=0\, .
\label{basta}
\end{equation}
Now we may alternatively contract equation (\ref{basta}) with
$C^{AB}$ or $\delta^L_C$ obtaining two relations on $ \stackrel{\circ}{P}_{AB} $
and   $\stackrel{\circ}{T}_{AB}$ which imply that they are separately
zero:
\begin{equation}
\stackrel{\circ}{P}_{AB}=\stackrel{\circ}{T}_{AB}=0\, ,
\end{equation}
which also imply, taking into account (\ref{basta})
\begin{equation}
\stackrel{\circ}{P}_{ABCD}=0\, .
\end{equation}
Thus
we have  reached the conclusion
\begin{eqnarray}
\label{42-27=0}
\stackrel{\circ}{P}_{ABCD|i}\partial_{\mu}\Phi^i\gamma^{\mu}\epsilon^D&=&0\nonumber\\
\stackrel{\circ}{P}_{AB|i}\partial_{\mu}\Phi^i\gamma^{\mu}\epsilon^B&=&0\\
\stackrel{\circ}{T}_{AB}&=&0 \label{tnot}
\end{eqnarray}
implying that $69$ out the $70$ scalar fields are actually
constant, while the only surviving central charge is that
associated with the singlet two-form $T$. Since $T_{AB}$ is a complex
combination of the electric and magnetic field strengths \ref{tab},
it is clear that eq. \ref{tnot} implies the vanishing of $54$ of the
quantized charges $p^{\Lambda \Sigma},q_{\Lambda \Sigma}$,the
surviving two charges transforming as a doublet of $O(1,1)$ according
to eq. (\ref{visto}).
The only non-trivial evolution equation relates $P$ and $T$ as follows:
\begin{equation}
\label{singleteqn1/2}
\left(
\hat P\partial_{\mu}\Phi\gamma^{\mu}-{3\over 2}
\mbox{\rm i}\, T^{(-)}_{\rho\sigma}\gamma^{\rho\sigma}\gamma^0 \right)\epsilon_A=0
\end{equation}
where we have set $P = \hat P d\Phi $  and $\Phi$ is the unique non trivial scalar
field parametrizing $O(1,1)$.
\par
In order to make this equation explicit we perform
the usual static ans\"atze. For the metric we set the ansatz \eqn{ds2U}.
The scalar fields are assumed to be radial dependent as generally
stated in eq. \eqn{phijey} and for the vector field strengths we
assume the ansatz of eq.\eqn{strenghtsans} which adapted to the
$E_{7(7)}$ notation reads as follows:
\begin{eqnarray}
F^{-\Lambda\Sigma}&=&{1 \over 4\pi}t^{\Lambda\Sigma}
\left(r\right)E^{\left(-\right)}\label{F}\\
t^{\Lambda\Sigma}\left(r\right)&=&
2\pi\left(g+\mbox{\rm i}\ell\left(r\right)\right)^{\Lambda\Sigma}
\label{smallt}
\end{eqnarray}
The anti self dual form $E^{\left(-\right)}$ was defined in eq.
\eqn{eaself}. Using   (\ref{defFpm}), (\ref{defG}),
(\ref{tab}),(\ref{eaself}),(\ref{F}) we have
\begin{equation}
\label{Tab}
T_{ab}^- = {\rm i}\ t^{\Lambda\Sigma}(r) E_{ab}^- C^{AB}
{\rm Im}{\cal N }_{\Lambda \Sigma,\Gamma\Delta}f^{\Gamma\Delta}_{~~AB} \, .
\end{equation}
A simple gamma matrix manipulation gives further
\begin{equation}
\label{gammaalgebra}
\gamma_{ab}\, E^{\mp}_{ab}=
2 \mbox{\rm i} {e^{2U}\over r^3}x^i\gamma^0\gamma^i\left(\pm 1+\gamma_5\over 2\right)
\end{equation}
and we arrive at the final equation
\begin{equation}
\label{dphidr}
\frac{d \Phi}{dr}= -\frac{\sqrt{3}}{4 }
\ell(r)^{\Lambda\Sigma}\, {\rm Im}{\cal N }_{\Lambda\Sigma|\Gamma\Delta}\,
f^{\Gamma\Delta}_{~~AB}\, \frac{e^U}{r^2}\, .
\label{finale}
\end{equation}
In eq. (\ref{finale}), we have set $p^{\Lambda\Sigma}=0$ since
reality of  the l.h.s. and of
$f_{AB}^{\Gamma\Delta} $(see eq. (\ref{f})) imply the vanishing of the
magnetic charge.
Furthermore, we have normalized the vielbein component of
the $Usp(8)$ singlet as follows
\begin{equation}
\label{P}
\hat P=4 \sqrt{3}
\end{equation}
which corresponds to normalizing the $Usp(8)$ vielbein as
\begin{equation}
\label{Pabcdnor}
P_{ABCD}^{\left({\it singlet}\right)} = {1\over 16}P C_{[AB} C_{CD]}=
{\sqrt{3}\over 4}C_{[AB}
C_{CD]}\, d \Phi\, .
\end{equation}
This choice agrees with the normalization of the scalar fields existing
in the current literature.
Let us now consider the gravitino equation \eqn{psisusy2}.
Computing the spin connention $\omega^a_{~b}$ from
equation (\ref{ds2U}), we find
\begin{eqnarray}
\omega^{0i}&=&{dU\over dr}{x^i\over r}
e^{U\left(r\right)}V^0\nonumber\\
\omega^{ij}&=&2{dU\over dr}\, {x_k\over r}\,
 \eta^{k[i}\, V^{j]} \, e^U
\end{eqnarray}
where $V^0=e^U \, dt$, $V^i = e^{-U} \, dx^i$.
Setting $\epsilon_A=e^{f\left(r\right)}\zeta_A$, where $\zeta_A$ is
a constant chiral spinor, we get
\begin{eqnarray}
\label{deltapsi}
&&\left\{{df\over dr}{x^i\over r}e^{f+U}\delta_A^BV^i+
\Omega_{A,\alpha}^{~B}\partial_i\Phi^{\alpha}e^fV^i\right.
\nonumber\\
&&\left.-{1\over 4}\left(2{dU\over dr}{x^i\over r}e^Ue^f
\left(\gamma^0\gamma^iV^0+\gamma^{ij}V_j\right)
\right)\delta_A^B \, + \,
\delta_A^B \,T^-_{ab}\gamma^{ab}\gamma^c\gamma^0 V_c\right\}\zeta_B=0
\end{eqnarray}
where we have used eqs.\eqn{psisusy2},(\ref{delepsi}), (\ref{usp8dec}).
This equation has two sectors; setting  to zero the coefficient
of $V^0$ or of $V^i\gamma^{ij}$ and tracing over the $A,B$ indices we find
two identical equations, namely:
\begin{equation}
{dU\over dr}=
-{1\over 8} \ell(r)^{\Lambda\Sigma}{e^U\over r^2}
C^{AB}{\rm Im}{\cal N }_{\Lambda \Sigma,\Gamma\Delta} f^{\Gamma\Delta}_{~~AB}.
\label{dudr}
\end{equation}
Instead, if we set  to zero the coefficient of $V^i$, we find
a differential equation for the function $f\left(r\right)$, which
is uninteresting for our purposes.
Comparing now equations (\ref{dphidr}) and (\ref{dudr}) we immediately find
\begin{equation}
\label{phiu}
\Phi= 2 \sqrt{3} \, U
\end{equation}

\subsection{Explicit computation of the Killing equations and of the reduced
Lagrangian in the $1/2$ case}

In order to compute the l.h.s. of eq.s
(\ref{finale}), (\ref{dudr}) and the lagrangian of the 1/2 model,
we need the explicit
form of the coset representative $\IL$ given in
equation (\ref{Lcoset}). This will also enable us to compute explicitly the
r.h.s. of equations (\ref{dphidr}), (\ref{dudr}).
In the present case the explicit form of $\IL$ can be retrieved by
exponentiating the $Usp\left(8\right)$ singlet generator.
As stated in equation (\ref{L-1dL}),
the scalar vielbein in the $Usp\left(28,28\right)$ basis is given by the off
diagonal block elements of $\IL^{-1}d\IL$, namely
\begin{equation}
\label{Pvielbein}
\IP =
\left(
\begin{array}{cc}
0&{\bar P}_{ABCD}\\
P_{ABCD}&0
\end{array}
\right).
\end{equation}
From equation (\ref{Pabcdnor}), we see that the $Usp\left(8\right)$
singlet corresponds to the generator
\begin{equation}
\label{kappa}
\IK ={\sqrt{3}\over 4}\left( \begin{array}{c|c}
                               0 & C^{[AB} C^{HL]}\\
                               \hline\\
                               C_{[CD}C_{RS]} & 0
                           \end{array}
                \right)
\end{equation}
and therefore, in order to construct the coset representative of the $O\left(1,1\right)$
subgroup of $E_{7(7)}$, we need only to exponentiate $\Phi\IK$.
Note that $\IK$ is a $Usp\left(8\right)$ singlet in the $\bf 70$
representation of $SU\left(8\right)$, but it acts non-trivially in the
$\bf 28$ representation of the quantized charges
$\left(q_{AB}, p^{AB}\right)$.
It follows that the various powers of $\IK$ are proportional to the
projection operators onto the irreducible $Usp(8)$
representations $\bf 1$ and $\bf 27$ of the
charges:
\begin{equation}
\label{P1}
\IP_1=\frac{1}{8} C^{AB}C_{RS}
\end{equation}
\begin{equation}
\label{P27}
\IP_{27}=(\delta_{RS}^{AB}- \frac{1}{8} C^{AB}C_{RS}).
\end{equation}
Straightforward exponentiation gives
\begin{eqnarray}
\exp(\Phi\IK)&=&\cosh\left({1\over 2\sqrt{3}}\Phi\right)\IP_{27}+{3\over 2}
\sinh\left({1\over 2\sqrt{3}}\Phi\right)\IP_{27}\IK\IP_{27}+\\
&&+\cosh\left({\sqrt{3}\over 2}\Phi\right)\IP_{1}+{1\over 2}
\sinh\left({\sqrt{3}\over 2}\Phi\right)\IP_{1}\IK\IP_{1}
\end{eqnarray}
Since we are interested only in the singlet subspace
\begin{equation}
\label{P1proj}
\IP_1\exp[\Phi\IK]\IP_1 = \cosh({\sqrt{3}\over 2}\Phi)\IP_1 +
\frac{1}{2} \sinh({\sqrt{3}\over 2}\Phi)
\IP_1\IK\IP_1
\end{equation}
\begin{equation}
\label{Lsinglet}
\IL_{singlet}= \frac{1}{8}\left( \begin{array}{c|c}
                \cosh ({\sqrt{3}\over 2}\Phi) C^{AB}C_{CD} &
                \sinh ({\sqrt{3}\over 2}\Phi) C^{AB} C^{FG}\\
                \hline\\
                 \sinh({\sqrt{3}\over 2}\Phi) C_{CD}C_{LM} &
                 \cosh({\sqrt{3}\over 2}\Phi) C_{CM}C^{FG}
                \end{array}
                \right).
\end{equation}
Comparing (\ref{Lsinglet}) with  the equation (\ref{Lcoset}),
we find \footnote{Note that
we are we are writing the coset matrix with the same pairs of
indices $AB, CD, \dots$ without distinction between the pairs
$\Lambda\Sigma$ and $AB$ as was done in sect.(\ref{41})}:
\begin{equation}
\label{f}
f=\frac{1}{8 \sqrt{2}} e^{{\sqrt{3}\over 2}\Phi} C^{AB}C_{CD}
\end{equation}
\begin{equation}
\label{h}
h=- \mbox{\rm i} \, \frac{1}{8 \sqrt{2}}\, e^{- {\sqrt{3}\over 2}\Phi} C_{AB}C_{CD}
\end{equation}
and hence, using ${\cal N}=hf^{-1}$, we find
\begin{equation}
\label{Nmatrix}
{\cal N}_{AB \, CD} = - \, \mbox{\rm i} \, {1\over 8}\, e^{-\sqrt{3}\Phi} C_{AB}C_{CD}
\end{equation}
so that we can compute the r.h.s. of (\ref{dphidr}), (\ref{dudr}).
Using the relation (\ref{phiu})  we find a single equation for the unknown
functions
$U\left(r\right)$, $\ell\left(r\right)=C_{\Lambda\Sigma}
\ell^{\Lambda\Sigma}\left(r\right)$
\begin{equation}
\label{dudr1}
\frac{dU}{dr}={1\over 8\sqrt{2}}{\ell\left(r\right)\over r^2}
\exp\left(-2U\right)
\end{equation}
At this point to solve the problem completely  we have to consider
also the second order field equation obtained from the lagrangian.
The bosonic supersymmetric
lagrangian of the $1/2$ preserving supersymmetry case is obtained
from equation (\ref{lN=8}) by substituting the values of $P_{ABCD}$
and ${\cal N}_{\Lambda\Sigma|\Gamma\Delta}$ given in equations
(\ref{Pabcdnor}) and (\ref{defN}) into equation (\ref{lN=8}).
We find
\begin{equation}
\label{lbose1/2}
{\cal L} = 2R- e^{- \sqrt{3} \Phi} F_{\mu \nu} F^{\mu \nu} +
\frac{1}{2} \partial_\mu \Phi \partial^\mu \Phi
\end{equation}
Note that this action has the general form of $0$--brane action in
$D=4$ (compare with eq.\eqn{paction}). Furthermore recalling eq.s
\eqn{critval1}, we see that the value of the parameter $a$ is
\begin{equation}
a = \sqrt{3}
\end{equation}
According to this we expect a solution where:
\begin{eqnarray}
U & = &-\frac{1}{4} \, \log \, H(r) \nonumber\\
\Phi &=& - \frac{\sqrt{3}}{2} \, \log \, H(r)\nonumber\\
\ell & = & 2 r^3\, \frac{d}{dr}  \left( H(r) \right)^{-\frac{1}{2}} =
k \, \times \, \left( H(r) \right)^{-\frac{3}{2}}
\label{indovin}
\end{eqnarray}
where $H(r)= 1+k/r$ denotes a harmonic function. In the next subsection,
by explicit calculation we show that this is indeed the BPS solution
we obtain.
\subsection{The $1/2$ solution}
The resulting field equations are
\\
{\underline {\sl Einstein equation:}}
\begin{equation}
\label{einstein1/2}
U'' + \frac{2}{r} U' - (U)^2 =
{1\over 4} (\Phi ')^2
\end{equation}
{\underline {\sl Maxwell equation:}}
\begin{equation}
\label{maxwell1/2}
\frac{d}{dr}(e^{-\sqrt{3}\Phi} \ell(r)) =0
\end{equation}
{\underline {\sl Dilaton equation:}}
\begin{equation}
\label{dilaton1/2}
\Phi'' + \frac{2}{r} \Phi' =-e^{-\sqrt{3}\Phi + 2 U}
        \ell(r)^2 \frac{1}{r^4}   \, .
\end{equation}
From Maxwell equations one immediately finds
\begin{equation}
\ell\left(r\right)=e^{\sqrt{3}\Phi\left(r\right)}.
\end{equation}
Taking into account (\ref{phiu}), the second order field equation and the
first order Killing spinor equation have the common solution
\begin{eqnarray}
\label{solution1/2}
U & = & -\frac{1}{4} \log \, H(x) \nonumber\\
\Phi     & = & - \frac{\sqrt{3}}{2} \log \, H(x)\nonumber\\
\ell       & = &  H(x)^{- \frac{3}{2}}
\end{eqnarray}
where:
\begin{equation}
H(x) \equiv 1 + \sum_{ i} \, \frac{k_i}{  {\vec x} - {\vec x}^0_i  }
\label{armonie}
\end{equation}
is a harmonic function describing
$0$--branes located at ${\vec x}^0_\ell$ for
$\ell=1, 2,\dots$, each brane carrying a charge $k_i$.
In particular for a single
$0$--brane we have:
\begin{equation}
H(x) = 1+ \frac{k}{r}
\label{unasola}
\end{equation}
and the solution reduces to the expected form \eqn{indovin}.
\par
Recall also from eq.\eqn{critval1} that for $a=\sqrt{3}$ we have
 $\Delta=4$. As it is shown in ref.\cite{mbrastelle} the parameter
 $\Delta$ is a dimensional reduction invariant so that we trace back
 the higher dimensional origin of the $1/2$ BPS black hole
 from such a value. If we interpret $N=8$ supergravity as the $T^7$
 compactification of M--theory we can compare our present four--dimensional
 $0$--brane solution with an eleven--dimensional $M2$--brane solution.
 They both have $\Delta=4$ and preserve $1/2$ of the $32$
 supersymmetry charges. We can identify our present black hole with
 the wrapping of the $M2$--brane on a $2$--cycle of the seven--torus.
\par
What we have shown is that the most
general $BPS$-saturated black hole preserving $1/2$ of the $N=8$
supersymmetry is actually described by the lagrangian
(\ref{lbose1/2}) with the solution given by (\ref{solution1/2}), in the
sense that any other solution with the same property can be obtained
from the present one by an $E_{7\left(7\right)}$ (U-duality) transformation.
\section{Detailed study of the $1/4$ case}
\label{seckilling1/4}
\setcounter{equation}{0}
\markboth{BPS BLACK HOLES IN SUPERGRAVITY: CHAPTER 5}
{5.6 DETAILED STUDY OF THE $1/4$ CASE}
Solutions preserving ${1\over 4}$ of $N=8$ supersymmetry have two
pairs of identical skew eigenvalues in the normal frame for the
central charges. In this case the stability subgroup preserving the
normal form is $O\left(5,5\right)$ with normalizer subgroup in
$E_{\left(7,7\right)}$ given by $SL\left(2,\IR\right)\times O\left(1,1\right)$
(see \cite{FGun}),
according to the decomposition
\begin{equation}
E_{\left(7,7\right)}\supset O\left(5,5\right)\times SL\left(2,\IR\right)\times O\left(1,1\right)=G_{stab}\times G_{norm}
\end{equation}
The relevant fields parametrize ${SL\left(2,\IR\right)\over U\left(1\right)}\times O\left(1,1\right)$ while the
surviving charges transform in the representation $\left({\bf 2,2}\right)$
of $SL\left(2,\IR\right)\times O\left(1,1\right)$. The group
$SL\left(2,\IR\right)$ rotates electric into electric and magnetic into magnetic
charges  while $O\left(1,1\right)$ mixes them. $O(1,1)$ is therefore
a true electromagnetic duality group.
\subsection{Killing spinor equations in the $1/4$ case: surviving
fields and charges.}
The holonomy subgroup $SU\left(8\right)$ decomposes in our
case as
\begin{equation}
SU\left(8\right) \rightarrow  Usp\left(4\right)
\times SU\left(4\right) \times  U\left(1\right)
\end{equation}
indeed in this case the killing spinors satisfy \eqn{urcunquart}
 where we recall the
index convention:
\begin{eqnarray}
&&A,B=1\dots 8 ~~~SU(8)~{\rm indices}\nonumber\\
&&a,b=1\dots 4 ~~~Usp(4)~~~{\rm indices} \nonumber\\
&&X,Y=5\dots 8 ~~~SU(4)~~~{\rm indices}
\end{eqnarray}
and $C_{ab}$ is the invariant metric of $Usp\left(4\right)$.
With respect to the holonomy subgroup
\newline $SU\left(4\right)\times Usp\left(4\right)$,
$P_{ABCD}$ and $T_{AB}$ appearing
in the equations (\ref{2}), (\ref{psisusy2}) decompose as follows:
\begin{eqnarray}
\label{7028decomp1/4}
{\bf 70}& \stackrel{Usp(4) \, \times \, SU(4)}{\longrightarrow}&
\left({\bf 1},{\bf 1}\right)\oplus\left({\bf 4},{\bf 4}\right)
\oplus\left({\bf 5},{\bf 6}\right)
\oplus\left({\bf 1},{\bf 6}\right)
\oplus\left({\bf\bar{4}},{\bf\bar{4}}\right)
\oplus\left({\bf\bar{1}},{\bf\bar{1}}\right)
\nonumber\\
{\bf 28}&\stackrel{Usp(4) \,
\times \, SU(4)}{\longrightarrow}&\left({\bf 1},{\bf 6}\right)
\oplus\left({\bf 4},{\bf 4}\right)\oplus\left({\bf 5},{\bf 1}\right)\oplus
\left({\bf 1},{\bf 1}\right)
\end{eqnarray}
We decompose eq. (\ref{2}) according to eq. (\ref{7028decomp1/4}).
We obtain:
\begin{eqnarray}
\delta\chi_{XYZ}&=&0 \\
\delta\chi_{aXY}&=&0 \\
\delta \stackrel{\circ}\chi_{abX} &=&
C^{ab}\delta\chi_{abX}= 0  \\
\delta\chi_{abc}&=&C_{[ab}\delta\chi_{c]} = 0.
\end{eqnarray}
\par
From $\delta\chi_{XYZ}=0$
we  immediately get:
\begin{equation}
P_{XYZa,\alpha}\partial_{\mu}\Phi^{\alpha}=0
\end{equation}
by means of which we recognize that $16$ scalar fields are actually constant
in the solution.
\par
From the reality condition
of the vielbein $P_{ABCD}$  (equation (\ref{realcondP}) )
we can also conclude
\begin{equation}
\label{16=0}
P_{Xabc}\equiv P_{X[a}C_{bc]}=0
\end{equation}
so that there are $16$ more scalar fields set to constants.
\par
From $\delta\chi_{aXY}=0$ we find
\begin{eqnarray}
\label{o111/4}
P_{XY,i}\partial_{\mu}\Phi^i\gamma^{\mu}\gamma^0\epsilon_a&=
&T_{XY\mu\nu}\gamma^{\mu\nu}\epsilon_a\\
\label{30=0}
\stackrel{\circ}{P}_{XYab,i}\partial_{\mu}\Phi^i&=&0
\end{eqnarray}
where we have set
\begin{equation}
\label{decompP1/4}
P_{XYab}=\stackrel{\circ}{P}_{XYab}+{1\over 4}C_{ab}P_{XY}
\end{equation}
Note that equation (\ref{30=0}) sets  $30$ extra scalar fields  to constant.
\par
From $\delta\chi_{Xab}=0$, using (\ref{16=0}), one  finds
that also $T_{Xa}=0$.

Finally, setting
\begin{eqnarray}
P_{abcd}&=&C_{[ab}C_{cd]}P\\
T_{ab}&=&\stackrel{\circ}{T}_{ab}+{1\over 4}C_{ab}T
\end{eqnarray}
the Killing spinor equation
$\delta\chi_{abc}\equiv C_{[ab}\delta\chi_{c]}=0$ yields:
\begin{eqnarray}
\stackrel{\circ}{T}_{ab}&=&0\\
\label{sl21/4}
P_{,i}\partial_{\mu}\Phi^i\gamma^{\mu}\gamma^0-{3\over
16}T_{\mu\nu}\gamma^{\mu\nu}\epsilon_a&=&0
\end{eqnarray}
Performing the gamma matrix algebra and using equation (\ref{gammaalgebra}),
the relevant evolution
equations (\ref{o111/4}), (\ref{sl21/4}) become
\begin{eqnarray}
\label{sl21/4svil}
P_{,i}{d\Phi^i\over dr}&=&\mbox{\rm i}{3\over
8}\left(p+\mbox{\rm i} \ell\left(r\right)\right)^{\Lambda\Sigma}
{\rm Im }{\cal N }_{ \Lambda \Sigma,\Gamma\Delta}f^{\Gamma\Delta}_{~~AB}
C^{AB}\, {e^U\over r^2}\nonumber\\
P_{XY,i}{d\Phi^i\over dr}&=& 2{\rm i }\left(p+\mbox{\rm i} \ell\left(r\right)
\right)^{\Lambda\Sigma}{\rm Im }
{\cal N }_{ \Lambda \Sigma,\Gamma\Delta}f^{\Gamma\Delta}_{~~XY}
\, {e^U\over r^2}
\end{eqnarray}
According to our previous discussion, $P_{XY,i}$ is the vielbein of the
coset $O\left(1,6\right)\over SU\left(4\right)$, which can be reduced to
depend on $6$ real fields $\Phi^i$ since, in force of the
$SU\left(8\right)$ pseudo--reality condition (\ref{realcondP}),
 $P_{XY,i}$  satisfies an analogous pseudo--reality condition.
 On the other hand $P_{,i}$ is the vielbein of $SL\left(2,\IR\right)
\over U\left(1\right)$, and it is intrinsically complex. Indeed
the $SU\left(8\right)$ pseudo--reality condition relates
the $SU\left(4\right)$ singlet $P_{XYZW}$ to the $Usp\left(4\right)$
singlet $P_{abcd}$. Hence $P_{,i}$ depends on a complex scalar field.
In conclusion we find that equations (\ref{sl21/4svil}) are evolutions
equations for $8$ real fields, the $6$ on which $P_{XY,i}$ depends
plus the $2$ real fields sitting in $P_{,i}$.
However, according to the discussion given above,
we expect that only three scalar fields,
parametrizing ${SL\left(2,\IR\right)\over U\left(1\right)}\times O\left(1,1\right)$
should be physically relevant.
To retrieve this number we note that
$O\left(1,1\right)$ is the subgroup of $O\left(1,6\right)$ which
commutes with the stability subgroup $O\left(5,5\right)$, and hence
also with its maximal compact subgroup $Usp\left(4\right)\times Usp\left(4\right)$.
Therefore out of the $6$ fields of
 $O\left(1,6\right)\over O\left(6\right)$ we restrict our
attention to the real field parametrizing $O\left(1,1\right)$,
whose corresponding vielbein is $C^{XY}P_{XY}=P_1d\Phi_1$. Thus
the second of equations (\ref{sl21/4svil}) can be  reduced to the
evolution equation for the single scalar field $\Phi_1$, namely:
\begin{equation}
\label{reduceo11}
P_{1}{d\Phi^1\over dr}= -2 \ell \left(r\right)^{\Lambda\Sigma}
{\rm Im}{\cal N }_{ \Lambda \Sigma,\Gamma\Delta}f^{\Gamma\Delta}_{~~XY}
\, {e^U\over r^2}
\end{equation}
In this equation we have set the magnetic charge
$p^{\Lambda\Sigma}=0$ since, as we show explicitly later, the quantity  $
{\rm Im}{\cal N }_{\Lambda \Sigma,\Gamma\Delta}f^{\Gamma\Delta}_{~~ab}$ is
actually real. Hence, since the left hand side of equation
(\ref{reduceo11}) is real, we are forced to set  the
corresponding magnetic charge to zero.
On the other hand, as we now show, inspection of the gravitino
Killing spinor equation, together with the first of equations
(\ref{sl21/4svil}),  further reduces the number of fields to two.
Indeed, let us consider the $\delta\psi_A=0$ Killing spinor equation.
The starting equation is the same as (\ref{deltapsi}), (\ref{tab}).
In the present case, however, the indices $A,B,\dots$ are $SU\left(8\right)$
indices, which have to be decomposed with respect to
$SU\left(4\right)\times Usp\left(4\right)\times U\left(1\right)$.
Then, from $\delta\Psi_X=0$, we obtain
\begin{equation}
\Omega_X^{~a}=0;~~~T_{Xa}=0
\end{equation}
From $\delta\psi_a=0$ we obtain an equation identical
to (\ref{deltapsi}) with $SU\left(8\right)$ indices
replaced by $SU\left(4\right)$ indices. With the same computations
performed in the $Usp\left(8\right)$ case we obtain the final
equation
\begin{equation}
\label{dudr1/4}
{d{  U}\over dr}=
-{1\over 4}\ell\left(r\right)^{\Lambda\Sigma}{e^U\over r^2}
C^{ab}{\rm Im}{\cal N }_{\Lambda \Sigma,\Gamma\Delta}
f^{\Gamma\Delta}_{~~ab}\, ,
\end{equation}
where we have taken into account
that $C^{ab}{\rm Im} {\cal N }_{\Lambda \Sigma,\Gamma\Delta}
f^{\Gamma\Delta}_{~~ab}$ must be real real,
implying the vanishing of the magnetic charge corresponding
to the singlet of
$U\left(1\right)\times SU\left(4\right)\times Usp\left(4\right)$.
Furthermore, since the right hand side of the
equation (\ref{sl21/4svil}) is proportional to the right
hand side of the gravitino equation, it
turns out that the vielbein $P_i$ must also be real.
Let us name $\Phi_2$ the scalar field appearing in left hand side
of the equation (\ref{sl21/4svil}),
and $P_2$ the corresponding vielbein component.
Equation (\ref{sl21/4svil}) can  be rewritten as:
\begin{equation}
\label{sl21/4fin}
P_2{d\Phi_2\over dr}=-{3\over
8} \ell \left(r\right)^{\Lambda\Sigma}
{\rm Im}{\cal N }_{ \Lambda \Sigma,\Gamma\Delta}f^{\Gamma\Delta}_{~~XY}
\,C^{XY}{e^U\over r^2}
\end{equation}
\par
In conclusion, we see that the most general model describing
$BPS$--saturated solutions preserving $1\over 4$ of $N=8$
supersymmetry is given, modulo $E_{7\left(7\right)}$ transformations,
in terms of two scalar fields and two electric charges.
\subsection{Derivation of the $1/4$ reduced lagrangian in Young basis}
\label{case1/4}
Our next step is to write down the lagrangian for this model.
This implies the construction of the coset representative
of ${SL\left(2,\IR\right)\over U\left(1\right)}\times O\left(1,1\right)$
in terms of which the kinetic matrix
of the vector fields and the $\sigma$--model metric of the scalar
fields is constructed.
\par
Once again we begin by considering such a construction in the Young
basis where the field strengths are labelled as antisymmetric tensors
and the $E_{7(7)}$ generators are written as $Usp(28,28)$ matrices.
\par
The basic steps in order to construct the desired lagrangian consist
of
\begin{enumerate}
\item Embedding of the appropriate $SL(2,\IR) \times O(1,1)$ Lie
algebra in the $Usp_Y(28,28)$ basis for the ${\bf 56}$ representation of
$E_{7(7)}$
\item Performing the explicit exponentiation of the two commuting
Cartan generators of the above algebra
\item Calculating the restriction of
the $\IL$ coset representative to the $4$--dimensional space spanned
by the $Usp(4) \times Usp(4)$ singlet field strengths and by their
magnetic duals
\item Deriving the restriction of
the matrix ${\cal N}_{\Lambda\Sigma}$ to the above
$4$--dimensional space.
\item Calculating the explicit form of the scalar vielbein
$P^{ABCD}$ and hence of the scalar kinetic terms.
\end{enumerate}
Let us begin with the first issue. To this effect we consider the
following two antisymmetric $8 \times 8$ matrices:
\begin{equation}
\varpi _{AB} = - \varpi_{BA} = \left (
\begin{array}{c|c}
C & 0 \\
\hline
0 & 0 \\
\end{array} \right) \quad ;  \quad \Omega _{AB} = - \Omega _{BA} = \left (
\begin{array}{c|c}
0 & 0 \\
\hline
0 & C \\
\end{array} \right)
\label{matriciotte}
\end{equation}
where each block is $4 \times 4$ and the non vanishing block $C$
satisfies
\footnote{The upper and lower matrices appearing in $\omega_{AB}$ and
in $\Omega_{AB}$ are actually the matrices $C_{ab}$, $a,b=1,\dots,4$
and $C_{XY}$ $X,Y=1,\dots,4$ used in the previous section.}:
\begin{equation}
C^T = - C \quad ; \quad C^2 = -\bfone
\end{equation}
The subgroup $Usp(4) \times Usp(4) \subset SU(8)$ is defined as the
set of unitary unimodular matrices that preserve simultaneously
$\varpi$ and $\Omega$:
\begin{equation}
A \, \in Usp(4) \, \times Usp(4) \subset SU(8) \quad \leftrightarrow
\quad  A^\dagger \varpi A =
 \varpi \quad \mbox{and}  \quad A^\dagger \Omega A = \Omega
\label{Usp42def}
\end{equation}
Obviously any other linear combinations of these two matrices is also
preserved by the same subgroup so that we can also consider:
\begin{equation}
\tau^\pm_{AB} \, \equiv \, \frac{1}{2}\left( \varpi_{AB} \pm
\Omega_{AB} \right) \,  = \, \left (
\begin{array}{c|c}
C & 0 \\
\hline
0 & \pm C \\
\end{array} \right)
\label{taumatte}
\end{equation}
Introducing also the matrices:
\begin{equation}
 \pi_{AB} \, = \, \left (
\begin{array}{c|c}
\bfone & 0 \\
\hline
0 &  0 \\
\end{array} \right)  \quad ; \quad \Pi_{AB} \, = \, \left (
\begin{array}{c|c}
0 & 0 \\
\hline
0 & \bfone\\
\end{array} \right)
\end{equation}
we have the obvious relations:
\begin{equation}
\pi_{AB} \, = \,  -  \varpi_{AC} \, \varpi_{CB} \quad ; \quad
 \Pi_{AB} \, = \,  -  \Omega_{AC} \, \Omega_{CB}
 \end{equation}
 In terms of these matrices we can easily construct the projection
 operators that single out from the ${\bf 28}$ of $SU(8)$ its
 $Usp(4) \times Usp(4)$ irreducible components according to:
 \begin{equation}
{\bf 28} \quad \stackrel{Usp(4) \times Usp(4)}{\Longrightarrow}\quad
{\bf (1,0)} \oplus {\bf (0,1)}  \oplus {\bf (5,0)}  \oplus {\bf (0,5)}
\oplus {\bf (4,4)}
\label{usp4deco}
\end{equation}
These projection operators are matrices mapping antisymmetric
$2$--tensors into antisymmetric $2$--tensors and  read  as
follows:
\begin{eqnarray}
\IP^{(1,0)}_{AB \  RS} & = & \frac{1}{4} \, \varpi_{AB} \,
\varpi_{RS} \nonumber \\
\IP^{(0,1)}_{AB \ RS} & = & \frac{1}{4} \, \Omega_{AB} \,
\Omega_{RS} \nonumber \\
\IP^{(5,0)}_{AB \ RS} & = & \frac{1}{2} \,\left(
\pi_{AR} \, \pi_{BS} \, - \, \pi_{AS} \, \pi_{BR} \right )\, - \,
\frac{1}{4} \, \varpi_{AB} \, \varpi_{RS}\nonumber \\
\IP^{(0,5)}_{AB \ RS} & = & \frac{1}{2} \,\left(
\Pi_{AR} \, \Pi_{BS} \, - \, \Pi_{AS} \, \Pi_{BR} \right )\, - \,
\frac{1}{4} \, \Omega_{AB} \, \Omega_{RS}\nonumber \\
\IP^{(4,4)}_{AB \ RS} & = & \frac{1}{8} \,\left(
\pi_{AR} \, \Pi_{BS} \,+ \,\Pi_{AR} \, \pi_{BS} \,  - \, \pi_{AS} \, \Pi_{BR}
\, - \, \Pi_{AS} \, \pi_{BR} \right)
\label{proiettori}
\end{eqnarray}
We also introduce the following shorthand notations:
\begin{eqnarray}
\ell^{AB}_{RS} & \equiv & \frac{1}{2} \,\left(
\pi_{AR} \, \pi_{BS} \, - \, \pi_{AS} \, \pi_{BR} \right )\,
\nonumber \\
L^{AB}_{RS} & \equiv &\frac{1}{2} \,\left(
\Pi_{AR} \, \Pi_{BS} \, - \, \Pi_{AS} \, \Pi_{BR} \right )\nonumber\\
U^{ABCD} & \equiv & \varpi^{[AB} \, \varpi^{CD]} \, = \,
\frac{1}{3} \left [ \varpi^{AB} \, \varpi^{CD} + \varpi^{AC} \, \varpi^{DB}\, +
\, \varpi^{AD} \, \varpi^{BC} \right] \nonumber \\
W^{ABCD} & \equiv & \Omega^{[AB} \, \Omega^{CD]} \, = \,
\frac{1}{3} \left [ \Omega^{AB} \, \Omega^{CD} + \Omega^{AC} \, \Omega^{DB}
\,+ \,  \Omega^{AD} \, \Omega^{BC} \right] \nonumber \\
Z^{ABCD}& \equiv & \varpi^{[AB} \, \Omega^{CD]} \, = \,
 \frac{1}{6} \Bigl [ \varpi^{AB} \, \Omega^{CD} + \varpi^{AC} \, \Omega^{DB}
\, \varpi^{AD} \, \Omega^{BC} \nonumber \\
&& + \, \Omega^{AB} \, \varpi^{CD} + \Omega^{AC} \, \varpi^{DB} \, +
\, \Omega^{AD} \, \varpi^{BC} \, \Bigr ]
\label{shortbread}
\end{eqnarray}
Then by direct calculation we can verify the following relations:
\begin{eqnarray}
Z_{ABRS} \, Z_{RSUV} & = & \frac{4}{9} \left( \IP^{(1,0)}_{AB \ UV}
\, + \, \IP^{(0,1)}_{AB \ UV} \, + \, \IP^{(4,4)}_{AB \  UV} \right )
\nonumber \\
U_{ABRS} \, U_{RSUV} & = & \frac{4}{9} \, \ell^{AB}_{UV} \nonumber \\
W_{ABRS} \, W_{RSUV} & = & \frac{4}{9} \, L^{AB}_{UV} \nonumber \\
\label{relazie2}
\end{eqnarray}
Using the above identities we can write
the explicit embedding of the relevant
$SL(2,\IR) \times O(1,1)$ Lie algebra
into the $E_{7(7)}$ Lie algebra, realized in the Young basis, namely
in terms of $Usp(28,28)$ matrices. Abstractly we have:
\begin{equation}
\begin{array}{ccc}
 \mbox{$SL(2,\IR)$ algebra} & \longrightarrow & \cases { \left[ L_+ \, , \, L_-
 \right] = 2 \, L_0 \cr
  \left[ L_0\, , \, L_\pm
 \right] = \pm \, L_\pm \cr } \\
 \null & \null & \null \\
 \mbox{$O(1,1)$ algebra} & \longrightarrow & {\cal C} \\
 \null & \null & \null \\
 \mbox{and they commute} & \null & \left[ {\cal C} \, , \, L_\pm \right] =
 \left[ {\cal C} \, , \, L_0 \right] = 0 \\
\end{array}
\end{equation}
The corresponding $E_{7(7)}$ generators in the ${\bf 56}$ Young basis
representation are:
\begin{eqnarray}
L_0 & = &  \, \left( \begin{array}{c|c}
 0 & \frac{3}{4}\left( U^{ABFG} + W^{ABFG}\right) \\
 \hline
 \frac{3}{4}\left( U_{LMCD} + W_{LMCD}\right) & 0 \\
 \end{array} \right) \nonumber\\
 L_\pm & = &  \, \left( \begin{array}{c|c}
 \pm \frac{\rm i}{2} \left( \ell^{AB}_{CD} - L^{AB}_{CD}\right) &
 \mbox{\rm i}\, \frac{3}{4}\left( U^{ABFG} - W^{ABFG}\right) \\
 \hline
 -\mbox{\rm i}\, \frac{3}{4}\left( U_{LMCD} - W_{LMCD}\right) &
 \mp \frac{\rm i}{2} \left( \ell^{LM}_{FG} - L^{LM}_{FG}\right) \\
 \end{array} \right) \nonumber\\
 {\cal C}& = & \, \left( \begin{array}{c|c}
 0 & \frac{3}{4}Z^{ABFG} \\
 \hline
 \frac{3}{4}Z_{LMCD} & 0 \\
 \end{array} \right)
\end{eqnarray}
The  non--compact Cartan subalgebra of $SL(2, \IR) \times O(1,1)$, spanned by
$L_0 \, , \, {\cal C}$ is a $2$--di\-men\-sio\-nal subalgebra of the full
$E_{7(7)}$ Cartan subalgebra. As such this abelian algebra is also a subalgebra
of the $70$--dimensional solvable Lie algebra $Solv_7$ defined in eq.
\eqn{defsolv7}. The scalar fields associated with $L_0$ and ${\cal C}$
are the two dilatons  parametrizing the reduced bosonic lagrangian we want to construct.
Hence our  programme is to construct the coset representative:
\begin{equation}
\IL( \Phi_1, \Phi_2) \equiv \exp \left[ \Phi_1 {\cal C}+ \Phi_2 L_0 \right]
\label{cosrepdef}
\end{equation}
and consider its restriction to the $4$--dimensional space spanned by
the $Usp(4) \times Usp(4)$ singlets $\varpi_{AB}$ and $\Omega_{AB}$.
Using the definitions \eqn{proiettori} and \eqn{shortbread} we can
easily verify that:
\begin{eqnarray}
\IP^{(1,0)}_{AB \ RS} \, \frac{3}{4} \, Z^{RSUV}
\, \IP^{(1,0)}_{UV \ PQ} \, &=& 0
\nonumber\\
\IP^{(0,1)}_{AB \ RS} \,\frac{3}{4} \, Z^{RSUV}
\, \IP^{(0,1)}_{UV \ PQ} \, &=& 0
\nonumber\\
\IP^{(1,0)}_{AB \ RS} \,\frac{3}{4} \, Z^{RSUV}
\, \IP^{(0,1)}_{UV \ PQ} \, &=&
\frac{1}{8} \varpi_{AB} \, \Omega_{PQ} \nonumber\\
\IP^{(0,1)}_{AB \ RS} \,\frac{3}{4} \, Z^{RSUV}
\, \IP^{(1,0)}_{UV \ PQ} \, &=&
\frac{1}{8} \Omega_{AB} \, \varpi_{PQ} \nonumber\\
\label{Cgenrest1}
\end{eqnarray}
and similarly:
\begin{eqnarray}
\IP^{(1,0)}_{AB \ RS} \, \frac{3}{4} \, \left( U^{RSUV} +
W^{RSUV} \right)
\, \IP^{(1,0)}_{UV \ PQ} \, &=& \frac{1}{2} \IP^{(1,0)}_{AB \ PQ}
\nonumber\\
\IP^{(1,0)}_{AB \ RS} \, \frac{3}{4} \, \left( U^{RSUV} +
W^{RSUV} \right)
\, \IP^{(0,1)}_{UV \ PQ} \, &=& 0
\nonumber\\
\IP^{(0,1)}_{AB \ RS} \, \frac{3}{4} \, \left( U^{RSUV} +
W^{RSUV} \right)
\, \IP^{(0,1)}_{UV \ PQ} \, &=& \frac{1}{2} \IP^{(0,1)}_{AB \ PQ}
\nonumber\\
\IP^{(0,1)}_{AB \ RS} \,\frac{3}{4} \, Z^{RSUV}
\, \IP^{(1,0)}_{UV \vert PQ} \, &=&
0 \nonumber\\
\label{Cgenrest2}
\end{eqnarray}
This means that in the $4$--dimensional space spanned by the
$Usp(4) \times Usp(4)$ singlets, using also the
definition \eqn{taumatte} and the shorthand notation
\begin{equation}
\Phi^\pm = \frac{\Phi_2 \pm \Phi_1}{2}
\label{shortmilk}
\end{equation}
the coset representative can be written as follows
\begin{eqnarray}
 & \exp\left[\Phi_1 {\cal C} + \Phi_2 L_0  \right]  =& \nonumber\\
 &  \left( \begin{array}{c|c}
 \cosh \Phi^+\, \frac{1}{2} \, \tau^+_{AB}
 \, \tau^+_{CD} + \cosh \Phi^- \, \frac{1}{2} \, \tau^-_{AB}
 \, \tau^-_{CD} & \sinh \Phi^+ \, \frac{1}{2} \, \tau^+_{AB}
 \, \tau^+_{CD} + \sinh \Phi^-\, \frac{1}{2} \, \tau^-_{AB}
 \, \tau^-_{CD} \\
 \hline
 \sinh \Phi^+ \, \frac{1}{2} \, \tau^+_{AB}
 \, \tau^+_{CD} + \sinh \Phi^-\, \frac{1}{2} \, \tau^-_{AB}
 \, \tau^-_{CD}\ & \cosh \Phi^+ \, \frac{1}{2} \, \tau^+_{AB}
 \, \tau^+_{CD} + \cosh \Phi^- \, \frac{1}{2} \, \tau^-_{AB}
 \, \tau^-_{CD}\\
 \end{array} \right) &  \nonumber\\
 \label{cassettone}
\end{eqnarray}
Starting from eq.\eqn{cassettone} we can easily write down the
matrices $f_{AB \ CD}, h_{AB \ CD}$ and ${\cal N}_{AB \
CD}$. We immediately find:
\begin{eqnarray}
f_{AB \ CD} & = & \frac{1}{\sqrt{2}} \,
\left( \exp[\Phi^+]  \, \frac{1}{2} \, \tau^+_{AB} \, \tau^+_{CD}  \,
+ \,  \exp[\Phi^-]  \, \frac{1}{2} \, \tau^-_{AB} \, \tau^-_{CD}
\right) \nonumber \\
h_{AB \ CD} & = & -\frac{\rm i}{\sqrt{2}} \,
\left( \exp[-\Phi^+]  \, \frac{1}{2} \, \tau^+_{AB} \, \tau^+_{CD}  \,
+ \,  \exp[-\Phi^-]  \, \frac{1}{2} \, \tau^-_{AB} \, \tau^-_{CD}
\right) \nonumber \\
{\cal N}_{AB \ CD} & = & -\frac{\rm i}{4} \,
\left( \exp[-2 \, \Phi^+]  \, \frac{1}{2} \, \tau^+_{AB} \, \tau^+_{CD}  \,
+ \,  \exp[-2\,\Phi^-]  \, \frac{1}{2} \, \tau^-_{AB} \, \tau^-_{CD}
\right)
\label{fhNmat}
\end{eqnarray}
To complete our programme, the last point we have to deal with is the
calculation of the scalar vielbein $P^{ABCD}$. We have:
\begin{small}
\begin{eqnarray}
& \IL^{-1} (\Phi_1, \Phi_2) \, d \, \IL (\Phi_1, \Phi_2)  \, = \, & \nonumber \\
 & \frac{3}{4} \,\left( \begin{array}{c|c}
 0 & d \Phi_1 \,  Z^{ABFG}   \, + \, d\Phi_2 \,
 \left( U^{ABFG} + W^{ABFG} \right) \\
 \hline
 d \Phi_1 \, Z^{LMCD}   \, + \, d\Phi_2 \,
  \left( U^{LMCD} + W^{LMCD} \right)  & 0 \\
 \end{array} \right) & \nonumber \\
 \label{linv1f}
\end{eqnarray}
\end{small}
so that we get:
\begin{equation}
P^{ABCD} \, = \, d \Phi_1 \, \frac{3}{4} \, Z^{ABCD}   \, + \, d\Phi_2 \,
 \frac{3}{4} \, \left( U^{ABCD} + W^{ABCD} \right)
 \label{scalvielb}
\end{equation}
and with a straightforward calculation:
\begin{equation}
 P^{ABCD}_\mu \, P_{ABCD}^\mu \, =
  \frac{3}{2} \, \partial_\mu \Phi_1 \,\partial^\mu \Phi_1  \, + \,
 3 \, \partial_\mu \Phi_2 \, \partial^\mu \Phi_2 \,
\label{paperino}
\end{equation}
Hence recalling the normalizations of the supersymmetric $N=8$
lagrangian \eqn{lN=8}, and introducing the two $Usp(4) \times Usp(4)$
singlet electromagnetic fields:
\begin{equation}
\label{potentials}
A_\mu ^{AB} = \tau^+_{AB} \, \frac{1}{2 \sqrt{2}} \,{\cal A}^{1}_\mu \,  + \,
\tau^-_{AB} \, \frac{1}{2 \sqrt{2}} \,{\cal A}^{2}_\mu \, + \,
\mbox{26 non singlet fields}
\end{equation}
we get the following reduced Lagrangian:
\begin{eqnarray}
{\cal L}^{1/4}_{red}& = & \sqrt{ -g } \, \Bigl [ 2 \, R[g] \, + \,
\frac{1}{4} \,   \partial_\mu \Phi_1 \, \partial^\mu  \Phi_1
\, + \, \frac{1}{2} \,   \partial_\mu \Phi_2 \, \partial^\mu  \Phi_2\nonumber \\
 &&- \,  \exp\left[-\Phi_1 - \Phi_2 \right]
  \left( F^1_{\mu \nu}\right)^2    \,
  - \,  \exp\left[ \Phi_1 -\Phi_2\right] \left( F^2_{\mu \nu}\right)^2
\label{primalag}
\end{eqnarray}
Redefining:
\begin{equation}
\Phi_1 = \sqrt{2} \, h_1 \quad ; \quad \Phi_2 = h_2
\label{redefo}
\end{equation}
we obtain the final standard form for the reduced lagrangian
\begin{eqnarray}
{\cal L}^{1/4}_{red}& = & \sqrt{ -g } \, \Bigl [ 2 \, R[g] \, + \,
\frac{1}{2} \,   \partial_\mu h_1 \, \partial^\mu  h_1
\, + \, \frac{1}{2} \,   \partial_\mu h_2 \, \partial^\mu  h_2\nonumber \\
 &&- \,  \exp\left[-\sqrt{2} \, h_1 - h_2 \right]
  \left( F^1_{\mu \nu}\right)^2    \,
  - \,  \exp\left[\sqrt{2} \, h_1 - h_2 \right]
  \left( F^2_{\mu \nu}\right)^2
\label{eff1/4}
\end{eqnarray}

\subsection{Solution of the reduced field equations}
We can easily solve the field equations for the reduced lagrangian
\eqn{eff1/4}. The Einstein equation is
\begin{equation}
\label{eqeinstein1/4}
U''+{2 \over r}U'-\left(U'\right)^2=
{1 \over 4}\, \left(h_1^\prime \right)^2+ {1 \over 4}\left(h_2^\prime \right)^2,
\end{equation}
the Maxwell equations are
\begin{eqnarray}
\sqrt{2} \, h_1^\prime- h_2^\prime &=&-{\ell_2' \over \ell_2}\nonumber\\
-\sqrt{2} \, h_1^\prime- h_2^\prime &=&-{\ell_1' \over \ell_1},
\label{eqmaxwell1/4}
\end{eqnarray}
the scalar equations are
\begin{eqnarray}
h_1''+{2 \over r}h_1'&=&{1 \over \sqrt{2} r^4}
\left(e^{ \sqrt{2}h_1- h_2+2U}\, \ell_2^2
-e^{-\sqrt{2}h_1- h_2+2U}\, \ell_1^2\right)\nonumber\\
h_2''+{2 \over r}h_2'&=&{1 \over 2 r^4}
\left(e^{ \sqrt{2}h_1- h_2+2U}\, \ell_2^2
+e^{-\sqrt{2}h_1- h_2+2U}\, \ell_1^2\right).
\label{eqscalars1/4}
\end{eqnarray}
The solution of the Maxwell equations is
\begin{eqnarray}
\ell_1\left(r\right)&=&\ell_1e^{\sqrt{2} h_1+h_2}\nonumber\\
\ell_2\left(r\right)&=&\ell_2e^{-\sqrt{2} h_1+h_2}
\label{solcharges1/4}
\end{eqnarray}
where
\begin{eqnarray}
\ell_1&\equiv&\ell_1\left(\infty\right)\nonumber\\
\ell_2&\equiv&\ell_2\left(\infty\right)
\end{eqnarray}
\begin{eqnarray}
h_1&=& -{1 \over \sqrt{2} }\, \log \, {{H_1\over H_2}} \nonumber\\
h_2&=&-{1 \over 2}\, \log \, {H_1 H_2}\nonumber\\
U&=&-{1 \over 4}\ln{H_1 H_2}
\label{solfields1/4}
\end{eqnarray}
and
\begin{eqnarray}
H_1\left(r\right)&=&
1 + \sum_{i} \, \frac{k^1_i}
{ \ {\vec x} - {\vec x}^{(1)}_i \ }\nonumber\\
H_2\left(r\right)&=&
1 + \sum_{i} \, \frac{l^2_i}{ \ {\vec x} - {\vec x}^{(2)}_i \ }
\end{eqnarray}
are a pair of harmonic functions.
\par
We can summarize the form of $1/4$ solution writing:
\begin{eqnarray}
ds^2 &= &\left(H_1 \,H_2 \right)^{-1/2} \, dt^2 -
\left(H_1 \,H_2 \right)^{1/2} \left( dr^2 + r^2 d\Omega_2 \right)
\nonumber \\
h_1 &=& - \frac{1}{\sqrt{2}}\, \log \frac{H_1}{H_2} \nonumber \\
h_2 &=& - \frac{1}{2} \, \log \left[ H_1 \, H_2 \right]
\nonumber \\
\label{finalsolution}
F^{1,2} &=& -dt \, \wedge \, {\vec {dx}} \,\cdot \,
\frac{\vec \partial}{\partial x} \, \left( H_{1,2} \right)^{-1}
\end{eqnarray}
Let us now   observe that by specializing eq.s\eqn{finalsolution}
to the case where the two harmonic functions are equal
$H_1(r)=H_2(r)=H(r)$ we exactly match the case $a=1$, $\Delta=2$
of eq.\eqn{critval1}. Indeed for this choice we have $h_1=0$ and
the two vector field strengths can be identified since they are
identical. The metric, the surviving scalar field $h_2$ and the
electric field they all have the form predicted in eq.\eqn{critval1}.
Looking back at eq.\eqn{eff1/4} we see that the parameter $a$ for
$h_2$ is indeed $a=1$ once $h_1$ is suppressed.
This explains why we claimed that the $a=1$ $0$--brane corresponds to
the case of $1/4$--preserved supersymmetry. The generating solution
for $1/4$ BPS black--holes is more general since it involves two
harmonic functions but the near--horizon behaviour of all such
solutions is the that of $a=1$ model. Indeed near the horizon $r=0$
all harmonic functions are proportional.
\subsection{Comparison with the Killing spinor equations}
In this section we show that the Killing spinor equations (\ref{reduceo11}),
(\ref{sl21/4fin}) are identically satisfied by the solution
(\ref{finalsolution}) giving no further restriction on the harmonic
function $H_1,H_2$. Using the formalism developed in section
\ref{case1/4} the equations (\ref{reduceo11}), (\ref{sl21/4fin})
can be combined as follows:
\begin{equation}
\label{16/3}
{16\over 3}P_2{d\Phi_2\over dr}\pm P_1{d\Phi_1\over
dr}=-2 \ell \left(r\right)^{\Lambda\Sigma}\tau^{\left(\pm\right)AB}
{\rm Im}{\cal N }_{ \Lambda \Sigma,\Gamma\Delta}
f^{\Gamma\Delta}_{~~AB}{e^U\over r^2}
\end{equation}
where
\begin{eqnarray}
P_2d\Phi_2&=&{3\over 8}C^{ab}C^{cd}P_{abcd}={3\over 8}U^{ABCD}P_{ABCD}
={3\over 4}d\Phi_2 \nonumber\\
P_1d\Phi_1&=&C^{XY}C^{ab}P_{XYab}=Z^{ABCD}P_{ABCD}=2d\Phi_1
\end{eqnarray}
furthermore, using equations (\ref{fhNmat}), we have
\begin{equation}
{\rm Im}{\cal N }_{ \Lambda\Sigma,\Gamma\Delta}f^{\Gamma\Delta}_{~~AB}=
-{1\over 2\sqrt{2}}\left(e^{-\Phi_+}\tau^{\left(+\right)\Lambda\Sigma}
\tau^{\left(+\right)AB}+
e^{-\Phi_-}\tau^{\left(-\right)\Lambda\Sigma}\tau^{\left(-\right)AB}\right).
\end{equation}
Therefore, the equation (\ref{16/3}) becomes
\begin{equation}
4{d\Phi_2\over dr}\pm 2{d\Phi_1\over dr}=4
\sqrt{2}\tau^{\left(\pm\right)}_{\Lambda\Sigma}\ell^{\Lambda\Sigma}
{e^{U-\Phi_{\pm}}\over r^2}
\end{equation}
Using equation (\ref{potentials}) we also have
\begin{eqnarray}
\tau^{\left(+\right)}_
{\Lambda\Sigma}q^{\Lambda\Sigma}&=&{1\over\sqrt{2}}\ell_1\left(r\right)\nonumber\\
\tau^{\left(-\right)}_
{\Lambda\Sigma}q^{\Lambda\Sigma}&=&{1\over\sqrt{2}}\ell_2\left(r\right).
\end{eqnarray}
Comparing equation (\ref{finalsolution}) with (\ref{f}),
(\ref{eaself}), (\ref{smallt}), we obtain
\begin{eqnarray}
\ell _1&=&-r^2{H'_1\over H_1^2}\sqrt{H_1H_2}\nonumber\\
\ell _2&=&-r^2{H'_2\over H_2^2}\sqrt{H_1H_2}.
\end{eqnarray}
Using all these information, a straightforward computation shows that
the Killing spinor equations are identically satisfied.
\section{Detailed study of the $1/8$ case}
\setcounter{equation}{0}
\label{det1/8}
\markboth{BPS BLACK HOLES IN SUPERGRAVITY: CHAPTER 5}
{5.7 DETAILED STUDY OF THE $1/8$ CASE}
As previously emphasized, the most general $1/8$ black--hole solution
of $N=8$ supergravity
is, up to $U$--duality transformations, a solution of an $STU$ model
suitably
embedded in the original $N=8$ theory. Therefore, in dealing with the
$STU$ model we would like
to keep trace of this embedding. To this end, we shall use, as anticipated,
the mathematical tool
of SLA which in general provides a suitable and simple description of the
embedding of a
supergravity theory in a larger one. The SLA formalism is very useful
in order to give a
geometrical and a quasi easy characterization of the different dynamical
scalar fields belonging
to the solution. Secondly, it enables one to write down the somewhat
heavy first order differential
system of equations for all the fields and to compute all the geometrical
quantities appearing in
the effective supergravity theory in a clear and direct way.
Instead of considering the $STU$ model embedded in the whole $N=8$ theory
with scalar manifold
${\cal M}=E_{7(7)}/SU(8)$, it suffices to focus on its $N=2$ truncation
with scalar manifold
${\cal M}_{T^6/\ZZ_3}=[SU(3,3)/SU(3)\times U(3)]\times {\cal M}_{Quat}$
which describes the
classical
limit of type $IIA$ Supergravity compactified on $T^6/\ZZ_3$,
${\cal M}_{Quat}$ being the quaternionic
manifold $SO(4,1)/SO(4)$ describing $1$ hyperscalar. Within this
latter simpler
model we are going to construct the $N=2$ $STU$ model as a consistent
truncation. Indeed the embedding of the manifold ${\cal M}_{T^6/\ZZ_3}$
inside the $N=8$ has already been described in eq.s\eqn{Alekal2}
where we have shown the embedding of the corresponding solvable Lie
algebra into $Solv_3 \subset Solv_7$.
The embedding of the $STU$ scalar manifold
${\cal M}_{STU}=(SL(2,\IR)/U(1))^3$
inside ${\cal M}_{T^6/\ZZ_3}$  was described as an example
in section \ref{trexe} (see in particular eq.\eqn{stuembed}).
On the other hand in section \ref{gen1/8} we discussed how,
up to $H=SU(8)$ transformations, the $N=8$ central charge
which is an $8\times 8$ antisymmetric complex matrix can
always be brought to
its {\it normal} form in which it is skew--diagonal with complex
eigenvalues $Z,Z_i$, $i=1,2,3$
($|Z|>|Z_i|$). In order to do this one needs to make a suitable
$48$--parameter $SU(8)$  transformation on the central charge.
This transformation may be seen as
the result of a $48$--parameter $E_{7(7)}$ duality
transformation on the $56$ dimensional charge vector and on the
$70$ scalars which, in the expression of the central charge,
sets to zero $48$ scalars (24 vector scalars and 24 hyperscalars
from the $N=2$ point of view) and $48$ charges.
Taking into account that there are 16 scalars parametrizing
the submanifold $SO(4,4)/SO(4)\times SO(4)$, $SO(4,4)$ being the
centralizer
of the normal form, on which the eigenvalues of the central
charge do not depend at all, the central
charge, in its normal form will depend only on the $6$
scalars and $8$ charges defining an $STU$ model.
The isometry group of ${\cal M}_{STU}$ is $[SL(2,\IR)]^3$,
which is the
{\it normalizer} of the normal form, i.e.
the residual $U$--duality which can still act non trivially
on the $6$ scalars and $8$
charges while keeping the central
charge skew diagonalized. As we show in the sequel, the $6$ scalars of the
$STU$ model consist of $3$ axions $a_i$ and $3$ dilatons $p_i$,
whose exponential
$\exp{p_i}$ will be denoted by $-b_i$.\par
In the framework of the $STU$ model, the
eigenvalues $Z(a_i,b_i,p^\Lambda,q_\Lambda)$
and $Z_i(a_i,b_i,p^\Lambda,q_\Lambda)$ are, respectively
the local realization  on  moduli space of
the $N=2$ supersymmetry algebra central charge and of the $3$
{\it matter} central charges associated with the $3$ matter
vector fields (see eq.\eqn{zentrum} and eq.\eqn{zmatta}).
The BPS condition for a $1/8$ black--hole is that the ADM mass
should equal
the modulus of the central charge:
\begin{equation}
 M_{ADM}=|Z(a_i,b_i,p^\Lambda,q_\Lambda)|.
\end{equation}
As explained in section \ref{fissato}, at the horizon
the field dependent central charge $|Z|$ flows to its
minimum value:
\begin{eqnarray}
  |Z|_{min}(p^\Lambda,q_\Lambda)&=&
  |Z(a_i^{fix},b_i^{fix},p^\Lambda,q_\Lambda)|\nonumber\\
 0 & = & \frac{\partial}{\partial a_i }|Z|_{a=b=fixed} \, = \,
 \frac{\partial}{\partial a_i }|Z|_{a=b=fixed}
\end{eqnarray}
which is obtained by extremizing it with
respect to the $6$ moduli $a_i,b_i$. At the horizon the
other eigenvalues $Z_i$ vanish (see chapter \ref{chargcha}). The
value $|Z|_{min}$ is related to the Bekenstein Hawking entropy of the
solution and it is expressed in terms of the quartic
invariant of the $56$--representation of $E_{7(7)}$,
which in principle depends on all the $8$
charges of the $STU$ model (see section \ref{5param} and in particular
eq.\eqn{bella}).
Nevertheless there is a residual
$[U(1)]^3\in [SL(2,\IR)]^3$ acting on the $N=8$ central charge matrix
in its normal form. These  three gauge parameters
can be used to  reduce  the number
of charges appearing in the quartic invariant (entropy)from 8 to 5, as we
have already pointed out.
  We shall see how these 3 conditions may be implemented
on the 8 charges at the level of the first order BPS equations
in order to obtain the $5$ parameter
generating solution for the most general $1/8$ black--holes in $N=8$
supergravity. This generating
solution coincides with the solution generating the orbit of $1/2$
BPS black--holes in the
truncated $N=2$ model describing type $IIA$ supergravity compactified
on $T^6/\ZZ_3$. Therefore,
in the framework of this latter simpler model, we shall work out the
$STU$ model and construct
the set of second and first order differential equations defining our
solution.
\subsection{The $STU$ model in the $SU(3,3)/SU(3)
\times U(3)$ theory and solvable Lie algebras}
As it was shown in \cite{mp1} and already discussed in these
lectures the hyperscalars do not contribute
to the dynamics of our BPS
black--hole, therefore, in what follows,
all hyperscalars will be set to zero and we shall forget
about the quaternionic factor ${\cal M}_{Quat}$ in
${\cal M}_{T_6/Z_3}$. The latter will then be
the scalar manifold of an $N=2$ supergravity describing
$9$ vector multiplets coupled with the
graviton multiplet.
The $18$ real scalars span the manifold
${\cal M}_{T_6/Z_3}=SU(3,3)/SU(3)\times U(3)$, while
the $10$ electric and $10$ magnetic charges
associated with the $10$ vector fields transform under
duality in the ${\bf 20}$ (three times antisymmetric)
of $SU(3,3)$. As anticipated, in order to show how the
$STU$ scalar manifold ${\cal M}_{STU}$ is
embedded in ${\cal M}_{T_6/Z_3}$ we just have to use the SLA description.
Indeed it suffices to quote the results of section \ref{trexe} where
the embedding of $Solv_{STU}$ into $Solv(SU(3,3)/SU(3)\times U(3)$
was explicitly derived.
\subsection{First order differential equations: the algebraic
approach}
Now that the $STU$ model has been constructed out the original
$SU(3,3)/SU(3)\times U(3)$ model, we may address the problem of
writing down
the BPS first order equations. To this end we shall use the
geometrical intrinsic approach defined
in \cite{mp1} and eventually compare it with the Special
K\"ahler geometry formalism. \par
The system of first order differential equations in
the background fields is
obtained from the Killing spinor equations \eqn{kigrav},\eqn{kigaug},
with parameter as in eq.\eqn{quispi}. In our case we have that
$i=1,2,3$  labels the three matter vector fields, while here as in
any other $N=2$ theory $A,B=1,2$ are
the $SU(2)$ R-symmetry indices.
\par
Following the procedure defined in previous chapters and sections,
in order to obtain a system of first order differential equations
out of the
killing spinor conditions   we make the  ans\"atze
\eqn{ds2U} for the metric, \eqn{fedfd},\eqn{strenghtsans},
\eqn{tlamb}for the field strengths and we assume radial dependence
for the scalar fields.
We represent the scalars of the $STU$ model in terms of three complex
fields $\{z^i\}\equiv \{S,T,U\}$, parametrizing each of the three factors
$SL(2,\IR)/SO(2)$ in ${\cal M}_{STU}$.
\par
In terms of special K\"ahler geometry the first order equations were
already derived for a generic $N=2$ theory and they are given by eq.s
\eqn{zequa} and \eqn{Uequa}. We just have to particularize such
equations to our specific $STU$--model.
\par
In order to compute the explicit form of eqs.\eqn{zequa} and \eqn{Uequa}
in a geometrical intrinsic way \cite{mp1} we need to decompose the
$4$ vector fields into the graviphoton $F_{\mu\nu}^0$ and the matter vector
fields $F_{\mu\nu}^i$ in the same representation of the scalars $z^i$ with respect to the
isotropy group $H=[SO(2)]^3$. This decomposition is immediately
performed by computing the positive weights $\vec{v}^\Lambda$ of the ${\bf (2,2,2)}$ on the three
generators $\{\tilde{g}_i\}$ of $H$ combined in such a way
as to factorize in $H$ the automorphism group $H_{aut}=SO(2)$ of
the supersymmetry algebra generated by
$\lambda=\tilde{g}_1+\tilde{g}_2+\tilde{g}_3$ from the remaining
$H_{matter}=[SO(2)]^2=\{\tilde{g}_1-\tilde{g}_2,\tilde{g}_1-\tilde{g}_3\}$ generators acting non
trivially only on the matter fields.
The real and imaginary components of the graviphoton central charge $Z$ will be associated with
the weight, say $\vec{v}^0$
having vanishing  value on the generators of $H_{matter}$. The remaining
weights will define a representation ${\bf (2,1,1)}\oplus {\bf (1,2,1)}\oplus {\bf (1,1,2)}$ of $H$
in which the real and imaginary parts of the central charges $Z^i$ associated with $F^i_{\mu\nu}$
transform and will be denoted by $\vec{v}^i$, $i=1,2,3$.
This representation is the same as the one in which the $6$ real scalar components of
$z^i=a_i+{\rm i}b_i$ transform with respect to $H$. It is useful to define on the tangent space of
${\cal M}_{STU}$ curved indices $\alpha$
and rigid indices $\hat{\alpha}$, both running form $1$ to $6$. Using the solvable parametrization
of ${\cal M}_{STU}$, which defines real coordinates $\phi^\alpha$, the generators of
$Solv_{STU}=\{T^\alpha\}$
carry curved indices since they are parametrized by the coordinates, but do
not transform in a representation of the isotropy group. The compact generators
$\IK=Solv_{STU}+Solv_{STU}^\dagger$ of $[SL(2,\IR)]^3$ on the other hand transform
 in the ${\bf (2,1,1)}\oplus {\bf (1,2,1)}\oplus {\bf (1,1,2)}$ of $H$ and we can choose an
orthonormal basis (with respect to the trace) for $\IK$ consisting of the generators
$\IK^{\hat{\alpha}}=T^\alpha +T^{\alpha \dagger}$. These generators now carry the rigid index and
are in one to one correspondence with the real scalar fields $\phi^\alpha$.
There is a one to one correspondence between the non--compact matrices $\IK^{\hat{\alpha}}$ and
the eigenvectors $\vert v^i_{x,y}\rangle$ ($i=1,2,3$) which are
orthonormal bases (in different
spaces) of the same representation of $H$:
\begin{eqnarray}
\underbrace{\{\IK^1,\IK^2,\IK^3,\IK^4,\IK^5,\IK^6\}}_{\{ \IK^{\hat{\alpha}}\}}\,
 &\leftrightarrow\, &\underbrace{\{\vert v^1_{x}\rangle,\vert v^2_{y}\rangle,\vert v^3_{y}\rangle,
\vert v^1_{y}\rangle,\vert v^2_{x}\rangle,\vert v^3_{x}\rangle \}}_{\{\vert v^{\hat{\alpha}}
\rangle \}}
\end{eqnarray}
The relation between the real parameters $\phi^\alpha$ of the SLA and the real and imaginary parts
of the complex fields $z^i$ is:
\begin{eqnarray}
\{\phi^\alpha\}\, &\equiv& \{-2a_1,-2a_2,-2a_3,\log {(-b_1)},\log {(-b_2)},\log {(-b_3) },\}
\end{eqnarray}
Using the $Sp(8)_D$ representation of $Solv_{STU}$, we construct the coset representative $
\IL(\phi^\alpha)$ of ${\cal M}_{STU}$ and the vielbein $\IP_\alpha^{\hat{\alpha}}$ as follows:
\begin{eqnarray}
\IL(a_i,b_i)\,&=&\, \exp\left(T_\alpha \phi^\alpha\right)\,=\,\nonumber\\
&&\left(1-2a_1 g_1\right)\cdot \left(1-2a_2 g_2\right)\cdot \left(1-2a_3 g_3\right)\cdot
\exp{\left(\sum_i\log{(-b_i)}h_i\right)}\nonumber\\
\IP^{\hat{\alpha}}\,&=&\,\frac{1}{2\sqrt{2}}{\rm Tr}\left(\IK^{\hat{\alpha}}\IL^{-1}d\IL
\right)\,=\,\{-\frac{da_1}{2b_1},-\frac{da_2}{2b_2},-\frac{da_3}{2b_3},\frac{db_1}{2b_1},
\frac{db_2}{2b_2},\frac{db_3}{2b_3}\}\nonumber\\
\end{eqnarray}
The scalar kinetic term of the lagrangian is $\sum_{\hat{\alpha}}(\IP_{\hat{\alpha}})^2$.
The following relations between quantities computed in the solvable approach and Special K\"ahler
formalism hold:
\begin{eqnarray}
\left(\IP^{\alpha}_{\hat{\alpha}}\langle v^{\hat{\alpha}}\vert \IL^t \IC {\bf M}\right)\,&=&\,
\sqrt{2}\left(\matrix{{\rm Re}(g^{ij^\star}(\bar{h}_{j^\star\vert \Lambda})),-{\rm Re}(g^{ij^\star}
(\bar{f}_{j^\star}^\Sigma))\cr{\rm Im}(g^{ij^\star}(\bar{h}_{j
^\star\vert \Lambda})),-{\rm Im}(g^{ij^\star}(\bar{f}_{j^\star}^\Sigma)
) }\right)\nonumber\\
\left(\matrix{\langle v^{0}_y\vert \IL^t \IC {\bf M}\cr \langle v^{0}_x\vert \IL^t \IC
{\bf M}}\right)\,&=&\,\sqrt{2}\left(\matrix{{\rm Re}(M_{ \Lambda}),-{\rm Re}(L^\Sigma)\cr{\rm Im}
(M_{ \Lambda}),-{\rm Im}(L^\Sigma)} \right)
\label{secm2}
\end{eqnarray}
where in the first equation both sides are  $6\times 8$ matrix
in which the rows are labelled by
$\alpha$. The first three values of $\alpha$
correspond to the axions $a_i$, the last three to the dilatons $\log (-b_i)$.
The columns  are to be contracted with the vector consisting of the $8$
electric and magnetic charges $\vert \vec{Q}\rangle_{sc} =2\pi (p^\Lambda,q_\Sigma)$ in the {\it
special coordinate} symplectic gauge of ${\cal M}_{STU}$.
 In eqs. (\ref{secm2}) $\IC$ is the symplectic invariant matrix, while ${\bf M}$ is the symplectic
matrix relating the charge vectors in the $Sp(8)_D$ representation and in the   {\it special
coordinate} symplectic gauge:
\begin{eqnarray}
\vert \vec{Q}\rangle_{Sp(8)_D}\, &=&\, {\bf M}\cdot \vert \vec{Q}\rangle_{sc}\nonumber\\
{\bf M}\, &=&\, \left(\matrix{0 & 0 & 0 & 0  & 0 & 0 & 1 & 0\cr
1 & 0 & 0 & 0 &  0 & 0 & 0 & 0\cr
0 & 0 & 0 & 0 &  0 & 0 & 0 & 1\cr
0 & 0 & 0 & 0 &  0 & -1 & 0 & 0\cr
0 & 0 & -1 & 0 &  0 & 0 & 0 & 0\cr
0 & 0 & 0 & 0 &  1 & 0 & 0 & 0\cr
0 & 0 & 0 & -1 & 0 & 0 & 0 & 0\cr
0 & 1 & 0 & 0  & 0 & 0 & 0 & 0\cr }\right)\in Sp(8,\IR)
\label{santiddio}
\end{eqnarray}
Using eqs. (\ref{secm2}) it is now possible to write in a geometrically
intrinsic way the first order equations:
\begin{eqnarray}
\frac{d\phi^\alpha}{dr}\,&=&\,
\left(\mp\frac{e^{U}}{r^2}\right)\frac{1}{2\sqrt{2}\pi}\IP^\alpha_{\hat{\alpha}}\langle
v^{\hat{\alpha}}\vert \IL^t\IC {\bf M}\vert \vec{Q} \rangle_{sc}\nonumber \\
\frac{dU}{dr}\,&=&\,\left(\mp\frac{e^{U}}{r^2}\right) \frac{1}{2\sqrt{2}\pi}\langle
v^0_y\vert \IL^t\IC {\bf M}\vert \vec{Q} \rangle_{sc}\nonumber \\
0\,&=&\,\langle v^0_x\vert \IL^t\IC {\bf M}\vert t \rangle_{sc}
\label{1ordeqs}
\end{eqnarray}
The full explicit form of eq.s (\ref{1ordeqs}) can be found
in Appendix A where, using eq.
(\ref{ncudierre}), everything is expressed in terms
of the quantized moduli-independent charges
$(q_{\Lambda},p^{\Sigma})$.
The fixed values of the scalars at the horizon are obtained
by setting the
right hand side of the above equations
to zero and the result is consistent with the literature
(\cite{STUkallosh}):
\begin{eqnarray}
\label{scalfixn}
(a_1+{\rm i}b_1)_{fix}\,&=&\,\frac{p^\Lambda q_\Lambda -2 p^1q_1-{\rm i}
\sqrt{f(p,q)}}{2p^2p^3 - 2p^0 q_1}\nonumber\\
(a_2+{\rm i}b_2)_{fix}\,&=&\,\frac{p^\Lambda q_\Lambda -2 p^2q_2-{\rm i}
\sqrt{f(p,q)}}{2p^1p^3 - 2p^0 q_2}\nonumber\\
(a_3+{\rm i}b_3)_{fix}\,&=&\,\frac{p^\Lambda q_\Lambda -2 p^3q_3-{\rm i}
\sqrt{f(p,q)}}{2p^1p^2 - 2p^0 q_3}
\end{eqnarray}
where $f(p,q)$ is the $E_{7(7)}$ quartic invariant
$S^2(p,q)$ (see eq.s \eqn{bella} and \eqn{funf})
expressed as a function of all the
$8$ charges (and whose square root is the entropy of the solution):
\begin{equation}
S^2(p,q) \, = \,
f(p,q)\,=\,-(p^0q_0-p^1q_1+p^2q_2+p^3q_3)^2+4(p^2p^3-p^0q_1)(p^1q_0+q_2q_3)
\end{equation}
The last of eqs. (\ref{1ordeqs}) expresses the reality condition for $Z(\phi, p,q)$ and it amounts
to fix one of the three $SO(2)$ gauge symmetries of $H$ giving therefore a condition on the $8$
charges.
Without spoiling the generality (up to $U$--duality) of the black--hole
solution it is still possible to fix the remaining $[SO(2)]^2$ gauges in $H$
by imposing two conditions on  the phases of the $Z^i(\phi, p,q)$.
For instance we could require two  of the  $Z^i(\phi, p,q)$ to be imaginary.
This would imply two more conditions on the charges, leading to a
generating solution depending   only on $5$ parameters as we expect
it to be \cite{bala}.
Hence we can conclude with the following:
\bst
Since the radial evolution of the axion fields $a_i$ is related to the real
part of the corresponding central charge $Z^i(\phi, p,q)$
(see (\ref{zequa})),
up to $U$ duality transformations, the
{\bf 5 parameter generating solution}
will have {\bf 3 dilatons} and {\bf 1 axion}
evolving from their fixed value at the horizon to the
boundary value at infinity, and 2 constant axions whose
value is the corresponding fixed
one at the horizon ({\it double fixed}).
\est
\subsection{The first order equations: the special geometry approach}
The first order BPS equations may be equivalently formulated within a Special K\"ahler description
of the manifold ${\cal M}_{STU}$. In the {\it special coordinate} symplectic gauge, all the
geometrical quantities defined on   ${\cal M}_{STU}$ may be deduced form a cubic {\it prepotential}
$F(X)$:
\begin{eqnarray}
\{z^i\}\,=\,\{S,T,U\}\;&,&\;
\Omega(z)\,=\, \left(\matrix{X^\Lambda (z) \cr F_\Sigma (z)}\right)\nonumber \\
X^\Lambda (z)\,&=&\,\left(\matrix{1\cr S
\cr T \cr U}\right)\nonumber\\
F_\Sigma (z)\,&=&\,\partial_\Sigma F(X)\nonumber\\
{\cal K}(z,\bar{z})\,&=& -\log (8\vert {\rm Im}S {\rm Im}T{\rm Im}U\vert)\nonumber\\
g_{ij^\star}(z,\bar{z})\,=\,\partial_i\partial_{j^\star}{\cal K}(z,\bar{z})\,&=&\,
{\rm diag}\{ -{{\left( \bar{S} - S \right) }^{-2}},
  -{{\left( \bar{T} - T \right) }^{-2}},
  -{{\left( \bar{U} - U \right) }^{-2}}\}\nonumber\\
{\cal N}_{\Lambda\Sigma}\, &=&\,\bar{F}_{\Lambda\Sigma}+2{\rm i}\frac{{\rm Im}
F_{\Lambda\Omega}{\rm Im}F_{\Sigma\Pi}L^\Omega L^\Pi}{L^\Omega L^\Pi
{\rm Im}F_{\Omega\Pi}}\nonumber\\
F_{\Lambda\Sigma}(z)\, &=&\,\partial_\Lambda \partial_\Sigma F(X)\nonumber\\
F(X)\, &=&\,\frac{X^1 X^2 X^3}{X^0}
\end{eqnarray}
The covariantly holomorphic symplectic section $V(z,\bar{z})$ and its covariant derivative
$U_i(z,\bar{z})$ are:
\begin{eqnarray}
V(z,\bar{z})\, &=&\, \left(\matrix{L^\Lambda (z,\bar{z}) \cr M_\Sigma (z,\bar{z})}\right)\,=
\,e^{{\cal K}(z,\bar{z})/2}\Omega(z,\bar{z})\nonumber\\
U_i(z,\bar{z})\, &=&\,\left(\matrix{f_i^\Lambda (z,\bar{z}) \cr h_{i\vert \Sigma}
(z,\bar{z})}\right)\,=\,\nabla_i V(z,\bar{z})\,=\,(\partial_i+\frac{\partial_i {\cal K}}{2})
V(z,\bar{z})\nonumber\\
\bar{U}_{i^\star}(z,\bar{z})\, &=&\,\left(\matrix{\bar{f}_{i^\star}^\Lambda (z,\bar{z}) \cr
\bar{h}_{i^\star \vert \Sigma} (z,\bar{z})}\right)\,=\,\nabla_{i^\star} \bar{V}(z,\bar{z})\,=
\,(\partial_{i^\star}+\frac{\partial_{i^\star} {\cal K}}{2})\bar{V}(z,\bar{z})\nonumber\\
M_\Sigma (z,\bar{z})\, &=&\,{\cal N}_{\Sigma\Lambda}(z,\bar{z})L^\Lambda (z,\bar{z})\nonumber \\
h_{i\vert \Sigma} (z,\bar{z})\, &=&\,\bar{{\cal N}}_{\Sigma\Lambda}(z,\bar{z})f_i^\Lambda (z,\bar{z})
\end{eqnarray}
The real and imaginary part of ${\cal N}$ in terms of the real part $a_i$
and imaginary part $b_i$ of the complex scalars $z^i$ are:
{\small
\begin{eqnarray}
{\rm Re}{\cal N}\, &=&\,\left(\matrix{ 2\,{a1}\,{a2}\,
   {a3} & -\left( {a2}\,{a3}
      \right)  & -\left( {a1}\,{a3} \right)
      & -\left( {a1}\,{a2} \right)
      \cr -\left( {a2}\,{a3} \right)
      & 0 & {a3} & {a2} \cr -\left(
     {a1}\,{a3} \right)  & {a3
   } & 0 & {a1} \cr -\left( {a1}\,
     {a2} \right)  & {a2} & {a1
   } & 0 \cr  }\right)\nonumber\\
{\rm Im}{\cal N}\, &=&\,\left(\matrix{ {\frac{{{{a1}}^2}\,{b2}\,
       {b3}}{{b1}}} +
   {\frac{{b1}\,
       \left( {{{a3}}^2}\,{{{b2}}^2} +
         \left( {{{a2}}^2} +
            {{{b2}}^2} \right) \,{{{b3}}^2}
          \right) }{{b2}\,{b3}}} & -{
      \frac{{a1}\,{b2}\,{b3}}
     {{b1}}} & -{\frac{{a2}\,
       {b1}\,{b3}}{{b2}}} &
    -{\frac{{a3}\,{b1}\,{b2}}
     {{b3}}} \cr -{\frac{{a1}\,
       {b2}\,{b3}}{{b1}}} &
    {\frac{{b2}\,{b3}}{{b1}}} &
   0 & 0 \cr -{\frac{{a2}\,{b1}\,
       {b3}}{{b2}}} & 0 & {\frac{
      {b1}\,{b3}}{{b2}}} & 0 \cr
   -{\frac{{a3}\,{b1}\,{b2}}
     {{b3}}} & 0 & 0 & {\frac{{b1}\,
      {b2}}{{b3}}} \cr  }\right)
\label{Ngen}
\end{eqnarray}
}
Using the above defined quantities, the first order BPS equations can be
written in a complex notation   as in eq. (\ref{zequa}):
{\small
\begin{eqnarray}
\frac{dS}{dr}\,&=&\, \pm \left(\frac{e^{{\cal U}(r)}}{r^2}\right)
\sqrt{\vert \frac{{\rm Im}(S)}{2{\rm Im}(T)
{\rm Im}(U)}\vert}\Biggl [ {q_0} + \bar{U}\,{q_3} -
\bar{U}\,{p^2}\,S + {q_1}\,S   \nonumber\\
&& + \bar{T}\,\left( -\left( \bar{U}\,{p1} \right)  + {q_2} +
     \bar{U}\,{p^0}\,S - {p^3}\,S \right)\Biggr ] \nonumber\\
\frac{dT}{dr}\,&=&\, \pm \left(\frac{e^{{\cal U}(r)}}{r^2}\right)
\sqrt{\vert\frac{{\rm Im}(T)}{2{\rm Im}(S)
{\rm Im}(U)}\vert}\Biggl [
{q_0} + \bar{U}\,{q_3} - \bar{U}\,{p^1}\,T + {q_2}\,T \nonumber\\
&& + \bar{S}\,\left( -\left( \bar{U}\,{p^2} \right)  + {q_1} +
     \bar{U}\,{p^0}\,T - {p^3}\,T \right)\Biggr ] \nonumber\\
\frac{dU}{dr}\,&=&\, \pm \left(\frac{e^{{\cal U}(r)}}{r^2}\right)
\sqrt{\vert\frac{{\rm Im}(U)}{2{\rm Im}(S)
{\rm Im}(T)}\vert}\Biggl [
{q_0} + \bar{t}\,{q_2} - \bar{T}\,{p^1}\,U + {q_3}\,U \nonumber \\
&&+ \bar{S}\,\left( -\left( \bar{T}\,{p^3} \right)  + {q_1} +
     \bar{T}\,{p^0}\,U - {p^2}\,U \right) \Biggr ] \nonumber\\
\frac{d  {\cal U}}{dr}\,&=&\, \pm
\left(\frac{e^{{\cal U}(r)}}{r^2}\right)\left(\frac{1}{2\sqrt{2}
(\vert{\rm Im}(S) {\rm Im}(T){\rm Im}(U)\vert)^{1/2}} \right)
[{q_0} + {S}\,\left( {T}\,{U}\,{p^0} -
     {U}\,{p^2} - {T}\,{p^3} + {q_1} \right)    \nonumber\\
 && + {T}\,\left( -( {U}\,{p^1})  + {q_2} \right)  +
  {U}\,{q_3}]
\label{eq_123}
\end{eqnarray}
}
Note that to avoid confusion in the above equations the real function
$U(r)$ appearing in the metric has been renamed ${\cal U}(r)$ so that
it should not be confused with the complex scalar field $U$. The
naming $STU$ for the three complex scalar fields of the $SL(2,\IR)^3$
model has by now become so traditional that it could not be avoided: in
the literature such an $N=2$ theory is universally referred to
as the $STU$--model. On the other hand the name $U$
for the scalar function in the ansatz  \eqn{ds2U} for the
extremum $4D$ black--hole is also universally adopted so that the
conflict of notation limited to eq.\eqn{eq_123} could be removed only
at the price of using unfamiliar notations throughout all the lectures.
\par
The central charge $Z(z,\bar{z},p,q)$ being given by:
\begin{eqnarray}
Z(z,\bar{z},p,q)\,&=&\, -\left(\frac{1}{2\sqrt{2}(\vert{\rm Im}(S)
{\rm Im}(T){\rm Im}(U)\vert)^{1/2}}\right)
[{q_0} + {S}\,\left( {T}\,{U}\,{p^0} -
     {U}\,{p^2} - {T}\,{p^3} + {q_1} \right)  + \nonumber\\
& &  {T}\,\left( -\left( {U}\,{p^1} \right)  + {q_2} \right)  +
  {U}\,{q_3}]
\end{eqnarray}
\subsection{The solution: preliminaries and comments on the most general one}
In order to find the solution of the $STU$ model we need
also the equations of motion that must
be satisfied together with the first order ones.
We go on using the Special K\"ahler formalism in
order to let the comparison with various papers being more immediate.
Let us first compute the field
equations for the scalar fields $z_i$,
which can be obtained from an $N=2$ pure supergravity action
coupled to 3 vector multiplets. From the action of eq.\eqn{ungausugra}
we get:
\noindent
\underline{Maxwell's equations :}\par
The field equations for the vector fields and the Bianchi identities
read:
\begin{eqnarray}
\partial_\mu \left(\sqrt{-g}\tilde{G}^{\mu\nu}\right)\,&=&\,0\nonumber\\
\partial_\mu \left(\sqrt{-g}\tilde{F}^{\mu\nu}\right)\,&=&\,0
\end{eqnarray}
Using the ansatz (\ref{strenghtsans}) the second equation is automatically fulfilled while the first
equation,
as it was anticipated before, requires the quantized electric charges
$q_\Lambda$ defined by eq. (\ref{ncudierre}) to be $r$-independent
(eq. (\ref{prione})).\\
\underline{Scalar equations :}\par
varying with respect to $z^i$ one gets:
\begin{eqnarray}
&&-\frac{1}{\sqrt{-g}}\partial_\mu\left(\sqrt{-g}g^{\mu\nu}g_{ij^\star}
\partial_\nu \bar{z}^{j^\star}
\right)+\partial_i (g_{kj^\star})\partial_\mu z^k
\partial_\nu\bar{z}^{j^\star} g^{\mu\nu}+\nonumber\\
&& (\partial_i{\rm Im}{\cal N}_{\Lambda\Sigma})F^{\Lambda}_{\cdot\cdot}
F^{\Sigma\vert\cdot\cdot}+
(\partial_i{\rm  Re}{\cal N}_{\Lambda\Sigma})F^{\Lambda}_{\cdot\cdot}
\tilde{F}^{\Sigma\vert\cdot
\cdot}\,=\,0
\end{eqnarray}
which, once projected onto  the real and imaginary parts of both sides, read:
\begin{eqnarray}
\frac{e^{2\cal {U}}}{4b_i^2}\left(a_i^{\prime\prime}+2\frac{a_i^{\prime}}{r}-2\frac{a_i^{\prime}
b_i^{\prime}}{b_i}\right)\,&=&\,-\frac{1}{2}\left((\partial_{a_i}{\rm Im}{\cal N}_{\Lambda\Sigma})
F^{\Lambda}_{\cdot\cdot}F^{\Sigma\vert\cdot\cdot}+(\partial_{
a_i}{\rm  Re}{\cal
N}_{\Lambda\Sigma})F^{\Lambda}_{\cdot\cdot}\tilde{F}^{\Sigma\vert\cdot\cdot}\right)
\nonumber\\
\frac{e^{2\cal {U}}}{4b_i^2}\left(b_i^{\prime\prime}+2\frac{b_i^{\prime }}{r}+
\frac{(a_i^{\prime 2}-b_i^{\prime2})}{b_i}\right)\,&=&\,-\frac{1}{2}\left((\partial_{b_i}{\rm
Im}{\cal
N}_{\Lambda\Sigma})F^{\Lambda}_{\cdot\cdot}F^{\Sigma\vert\cdot\cdot}+(\partial_{b_i}{\rm  Re}
{\cal N}_{\Lambda\Sigma})F^{\Lambda}_{\cdot\cdot}
\tilde{F}^{\Sigma\vert\cdot\cdot}\right)\nonumber\\
\label{scaleq}
\end{eqnarray}
\underline{Einstein equations :}\par
 Varying the action (\ref{ungausugra}) with respect to the metric we obtain the
following equations:
\begin{eqnarray}
R_{MN}\,&=&\, -g_{ij^\star}\partial_M z^i\partial_N\bar{z}^{ j^\star}+S_{MN}\nonumber\\
S_{MN}\,&=&\,-2{\rm Im}{\cal N}_{\Lambda\Sigma}\left(F^\Lambda_{M\cdot}F^{\Sigma\vert\cdot}_{N}-
\frac{1}{4}g_{MN}F^\Lambda_{\cdot\cdot}F^{\Sigma\vert\cdot\cdot}\right)+\nonumber\\
&&-2{\rm Re}{\cal N}_{\Lambda\Sigma}\left(F^\Lambda_{M\cdot}\tilde{F}^{\Sigma\vert\cdot}_{N}-
\frac{1}{4}g_{MN}F^\Lambda_{\cdot\cdot}\tilde{F}^{\Sigma\vert\cdot\cdot}\right)
\label{eineq}
\end{eqnarray}
Projecting on the components $(M,N)=({\underline{0}},{\underline{0}})$ and
$(M,N)=({\underline{a}},{\underline{b}})$, respectively, these equations can be written in the
following way:
\begin{eqnarray}
{\cal U}^{\prime\prime}+\frac{2}{r}{\cal U}^\prime\,&=&\,-2e^{-2{\cal U}}S_{{\underline{0}}
{\underline{0}}}\nonumber\\
({\cal U}^\prime)^2+\sum_i\frac{1}{4b_i^2}\left((b_i^\prime)^2+(a_i^\prime)^2\right)\,&=&\,
-2e^{-2{\cal U}}S_{{\underline{0}}{\underline{0}}}
\label{2eqeinformern}
\end{eqnarray}
where:
\begin{equation}
S_{{\underline{0}}{\underline{0}}}\,=\, -\frac{2e^{4{\cal U}}}{(8\pi)^2 r^4}
{\rm Im}{\cal N}_{\Lambda\Sigma}(p^\Lambda p^\Sigma+\ell (r)^\Lambda \ell (r)^\Sigma)
\end{equation}
In order to solve these equations one would need to make explicit the right hand side expression in
terms of scalar fields $a_i$,$b_i$ and quantized charges $(p^{\Lambda},q_{\Sigma})$. In order to do
that, one has to consider the ansatz for the field strengths (\ref{strenghtsans}) substituting to
the moduli-dependent charges $\ell_{\Lambda}(r)$ appearing in the previous equations their
expression
in terms of the quantized charges obtained by inverting
eq.(\ref{ncudierre}):
\begin{eqnarray}
\hskip -3pt \ell^{\Lambda} (r) &=& {\rm Im}{\cal N}^{-1\vert \Lambda\Sigma}\left(
q_{\Sigma}-{\rm Re}{\cal N}_{\Sigma\Omega}p^\Omega\right)
\label{qrgen}
\end{eqnarray}
Using now the expression for the matrix ${\cal N}$ in eq. (\ref{Ngen}),
one can
find the explicit expression of the scalar fields equations of motion written in terms of the
quantized $r$-independent charges. They can be found in Appendix A.
In Appendix A we report the full explicit expression of the equations of motion for both the scalars
and the metric. Differently from what stated in \cite{stu-}, in order to find the 5 parameter
generating solution of the $STU$ model, it is not sufficient to substitute to each
charge, in the scalar fixed values of eq.(\ref{scalfixn}), a corresponding harmonic function
($q_i \rightarrow H_i=1+q_i/r$).
As already explained, the generating solution should
depend on 5 parameters and 4 harmonic
functions, as in \cite{stu+}. In particular, as explained above, 2 of the 6 scalar fields
parametrizing the $STU$ model, namely 2 axion fields, should be taken to be constant.
Therefore, in order to find the generating solution one as to solve the two systems of eq.s
(\ref{mammamia}) (first order) and (\ref{porcodue}) (second order) explicitly putting as an external
input the information on the constant nature of 2 of the 3 axion fields. As it is evident from
the above quoted system of eq.s,
it is quite difficult to give a not double extreme
solution of the combined system
that is both explicit and manageable.
It is however   work in progress for a forthcoming paper \cite{bft}.
\subsection{The solution: a simplified case, namely $S=T=U$}
In order to find a fully explicit solution we can deal with, let us consider the particular case where
$S=T=U$. Although simpler, this solution encodes all non-trivial aspects of the most general one:
it is regular, i.e. has non-zero entropy, and the scalars do evolve, i.e. it is an extreme but
{\em not} double extreme solution. First of all let us notice that eq.s (\ref{mammamia}) remain
invariant if the same set of permutations are performed on the triplet of subscripts $(1,2,3)$
in both the fields and the charges. Therefore the solution $S=T=U$ implies the positions
$q_1=q_2=q_3\equiv q$ and $p^1=p^2=p^3\equiv p$ on the charges and
therefore it will correspond
to a solution with (apparently only) $4$ independent charges $(p^0,p,q_0,q)$.
According to this identification, what we do expect now, is to find a solution which depends on
(apparently) only 3 charges and 2 harmonic functions.
Notice that this is not simply an axion--dilaton black--hole: such
a solution would  have a vanishing  entropy differently from our case.
 The fact that we have just one
complex field in our solution
is because the three complex fields are taken to be equal in value.
The equations (\ref{mammamia}) simplify in the following way:
\begin{eqnarray}
\label{sfirst}
\frac{da}{dr}\,&=&\,\pm \left(\frac{e^{{\cal U}(r)}}{r^2}\right)\frac{1}{\sqrt{-2b}}
({bq} - 2\,{ab}\,p + \left( {a^2}\,b + {b^3} \right) \,{p^0})\nonumber\\
\frac{db}{dr}\,&=&\,\pm \left(\frac{e^{{\cal U}(r)}}{r^2}\right)\frac{1}{\sqrt{-2b}}
(3\,{aq} - \left( 3\,{a^2} + {b^2} \right) \,p + \left( {a^3} + a\,{b^2} \right) \,{p^0} +
  {q_0})\nonumber\\
\frac{d\cal {U}}{dr}\,&=&\, \pm \left(\frac{e^{{\cal U}(r)}}{r^2}\right)\left(\frac{1}{2\sqrt{2}
(- b)^{3/2}}\right) (3\,{aq} - \left( 3\,{a^2} - 3\,{b^2} \right) \,p +
  \left( {a^3} - 3\,a\,{b^2} \right) \,{p^0} + {q_0})\nonumber\\
0\,&=&\, 3\,{bq} - 6\,{ab}\,p + \left( 3\,{a^2}\,b - {b^3} \right) \,{p^0}
\end{eqnarray}
where $a\equiv a_i\,,\,b\equiv b_i\;(i=1,2,3)$.
In this case the fixed values for the scalars $a,b$ are:
\begin{eqnarray}
\label{scalfixs3}
a_{fix}\,&=&\, {\frac{p\,q + {p^0}\,{q_0}}
   {2\,{p^2} - 2\,{p^0}\,q}}\nonumber\\
b_{fix}\,&=&\,-\,\frac{\sqrt{f(p,q,p^0,q_0)}}{2(p^2-p^0q)}\nonumber\\
\mbox{where}\;f(p,q,p^0,q_0)\,&=&\,3\,{p^2}\,{q^2} + 4\,{p^3}\,{q_0} -
  6\,p\,{p^0}\,q\,{q_0} -
  {p^0}\,\left( 4\,{q^3} +
     {p^0}\,{{{q_0}}^2} \right)
\end{eqnarray}
Computing the central charge at the fixed point $Z_{fix}(p,q,p^0,q_0)=
Z(a_{fix},b_{fix},p,q,p^0,q_0)$ one finds:
\begin{eqnarray}
Z_{fix}(p,q,p^0,q_0)\,&=&\,\vert Z_{fix}\vert e^{\theta}\nonumber\\
\vert Z_{fix}(p,q,p^0,q_0)\vert\,&=&\, f(p,q,p^0,q_0)^{1/4}\nonumber\\
\sin\theta\,&=&\,\frac{p^0f(p,q,p^0,q_0)^{1/2}}{2(p^2-qp^0)^{3/2}}\nonumber\\
\cos\theta\,&=&\,{\frac{-2\,{p^3} + 3\,p\,{p^0}\,q +
     {{{p^0}}^2}\,{q_0}}{2\,{{\left( {p^2} - {p^0}\,q \right) }^{{3/2}}}}}
\label{components}
\end{eqnarray}
The value of the $U$--duality group quartic invariant (whose square root is the entropy) is:
\begin{eqnarray}
S^2(p,q,p^0,q_0)\,&=&\,\vert Z_{fix}(p,q,p^0,q_0)\vert^4\,=\,f(p,q,p^0,q_0)
\end{eqnarray}
We see form eqs.(\ref{components}) that in order for $Z_{fix}$ to be real and the entropy to be
non vanishing the only possibility is $p^0=0$ corresponding to $\theta=\pi$. It is in fact necessary
that $\sin\theta=0$ while keeping $f\not= 0$. We are therefore left with 3 independent charges
($q,p,q_0$), as anticipated.
\subsubsection{Solution of the $1^{st}$ order equations}
Setting $p^0=0$ the fixed values of the scalars and the quartic invariant become:
\begin{eqnarray}
\label{fixeds3}
a_{fix}\,&=&\, \frac{q}{2p}\nonumber\\
b_{fix}\,&=&\,-\,\frac{\sqrt{3q^2+4q_0 p}}{2p}\nonumber\\
I_4\,&=&\, (3q^2p^2+4q_0 p^3)
\end{eqnarray}
From the last of eq.s (\ref{sfirst}) we see that in this case the axion is double fixed, namely
does not evolve, $a\equiv a_{fix}$ and the reality condition for the central charge
is fulfilled for any $r$. Of course, also the axion equation is fulfilled and therefore
we are left with two axion--invariant equations for $b$ and $\cal {U}$:
\begin{eqnarray}
\frac{db}{dr}\,&=&\, \pm\frac{e^{\cal U}}{r^2\sqrt{2b}}(q_0+\frac{3q^2}{4p}-b^2p)\nonumber\\
\frac{d\cal {U}}{dr}\,&=&\, \pm\frac{e^{\cal U}}{r^2 (2b)^{3/2}}(q_0+\frac{3q^2}{4p}+3b^2p)
\label{eqbU}
\end{eqnarray}
which admit the following solution:
\begin{eqnarray}
\label{k1ek2}
b(r)\,&=&\,-\,\sqrt{\frac{(A_1+k_1/r)}{(A_2+k_2/r)}}\nonumber\\
e^{\cal U}\,&=&\,\left((A_2+\frac{k_2}{r})^3(A_1+k_1/r)\right)^{-1/4}\nonumber\\
k_1\,&=&\,\pm\frac{\sqrt{2}(3q^2+4q_0p)}{4p}\nonumber\\
k_2\,&=&\,\pm\sqrt{2}p
\end{eqnarray}
In the limit $r\rightarrow 0$:
\begin{eqnarray}
b(r)&\rightarrow&-\,\left(\frac{k_1}{k_2}\right)^{1/2}\,=\,b_{fix}\nonumber\\
e^{{\cal U}(r)}&\rightarrow& \,r\,(k_1k_2^3)^{-1/4}\,=\,r\,f^{-1/4} \nonumber
\end{eqnarray}
as expected, and the only undetermined constants are $A_1\,,\,A_2$. In order for the solution to be
asymptotically minkowskian it is necessary that $(A_1\,A_2^3)^{-1/4}=1$. There is then just one
undetermined parameter which is fixed by the asymptotic value of the dilaton $b$. We choose for
simplicity it to be $-1$, therefore $A_1=1\,,\,A_2=1$. This choice is arbitrary in the sense that
the different value of $b$ at infinity the different universe ($\equiv$black--hole solution), but
with the same entropy. Summarizing, before considering the eq.s of
motion, the solution is:
\begin{eqnarray}
\label{sol}
a\,&=&\,a_{fix}\,=\,\frac{q}{2p}\nonumber \\
b\,&=&\,-\,\sqrt{\frac{(1+k_1/r)}{(1+k_2/r)}}\nonumber\\
e^{\cal U}\,&=&\,\left[(1+k_1/r)(1+k_2/r)^3\right]^{-1/4}
\label{sol1/8}
\end{eqnarray}
with $k_1$ and $k_2$ given in (\ref{k1ek2}).
\subsubsection{Solution of the $2^{st}$ order equations}
In the case $S=T=U$ the structure of the $\cal N$ matrix (\ref{Ngen}) and of the field strengths
reduces considerably. For the period matrices one simply obtains:
\begin{equation}
{\rm Re}{\cal N}=\left(\matrix{ 2\,{a^3} & -{a^2} & -{a^2} & -{a^2} \cr
    -{a^2} & 0 & a & a \cr -{a^2} & a & 0 & a \cr
    -{a^2} & a & a & 0 \cr  }\right)\;,\;
{\rm Im}{\cal N}=\left(\matrix{ 3\,{a^2}\,b + {b^3} & -\left( a\,b \right)  & -\left(
     a\,b \right)  & -\left( a\,b \right)  \cr -\left( a\,b
      \right)  & b & 0 & 0 \cr -\left( a\,b \right)
      & 0 & b & 0 \cr -\left( a\,b \right)  & 0 & 0 & b \cr  }\right)
\end{equation}
while the dependence of $\ell^{\Lambda}(r)$ from the quantized charges simplifies to:
\begin{eqnarray}
\ell^\Lambda (r)\, &=&\, \left(\matrix{{\frac{-3\,{a^2}\,p + 3\,a\,q + {q_0}}{{b^3}}}\cr
{\frac{-3\,{a^3}\,p + {b^2}\,q +
     3\,{a^2}\,q + a\,\left( -2\,{b^2}\,p + {q_0} \right) }{{b^3}}}\cr
{\frac{-3\,{a^3}\,p + {b^2}\,q +
     3\,{a^2}\,q + a\,\left( -2\,{b^2}\,p + {q_0} \right) }{{b^3}}}\cr
{\frac{-3\,{a^3}\,p + {a^4}\,{p^0} + {b^2}\,q +
     3\,{a^2}\,q + a\,\left( -2\,{b^2}\,p + {q_0} \right) }{{b^3}}}}\right)
\label{qr}
\end{eqnarray}
Inserting (\ref{qr}) in the expressions (\ref{strenghtsans}) and  substituting the result in
the eq.s of motion (\ref{scaleq}) one finds:
{\small
\begin{eqnarray}
\left(a^{\prime\prime}-2\frac{a^{\prime}b^{\prime}}{b}+2\frac{a^{\prime}}{r}\right)\,&=&\,
            0\nonumber\\
\left(b^{\prime\prime}+2\frac{b^{\prime }}{r}+\frac{(a^{\prime 2}-b^{\prime2})}{b}\right)\,&=&\,
-\frac{{b^2}\,{e^{2\,\cal {U}}}
              \,( {p^2} - \,\frac{(-3\,{a^2}\,p + 3\,a\,q + q_0)^2}{b^6} \, )
              \, }{{r^4}}
\end{eqnarray}
}
The equation for $a$ is automatically fulfilled by our solution (\ref{sol}). The equation for $b$
is fulfilled as well and both sides are equal to:
\begin{eqnarray}
{\frac{\left( {k_2} \,-\,{k_1} \right) \,
     {e^{4\,\cal {U}}}\,\left( {k_1} + {k_2} +
{\frac{2\,{k_1}\,{k_2}}{r}}
       \right) }{2\,b\,{r^4}}} \nonumber
\end{eqnarray}
If $\left( {k_2} - {k_1} \right)=0$ both sides are separately equal to $0$
which corresponds to the double fixed solution already
found in \cite{STUkallosh}.
Let us now consider the Einstein's equations.
From equations (\ref{2eqeinformern}) we obtain in our
simpler case the following ones:
\begin{eqnarray}
{\cal U}^{\prime\prime}+\frac{2}{r}{\cal U}^\prime\,&=&\,({\cal U}^\prime)^2+\frac{3}{4b^2}
\left((b^\prime)^2+
(a^\prime)^2\right)\nonumber\\
{\cal U}^{\prime\prime}+\frac{2}{r}{\cal U}^\prime\,&=&\,-2e^{-2{\cal U}}S_{{\underline{0}}
{\underline{0}}}
\label{2eqeinn}
\end{eqnarray}
The first of eqs.(\ref{2eqeinn}) is indeed fulfilled by our ansatz. Both sides are equal to:
\begin{eqnarray}
{\frac{3\,{{\left( k_2 - k_1 \right)
          }^2}}{16\,r^4{{\left( H_1\right) }^2}\,
     {{\left( H_2\right) }^2}}}
\end{eqnarray}
Again, both sides are separately zero in the double-extreme case
$\left( k_2 - k_1\right)=0$.
The second equation is fulfilled, too, by our ansatz and again both sides are zero in the
double-extreme case.
Therefore we can conclude with the
\bst
Eq.\eqn{sol1/8} yields a $\frac{1}{8}$ supersymmetry preserving solution of
N=8 supergravity that is {\bf not double extreme} and has a {\bf finite
entropy}:
\begin{eqnarray}
S_{BH}\,=\,\frac{1}{4\pi} \left(3 \, p^2 \, q^2 + 4 \, q_0 p^3 \right)^{1/2}
\end{eqnarray}
depending on three of the $5$ truly independent charges
\est
On the other hand if in eq.\eqn{sol1/8} we set $k_1=k_2$ we see that
both the dilaton and the axion are set to  a constant value equal to
the fixed value. At the same time the form of the metric becomes that
of an $a=0$ $0$--brane solution as described in eq. \eqn{critval1}.
As  the critical values $a=\sqrt{3}$ and $a=1$
are respectively  associated with $1/2$ and $1/4$ supersymmetry preserving
black--holes, in the same way the last critical value $a=0$
is associated with $1/8$ preserving black--holes. This case is the
only one where scalar fields are finite at the horizon and
correspondingly the entropy can be non--vanishing and finite.
\appendix
\chapter{The full set of first
and second order differential equations for the $STU$--model}
\markboth{BPS BLACK HOLES IN SUPERGRAVITY: APPENDIX A}{A. Explicit Differential
Equations}
\label{appendiceB}
\setcounter{equation}{0}
\addtocounter{section}{1}
Setting $z^i=a_i+{\rm i}b_i$ eqs.(\ref{eq_123}) can be rewritten in the form:
{\small
\begin{eqnarray}
\label{mammamia}
\frac{da_1}{dr}\,&=&\,\pm \frac{e^{{\cal U}(r)}}{r^2}\sqrt{- \frac{b_1}{2b_2b_3}}
[-{b_1q_1} + {b_2q_2} + {b_3q_3} +
  \Bigl ( -\left( {a_2}\,{a_3}\,{b_1} \right)    \nonumber\\
    && + {a_1}\,{a_3}\,{b_2} + {a_1}\,{a_2}\,{b_3} +
     {b_1}\,{b_2}\,{b_3} \Bigr) \,{p^0} + \nonumber \\
&&  + \left( -\left( {a_3}\,{b_2} \right)  - {a_2}\,{b_3} \right) \,
   {p^1} + \left( {a_3}\,{b_1} - {a_1}\,{b_3} \right) \,
   {p^2} \nonumber\\
&&+ \left( {a_2}\,{b_1} - {a_1}\,{b_2} \right) \,
   {p^3}] \nonumber\\
\frac{db_1}{dr}\,&=&\,\pm \frac{e^{{\cal U}(r)}}{r^2}\sqrt{- \frac{b_1}{2b_2b_3}}
[{a_1q_1} + {a_2q_2} + {a_3q_3} \nonumber\\
&&+
  \left( {a_1}\,{a_2}\,{a_3} + {a_3}\,{b_1}\,{b_2} +
     {a_2}\,{b_1}\,{b_3} - {a_1}\,{b_2}\,{b_3} \right) \,
   {p^0} + \nonumber \\
&& + \left( -\left( {a_2}\,{a_3} \right)  + {b_2}\,{b_3}
      \right) \,{p^1}
 - \left( {a_1}\,{a_3} + {b_1}\,{b_3} \right)
     \,{p^2} \nonumber\\
&& - \left( {a_1}\,{a_2} + {b_1}\,{b_2} \right) \,
   {p^3} + {q_0}] \nonumber\\
\frac{da_2}{dr}\,&=&\, (1,2,3) \rightarrow (2,1,3) \nonumber\\
\frac{db_2}{dr}\,&=&\, (1,2,3) \rightarrow (2,1,3) \nonumber\\
\frac{da_3}{dr}\,&=&\, (1,2,3) \rightarrow (3,2,1) \nonumber\\
\frac{db_3}{dr}\,&=&\, (1,2,3) \rightarrow (3,2,1) \nonumber\\
\frac{d\cal {U}}{dr}\,&=&\, \pm \frac{e^{{\cal U}(r)}}{r^2}\frac{1}{2\sqrt{2}
(- b_1b_2b_3)^{1/2}}[{a_1q_1} + {a_2q_2} + {a_3q_3} \nonumber\\
&&+
  \left( {a_1}\,{a_2}\,{a_3} - {a_3}\,{b_1}\,{b_2} -
     {a_2}\,{b_1}\,{b_3} - {a_1}\,{b_2}\,{b_3} \right) \,
   {p^0} + \nonumber \\
&& - \left( {a_2}\,{a_3} - {b_2}\,{b_3} \right) \,
   {p^1} - \left( {a_1}\,{a_3} - {b_1}\,{b_3} \right) \,
   {p^2} - \left( {a_1}\,{a_2} - {b_1}\,{b_2} \right) \,
   {p^3} + {q_0}] \nonumber\\
0\,&=&\,{b_1q_1} + {b_2q_2} + {b_3q_3} +
  \left( {a_2}\,{a_3}\,{b_1} + {a_1}\,{a_3}\,{b_2} +
     {a_1}\,{a_2}\,{b_3} - {b_1}\,{b_2}\,{b_3} \right) \,
   {p^0}\nonumber\\
&&- \left( {a_3}\,{b_2} + {a_2}\,{b_3} \right) \,
   {p^1}\nonumber\\
&& - \left( {a_3}\,{b_1} + {a_1}\,{b_3} \right) \,
   {p^2} - \left( {a_2}\,{b_1} - {a_1}\,{b_2} \right) \,
   {p^3}
\end{eqnarray}
}
The explicit form of the equations of motion for the most general case is:
{\small
\begin{eqnarray}
\underline{\mbox{Scalar equations :}}\quad\quad \quad\quad&&\nonumber\\
\left(a_1^{\prime\prime}-2\frac{a_1^{\prime}b_1^{\prime}}{b_1}+
2\frac{a_1^{\prime}}{r}\right)\,&=&\,
\frac{-2\,{b_1}\,{e^{2\,{\cal U}}}}{{r^4}}\,
     [ {a_1}\,{b_2}\,{b_3}\,( {{{p^1}}^2} - {{{\ell (r)_1}}^2} ) \nonumber\\
&&+ {b_2}\,( -( {b_3}\,{p^1}\,{p^2} )
              + {b_3}\,{\ell (r)_1}\,{\ell (r)_2} )  + \nonumber \\
&&      + {b_1}\,( -2\,{a_2}\,{a_3}\,{p^1}\,
           {\ell (r)_1} + {a_3}\,{p^3}\,{\ell (r)_1} +
           {a_2}\,{p^4}\,{\ell (r)_1} +
          {a_3}\,{p^1}\,{\ell (r)_3} + \nonumber \\
&&        -  {p^4}\,{\ell (r)_3} +
          {a_2}\,{p^1}\,{\ell (r)_4} - {p^3}\,{\ell (r)_4}
           )  ]  \nonumber\\
\left(b_1^{\prime\prime}+2\frac{b_1^{\prime }}{r}+
\frac{(a_1^{\prime 2}-b_1^{\prime2})}{b_1}\right)\,&=&\,
 -\frac{{e^{2\,{\cal U}}}}{{b_2}\,{b_3}\,{r^4}}\,[ -( {{{a_1}}^2}\,{{{b_2}}^2}\,
           {{{b_3}}^2}\,{{{p^1}}^2} )  +
        {{{b_1}}^2}\,{{{b_2}}^2}\,{{{b_3}}^2}\,
         {{{p^1}}^2} \nonumber\\
&&         + 2\,{a_1}\,{{{b_2}}^2}\,
         {{{b_3}}^2}\,{p^1}\,{p^2} + \nonumber \\
&&       - {{{b_2}}^2}\,{{{b_3}}^2}\,{{{p^2}}^2} +
        {{{b_1}}^2}\,{{{b_3}}^2}\,{{{p^3}}^2} +
        {{{b_1}}^2}\,{{{b_2}}^2}\,{{{p^4}}^2}\nonumber\\
&&      + {{{a_1}}^2}\,{{{b_2}}^2}\,{{{b_3}}^2}\,
         {{{\ell (r)_1}}^2} + \nonumber \\
&&       - {{{b_1}}^2}\,{{{b_2}}^2}\,
         {{{b_3}}^2}\,{{{\ell (r)_1}}^2} +
        {{{a_3}}^2}\,{{{b_1}}^2}\,{{{b_2}}^2}\,
         ( {{{p^1}}^2} - {{{\ell (r)_1}}^2} )\nonumber\\
&&       + {{{a_2}}^2}\,{{{b_1}}^2}\,{{{b_3}}^2}\nonumber \\
&&         ( {{{p^1}}^2} - {{{\ell (r)_1}}^2} )
        - 2\,{a_1}\,{{{b_2}}^2}\,{{{b_3}}^2}\,{\ell (r)_1}\,
         {\ell (r)_2} +
          {{{b_2}}^2}\,{{{b_3}}^2}\,
         {{{\ell (r)_2}}^2} + \nonumber \\
&&       - {{{b_1}}^2}\,{{{b_3}}^2}\,
         {{{\ell (r)_3}}^2} +
          2\,{a_2}\,{{{b_1}}^2}\,
         {{{b_3}}^2}\,( -( {p^1}\,{p^3} )  +
           {\ell (r)_1}\,{\ell (r)_3} ) + \nonumber \\
&&       - {{{b_1}}^2}\,{{{b_2}}^2}\,{{{\ell (r)_4}}^2}
       + 2\,{a_3}\,{{{b_1}}^2}\,{{{b_2}}^2}\,
         ( -( {p^1}\,{p^4} )  +
           {\ell (r)_1}\,{\ell (r)_4} )  ] \,
      \nonumber\\
\left(a_2^{\prime\prime}-2\frac{a_2^{\prime}b_2^{\prime}}{b_2}
+2\frac{a_2^{\prime}}{r}\right)\,&=&\,
 (1,2,3) \rightarrow (2,1,3) \nonumber\\
\left(b_2^{\prime\prime}+2\frac{b_2^{\prime }}{r} +
\frac{(a_2^{\prime
2}-b_2^{\prime2})}{b_2}\right)\,&=&\,
 (1,2,3) \rightarrow (2,1,3)  \nonumber\\
\left(a_3^{\prime\prime}-2\frac{a_3^{\prime}b_3^{\prime}}{b_3}
+2\frac{a_3^{\prime}}{r}\right)\,&=&\,
 (1,2,3) \rightarrow (3,2,1)    \nonumber\\
\left(b_3^{\prime\prime}+2\frac{b_3^{\prime }}{r}+
\frac{(a_3^{\prime
2}-b_3^{\prime2})}{b_3}\right)\,&=&\,
 (1,2,3) \rightarrow (3,2,1) \nonumber \\
\underline{\mbox{Einstein equations :}}\quad\quad\quad\quad &&\nonumber\\
{\cal U}^{\prime\prime}+\frac{2}{r}{\cal U}^\prime\,&=&\,-2e^{-2{\cal U}}
S_{00} \nonumber\\
({\cal U}^\prime)^2+\sum_i\frac{1}{4b_i^2}\left((b_i^\prime)^2
+(a_i^\prime)^2\right)\,&=&\,
2e^{-2{\cal U}}S_{00}
\label{porcodue}
\end{eqnarray}
}
where the quantity $S_{00}$ on the right hand side of the Einstein eqs.
has the following form:
{\small
\begin{eqnarray}
S_{00}\, &=&\,\frac{{e^{4\,{\cal U}}}}{4\,{b_1}\,{b_2}\,{b_3}\,{r^4}}\,
( {{{a_1}}^2}\,{{{b_2}}^2}\,{{{b_3}}^2}\,
        {{{p_1}}^2} + {{{b_1}}^2}\,{{{b_2}}^2}\,
        {{{b_3}}^2}\,{{{p_1}}^2} -
       2\,{a_1}\,{{{b_2}}^2}\,{{{b_3}}^2}\,{p_1}\,
        {p_2} + {{{b_2}}^2}\,{{{b_3}}^2}\,{{{p_2}}^2} +
       {{{b_1}}^2}\,{{{b_3}}^2}\,{{{p_3}}^2} + \nonumber \\
&&      + {{{b_1}}^2}\,{{{b_2}}^2}\,{{{p_4}}^2} +
       {{{a_1}}^2}\,{{{b_2}}^2}\,{{{b_3}}^2}\,
        {{{\ell (r)_1}}^2} + {{{b_1}}^2}\,{{{b_2}}^2}\,
        {{{b_3}}^2}\,{{{\ell (r)_1}}^2} +
       {{{a_3}}^2}\,{{{b_1}}^2}\,{{{b_2}}^2}\,
        ( {{{p_1}}^2} + {{{\ell (r)_1}}^2} ) + \nonumber \\
&&     + {{{a_2}}^2}\,{{{b_1}}^2}\,{{{b_3}}^2}\,
        ( {{{p_1}}^2} + {{{\ell (r)_1}}^2} )  -
       2\,{a_1}\,{{{b_2}}^2}\,{{{b_3}}^2}\,{\ell (r)_1}\,
        {\ell (r)_2} + {{{b_2}}^2}\,{{{b_3}}^2}\,
        {{{\ell (r)_2}}^2} + {{{b_1}}^2}\,{{{b_3}}^2}\,
        {{{\ell (r)_3}}^2} + \nonumber \\
&&       - 2\,{a_2}\,{{{b_1}}^2}\,
        {{{b_3}}^2}\,( {p_1}\,{p_3} +
          {\ell (r)_1}\,{\ell (r)_3} )  +
       {{{b_1}}^2}\,{{{b_2}}^2}\,{{{\ell (r)_4}}^2} -
       2\,{a_3}\,{{{b_1}}^2}\,{{{b_2}}^2}\,
        ( {p_1}\,{p_4} + {\ell (r)_1}\,{\ell (r)_4} ))
\end{eqnarray}
}
The explicit expression of the $\ell_{\Lambda}(r)$ charges
in terms of the quantized ones is
computed from eq. (\ref{qrgen}):
\begin{equation}
  \ell_{\Lambda} (r) \,=\,
  \frac{ 1} {{b_1}\, {b_2}\,{b_3}} \,\left(
 { \begin{array}{l}
\Biggl [  \,{q_1} + {a_1}\,
      \left( {a_2}\,{a_3}\,{p^1} -
        {a_3}\,{p^3} - {a_2}\,{p^4} + {q_2}
         \right) \\
         + {a_2}\,\left( -\left( {a_3}\,{p^2} \right)
            + {q_3} \right)  + {a_3}\,{q_4}\Biggr ] \\
            \null \\
            \hline
            \null \\
  \Biggl [ {{a_1}}^2 \,\left( {a_2}\,{a_3}\,{p^1} -
        {a_3}\,{p^3} - {a_2}\,{p^4} + {q_2}
         \right)  + {{{b_1}}^2}\,
      \left( {a_2}\,{a_3}\,{p^1} -
        {a_3}\,{p^3} - {a_2}\,{p^4} + {q_2}
         \right)  \\
         + {a_1}\,\left( {q_1} +
        {a_2}\,\left( -\left( {a_3}\,{p^2} \right)  +
           {q_3} \right)  + {a_3}\,{q_4} \right)  \Biggr ] \\
           \null \\
           \hline
           \null \\
 \Biggl[ {a_1}\,\left( {{{a_2}}^2}\,
         \left( {a_3}\,{p^1} - {p^4} \right)  +
        {{{b_2}}^2}\,\left( {a_3}\,{p^1} - {p^4}
           \right)  + {a_2}\,
         \left( -\left( {a_3}\,{p^3} \right)  + {q_2} \right)
         \right)  \\
         + {{{a_2}}^2}\,
      \left( -\left( {a_3}\,{p^2} \right)  + {q_3} \right)  +
     {{{b_2}}^2}\,\left( -\left( {a_3}\,{p^2} \right)  +
        {q_3} \right)  + {a_2}\,
      \left( {q_1} + {a_3}\,{q_4} \right)  \Biggr ] \\
      \null \\
      \hline
      \null \\
  \Biggl [ {a_3}\,{q_1} +
     {a_1}\,\left( -\left( {{{a_3}}^2}\,{p^3} \right)  -
        {{{b_3}}^2}\,{p^3} +
        {a_2}\,\left( {{{a_3}}^2}\,{p^1} +
           {{{b_3}}^2}\,{p^1} - {a_3}\,{p^4} \right)  +
         {a_3}\,{q_2} \right) \\
         -
     {a_2}\,\left( {{{a_3}}^2}\,{p^2} +
        {{{b_3}}^2}\,{p^2} - {a_3}\,{q_3} \right)  +
     {{{a_3}}^2}\,{q_4} + {{{b_3}}^2}\,{q_4} \Biggr ] \\
 \end{array}  }
 \right )
\end{equation}

\end{document}